\documentclass[acmlarge,dvipsnames]{acmart}

\usepackage{booktabs} 
\usepackage{gensymb}
\usepackage{multirow}
\usepackage{xspace}
\usepackage{algorithm}
\usepackage[noend]{algpseudocode}
\usepackage{algorithmicx}
\usepackage[binary-units=true]{siunitx}
\usepackage{fixltx2e}
\usepackage{pifont}
\usepackage{hhline}
\usepackage{listings}
\usepackage{enumitem}
\usepackage{xstring}
\usepackage[utf8]{inputenc}
\usepackage{array}
\usepackage{graphicx}
\usepackage{makecell}
\usepackage{colortbl}
\usepackage{wrapfig}
\usepackage{subcaption}
\usepackage[us,12hr]{datetime}

\renewcommand\footnotetextcopyrightpermission[1]{}
\setcopyright{none}

\newcommand{\ignore}[1]{}
\newcommand{\chI}[0]{}
\newcommand{\chII}[0]{}
\newcommand{\chIII}[0]{}
\newcommand{\chIV}[0]{}
\newcommand{\todo}[1][]{}
\newcommand{\chV}[0]{}
\newcommand{\chVI}[0]{}
\newcommand{\chVII}[0]{}
\newcommand{\chVIII}[0]{}

\newcommand{\paratitle}[1]{\vspace{8pt}\textbf{#1.}}

\newcommand{\incircle}[1]{%
    \IfEqCase{#1}{%
        {1}{\ding{182}}%
        {2}{\ding{183}}%
        {3}{\ding{184}}%
        {4}{\ding{185}}%
        {5}{\ding{186}}%
        {6}{\ding{187}}%
        {7}{\ding{188}}%
        {8}{\ding{189}}%
    }[\PackageError{incircle}{Undefined option to incircle: #1}{}]%
}%

\newcommand{\squishlist}{
   \begin{itemize}
   \itemsep 0pt
}

\newcommand{\squishend}{
    \end{itemize}
}

\newcolumntype{L}[1]{>{\raggedright\let\newline\\\arraybackslash\hspace{0pt}}m{#1}}
\newcolumntype{C}[1]{>{\centering\let\newline\\\arraybackslash\hspace{0pt}}m{#1}}
\newcolumntype{R}[1]{>{\raggedleft\let\newline\\\arraybackslash\hspace{0pt}}m{#1}}

\begin{document}
\title[Errors in Flash-Memory-Based Solid-State Drives: Analysis, Mitigation,
and Recovery]{Errors in Flash-Memory-Based Solid-State Drives: \\
Analysis, Mitigation, and Recovery}

\author{Yu Cai}
\author{Saugata Ghose}
\affiliation{%
  \institution{Carnegie Mellon University}
}

\author{Erich F. Haratsch}
\affiliation{%
  \institution{Seagate Technology}
}

\author{Yixin Luo}
\affiliation{%
  \institution{Carnegie Mellon University}
}

\author{Onur Mutlu}
\affiliation{%
  \institution{ETH Z{\"u}rich and Carnegie Mellon University}
}

\renewcommand{\shortauthors}{Y. Cai et al.}


\begin{abstract}

NAND flash memory is ubiquitous in everyday life
today because its capacity has continuously increased and cost
has continuously decreased over decades. This positive growth
is a result of two key trends: (1)~effective process technology
scaling; and (2)~multi-level (e.g., MLC, TLC) cell data coding.
Unfortunately, the reliability of raw data stored in flash memory
has also continued to become more difficult to ensure, because
these two trends lead to (1)~fewer electrons in the flash memory
cell floating gate to represent the data; and (2)~larger cell-to-cell
interference and disturbance effects. Without mitigation,
worsening reliability can reduce the lifetime of NAND flash
memory. As a result, flash memory controllers in solid-state drives
(SSDs) have become much more sophisticated: they incorporate
many effective techniques to ensure the correct interpretation of
noisy data stored in flash memory cells.

In this \chI{chapter}, we review
recent advances in SSD error characterization, mitigation, and
data recovery techniques for reliability and lifetime improvement.
We provide rigorous experimental data from state-of-the-art MLC
and TLC NAND flash devices on various types of flash memory
errors, to motivate the need for such techniques. Based on the
understanding developed by the experimental characterization,
we describe several mitigation and recovery techniques, including
(1)~cell-to-cell interference mitigation; 
(2)~optimal multi-level cell sensing; 
(3)~error correction using state-of-the-art algorithms
and methods; and 
(4)~data recovery when error correction fails.
We quantify the reliability improvement provided by each of
these techniques. Looking forward, we briefly discuss how flash
memory and these techniques could evolve into the future.

\end{abstract}

\maketitle



\label{sec:intro}

Solid-state drives (SSDs) are widely used in computer
systems today as a primary method of data storage. In comparison
with magnetic hard drives, the previously dominant
choice for storage, SSDs deliver significantly higher
read and write performance, with orders of magnitude of
improvement in random-access input/output (I/O) operations,
and are resilient to physical shock, while requiring a
smaller form factor and consuming less static power. SSD
capacity (i.e., storage density) and cost-per-bit have been
improving steadily in the past two decades, which has led
to the widespread adoption of SSD-based data storage in
most computing systems, from mobile consumer 
devices~\cite{R51, R96} to enterprise data centers~\cite{R48, R49, R50, R83, R97}.

The first major driver for the improved SSD capacity
and cost-per-bit has been \emph{manufacturing process scaling},
which has increased the number of flash memory cells
within a fixed area. Internally, commercial SSDs are made
up of NAND flash memory chips, which provide nonvolatile
memory storage (i.e., the data stored in NAND flash is
correctly retained even when the power is disconnected) using
\emph{floating-gate (FG) transistors}~\cite{R46, R47, R171} or 
\emph{charge trap transistors}~\cite{R105, R172}. In this paper, we mainly focus on 
floating-gate transistors, since they are the most common transistor
used in today's flash memories.
A floating-gate transistor constitutes
a flash memory cell. It can encode one or more bits of
digital data, which is represented by the level of charge stored
inside the transistor's \emph{floating gate}. The transistor traps charge
within its floating gate, which dictates the \emph{threshold voltage}
level
at which the transistor turns on. The threshold voltage level of
the floating gate is used to determine the value of the digital
data stored inside the transistor. When manufacturing process
scales down to a smaller technology node, the size of each flash
memory cell, and thus the size of the transistor, decreases,
which in turn reduces the amount of charge that can be trapped
within the floating gate. Thus, process scaling increases storage
density by enabling more cells to be placed in a given area, but
it also causes reliability issues, which are the focus of this paper.

The second major driver for improved SSD capacity has
been the use of a single floating-gate transistor to represent \emph{more
than} one bit of digital data. Earlier NAND flash chips stored a
single bit of data in each cell (i.e., a single floating-gate transistor),
which was referred to as single-level cell (SLC) NAND
flash. Each transistor can be set to a specific threshold voltage
within a fixed range of voltages. SLC NAND flash divided this
fixed range into two \emph{voltage windows}, where one window represents
the bit value 0 and the other window represents the bit
value 1. Multi-level cell (MLC) NAND flash was commercialized
in the last two decades, where the same voltage range is
instead divided into \emph{four} voltage windows that represent each
possible 2-bit value (00, 01, 10, and 11). Each voltage window
in MLC NAND flash is therefore much smaller than a voltage
window in SLC NAND flash. This makes it more difficult to
identify the value stored in a cell. More recently, triple-level
cell (TLC) flash has been commercialized~\cite{R65, R183}, which
further divides the range, providing \emph{eight} voltage windows to
represent a 3-bit value. Quadruple-level cell (QLC) flash, storing
a 4-bit value per cell, is currently being developed~\cite{R184}.
Encoding more bits per cell increases the capacity of the SSD
without increasing the chip size, yet it also decreases reliability
by making it more difficult to correctly store and read the bits.

The two major drivers for the higher capacity, and thus
the ubiquitous commercial success, of flash memory as a storage
device, are also major drivers for its reduced reliability
and are the causes of its scaling problems. As the amount of
charge stored in each NAND flash cell decreases, the voltage
for each possible bit value is distributed over a wider voltage
range due to greater process variation, and the \emph{margins} (i.e.,
the width of the gap between neighboring voltage windows)
provided to ensure the raw reliability of NAND flash chips
have been diminishing, leading to a greater probability of flash
memory errors with newer generations of SSDs. NAND
flash memory errors can be induced by a variety of sources~\cite{R32}, including flash cell wearout~\cite{R32, R33, R42}, errors
introduced during programming~\cite{R35, R40, R42, R53}, interference
from operations performed on adjacent cells~\cite{R20, R26, R27, R35, R36, R38, R55, R62}, and data retention
issues due to charge leakage~\cite{R20, R32, R34, R37, R39}.

To compensate for this, SSDs employ sophisticated error-correc\-ting
codes (ECCs) within their controllers. An SSD controller
uses the ECC information stored alongside a piece of
data in the NAND flash chip to detect and correct a number
of \emph{raw bit errors} (i.e., the number of errors experienced before
correction is applied) when the piece of data is read out. The
number of bits that can be corrected for every piece of data is
a fundamental tradeoff in an SSD. A more sophisticated ECC
can tolerate a larger number of raw bit errors, but it also consumes
greater area overhead and latency. Error characterization
studies~\cite{R20, R32, R33, R42, R53, R62} have found that,
due to NAND flash wearout, the probability of raw bit errors
increases as more \emph{program/erase (P/E) cycles} (i.e., \emph{write accesses},
or \emph{writes}) are performed to the drive. The raw bit error rate
eventually exceeds the maximum number of errors that can be
corrected by ECC, at which point data loss occurs~\cite{R37, R44, R48, R49}. The \emph{lifetime} of a NAND-flash-memory-based SSD is
determined by the number of P/E cycles that can be performed
successfully while avoiding data loss for a minimum \emph{retention
guarantee}
(i.e., the required minimum amount of time, after
being written, that the data can still be read out without uncorrectable
errors).

The decreasing raw reliability of NAND flash memory
chips has drastically impacted the lifetime of commercial
SSDs. For example, older SLC NAND-flash-based SSDs were
able to withstand 150,000 P/E cycles (writes) to each flash
cell, but contemporary 1x-nm (i.e., \SIrange{15}{19}{\nano\meter}) process-based
SSDs consisting of MLC NAND flash can sustain only 3,000
P/E cycles~\cite{R53, R60, R81}. With the raw reliability of a flash
chip dropping so significantly, approaches to mitigating reliability
issues in NAND-flash-based SSDs have been the focus
of an important body of research. A number of solutions
have been proposed to increase the lifetime of contemporary
SSDs, ranging from changes to the low-level device behavior
(e.g., \cite{R33, R38, R40, R72}) to making SSD controllers
much more intelligent in dealing with individual flash memory
chips (e.g., \cite{R34, R36, R37, R39, R41, R42, R43, R45, R65}).
In addition, various mechanisms have been developed to successfully
recover data in the event of data loss that may occur
during a read operation to the SSD (e.g., \cite{R37, R38, R45}).

In this \chI{chapter}, we provide a comprehensive overview of the
state of flash-memory-based SSD reliability, with a focus on
(1)~fundamental causes of flash memory errors, backed up by
(2)~quantitative error data collected from real state-of-the-art
flash memory devices, and (3)~sophisticated error mitigation
and data recovery techniques developed to tolerate, correct,
and recover from such errors. To this end, we first discuss the
architecture of a state-of-the-art SSD, and describe mechanisms
used in a commercial SSD to reduce the probability of data loss
(Section~\ref{sec:ssdarch}). Next, we discuss the low-level behavior of the
underlying NAND flash memory chip in an SSD, to illustrate
fundamental reasons why errors can occur in flash memory
(Section~\ref{sec:flash}). We then discuss the root causes of these errors,
quantifying the impact of each error source using experimental
characterization data collected from real NAND flash memory
chips (Section~\ref{sec:errors}). For each of these error sources, we describe
various state-of-the-art mechanisms that mitigate the induced
errors (Section~\ref{sec:mitigation}). We next examine several error recovery
flows to successfully extract data from the SSD in the event of
data loss during a read operation (Section~\ref{sec:correction}). Then, we look to
the future to foreshadow how the reliability of SSDs might be
affected by emerging flash memory technologies (Section~\ref{sec:3d}).
Finally, we briefly examine how other memory technologies
(such as DRAM, which is used prominently in a modern SSD,
and emerging nonvolatile memory) suffer from similar reliability
issues to SSDs (Section~\ref{sec:othermem}).


\section{State-of-the-Art SSD Architecture}
\label{sec:ssdarch}

\begin{figure}[h]
  \centering
  \includegraphics[width=0.75\columnwidth]{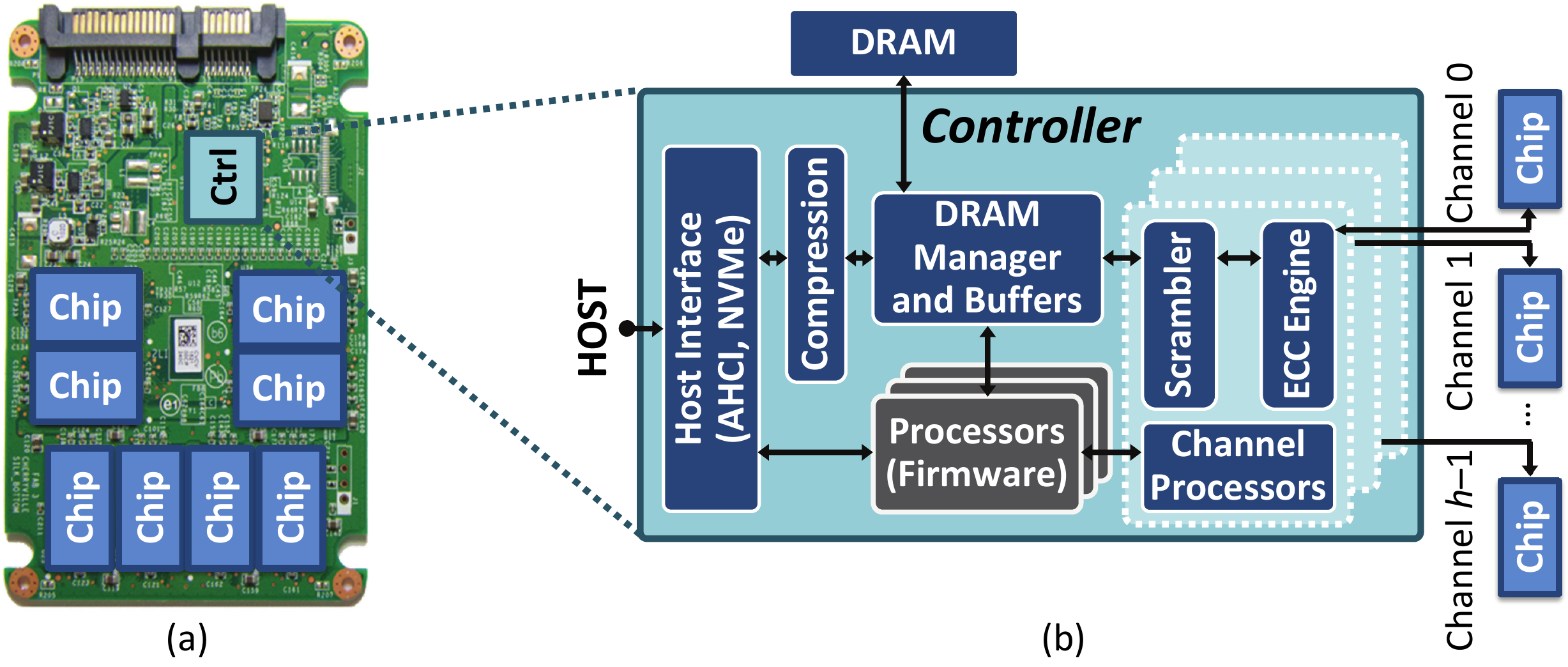}%
  \vspace{-8pt}%
  \caption{(a)~SSD system architecture, showing controller (Ctrl)
and chips. (b)~Detailed view of connections between controller
components and chips. \chI{Adapted from \cite{cai.arxiv17}.}}%
  \label{fig:F1}%
\end{figure}

In order to understand the root causes of reliability issues
within SSDs, we first provide an overview of the system architecture
of a state-of-the-art SSD. The SSD consists of a group of
NAND flash memories (or \emph{chips}) and a \emph{controller}, as shown in
Figure~\ref{fig:F1}. A host computer communicates with the SSD through a
high-speed host interface (e.g., \chIII{AHCI, NVMe}; \chIV{see Section~\ref{sec:ssdarch:ctrl:scheduling}}), which
connects to the SSD controller. The controller is then connected
to each of the NAND flash chips via memory \emph{channels}.

\subsection{Flash Memory Organization}
\label{sec:ssdarch:flash}

\begin{figure}[h]
  \centering
  \includegraphics[width=0.65\columnwidth]{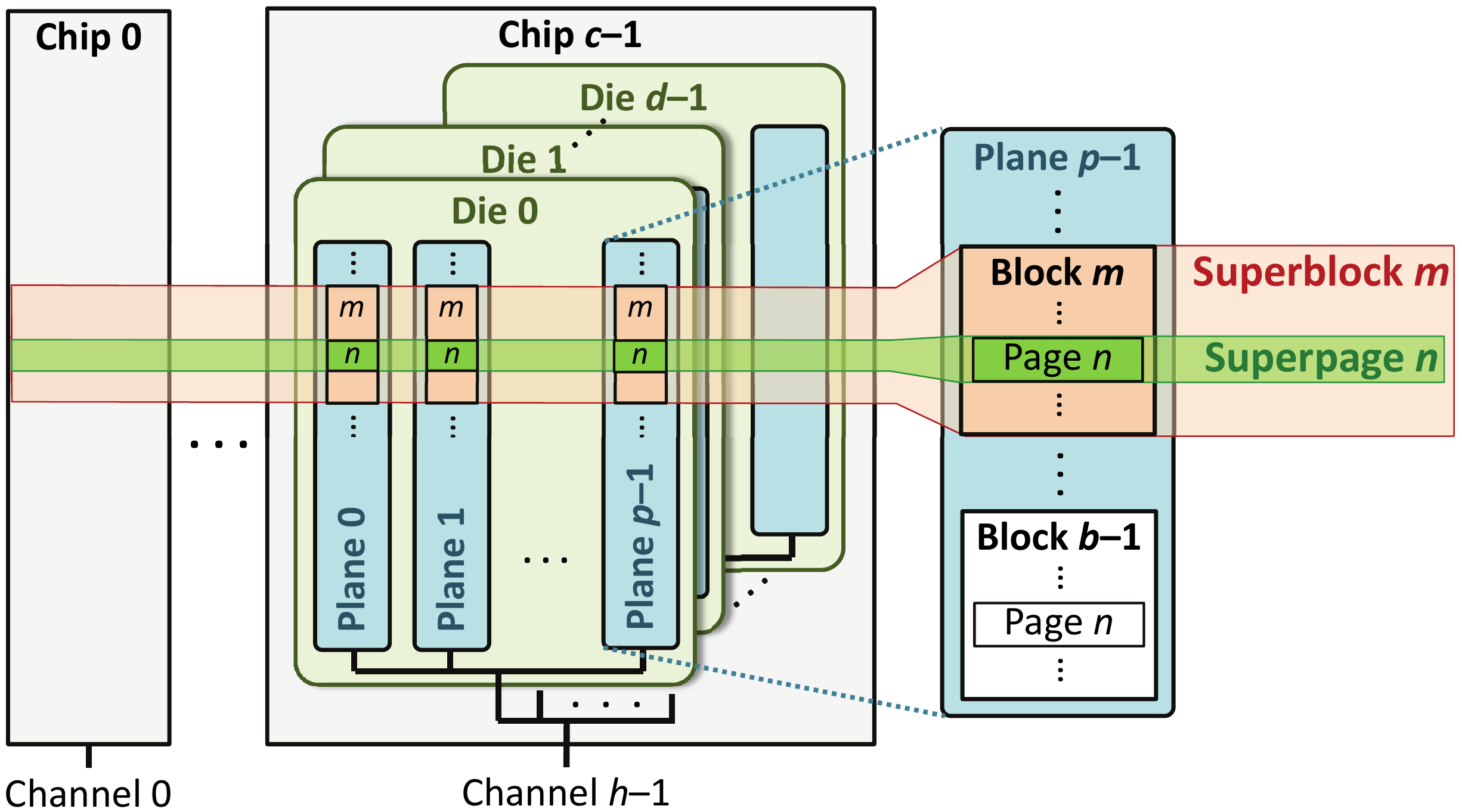}%
  \vspace{-5pt}%
  \caption{Flash memory organization. \chI{Reproduced from \cite{cai.arxiv17}.}}%
  \label{fig:F2}%
\end{figure}

Figure~\ref{fig:F2} shows an example of how NAND flash memory is
organized within an SSD. The flash memory is spread across
multiple flash chips, where each chip contains one or more
flash \emph{dies}, which are individual pieces of silicon wafer that
are connected together to the pins of the chip. Contemporary
SSDs typically have 4--16 chips per SSD, and can have as many
as 16 dies per chip. Each chip is connected to one or more
physical memory channels, and these memory channels are
not shared across chips. A flash die operates independently
of other flash dies, and contains between one and four \emph{planes}.
Each plane contains hundreds to thousands of flash \emph{blocks}.
Each block is a 2D array that contains hundreds of rows of
flash cells (typically 256--1024 rows) where the rows store
contiguous pieces of data. Much like banks in a multi-bank
memory (e.g., DRAM banks~\cite{R84, R85, R99, R101, R102, R108, R193, R194, R195, R196}), 
the planes can execute flash operations
in parallel, but the planes within a die share a single set of data
and control buses~\cite{R185}. Hence, an operation can be started in
a different plane in the same die in a pipelined manner, every
cycle. Figure~\ref{fig:F2} shows how blocks are organized within chips
across multiple channels. In the rest of this work, without
loss of generality, we assume that a chip contains a single die.

Data in a block is written at the unit of a \emph{page}, which is
typically between 8 and \SI{16}{\kilo\byte} in size in NAND flash memory.
All read and write operations are performed at the granularity
of a page. Each block typically contains hundreds of pages.
Blocks in each plane are numbered with an ID that is unique
within the plane, but is shared across multiple planes. Within
the block, each page is numbered in sequence. The controller
firmware groups blocks with the same ID number across
multiple chips and planes together into a \emph{superblock}. Within
each superblock, the pages with the same page number are
considered a \emph{superpage}. The controller \emph{opens} one superblock
(i.e., an empty superblock is selected for write operations) at a
time, and typically writes data to the NAND flash memory one
superpage at a time to improve sequential read/write performance
and make error correction efficient, since some parity
information is kept at superpage granularity (see \chIV{Section~\ref{sec:ssdarch:ctrl:parity}}).
Having the ability to write to all of the pages in a superpage
simultaneously, the SSD can fully exploit the internal parallelism
offered by multiple planes/chips, which in turn maximizes
write throughput.

\subsection{Memory Channel}
\label{sec:ssdarch:channel}

Each flash memory channel has its own data and control
connection to the SSD controller, much like a main
memory channel has to the DRAM 
controller\chV{~\cite{R99, R100, R102, R108, R197, R198, R199, R200, R201, 
R157, kim.micro10, hassan.hpca16, subramanian.hpca13, ipek.isca08, ghose.isca13}}. 
The connection for each channel
is typically an 8- or 16-bit wide bus between the controller
and one of the flash memory chips~\cite{R185}. Both data and flash
commands can be sent over the bus.

Each channel also contains its own control signal pins
to indicate the type of data or command that is on the bus.
The \emph{address latch enable} (ALE) pin signals that the controller
is sending an address, while the \emph{command latch enable}
(CLE) pin signals that the controller is sending a flash
command. Every rising edge of the \emph{write enable} (WE) signal
indicates that the flash memory should write the piece of
data currently being sent on the bus by the SSD controller.
Similarly, every rising edge of the \emph{read enable} (RE) signal
indicates that the flash memory should send the next piece
of data from the flash memory to the SSD controller.

Each flash memory die connected to a memory channel
has its own \emph{chip enable} (CE) signal, which selects the die
that the controller currently wants to communicate with.
On a channel, the bus broadcasts address, data, and flash
commands to all dies within the channel, but only the die
whose CE signal is active reads the information from the bus
and executes the corresponding operation.

\subsection{SSD Controller}
\label{sec:ssdarch:ctrl}

The SSD controller, shown in Figure~\ref{fig:F1}b, is responsible for
\chIII{\chIV{(1)}~handling I/O requests received from the host, 
\chIV{(2)~ensuring data integrity and efficient storage, and (3)~}managing the 
underlying NAND flash memory.}
To perform these
tasks, the controller runs firmware, which is often referred
to as the \emph{flash translation layer} (FTL). FTL tasks are executed
on one or more embedded processors that exist inside
the controller. The controller has access to DRAM, which
can be used to store various controller metadata (e.g., how
host memory addresses map to physical SSD addresses) and
to cache relevant (e.g., frequently accessed) SSD pages~\cite{R48, R161}. 

When the controller handles I/O requests, it performs
a number of operations on \chIII{both the requests and the data. For requests,
the controller \emph{schedules} them in a manner that ensures correctness and provides
high/reasonable performance. For data, the controller \emph{scrambles}
the data to improve raw bit error rates, performs \emph{ECC
encoding/decoding}, and in some cases \chIV{\emph{compresses/decompresses}
and/or \emph{encrypts/decrypts}} the
data and employs \emph{superpage-level data parity}.
To manage the NAND flash memory, the controller runs \emph{firmware} that
maps host data to physical NAND flash pages, performs \emph{garbage
collection} on flash pages that have been invalidated, applies
\emph{wear leveling} to evenly distribute the impact of writes on NAND flash
reliability across all pages, and manages bad NAND flash blocks.} We briefly
examine the various tasks of the SSD controller.

\subsubsection{\chI{Scheduling Requests}}
\label{sec:ssdarch:ctrl:scheduling}

\chI{The controller receives I/O requests over a \emph{host controller interface}
\chII{(shown as \emph{Host Interface} in Figure~\ref{fig:F1}b)},
which consists of a system I/O bus and the protocol used to communicate along 
the bus.  When an application running on the host \chII{system} needs to access the
SSD, it generates an I/O request, which is \chII{sent by the host} over the host controller
interface.  The SSD controller receives the I/O request, and inserts the 
request into a queue.  The controller uses a \emph{scheduling policy} to 
determine the order in which the controller processes the requests that 
\chII{are} in the queue.  \chII{The controller then sends the request selected for
scheduling to}
the FTL \chIII{(part of the \emph{Firmware} shown in Figure~\ref{fig:F1}b)}.}

\chI{The host controller interface determines how requests are sent to the SSD
and how the requests are queued for scheduling.
Two of the most common host controller interfaces used by \chII{modern} SSDs are
the Advanced Host Controller Interface (AHCI)~\cite{ahci.1.3.1.spec} and 
NVM Express (NVMe)~\cite{nvme.1.3.spec}.
AHCI builds upon the Serial Advanced Technology Attachment (SATA) system bus
protocol~\cite{sata.3.3.spec}, which was originally designed to connect the host
system to magnetic hard disk drives.  AHCI allows the host to use advanced
features with SATA, such as \emph{native command queuing} (NCQ).  When an
application executing on the host generates an I/O request, the application 
sends the request to the operating system (OS).  The OS sends the request over
the SATA bus to the SSD controller, and the controller adds the request to a
single \emph{command queue}.  NCQ allows the controller to schedule the queued
I/O requests in a different order than the order in which requests were 
received (i.e., requests are scheduled \emph{out of order}).  As a result, 
the controller can choose requests from the queue \chIII{in a manner that
maximizes} the overall SSD
performance (e.g., \chII{a \chIII{younger} request can be scheduler earlier than
an
older request that requires access to a \chIII{plane} that is \chIII
{occupied with} serving another
request}).
A major drawback of AHCI and SATA is the limited throughput they enable 
for SSDs~\cite{xu.systor15}, as the protocols were originally designed to match the much
lower throughput of magnetic hard disk drives.
\chII{For example, a modern magnetic hard drive has a sustained read throughput of \SI{300}{\mega\byte\per\second}~\cite{seagate.cheetah15k.spec},
whereas a modern SSD has a read throughput of \SI{3500}{\mega\byte\per\second}~\cite{samsung.960pro.spec}.}
However, AHCI and SATA are widely deployed in modern \chIII{computing systems},
and \chIII{they currently} remain
a common choice for the SSD host controller interface.}

\chI{To alleviate the throughput bottleneck of AHCI and SATA, many manufacturers
\chIII{have started adopting} host controller interfaces that use the PCI
Express (PCIe) system
bus~\cite{pcie.3.1a.spec}.  A popular standard interface for the PCIe bus is the NVM
Express (NVMe) interface~\cite{nvme.1.3.spec}.  Unlike AHCI, which requires an
application to send I/O requests through the OS, NVMe directly exposes multiple
SSD I/O queues to the applications executing on the host.  By directly exposing 
the queues to the applications, NVMe simplifies the software I/O stack, 
eliminating most OS involvement\chII{~\cite{xu.systor15}}, which in turn reduces communication overheads.
An SSD using the NVMe interface maintains a separate set of queues for
\emph{each} application (as opposed to the single
queue \chII{used for all applications with} AHCI) \chIII{within the host interface}.  With more
queues, the controller \chIII{(1)~has a larger number of requests to select from
during scheduling, increasing its ability to utilize idle resources (i.e., 
\emph{channels}, \emph{dies}, \emph{planes}; see \chIV{Section~\ref{sec:ssdarch:flash}}); and
(2)~can more easily manage and control} the amount of interference that
an application experiences from
other concurrently-executing applications. Currently, NVMe is used by modern
SSDs that are designed \chIII{mainly} for high-performance systems (e.g.,
enterprise servers, data centers\chIII{~\cite{xu.sigmetrics15,xu.systor15}}).}

\subsubsection{Flash Translation Layer}
\label{sec:ssdarch:ctrl:ftl}

The main duty of the FTL \chIII{(which is part of the \emph{Firmware} shown in Figure~\ref{fig:F1})} is to
manage the mapping of \emph{logical addresses} (i.e., the address
space utilized by the host) to \emph{physical addresses} in the
underlying flash memory (i.e., the address space for actual
locations where the data is stored, visible only to the SSD
controller) for each page of data~\cite{R1, R2}. By providing this
indirection between address spaces, the FTL can \emph{remap} the
logical address to a different physical address (i.e., move
the data to a different physical address) \emph{without} notifying
the host. Whenever a page of data is written to by the host
or moved for underlying SSD maintenance operations (e.g.,
\chI{garbage collection~\cite{R3, R4}; see 
\chIV{Section~\ref{sec:ssdarch:ctrl:gc}}}), the old data (i.e., the
physical location where the overwritten data resides) is simply
marked as invalid in the physical block's \emph{metadata}, and
the new data is written to a page in the flash block that is
currently open for writes (see Section~\ref{sec:flash:pgmerase} for more detail
on how writes are performed).

The FTL is also responsible for \emph{wear leveling}, to ensure that
all of the blocks within the SSD are evenly worn out~\cite{R3, R4}.
By evenly distributing the \emph{wear} (i.e., the number of P/E cycles
that take place) \emph{across} different blocks, the SSD controller
reduces the heterogeneity of the amount of wearout across
these blocks, \chIII{thereby} extending the lifetime of the device. \chIII{The
wear-leveling
algorithm is} invoked when the current block that
is being written to is full (i.e., no more pages in the block are
available to write to), and \chIII{it enables} the controller \chIII{to select}
a new block \chIII{from the \chIV{\emph{free list}} to direct the future writes to}. The wear-leveling
algorithm dictates
which of the blocks from the free list is selected. One
simple approach is to select the block in the free list with the
lowest number of P/E cycles to minimize the variance of the
wearout amount across blocks, though many algorithms have
been developed for wear leveling~\cite{R98, R203}.

\subsubsection{\chI{Garbage Collection}}
\label{sec:ssdarch:ctrl:gc}

\chI{When the host issues a write request to a logical address stored in the
SSD, the SSD controller performs the write \emph{out of place} (i.e., \chIII
{the} updated
version of the \chIII{page data} is written to a different physical page in the
NAND flash
memory), because in-place updates cannot be performed 
(see Section~\ref{sec:flash:pgmerase}).  The old physical page is marked as
\emph{invalid} when the out-of-place write completes.
\chII{\emph{Fragmentation}
refers to the waste of space within a block \chIII{due to the presence of
invalid pages}.
In a \emph{fragmented} block, \chIII{a fraction of} the pages are
invalid, but these pages are unable to store new data until the
page is erased.  Due to circuit-level limitations, the controller can perform
erase operations \chIII{\chIV{only at} the granularity of} an \emph{entire block} (see 
Section~\ref{sec:flash:pgmerase} for details).  As a result, until a fragmented
block is erased, the block wastes physical space within the SSD.  Over time,
if fragmented blocks are not erased, the SSD will run out of pages that it
can write new data to.}} \chIII{The problem becomes especially severe if the
blocks are highly fragmented (i.e., a large fraction of the pages within a
block are invalid).}

\chI{To reduce the \chIII{negative} impact of fragmentation \chIII{on usable
SSD storage space}, the FTL periodically performs \chII{a \chIV{process
called}}
\emph{garbage collection}.  
\chII{Garbage collection finds highly-fragmented flash blocks in the SSD
and recovers the wasted space due to invalid pages.}
\chII{The basic garbage collection algorithm}~\cite{R3, R4}
(1)~identifies the \chII{highly-fragmented blocks} (which we call the
\emph{selected blocks}),
(2)~migrates any valid pages in a selected block (i.e., each valid page is 
written to a new block, \chII{its virtual-to-physical address mapping is updated, 
and the page in the selected block is marked as invalid}),
(3)~erases each selected block (see Section~\ref{sec:flash:pgmerase}),
and (4)~adds a pointer to each selected block into \chIII{the} \emph{free list}
within the FTL.
The \chII{garbage collection} algorithm typically selects blocks with the
\chIII{highest number of invalid pages}.
When the controller needs a new block to write pages to, it selects one of the
blocks currently in the free list.}

\chI{\chII{We briefly discuss five optimizations that prior works propose to
improve the performance and/or efficiency of garbage collection~\cite{R4, qin.dac11, han.uic06, he.eurosys17, R185, choudhuri.codes08,
wu.fast12, R1, R41}}.
First, the \chII{garbage collection} algorithm can be optimized to determine the most efficient frequency to invoke
garbage collection~\cite{R4, qin.dac11}, as performing garbage collection too
frequently can delay I/O requests from the host, while not performing garbage
collection frequently enough can cause the controller to stall when there are
no blocks available in the free list.
Second, the algorithm can be optimized to select blocks \chIII{in a way that
reduces} the number
of page copy and erase operations required each time the garbage collection 
algorithm is invoked~\cite{han.uic06, qin.dac11}.
Third, some works reduce the latency of garbage collection by using multiple channels
to \chII{perform garbage collection on} multiple blocks in parallel~\cite{he.eurosys17, R185}.
Fourth, the FTL can minimize the latency of I/O requests from the host by
pausing erase and copy operations \chIII{that are} being performed for garbage
collection, in
order to service the host requests immediately~\cite{choudhuri.codes08, wu.fast12}.
Fifth, \chII{pages can be grouped together such that all of the pages within a
block become invalid around the same time~\cite{R1, he.eurosys17, R41}.
For example, the controller can group pages with (1)~a similar degree of 
\emph{write-hotness} (i.e., the frequency at which a page is updated; see 
Section~\ref{sec:mitigation:hotcold}) or (2)~a similar} \emph{death time} (i.e., the time at 
which a page is \chIII{overwritten}).}
\chII{Garbage collection remains an active area of research.}

\subsubsection{Flash Reliability Management}
\label{sec:ssdarch:ctrl:reliability}

The SSD controller
performs many background optimizations that improve
flash reliability. These flash reliability management techniques,
as we will discuss in more detail in Section~\ref{sec:mitigation},
can effectively improve flash lifetime at a very low cost,
since the optimizations are usually performed during idle
times, when the interference with the running workload
is minimized. These management techniques sometimes
require small metadata storage in memory (e.g., for storing
\chIII{the near-optimal} read reference voltages~\cite{R37, R38, R42}), or
require a timer (e.g., for triggering refreshes in 
time~\cite{R34, R39}).

\subsubsection{Compression}
\label{sec:ssdarch:ctrl:compression}

Compression can reduce the size of the
data written to minimize the number of flash cells worn out
by the original data. Some controllers provide compression,
as well as decompression, which reconstructs the original
data from the compressed data stored in the flash 
memory~\cite{R5, R6}. The controller may contain a \emph{compression engine},
which, for example, performs the LZ77 or LZ78 algorithms.
Compression is optional, as some types of data being stored
by the host (e.g., JPEG images, videos, encrypted files, files
that are already compressed) may not be compressible.

\subsubsection{Data Scrambling and Encryption}
\label{sec:ssdarch:ctrl:scrambling}
\label{sec:ssdarch:ctrl:encryption}

The occurrence of
errors in flash memory is highly dependent on the data values
stored into the memory cells~\cite{R32, R35, R36}. To reduce
the dependence of the error rate on data values, an SSD
controller first scrambles the data before writing it into the
flash chips~\cite{R7, R8}. The key idea of scrambling is to probabilistically
ensure that the actual value written to the SSD
contains an equal number of randomly distributed zeroes
and ones, thereby minimizing any data-dependent behavior.
Scrambling is performed using a reversible process, and
the controller \emph{descrambles} the data stored in the SSD during
a read request. The controller employs a \emph{linear feedback shift
register} (LFSR) to perform scrambling and descrambling.
An $n$-bit LFSR generates $2^{n-1}$ bits worth of pseudo-random
numbers without repetition. For each page of data to be written,
the LFSR can be seeded with the \emph{logical} address of that
page, so that the page can be correctly descrambled even if
maintenance operations (e.g., garbage collection) migrate
the page to another physical location, as the logical address
is unchanged. (This also reduces the latency of maintenance
operations, as they do not need to descramble and rescramble
the data when a page is migrated.) The LFSR then generates
a pseudo-random number based on the seed, which is
then XORed with the data to produce the scrambled version
of the data. As the XOR operation is reversible, the same
process can be used to descramble the data.

In addition to the data scrambling employed to minimize
data value dependence, several SSD controllers
include data encryption hardware~\cite{R167, R168, R170}. An
SSD that contains data encryption hardware within its
controller is known as a \emph{self-encrypting drive} (SED). In the
controller, data encryption hardware typically employs
AES encryption~\cite{R168, R169, R170, R204}, which performs multiple
rounds of substitutions and permutations to the unencrypted
data in order to encrypt it. AES employs a separate
key for each round~\cite{R169, R204}. In an SED, the controller
contains hardware that generates the AES keys for each
round, and performs the substitutions and permutations
to encrypt or decrypt the data using dedicated 
hardware~\cite{R167, R168, R170}.

\subsubsection{Error-Correcting Codes}
\label{sec:ssdarch:ctrl:ecc}

ECC is used to detect and correct
the raw bit errors that occur within flash memory. A
host writes a page of data, which the SSD controller splits
into one or more chunks. For each chunk, the controller
generates a \emph{codeword}, consisting of the chunk and a correction
code. The strength of protection offered by ECC
is determined by the \emph{coding rate}, which is the chunk size
divided by the codeword size. A higher coding rate provides
weaker protection, but consumes less storage, representing
a key reliability tradeoff in SSDs.

The ECC algorithm employed (typically BCH~\cite{R9, R10, R92, R93} 
or LDPC\chV{~\cite{R9, R11, R94, R95, gallager.tit62}}; see Section~\ref{sec:correction}),
as well as the length of the codeword and the coding rate,
determine the total \emph{error correction capability}, i.e., the
maximum number of raw bit errors that can be corrected
by ECC. ECC engines in contemporary SSDs are able to
correct data with a relatively high raw bit error rate (e.g.,
between $10^{-3}$ and $10^{-2}$~\cite{R110}) and return data to the host at
an error rate that meets traditional data storage reliability
requirements (e.g., a post-correction error rate of $10^{-15}$ in
the JEDEC standard~\cite{R12}). The \emph{error correction failure rate}
($P_{ECFR}$) of an ECC implementation, with a codeword length
of $l$ where the codeword has an error correction capability
of $t$ bits, can be modeled as:
\begin{equation}
P_{ECFR} = \sum_{k=t+1}^{l}  \binom{l}{k} (1 - \text{BER})^{(l-k)} \text{BER}^{k}
\label{eq:E1}
\end{equation}
where BER is the bit error rate of the NAND flash memory.
We assume in this equation that errors are independent and
identically distributed.

In addition to the ECC information, a codeword contains
cyclic redundancy checksum (CRC) parity information~\cite{R161}.
When data is being read from the NAND flash memory,
there may be times when the ECC algorithm incorrectly
indicates that it has successfully corrected all errors in the
data, when uncorrected errors remain. To ensure that incorrect
data is not returned to the user, the controller performs
a CRC check in hardware to verify that the data is error 
free~\cite{R161, R205}.

\subsubsection{Data Path Protection}
\label{sec:ssdarch:ctrl:datapathprotection}

\begin{figure}[h]
  \centering
  \includegraphics[width=0.65\columnwidth]{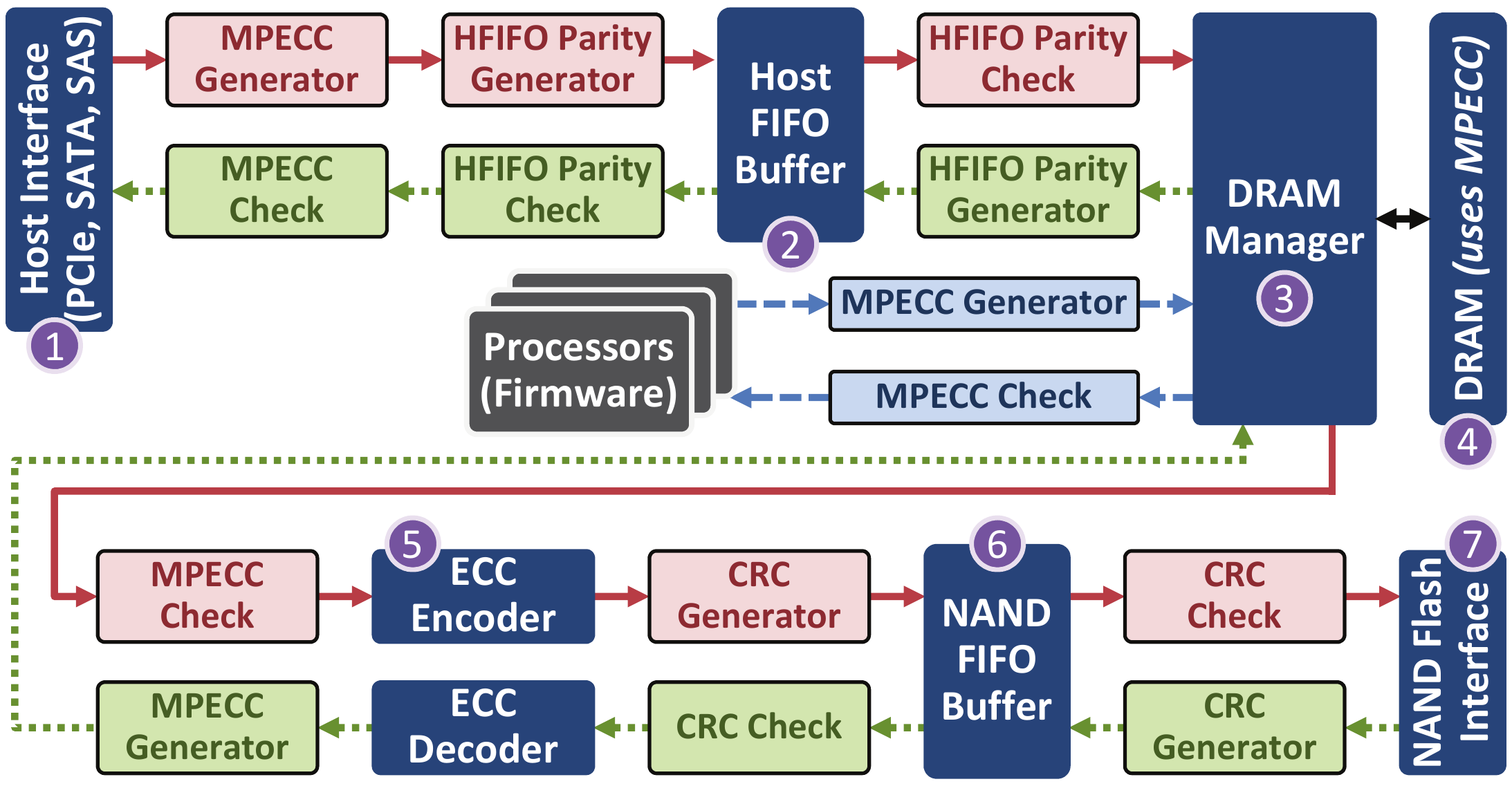}%
  \vspace{-5pt}%
  \caption{Data path protection employed within the controller. \chI{Reproduced from \cite{cai.arxiv17}.}}%
  \label{fig:F3}%
\end{figure}

In addition to protecting the
data from raw bit errors within the NAND flash memory,
newer SSDs incorporate error detection and correction
mechanisms throughout the SSD controller, in order to
further improve reliability and data integrity~\cite{R161}. These
mechanisms are collectively known as \emph{data path protection},
and protect against errors that can be introduced by the various
SRAM and DRAM structures that exist within the SSD.\footnote{See 
Section~\ref{sec:othermem} for a discussion on the possible types of errors that
can be present in DRAM.}
Figure~\ref{fig:F3} illustrates the various structures within the controller
that employ data path protection mechanisms. There are
three data paths that require protection: (1)~the path for data
written by the host to the flash memory, shown as a red solid
line in Figure~\ref{fig:F3}; (2)~the path for data read from the flash memory
by the host, shown as a green dotted line; and (3)~the path
for metadata transferred between the firmware (i.e., FTL)
processors and the DRAM, shown as a blue dashed line.

In the write data path of the controller (the red solid
line shown in Figure~\ref{fig:F3}), data received from the host interface
(\incircle{1} in the figure) is first sent to a host FIFO buffer (\incircle{2}).
Before the data is written into the host FIFO buffer, the data
is appended with \emph{memory protection ECC} (MPECC) and
\emph{host FIFO buffer} (HFIFO) parity~\cite{R161}. The MPECC parity is
designed to protect against errors that are introduced when
the data is stored within DRAM (which takes place later
along the data path), while the HFIFO parity is designed
to protect against SRAM errors that are introduced when
the data resides within the host FIFO buffer. When the
data reaches the head of the host FIFO buffer, the controller
fetches the data from the buffer, uses the HFIFO parity
to correct any errors, discards the HFIFO parity, and sends
the data to the DRAM manager (\incircle{3}). The DRAM manager
buffers the data (which still contains the MPECC information)
within DRAM (\incircle{4}), and keeps track of the location of
the buffered data inside the DRAM. When the controller
is ready to write the data to the NAND flash memory, the
DRAM manager reads the data from DRAM. Then, the controller
uses the MPECC information to correct any errors,
and discards the MPECC information. The controller then
encodes the data into an ECC codeword (\incircle{5}), generates CRC
parity for the codeword, and then writes both the codeword
and the CRC parity to a NAND flash FIFO buffer (\incircle{6})~\cite{R161}.
When the codeword reaches the head of this buffer, the controller
uses CRC parity to \chIV{detect} any errors in the codeword,
and then dispatches the data to the flash interface (\incircle{7}),
which writes the data to the NAND flash memory. The read
data path of the controller (the green dotted line shown in
Figure~\ref{fig:F3}) performs the same procedure as the write data path,
but in reverse order~\cite{R161}.

Aside from buffering data along the write and read paths,
the controller uses the DRAM to store essential metadata,
such as the table that maps each host data address to a physical
block address within the NAND flash memory~\cite{R48, R161}.
In the metadata path of the controller (the blue dashed
line shown in Figure~\ref{fig:F3}), the metadata is often read from or written
to DRAM by the firmware processors. In order to ensure
correct operation of the SSD, the metadata must not contain
any errors. As a result, the controller uses memory protection
ECC (MPECC) for the metadata stored within DRAM~\cite{R130, R161}, 
just as it did to buffer data along the write and
read data paths. Due to the lower rate of errors in DRAM
compared to NAND flash memory (see Section~\ref{sec:othermem}), the
employed memory protection ECC algorithms are not as
strong as BCH or LDPC. We describe common ECC algorithms
employed for DRAM error correction in Section~\ref{sec:othermem}.

\subsubsection{Bad Block Management}
\label{sec:ssdarch:ctrl:badblocks}

Due to process variation or
uneven wearout, a small number of flash blocks may have
a much higher raw bit error rate (RBER) than an average
flash block. Mitigating or tolerating the RBER on these flash
blocks often requires a much higher cost than the benefit of
using them. Thus, it is more efficient to identify and record
these blocks as \emph{bad blocks}, and avoid using them to store
useful data. There are two types of bad blocks: \emph{original bad
blocks} (OBBs), which are defective due to manufacturing
issues (e.g., process variation), and \emph{growth bad blocks}
(GBBs), which fail during runtime~\cite{R91}.

The flash vendor performs extensive testing, known
as \emph{bad block scanning}, to identify OBBs when a flash chip
is manufactured~\cite{R106}. Initially, all blocks are kept in
the erased state, and contain the value 0xFF in each byte
(see Section~\ref{sec:flash:data}). Inside each OBB, the bad block scanning
procedure writes a specific data value (e.g., 0x00) to
a specific byte location within the block that indicates the
block status. A good block (i.e., a block without defects) is
not modified, and thus its block status byte remains at the
value 0xFF. When the SSD is powered up for the first time,
the SSD controller iterates through all blocks and checks
the value stored in the block status byte of each block. Any
block that does not contain the value 0xFF is marked as bad,
and is recorded in a \emph{bad block table} stored in the controller.
A small number of blocks in each plane are set aside as
\emph{reserved blocks} (i.e., blocks that are not used during normal
operation), and the bad block table automatically remaps
any operation originally destined to an OBB to one of the
reserved blocks. The bad block table remaps an OBB to a
reserved block in the same plane, to ensure that the SSD
maintains the same degree of parallelism when writing to a
superpage, thus avoiding performance loss. Less than 2\% of
all blocks in the SSD are expected to be OBBs~\cite{R162}.

The SSD identifies growth bad blocks during runtime by
monitoring the status of each block. Each superblock contains
a bit vector indicating which of its blocks are GBBs.
After each program or erase operation to a block, the SSD
reads the \emph{status reporting registers} to check the operation
status. If the operation has failed, the controller marks the
block as a GBB in the superblock bit vector. At this point,
the controller uses superpage-level parity to recover the data
that was stored in the GBB (see \chIV{Section~\ref{sec:ssdarch:ctrl:parity}}), and \emph{all data
in the superblock} is copied to a different superblock. The
superblock containing the GBB is then erased. When the
superblock is subsequently opened, blocks marked as GBBs
are \emph{not} used, but the remaining blocks can store new data.

\subsubsection{Superpage-Level Parity}
\label{sec:ssdarch:ctrl:parity}
\label{sec:ssdarch:ctrl:superpageparity}

\begin{figure}[h]
  \centering
  \includegraphics[width=0.65\columnwidth]{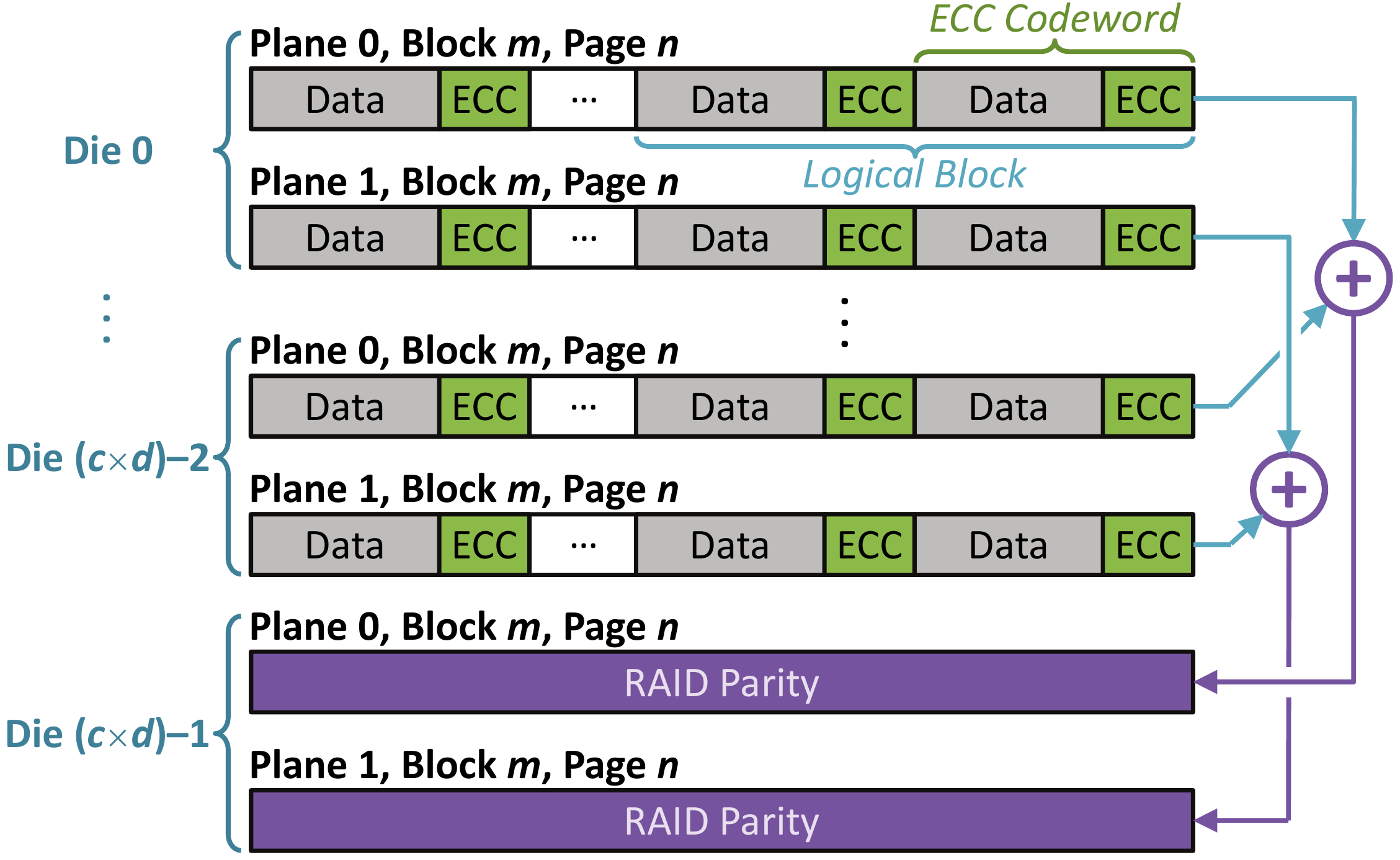}%
  \vspace{-5pt}%
  \caption{Example layout of ECC codewords, logical blocks, and
superpage-level parity for superpage \emph{n} in superblock \emph{m}. In this
example, we assume that a logical block contains two codewords. \chI{Reproduced from \cite{cai.arxiv17}.}}%
  \label{fig:F4}%
\end{figure}

In addition to ECC to protect
against bit-level errors, many SSDs employ RAID-like parity~\cite{R13, R14, R15, R16}. 
The key idea is to store parity information within
each superpage to protect data from ECC failures that occur
within a single chip or plane. Figure~\ref{fig:F4} shows an example of
how the ECC and parity information are organized within
a superpage. For a superpage that spans across multiple
chips, dies, and planes, the pages stored within one die or
one plane (depending on the implementation) are used to
store parity information for the remaining pages. Without
loss of generality, we assume for the rest of this section that
a superpage that spans $c$ chips and $d$ dies per chip stores parity
information in the pages of a single die (which we call
the \emph{parity die}), and that it stores user data in the pages of
the remaining $(c \times d) - 1$ dies. When all of the user data is
written to the superpage, the SSD controller XORs the data
together one plane at a time (e.g., in Figure~\ref{fig:F4}, all of the pages
in Plane~0 are XORed with each other), which produces the
parity data for that plane. This parity data is written to the
corresponding plane in the parity die, e.g., Plane~0 page in
Die $(c \times d) - 1$ in the figure.

The SSD controller invokes superpage-level parity when
an ECC failure occurs during a host software (e.g., OS, file
system) access to the SSD. The host software accesses data
at the granularity of a \emph{logical block} (LB), which is indexed
by a \emph{logical block address} (LBA). Typically, an LB is \SI{4}{\kilo\byte}
in size, and consists of several ECC codewords (which are
usually \SI{512}{\byte}B to \SI{2}{\kilo\byte} in size) stored consecutively within
a flash memory page, as shown in Figure~\ref{fig:F4}. During the LB
access, a read failure can occur for one of two reasons.
First, it is possible that the LB data is stored within a \emph{hidden}
GBB (i.e., a GBB that has not yet been detected and
excluded by the bad block manager). The probability of
storing data in a hidden GBB is quantified as $P_{HGBB}$. Note
that because bad block management successfully identifies
and excludes most GBBs, $P_{HGBB}$ is much lower than the
total fraction of GBBs within an SSD. Second, it is possible
that at least one ECC codeword within the LB has \emph{failed}
(i.e., the codeword contains an error that cannot be corrected
by ECC). The probability that a codeword fails is
$P_{ECFR}$ (see \chIV{Section~\ref{sec:ssdarch:ctrl:ecc}}). For an LB that contains $K$ ECC
codewords, we can model $P_{LBFail}$, the overall probability
that an LB access fails (i.e., the rate at which superpage-level
parity needs to be invoked), as:
\begin{equation}
P_{LBFail} = P_{HGBB} + [1 - P_{HGBB}] \times [1 - (1 - P_{ECFR})^K]
\label{eq:E2}
\end{equation}
In Equation~\ref{eq:E2}, $P_{LBFail}$ consists of (1)~the probability that an LB is
inside a hidden GBB (left side of the addition); and (2)~for
an LB that is not in a hidden GBB, the probability of any
codeword failing (right side of the addition).

When a read failure occurs for an LB in plane $p$, the SSD
controller reconstructs the data using the other LBs in the
same superpage. To do this, the controller reads the LBs
stored in plane $p$ in the other $(c \times d) - 1$ dies of the superpage,
including the LBs in the parity die. The controller
then XORs all of these LBs together, which retrieves the
data that was originally stored in the LB whose access failed.
In order to correctly recover the failed data, all of the LBs
from the $(c \times d) - 1$ dies must be correctly read. The overall
superpage-level parity failure probability $P_{parity}$ (i.e., the
probability that more than one LB contains a failure) for an
SSD with $c$ chips of flash memory, with $d$ dies per chip, can
be modeled as~\cite{R16}:
\begin{equation}
P_{parity} = P_{LBFail} \times \lbrack 1 - (1 - P_{LBFail})^{(c \times d) - 1} \rbrack
\label{eq:E3}
\end{equation}
Thus, by designating one of the dies to contain parity information
(in a fashion similar to RAID 4~\cite{R16}), the SSD can
tolerate the \emph{complete failure} of the superpage data in one die
without experiencing data loss during an LB access.

\subsection{Design Tradeoffs for Reliability}
\label{sec:ssdarch:reliability}

Several design decisions impact the SSD \emph{lifetime} (i.e.,
the duration of time that the SSD can be used within a
bounded probability of error without exceeding a given
performance overhead). To capture the tradeoff between
these decisions and lifetime, SSD manufacturers use the
following model:
\begin{equation}
\text{Lifetime (Years)} = \frac{\text{PEC} \times (1 + \text{OP})}{365 \times \text{DWPD} \times \text{WA} \times R_{compress}}
\label{eq:E4}
\end{equation}

In Equation~\ref{eq:E4}, the numerator is the total number of full drive writes
the SSD can endure (i.e., for a drive with an $X$-byte capacity,
the number of times $X$ bytes of data can be written). The number
of full drive writes is calculated as the product of PEC, the
total P/E cycle \emph{endurance} of each flash block (i.e., the number
of P/E cycles the block can sustain before its raw error rate
exceeds the ECC correction capability), and $1+\text{OP}$, where OP
is the \emph{overprovisioning factor} selected by the manufacturer.
Manufacturers overprovision the flash drive by providing
more physical block addresses, or PBAs, to the SSD controller
than the \emph{advertised capacity} of the drive, i.e., the number of
logical block addresses (LBAs) available to the operating system.
Overprovisioning improves performance and endurance,
by providing additional free space in the SSD so that maintenance
operations can take place without stalling host requests.
OP is calculated as:
\begin{equation}
\text{OP} = \frac{\text{PBA count} - \text{LBA count}}{\text{LBA count}}
\label{eq:E5}
\end{equation}

The denominator in Equation~\ref{eq:E4} is the number of full drive writes
per year, which is calculated as the product of days per year
(i.e., 365), DWPD, and the ratio between the total size of
the data written to flash media and the size of the data sent
by the host (i.e., $\text{WA} \times R_{compress}$). DWPD is the number of
full disk writes per day (i.e., the number of times per day the
OS writes the advertised capacity's worth of data). DWPD
is typically less than 1 for read-intensive applications, and
could be greater than 5 for write-intensive applications~\cite{R34}.
WA (\emph{write amplification}) is the ratio between the amount
of data written into NAND flash memory by the controller
over the amount of data written by the host machine. Write
amplification occurs because various procedures (e.g.,
garbage collection~\cite{R3, R4}; and remapping-based refresh,
Section~\ref{sec:mitigation:refresh}) in the SSD perform additional writes in the
background. For example, when garbage collection selects a
block to erase, the pages that are remapped to a new block
require background writes. $R_{compress}$, or the compression
ratio, is the ratio between the size of the compressed data
and the size of the uncompressed data, and is a function of
the entropy of the stored data and the efficiency of the compression
algorithms employed in the SSD controller. In Equation~\ref{eq:E4},
DWPD and $R_{compress}$ are largely determined by the workload
and data compressibility, and cannot be changed to optimize
flash lifetime. For controllers that do not implement
compression, we set R compress to 1. However, the SSD controller
can trade off other parameters between one another
to optimize flash lifetime. We discuss the most salient tradeoffs
next.

\begin{figure}[h]
  \centering
  \includegraphics[width=0.65\columnwidth]{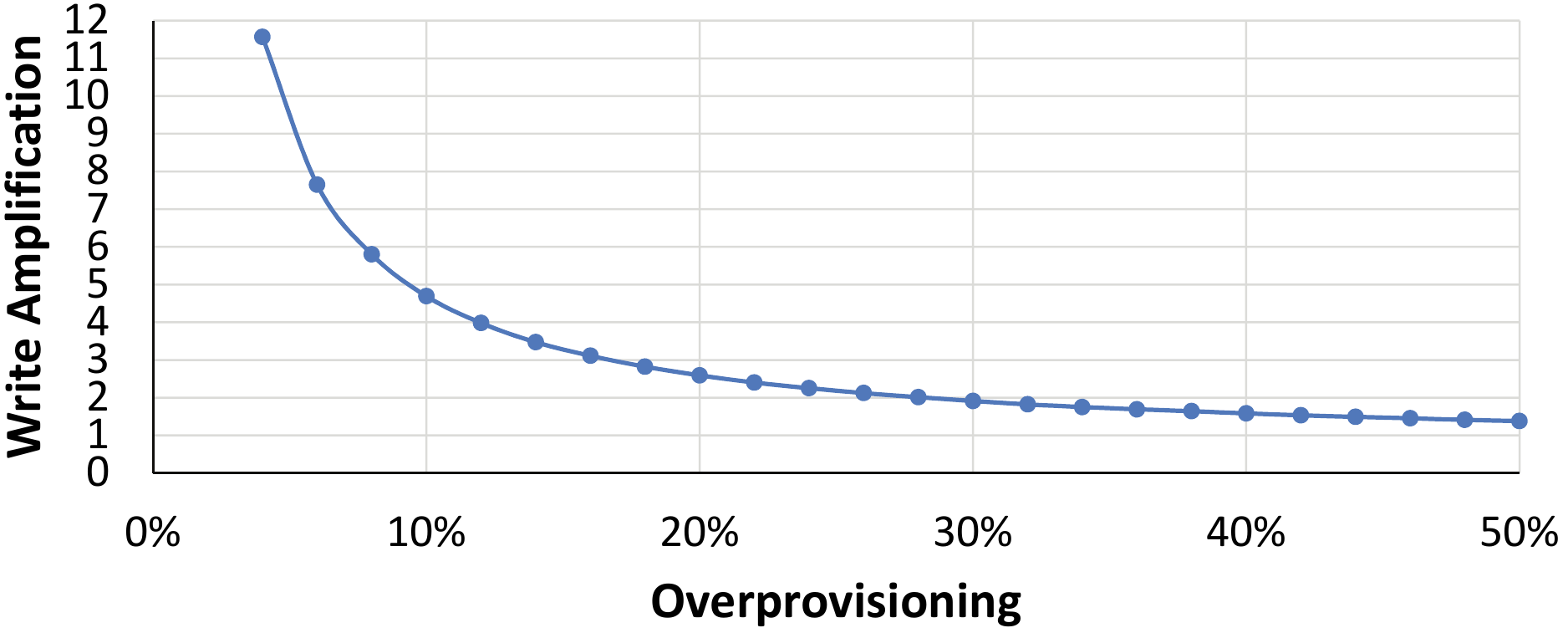}%
  \vspace{-5pt}%
  \caption{Relationship between write amplification (WA) and the
overprovisioning factor (OP). \chI{Reproduced from \cite{cai.arxiv17}.}}%
  \label{fig:F5}%
\end{figure}

\paratitle{Tradeoff Between Write Amplification and Overprovisioning}
As mentioned in \chIV{Section~\ref{sec:ssdarch:ctrl:gc}}, due to the
granularity mismatch between flash erase and program
operations, garbage collection occasionally remaps remaining
valid pages from a selected block to a new flash block,
in order to avoid block-internal fragmentation. This remapping
causes additional flash memory writes, leading to
\emph{write amplification}. In an SSD with more overprovisioned
capacity, the amount of write amplification decreases,
as the blocks selected for garbage collection are older
and tend to have fewer valid pages. For a greedy garbage collection
algorithm and a random-access workload, the correlation
between WA and OP can be calculated~\cite{R17, R18}, as
shown in Figure~\ref{fig:F5}. In an ideal SSD, both WA and OP should
be minimal, i.e., WA = 1 and OP = 0\%, but in reality there
is a tradeoff between these parameters: when one increases,
the other decreases. As Figure~\ref{fig:F5} shows, WA can be reduced by
increasing OP, and with an infinite amount of OP, WA converges
to 1. However, the reduction of WA is smaller when
OP is large, resulting in diminishing returns.

In reality, the relationship between WA and OP is also a
function of the storage space utilization of the SSD. When the
storage space is \emph{not} fully utilized, many more pages are available,
reducing the need to invoke garbage collection, and thus
WA can approach 1 without the need for a large amount of OP.

\paratitle{Tradeoff Between P/E Cycle Endurance and Overprovisioning}
PEC and OP can be traded against each
other by adjusting the amount of redundancy used for error
correction, such as ECC and superpage-level parity (as discussed
in \chIV{Section~\ref{sec:ssdarch:ctrl:parity}}). As the error correction capability
increases, PEC increases because the SSD can tolerate the
higher raw bit error rate that occurs at a higher P/E cycle
count. However, this comes at a cost of reducing the amount
of space available for OP, since a stronger error correction
capability requires higher redundancy (i.e., more space).
Table~\ref{tbl:T1} shows the corresponding OP for four different error
correction configurations for an example SSD with \SI{2.0}{\tera\byte}
of advertised capacity and \SI{2.4}{\tera\byte} (20\% extra) of physical
space. In this table, the top two configurations use ECC-1
with a coding rate of 0.93, and the bottom two configurations
use ECC-2 with a coding rate of 0.90, which has higher
redundancy than ECC-1. Thus, the ECC-2 configurations
have a lower OP than the top two. ECC-2, with its higher
redundancy, can correct a greater number of raw bit errors,
which in turn increases the P/E cycle endurance of the SSD.
Similarly, the two configurations with superpage-level parity
have a lower OP than configurations without superpage-level
parity, as parity uses a portion of the overprovisioned
space to store the parity bits.

\begin{table}[h]
\centering
\small
\setlength{\tabcolsep}{0.5em}
\caption{Tradeoff between strength of error correction configuration
and amount of SSD space left for overprovisioning.}
\label{tbl:T1}
\vspace{-5pt}%
\begin{tabular}{|c|c|}
\hline
\textbf{Error Correction Configuration} & \textbf{Overprovisioning Factor} \\ \hhline{|=|=|}
ECC-1 (0.93), no superpage-level parity & 11.6\% \\ \hline
ECC-1 (0.93), with superpage-level parity & 8.1\% \\ \hline
ECC-2 (0.90), no superpage-level parity & 8.0\% \\ \hline
ECC-2 (0.90), with superpage-level parity & 4.6\% \\ \hline
\end{tabular}
\end{table}

When the ECC correction strength is increased, the
amount of overprovisioning in the SSD decreases, which
in turn increases the amount of write amplification that
takes place. Manufacturers must find and use the correct
tradeoff between ECC correction strength and the overprovisioning
factor, based on which of the two is expected
to provide greater reliability for the target applications of
the SSD.


\section{NAND Flash Memory Basics}
\label{sec:flash}

A number of underlying properties of the NAND flash
memory used within the SSD affect SSD management,
performance, and reliability~\cite{R20, R22, R24}. In this section,
we present a primer on NAND flash memory and its
operation, to prepare the reader for understanding our
further discussion on error sources (Section~\ref{sec:errors}) and mitigation
mechanisms (Section~\ref{sec:mitigation}). Recall from Section~\ref{sec:ssdarch:flash}
that within each plane, flash cells are organized as multiple
2D arrays known as flash blocks, each of which
contains multiple pages of data, where a page is the granularity
at which the host reads and writes data. We first
discuss how data is stored in NAND flash memory. We
then introduce the three basic operations supported by
NAND flash memory: read, program, and erase.

\subsection{Storing Data in a Flash Cell}
\label{sec:flash:data}

\begin{figure}[h]
  \centering
  \includegraphics[width=0.4\columnwidth]{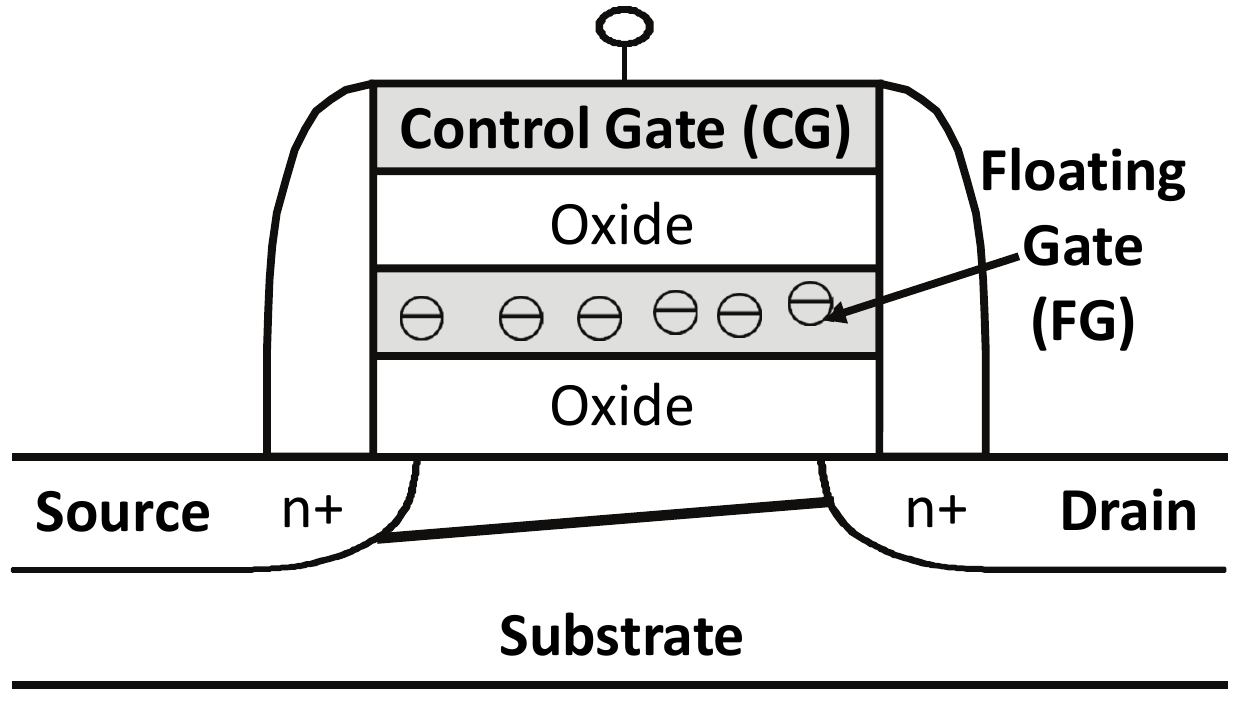}%
  \vspace{-3pt}%
  \caption{Flash cell (i.e., floating-gate transistor) cross section. \chI{Reproduced from \cite{cai.arxiv17}.}}%
  \label{fig:F6}%
\end{figure}

NAND flash memory stores data as the \emph{threshold voltage}
of each flash cell, which is made up of a \emph{floating-gate
transistor}. Figure~\ref{fig:F6} shows a cross section of a floating-gate
transistor. On top of a flash cell is the \emph{control gate} (CG) and
below is the floating gate (FG). The floating gate is insulated
on both sides, on top by an inter-poly oxide layer and at the
bottom by a tunnel oxide layer. As a result, the electrons
programmed on the floating gate do not discharge even
when flash memory is powered off.

For \emph{single-level cell} (SLC) NAND flash, each flash cell
stores a 1-bit value, and can be programmed to one of two
threshold voltage states, which we call the ER and P1 states.
\emph{Multi-level cell} (MLC) NAND flash stores a 2-bit value in each
cell, with four possible states (ER, P1, P2, and P3), and \emph{triple-level
cell} (TLC) NAND flash stores a 3-bit value in each cell
with eight possible states (ER, P1--P7). Each state represents
a different value, and is assigned a \emph{voltage window} within
the range of all possible threshold voltages. Due to variation
across \emph{program} operations, the threshold voltage of flash cells
programmed to the same state is initially distributed across
this voltage window.

\begin{figure}[h]
  \centering
  \includegraphics[width=0.75\columnwidth]{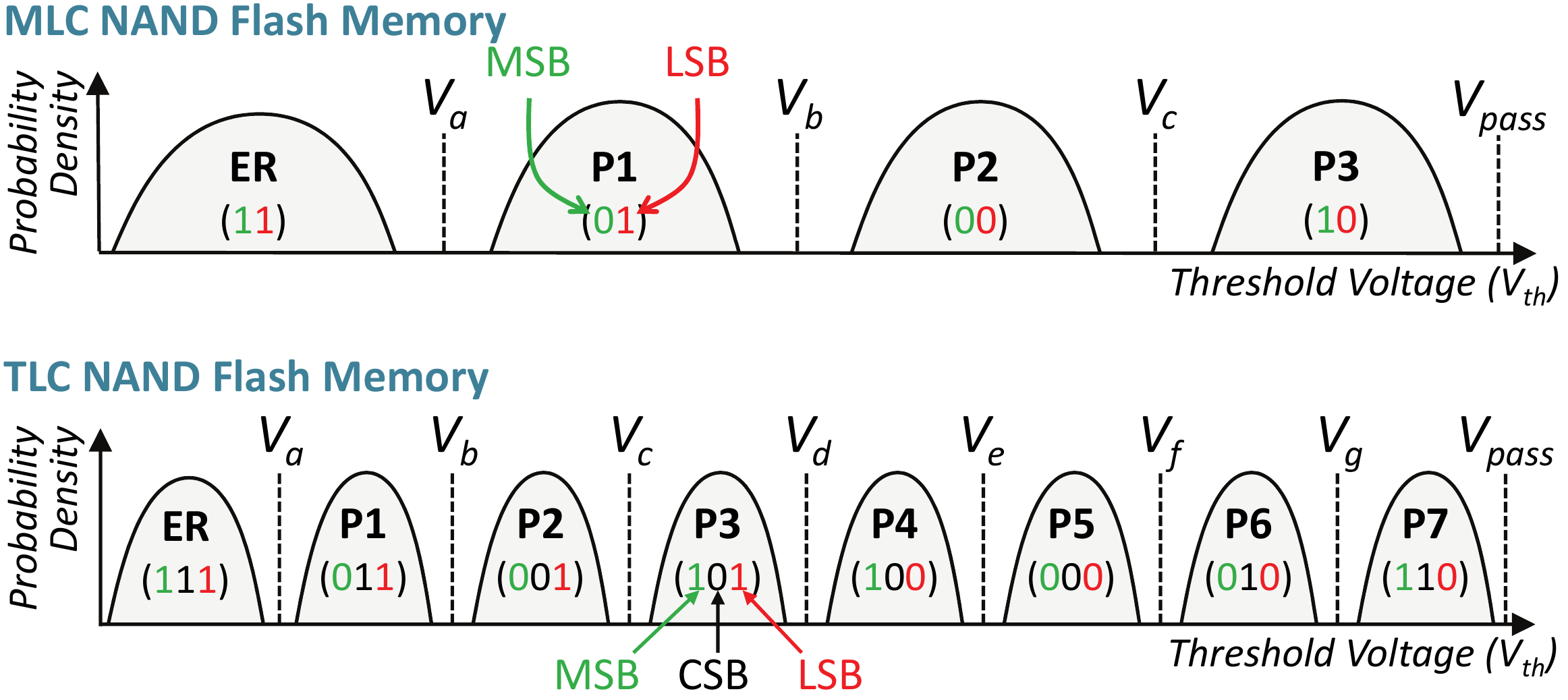}%
  \vspace{-4pt}%
  \caption{Threshold voltage distribution of MLC (top) and TLC (bottom)
NAND flash memory. \chI{Reproduced from \cite{cai.arxiv17}.}}%
  \label{fig:F7}%
\end{figure}

Figure~\ref{fig:F7} illustrates the threshold voltage distribution of
MLC (top) and TLC (bottom) NAND flash memories. The
x-axis shows the threshold voltage ($V_{th}$), which spans a certain
voltage range. The y-axis shows the probability density
of each voltage level across all flash memory cells. The
threshold voltage distribution of each threshold voltage
state can be represented as a probability density curve that
spans over the state's voltage window.

We label the distribution curve for each state with the
name of the state and a corresponding bit value. Note that
some manufacturers may choose to use a different mapping
of values to different states. The bit values of adjacent
states are separated by a Hamming distance of 1. We break
down the bit values for MLC into the most significant bit
(MSB) and least significant bit (LSB), while TLC is broken
down into the MSB, the center significant bit (CSB), and
the LSB. The boundaries between neighboring threshold
voltage windows, which are labeled as $V_a$, $V_b$, and $V_c$ for the
MLC distribution in Figure~\ref{fig:F7}, are referred to as \emph{read reference
voltages}. These voltages are used by the SSD controller to
identify the voltage window (i.e., state) of each cell upon
reading the cell.

\subsection{Flash Block Design}
\label{sec:flash:block}

\begin{figure}[h]
  \centering
  \includegraphics[width=0.5\columnwidth]{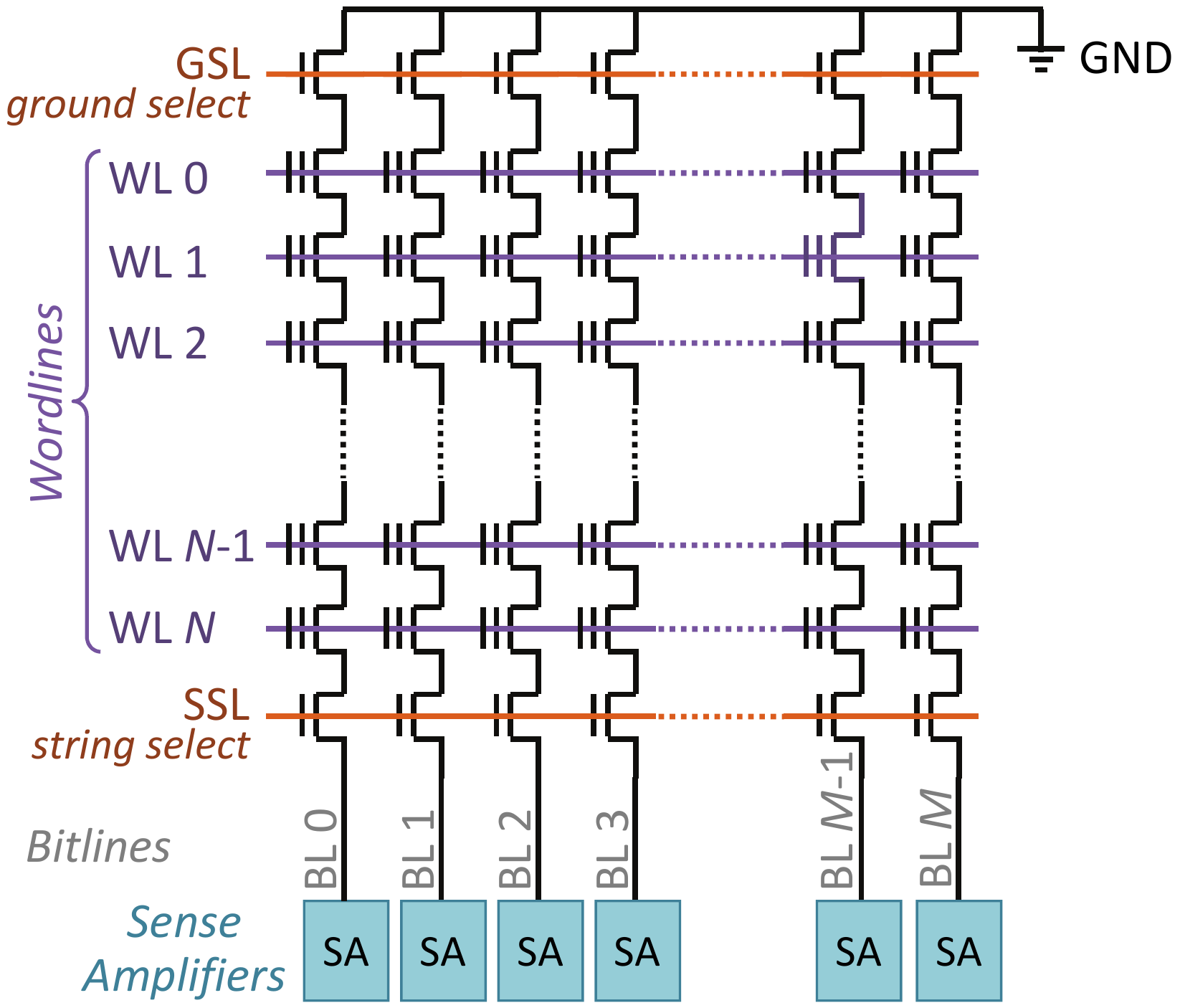}%
  \vspace{-3pt}%
  \caption{Internal organization of a flash block. \chI{Reproduced from \cite{cai.arxiv17}.}}%
  \label{fig:F8}%
\end{figure}

Figure~\ref{fig:F8} shows the high-level internal organization of a
NAND flash memory block. Each block contains multiple
rows of cells (typically 128--512 rows). Each row of cells is
connected together by a common \emph{wordline} (WL, shown horizontally
in Figure~\ref{fig:F8}), typically spanning 32K--64K cells. All of
the cells along the wordline are logically combined to form
a page in an SLC NAND flash memory. For an MLC NAND
flash memory, the MSBs of all cells on the same wordline are
combined to form an \emph{MSB page}, and the LSBs of all cells on
the wordline are combined to form an \emph{LSB page}. Similarly,
a TLC NAND flash memory logically combines the MSBs
on each wordline to form an MSB page, the CSBs on each
wordline to form a \emph{CSB page}, and the LSBs on each wordline
to form an LSB page. In MLC NAND flash memory, each
flash block contains 256--1024 flash pages, each of which
are typically \SIrange{8}{16}{\kilo\byte} in size.

Within a block, all cells in the same column are connected
in series to form a \emph{bitline} (BL, shown vertically in
Figure~\ref{fig:F8}) or \emph{string}. All cells in a bitline share a common ground
(GND) on one end, and a common \emph{sense amplifier} (SA) on
the other for reading the threshold voltage of one of the cells
when decoding data. Bitline operations are controlled by
turning the \emph{ground select line} (GSL) and \emph{string select line}
(SSL) transistor of each bitline on or off. The SSL transistor
is used to enable operations on a bitline, and the GSL
transistor is used to connect the bitline to ground during a
read operation~\cite{R103}. The use of a common bitline across
multiple rows reduces the amount of circuit area required
for read and write operations to a block, improving storage
density.

\subsection{Read Operation}
\label{sec:flash:read}

Data can be read from NAND flash memory by applying
read reference voltages onto the control gate of each cell, to
sense the cell's threshold voltage. To read the value stored
in a single-level cell, we need to distinguish only the state
with a bit value of 1 from the state with a bit value of 0.
This requires us to use only a single read reference voltage.
Likewise, to read the LSB of a multi-level cell, we need to
distinguish only the states where the LSB value is 1 (ER and
P1) from the states where the LSB value is 0 (P2 and P3),
which we can do with a single read reference voltage ($V_b$ in
the top half of Figure~\ref{fig:F7}). To read the MSB page, we need to distinguish
the states with an MSB value of 1 (ER and P3) from
those with an MSB value of 0 (P1 and P2). Therefore, we
need to determine whether the threshold voltage of the cell
falls between $V_a$ and $V_c$, requiring us to apply each of these
two read reference voltages (which can require up to two
consecutive read operations) to determine the MSB.

Reading data from a triple-level cell is similar to the data
read procedure for a multi-level cell. Reading the LSB for TLC
again requires applying only a single read reference voltage
($V_d$ in the bottom half of Figure~\ref{fig:F7}). Reading the CSB requires two
read reference voltages to be applied, and reading the MSB
requires four read reference voltages to be applied.

\begin{figure}[h]
  \centering
  \includegraphics[width=0.65\columnwidth]{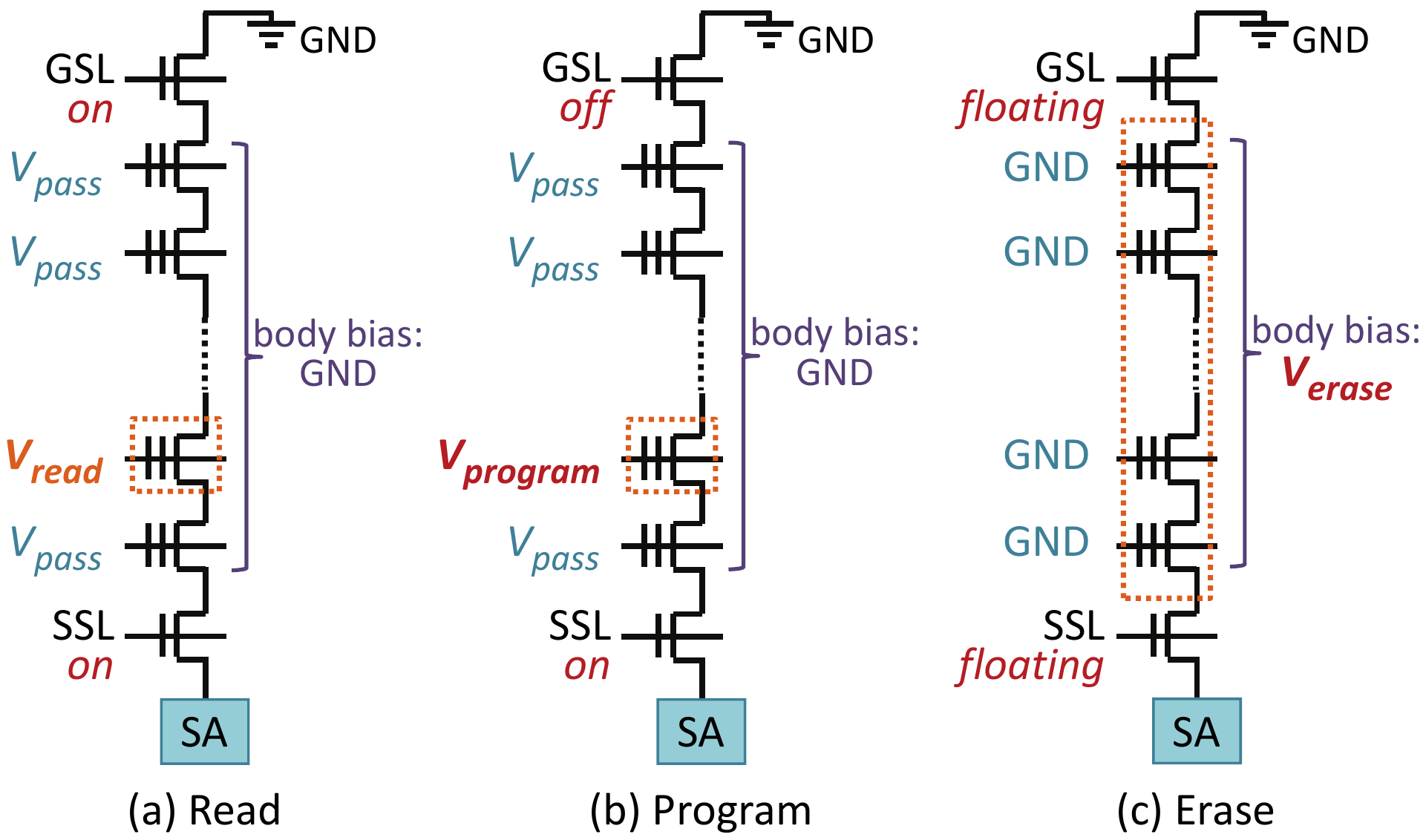}%
  \vspace{-3pt}%
  \caption{Voltages applied to flash cell transistors on a bitline to
perform (a)~read, (b)~program, and (c)~erase operations. \chI{Reproduced from \cite{cai.arxiv17}.}}%
  \label{fig:F9}%
\end{figure}

As Figure~\ref{fig:F8} shows, cells from multiple wordlines (WL in the
figure) are connected in series on a \emph{shared} bitline (BL) to the
sense amplifier, which drives the value that is being read from
the block onto the memory channel for the plane. In order to
read from a single cell on the bitline, \emph{all of the other cells} (i.e.,
\emph{unread} cells) on the same bitline must be switched on to allow
the value that is being read to propagate through to the sense
amplifier. The NAND flash memory achieves this by applying
the \emph{pass-through voltage} onto the wordlines of the unread cells,
as shown in Figure~\ref{fig:F9}a. When the pass-through voltage (i.e., the
maximum possible threshold voltage $V_{pass}$) is applied to a flash
cell, the source and the drain of the cell transistor are connected,
regardless of the voltage of the floating gate. Modern
flash memories guarantee that all \emph{unread} cells are \emph{passed through}
to minimize errors during the read operation~\cite{R38}.

\subsection{Program and Erase Operations}
\label{sec:flash:pgmerase}

The threshold voltage of a floating-gate transistor is controlled
through the injection and ejection of electrons through
the tunnel oxide of the transistor, which is enabled by the
Fowler–Nordheim (FN) tunneling effect~\cite{R21, R24, R28}. The
tunneling current ($J_{FN}$)~\cite{R22, R28} can be modeled as:
\begin{equation}
J_{FN} = \alpha_{FN} E_{ox}^2 e^{-\beta_{FN} / E_{ox}}
\label{eq:E6}
\end{equation}
In Equation~\ref{eq:E6}, $\alpha_{FN}$ and $\beta_{FN}$ are constants, and $E_{ox}$ is the electric field
strength in the tunnel oxide. As Equation~\ref{eq:E6} shows, $J_{FN}$ is exponentially
correlated with $E_{ox}$.

During a program operation, electrons are injected into
the floating gate of the flash cell from the substrate when
applying a high positive voltage to the control gate (see Figure~\ref{fig:F6}
for a diagram of the flash cell). The pass-through voltage is
applied to all of the other cells on the same bitline as the
cell that is being programmed as shown in Figure~\ref{fig:F9}b. When
data is programmed, charge is transferred into the floating
gate through FN tunneling by repeatedly pulsing the programming
voltage, in a procedure known as \emph{incremental
step-pulse programming} (ISPP)~\cite{R20, R23, R24, R25}. During
ISPP, a high programming voltage ($V_{program}$) is applied for
a very short period, which we refer to as a \emph{step-pulse}. ISPP
then verifies the current voltage of the cell using the voltage
$V_{verify}$. ISPP repeats the process of applying a step-pulse and
verifying the voltage until the cell reaches the desired target
voltage. In the modern all-bitline NAND flash memory,
all flash cells in a single wordline are programmed concurrently.
During programming, when a cell along the wordline
reaches its target voltage but other cells have yet to reach
their target voltage, ISPP \emph{inhibits} programming pulses to
the cell by turning off the SSL transistor of the cell's bitline.

In SLC NAND flash and older MLC NAND flash, \emph{one-shot
programming} is used, where all of the ISPP step-pulses
required to program a cell are applied back to back until all
cells in the wordline are fully programmed. One-shot programming
does \emph{not} interleave the program operations to
a wordline with the program operations to another wordline.
In newer MLC NAND flash, the lack of interleaving
between program operations can introduce a significant
amount of cell-to-cell program interference on the cells of
immediately-adjacent wordlines (see Section~\ref{sec:errors:celltocell}).

\begin{figure}[h]
  \centering
  \includegraphics[width=0.55\columnwidth]{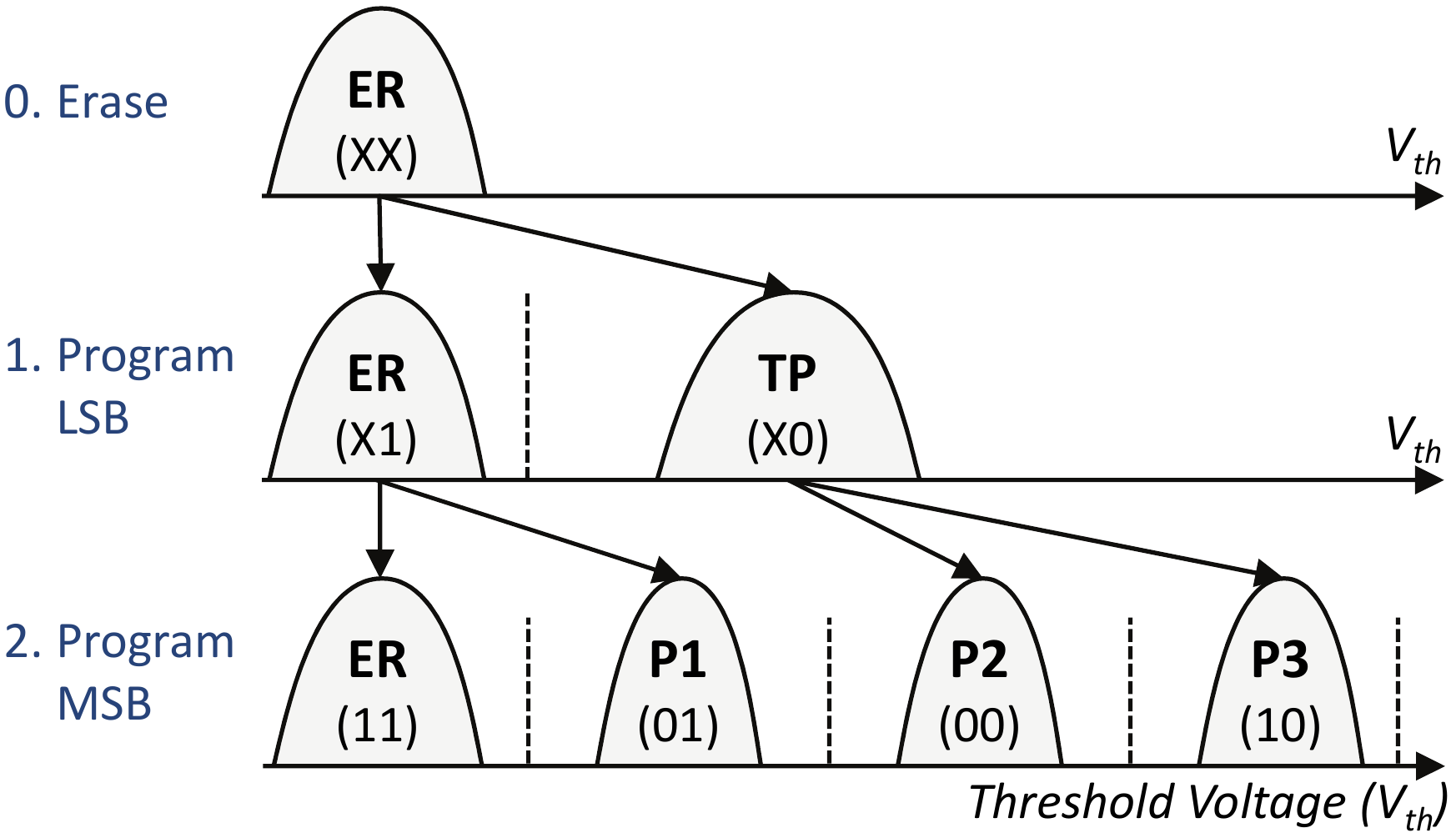}%
  \vspace{-5pt}%
  \caption{Two-step programming algorithm for MLC flash. \chI{Reproduced from \cite{cai.arxiv17}.}}%
  \label{fig:F10}%
\end{figure}

To reduce the impact of program interference, the controller
employs \emph{two-step programming} for sub-\SI{40}{\nano\meter} MLC
NAND flash~\cite{R26, R35}: it first programs the LSBs into the
erased cells of an unprogrammed wordline, and then programs
the MSBs of the cells using a separate program operation~\cite{R26, R27, R33, R40}. 
Between the programming of the
LSBs and the MSBs, the controller programs the LSBs of
the cells in the wordline immediately above~\cite{R26, R27, R33, R40}. 
Figure~\ref{fig:F10} illustrates the two-step programming algorithm.
In the first step, a flash cell is \emph{partially programmed}
based on its LSB value, either staying in the ER state if the
LSB value is 1, or moving to a temporary state (TP) if the LSB
value is 0. The TP state has a mean voltage that falls between
states P1 and P2. In the second step, the LSB data is first
read back into an internal buffer register within the flash
chip to determine the cell's current threshold voltage state,
and then further programming pulses are applied based on
the MSB data to increase the cell's threshold voltage to fall
within the voltage window of its final state. Programming
in MLC NAND flash is discussed in detail in \cite{R40} and \cite{R33}.

\begin{figure}[h]
  \centering
  \includegraphics[width=0.7\columnwidth]{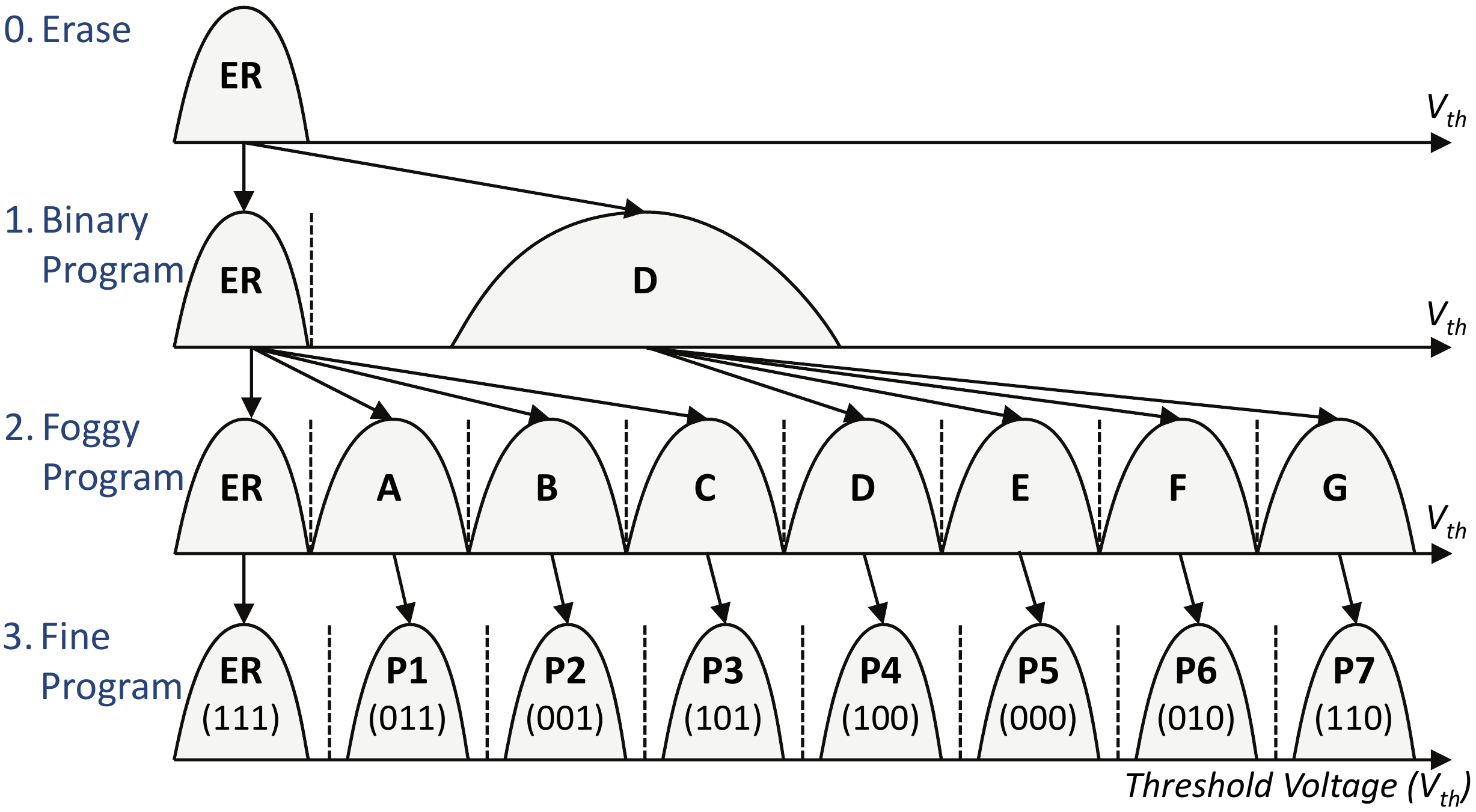}%
  \vspace{-5pt}%
  \caption{Foggy-fine programming algorithm for TLC flash. \chI{Reproduced from \cite{cai.arxiv17}.}}%
  \label{fig:F11}%
\end{figure}

TLC NAND flash takes a similar approach to the two-step
programming of MLC, with a mechanism known as
\emph{foggy-fine programming}~\cite{R19}, which is illustrated in Figure~\ref{fig:F11}.
The flash cell is first partially programmed based on its LSB
value, using a \emph{binary} programming step in which very large
ISPP step-pulses are used to significantly increase the voltage
level. Then, the flash cell is partially programmed again based
on its CSB and MSB values to a new set of temporary states
(these steps are referred to as \emph{foggy} programming, which uses
smaller ISPP step-pulses than binary programming). Due to
the higher potential for errors during TLC programming as a
result of the narrower voltage windows, all of the programmed
bit values are buffered after the binary and foggy programming
steps into SLC buffers that are reserved in each chip/plane. 
Finally, \emph{fine} programming takes place, where these bit
values are read from the SLC buffers, and the smallest ISPP
step-pulses are applied to set each cell to its final threshold
voltage state. The purpose of this last fine programming step
is to fine tune the threshold voltage such that the threshold
voltage distributions are tightened (bottom of Figure~\ref{fig:F11}).

Though programming sets a flash cell to a specific
threshold voltage using programming pulses, the voltage
of the cell can drift over time after programming. When no
external voltage is applied to any of the electrodes (i.e., CG,
source, and drain) of a flash cell, an electric field still exists
between the FG and the substrate, generated by the charge
present in the FG. This is called the \emph{intrinsic electric field}~\cite{R22}, 
and it generates \emph{stress-induced leakage current} (SILC)~\cite{R24, R29, R30}, 
a weak tunneling current that leaks charge
away from the FG. As a result, the voltage that a cell is programmed
to may not be the same as the voltage read for that
cell at a subsequent time.

In NAND flash, a cell can be reprogrammed with new
data \emph{only after} the existing data in the cell is erased. This is
because ISPP can only \emph{increase} the voltage of the cell. The
erase operation resets the threshold voltage state of \emph{all cells
in the flash block} to the ER state. During an erase operation,
electrons are ejected from the FG of the flash cell into
the substrate by inducing a high negative voltage on the cell
transistor. The negative voltage is induced by setting the CG
of the transistor to GND, and biasing the transistor body
(i.e., the substrate) to a high voltage ($V_{erase}$), as shown in
Figure~\ref{fig:F9}c. Because all cells in a flash block share a common
transistor substrate (i.e., the bodies of all transistors in the
block are connected together), a flash block must be erased
in its entirety~\cite{R103}.


\section{NAND Flash Error Characterization}
\label{sec:errors}

\begin{figure}[h]
  \centering
  \includegraphics[width=0.65\columnwidth]{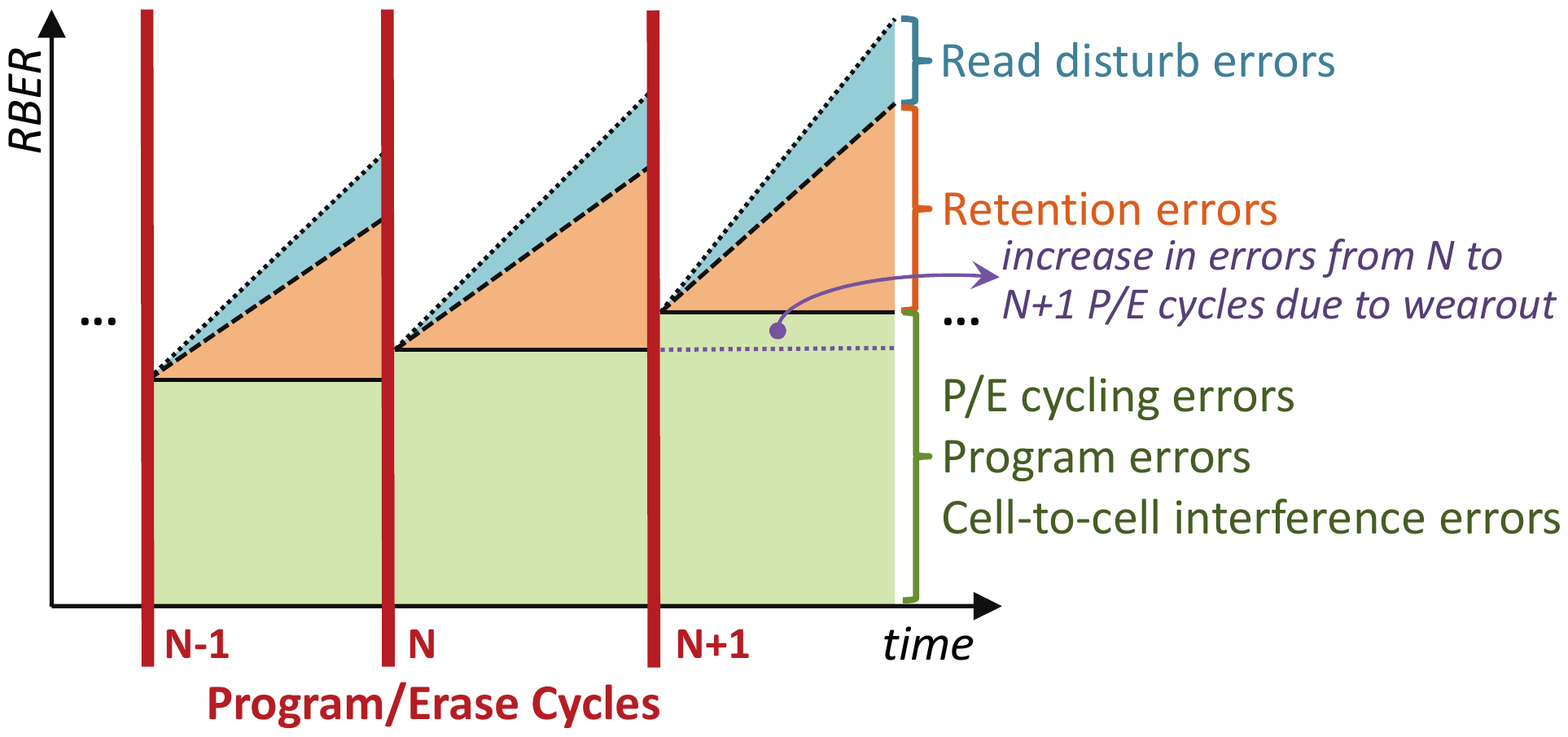}%
  \vspace{-5pt}%
  \caption{Pictorial depiction of errors accumulating within a NAND
flash block as P/E cycle count increases. \chI{Reproduced from \cite{cai.arxiv17}.}}%
  \label{fig:F12}%
\end{figure}

Each block in NAND flash memory is used in a cyclic fashion,
as is illustrated by the observed raw bit error rates seen
over the lifetime of a flash memory block in Figure~\ref{fig:F12}. At the
beginning of a \emph{cycle}, known as a \emph{program/erase (P/E) cycle},
an erased block is \emph{opened} (i.e., selected for programming).
Data is then programmed into the open block one page at
a time. After all of the pages are programmed, the block is
closed, and none of the pages can be reprogrammed until
the whole block is erased. At any point before erasing, read
operations can be performed on a \emph{valid} programmed page
(i.e., a page containing data that has not been modified
by the host). A page is marked as invalid when the data
stored at that page's logical address by the host is modified.
As ISPP can only inject more charge into the floating gate
but cannot remove charge from the gate, it is not possible
to modify data to a new arbitrary value \emph{in place} within
existing NAND flash memories. Once the block is erased,
the P/E cycling behavior repeats until the block is \emph{worn out}
(i.e., the block can no longer avoid data loss over the course
of the minimum data retention period guaranteed by the
manufacturer). Although the 5x-nm (i.e., \SIrange{50}{59}{\nano\meter})
generation of MLC NAND flash could endure \textasciitilde 10,000 P/E
cycles per block before being worn out, modern 1x-nm
(i.e., \SIrange{15}{19}{\nano\meter}) MLC and TLC NAND flash can endure
only \textasciitilde 3,000 and \textasciitilde 1,000 P/E cycles per block, 
respectively~\cite{R53, R60, R81, R86}.

\begin{figure}[h]
  \centering
  \includegraphics[width=0.75\columnwidth]{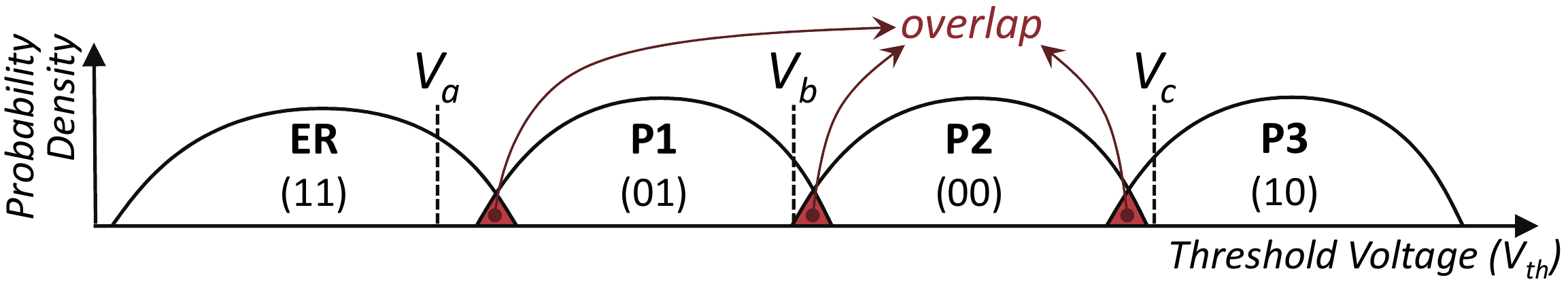}%
  \vspace{-5pt}%
  \caption{Threshold voltage distribution shifts and widening can
cause the distributions of two neighboring states to overlap with
each other (compare to Figure~\ref{fig:F7}), leading to read errors. \chI{Reproduced from \cite{cai.arxiv17}.}}%
  \label{fig:F13}%
\end{figure}

As shown in Figure~\ref{fig:F12}, several different types of errors can
be introduced at any point during the P/E cycling process:
\emph{P/E cycling errors}, \emph{program errors}, errors due to \emph{cell-to-cell program
interference}, \emph{data retention errors}, and errors due to \emph{read
disturb}. As discussed in Section~\ref{sec:flash:data}, the threshold voltage
of flash cells programmed to the same state is distributed
across a voltage window due to variation across program
operations and across different flash cells. Several types of
errors introduced during the P/E cycling process, such as
data retention and read disturb, cause the threshold voltage
distribution of each state to shift and widen. Due to the shift
and widening, the tails of the distributions of each state can
enter the margin that originally existed between each of the
two neighboring states' distributions. Thus, the threshold
voltage distributions of different states can start overlapping,
as shown in Figure~\ref{fig:F13}. When the distributions overlap
with each other, the read reference voltages can no longer
correctly identify the state of some flash cells in the overlapping
region, leading to \emph{raw bit errors} during a read operation.

In this section, we discuss the causes of each type of error
in detail, and characterize the impact that each error type
has on the amount of raw bit errors occurring within NAND
flash memory. We use an FPGA-based testing platform~\cite{R31}
to characterize state-of-the-art TLC NAND flash chips. We
use the read-retry operation present in NAND flash devices
to accurately read the cell threshold voltage~\cite{R33, R34, R35, R36, R37, R38, R42, R52, R107} 
(for a detailed description of the read-retry operation,
see Section~\ref{sec:mitigation:retry}). As absolute threshold voltage values
are proprietary information to flash vendors, we present our
results using normalized voltages, where the nominal maximum
value of $V_{th}$ is equal to 512 in our normalized scale,
and where 0 represents GND. We also describe characterization
results and observations for MLC NAND flash chips.
These MLC NAND results are taken from our prior works~\cite{R32, R33, R34, R35, R36, R37, R38, R39, R40, R42, R177}, 
which provide more detailed error characterization
results and analyses. To our knowledge, this paper
provides the first experimental characterization and analysis
of errors in real \emph{TLC} NAND flash memory chips.

We later discuss mitigation techniques for these flash
memory errors in Section~\ref{sec:mitigation}, and provide procedures to
recover in the event of data loss in Section~\ref{sec:correction}.

\subsection{P/E Cycling Errors}
\label{sec:errors:pe}

A P/E cycling error occurs when either (1)~an erase operation
fails to reset a cell to the ER state; or (2)~when a program
operation fails to set the cell to the desired target state.
P/E cycling errors occur because electrons become trapped
in the tunnel oxide after stress from repeated P/E cycles.
Errors due to such electron trapping (which we refer to as
\emph{P/E cycling noise}) continue to accumulate over the lifetime
of a NAND flash block. This behavior is called \emph{wearout},
and it refers to the phenomenon where, as more writes are
performed to a block, there are a greater number of raw bit
errors that must be corrected, exhausting more of the fixed
error correction capability of the ECC (see \chIV{Section~\ref{sec:ssdarch:ctrl:ecc}}).

\begin{figure}[h]
  \centering
  \includegraphics[width=0.7\columnwidth]{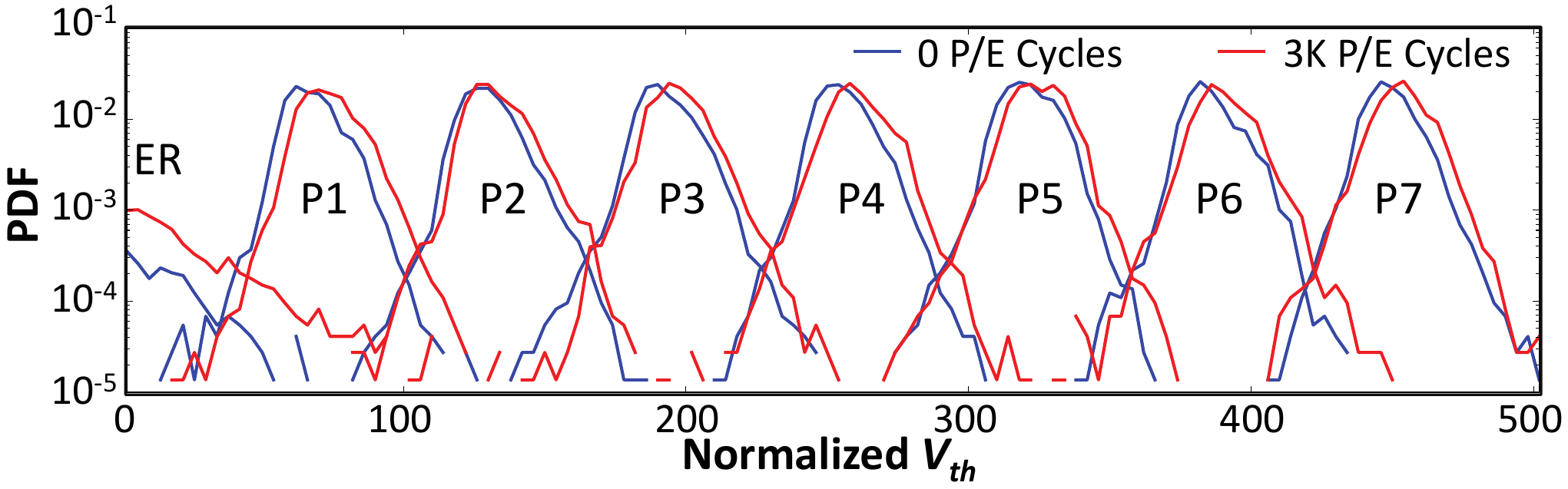}%
  \vspace{-5pt}%
  \caption{Threshold voltage distribution of TLC NAND flash memory
after 0 P/E cycles and 3,000 P/E cycles. \chI{Reproduced from \cite{cai.arxiv17}.}}%
  \label{fig:F14}%
\end{figure}

Figure~\ref{fig:F14} shows the threshold voltage distribution of TLC
NAND flash memory after 0 P/E cycles and after 3,000 P/E
cycles, without any retention or read disturb errors present
(which we ensure by reading the data \emph{immediately} after
programming). The mean and standard deviation of each
state's distribution are provided in Table~\ref{tbl:T4} in the Appendix
(for other P/E cycle counts as well). We make two observations
from the two distributions. First, as the P/E cycle
count increases, each state's threshold voltage distribution
systematically (1)~shifts to the right and (2)~becomes wider.
Second, the amount of the shift is greater for lower-voltage
states (e.g., the ER and P1 states) than it is for higher-voltage
states (e.g., the P7 state).

The threshold voltage distribution shift occurs because
as more P/E cycles take place, the quality of the tunnel
oxide degrades, allowing electrons to tunnel through the
oxide more easily~\cite{R58}. As a result, if the same ISPP conditions
(e.g., programming voltage, step-pulse size, program
time) are applied throughout the lifetime of the NAND flash
memory, more electrons are injected during programming
as a flash memory block wears out, leading to higher threshold
voltages, i.e., the right shift of the distribution. The distribution
of each state widens due to the process variation
present in (1)~the wearout process, and (2)~the cell's structural
characteristics. As the distribution of each voltage state
widens, more overlap occurs between neighboring distributions,
making it less likely for a read reference voltage to
determine the correct value of the cells in the overlapping
regions, which leads to a greater number of raw bit errors.

The threshold voltage distribution trends we observe here
for TLC NAND flash memory trends are similar to trends
observed previously for MLC NAND flash memory~\cite{R32, R33, R42, R53}, 
although the MLC NAND flash characterizations
reported in past studies span up to a larger P/E cycle count than
the TLC experiments due to the greater endurance of MLC
NAND flash memory. More findings on the nature of wearout
and the impact of wearout on NAND flash memory errors and
lifetime can be found in our prior work~\cite{R32, R33, R42, R177}.

\subsection{Program Errors}
\label{sec:errors:pgm}

\begin{figure}[h]
  \centering
  \includegraphics[width=0.65\columnwidth]{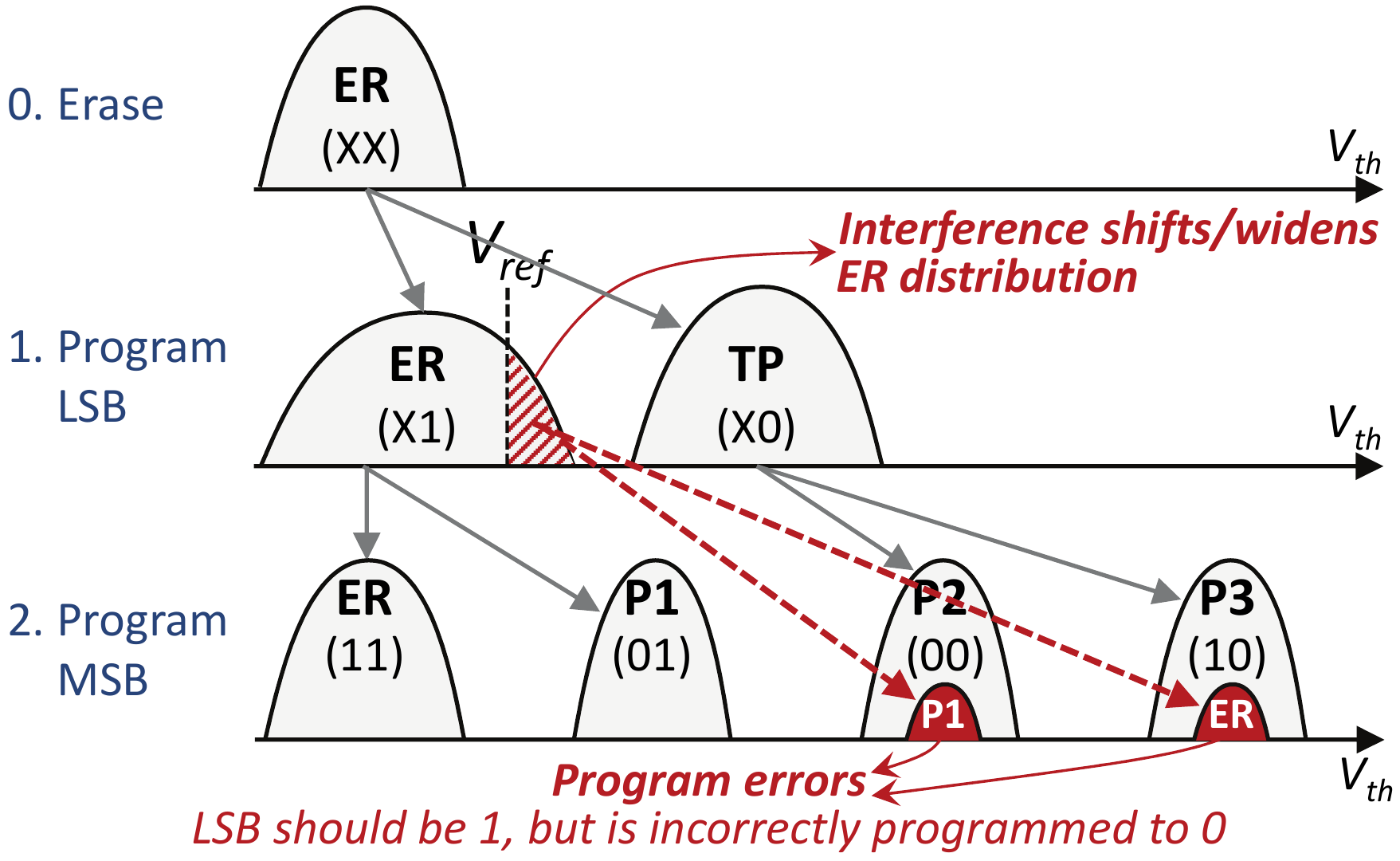}%
  \vspace{-5pt}%
  \caption{Impact of program errors during two-step programming on
cell threshold voltage distribution. \chI{Reproduced from \cite{cai.arxiv17}.}}%
  \label{fig:F15}%
\end{figure}

Program errors occur when data read directly from the
NAND flash array contains errors, and the erroneous values
are used to program the new data. Program errors occur in two
major cases: (1)~partial programming during two-step or foggy-fine
programming, and (2)~\emph{copyback} (i.e., when data is copied
inside the NAND flash memory during a maintenance operation)~\cite{R109}. 
During two-step programming for MLC NAND
flash memory (see Figure~\ref{fig:F10}), in between the LSB and MSB programming
steps of a cell, threshold voltage shifts can occur
on the partially-programmed cell. These shifts occur because
several other read and program operations to cells in \emph{other}
pages within the same block may take place, causing interference
to the partially-programmed cell. Figure~\ref{fig:F15} illustrates
how the threshold distribution of the ER state widens and
shifts to the right after the LSB value is programmed (step~1
in the figure). The widening and shifting of the distribution
causes some cells that were originally partially programmed
to the ER state (with an LSB value of 1) to be misread as being
in the TP state (with an LSB value of 0) during the \emph{second}
programming step (step~2 in the figure). As shown in Figure~\ref{fig:F15},
the misread LSB value leads to a program error when the
final cell threshold voltage is programmed~\cite{R40, R42, R53}.
Some cells that should have been programmed to the P1 state
(representing the value 01) are instead programmed to the
P2 state (with the value 00), and some cells that should have
been programmed to the ER state (representing the value 11)
are instead programmed to the P3 state (with the value 10).

The incorrect values that are read before the second programming
step are \emph{not} corrected by ECC, as they are read
directly inside the NAND flash array, without involving the
controller (where the ECC engine resides). Similarly, during
foggy-fine programming for TLC NAND flash (see Figure~\ref{fig:F11}),
the data may be read incorrectly from the SLC buffers used to
store the contents of partially-programmed wordlines, leading
to errors during the fine programming step. Program errors
occur during \emph{copyback}~\cite{R109} when valid data is read out from
a block during maintenance operations (e.g., a block about to
be garbage collected) and reprogrammed into a new block, as
copyback operations do \emph{not} go through the SSD controller.

Program errors that occur during partial programming
predominantly shift data from lower-voltage states to higher-voltage
states. For example, in MLC NAND flash, program
errors predominantly shift data that should be in the ER state
(11) into the P3 state (10), or data that should be in the P1 state
(01) into the P2 state (00)~\cite{R40}. This occurs because MSB programming
can only \emph{increase} (and not reduce) the threshold
voltage of the cell from its partially-programmed voltage
(and thus cannot move a multi-level cell that should be
in the P3 state into the ER state, or one that should be in
the P2 state into the P1 state). TLC NAND flash is much
less susceptible to program errors than MLC NAND flash,
as the data read from the SLC buffers in TLC NAND flash
has a much lower error rate than data read from a partially-programmed
MLC NAND flash wordline~\cite{R202}.

\sloppypar
From a rigorous experimental characterization of modern
MLC NAND flash memory chips~\cite{R40}, we find that program
errors occur primarily due to two types of errors affecting the
partially-programmed data. First, cell-to-cell program interference
(Section~\ref{sec:errors:celltocell}) on a partially-programmed wordline is
no longer negligible in newer NAND flash memory compared
to older NAND flash memory, due to manufacturing process
scaling. As flash cells become smaller and are placed closer to
each other, cells in partially-programmed wordlines become
more susceptible to bit flips. Second, partially-programmed
cells are more susceptible to read disturb errors than fully-program\-med
cells (Section~\ref{sec:errors:readdisturb}), as the threshold voltages
stored in these cells are no more than approximately half of
$V_{pass}$~\cite{R40}, and cells with lower threshold voltages are more
likely to experience read disturb errors.

More findings on the nature of program errors and the
impact of program errors on NAND flash memory lifetime
can be found in our prior work~\cite{R40, R42}.

\subsection{Cell-to-Cell Program Interference Errors}
\label{sec:errors:celltocell}

Program interference refers to the phenomenon where the
programming of a flash cell induces errors on adjacent flash
cells within a flash block~\cite{R35, R36, R55, R61, R62}. The interference
occurs due to \emph{parasitic capacitance coupling} between
these cells. As a result, when the threshold voltage of an adjacent
flash cell increases, the threshold voltage of the \emph{victim
cell} increases as well. The unintended threshold voltage shifts
can eventually move a cell into a different state than the one it
was originally programmed to, leading to a bit error.

\begin{figure}[h]
  \centering
  \includegraphics[width=0.55\columnwidth]{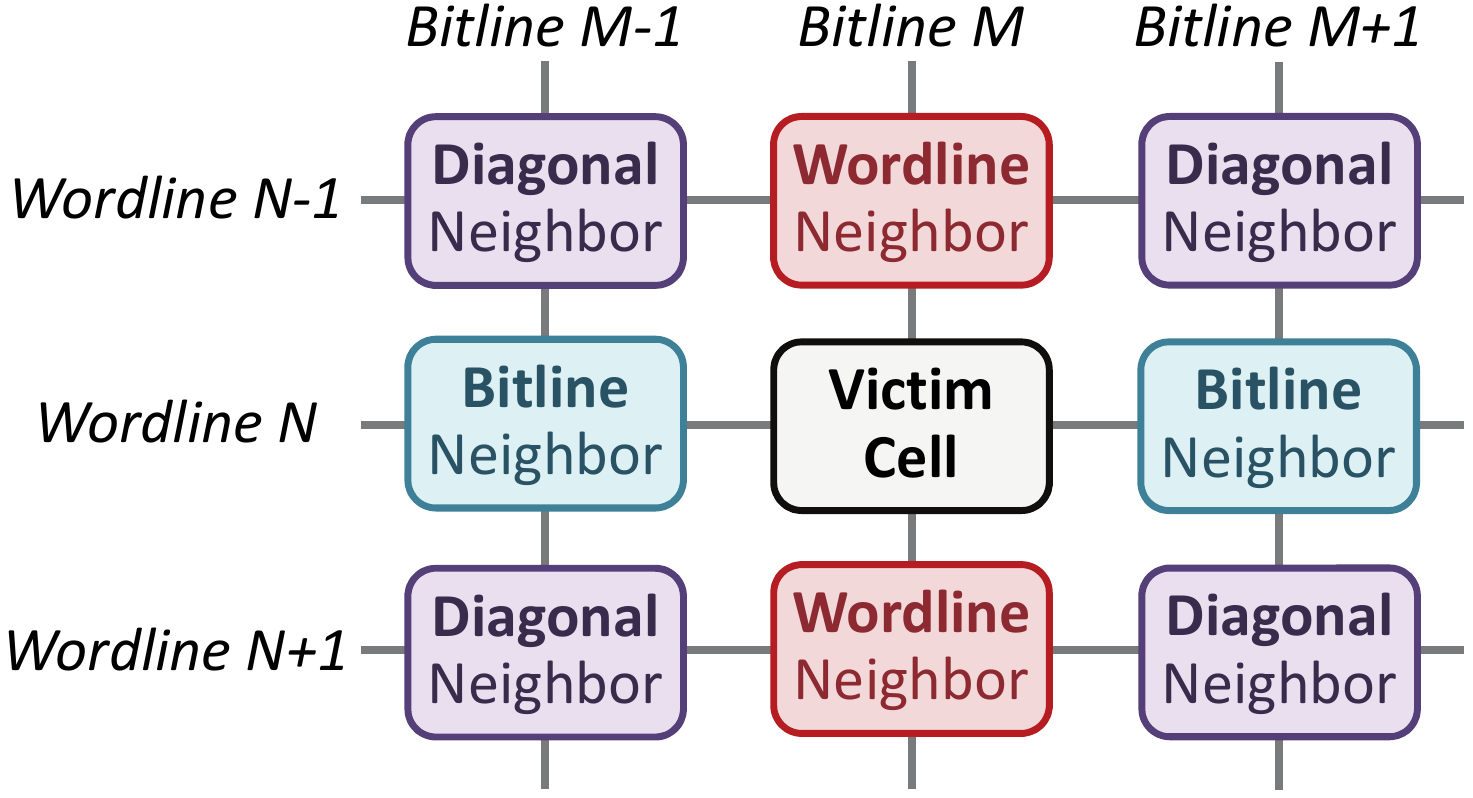}%
  \vspace{-5pt}%
  \caption{Immediately-adjacent cells that can induce program
interference on a victim cell that is on wordline~\emph{N} and bitline~\emph{M}. 
\chI{Reproduced from \cite{cai.arxiv17}.}}%
  \label{fig:F16}%
\end{figure}

We have shown, based on our experimental analysis of
modern MLC NAND flash memory chips, that the threshold
voltage change of the victim cell can be accurately modeled
as a linear combination of the threshold voltage changes of
the adjacent cells when they are programmed, using linear
regression with least-square-error estimation~\cite{R35, R36}.
The cells that are physically located immediately next to the
victim cell (called the \emph{immediately-adjacent cells}) are the
major contributors to the cell-to-cell interference of a victim
cell~\cite{R35}. Figure~\ref{fig:F16} shows the eight immediately-adjacent cells
for a victim cell in 2D planar NAND flash memory.

The amount of interference that program operations to
the immediately-adjacent cells can induce on the victim cell
is expressed as:
\begin{equation}
\Delta V_{victim} = \sum_{X} K_X \Delta V_X
\label{eq:E7}
\end{equation}
where $\Delta V_{victim}$ is the change in voltage of the victim cell
due to cell-to-cell program interference, $K_X$ is the \emph{coupling
coefficient} between cell $X$ and the victim cell, and $\Delta V_{X}$ is
the threshold voltage change of cell $X$ during programming.
Table~\ref{tbl:T2} lists the coupling coefficients for both 2y-nm and
1x-nm NAND flash memory. We make two key observations
from Table~\ref{tbl:T2}. First, we observe that the coupling coefficient
is greatest for wordline neighbors (i.e., immediately-adjacent
cells on the same bitline, but on a neighboring
wordline)~\cite{R35}. The coupling coefficient is directly related
to the effective capacitance $C$ between cell $X$ and the victim
cell, which can be calculated as:
\begin{equation}
C = \varepsilon S / d
\label{eq:E8}
\end{equation}
where $\varepsilon$ is the permittivity, $S$ is the effective cell area of cell
$X$ that faces the victim cell, and $d$ is the distance between the
cells. Of the immediately-adjacent cells, the wordline neighbor
cells have the greatest coupling capacitance with the victim
cell, as they likely have a large effective facing area to,
and a small distance from, the victim cell compared to other
surrounding cells. Second, we observe that the coupling
coefficient grows as the feature size decreases~\cite{R35, R36}.
As NAND flash memory process technology scales down
to smaller feature sizes, cells become smaller and get closer
to each other, which increases the effective capacitance
between them. As a result, at smaller feature sizes, it is easier
for an immediately-adjacent cell to induce program interference
on a victim cell. We conclude that (1)~the program interference
an immediately-adjacent cell induces on a victim cell
is primarily determined by the distance between the cells and
the immediately-adjacent cell's effective area facing the victim
cell; and (2)~the wordline neighbor cell causes the highest
such interference, based on empirical measurements.

\begin{table}[h]
\centering
\small
\setlength{\tabcolsep}{1.45em}
\caption{Coupling coefficients for immediately-adjacent cells.}
\label{tbl:T2}
\vspace{-5pt}%
\begin{tabular}{|c||c|c|c|}
\hline
\textbf{Process} & \textbf{Wordline} & \textbf{Bitline} & \textbf{Diagonal} \\
\textbf{Technology} & \textbf{Neighbor} & \textbf{Neighbor} & \textbf{Neighbor} \\ \hhline{|=#=|=|=|}
\textbf{2y-nm} & 0.060 & 0.032 & 0.012 \\ \hline
\textbf{1x-nm} & 0.110 & 0.055 & 0.020 \\ \hline
\end{tabular}
\end{table}

Due to the order of program operations performed in
NAND flash memory, many immediately-adjacent cells do
\emph{not} end up inducing interference after a victim cell is fully
programmed (i.e., once the victim cell is at its target voltage).
In modern all-bitline NAND flash memory, all flash cells on
the same wordline are programmed at the same time, and
wordlines are fully programmed sequentially (i.e., the cells
on wordline $i$ are fully programmed before the cells on wordline
$i+1$). As a result, an immediately-adjacent cell on the
wordline below the victim cell or on the same wordline as the
victim cell does \emph{not} induce program interference on a fully-programmed
victim cell. Therefore, the major source of program
interference on a fully-programmed victim cell is the
programming of the wordline immediately above it.

\begin{figure}[h]
  \centering
  \includegraphics[width=0.7\columnwidth]{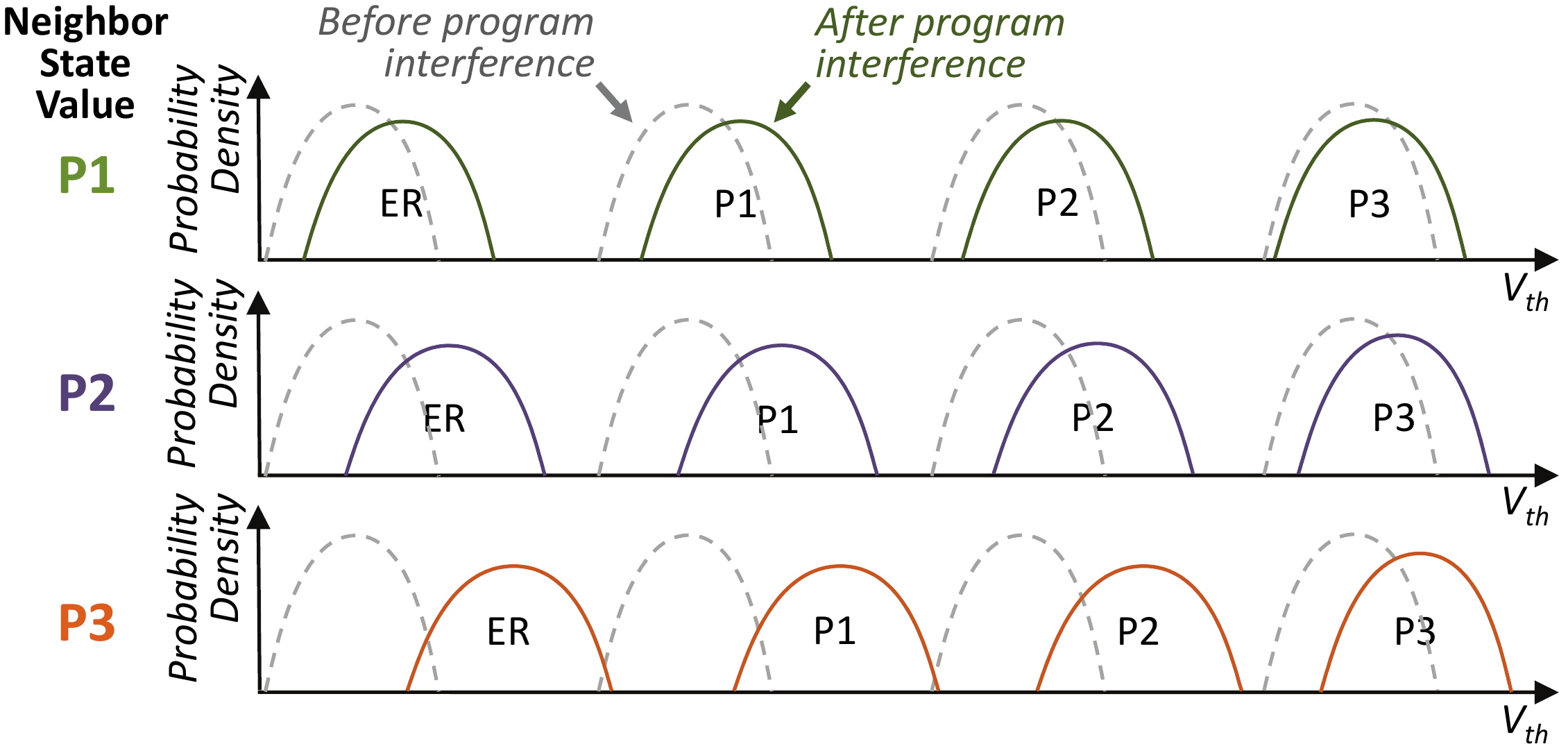}%
  \vspace{-5pt}%
  \caption{Impact of cell-to-cell program interference on a victim
cell during one-shot programming, depending on the value its
neighboring cell is programmed to. \chI{Reproduced from \cite{cai.arxiv17}.}}%
  \label{fig:F17}%
\end{figure}

Figure~\ref{fig:F17} shows how the threshold voltage distribution of
a victim cell shifts when different values are programmed
onto its immediately-adjacent cells in the wordline above
the victim cell for MLC NAND flash, when one-shot programming
is used. The amount by which the victim cell
distribution shifts is directly correlated with the number
of programming step-pulses applied to the immediately-adjacent
cell. That is, when an immediately-adjacent cell
is programmed to a higher-voltage state (which requires
more step-pulses for programming), the victim cell distribution
shifts further to the right~\cite{R35}. When an immediately-adjacent
cell is set to the ER state, no step-pulses are applied,
as an unprogrammed cell is already in the ER state. Thus, no
interference takes place. Note that the amount by which a
fully-programmed victim cell distribution shifts is different
when two-step programming is used, as a fully-programmed
cell experiences interference from only one of the two programming
steps of a neighboring wordline~\cite{R40}.

More findings on the nature of cell-to-cell program
interference and the impact of cell-to-cell program interference
on NAND flash memory errors and lifetime can be
found in our prior work~\cite{R35, R36, R40, R177}.

\subsection{Data Retention Errors}
\label{sec:errors:retention}

Retention errors are caused by charge leakage over time
after a flash cell is programmed, and are the dominant source
of flash memory errors, as demonstrated previously~\cite{R20, R32, R34, R37, R39, R56}. 
As flash memory process technology
scales to smaller feature sizes, the capacitance of a flash
cell, and the number of electrons stored on it, decreases.
State-of-the-art (i.e., 1x-nm) MLC flash memory cells can
store only \textasciitilde 100 electrons~\cite{R81}. Gaining or losing several electrons
on a cell can significantly change the cell's voltage level
and eventually alter its state. Charge leakage is caused by the
unavoidable trapping of charge in the tunnel oxide~\cite{R37, R57}.
The amount of trapped charge increases with the electrical
stress induced by repeated program and erase operations,
which degrade the insulating property of the oxide.

Two failure mechanisms of the tunnel oxide lead to retention
loss. \emph{Trap-assisted tunneling} (TAT) occurs because the
trapped charge forms an electrical tunnel, which exacerbates
the weak tunneling current, SILC (see Section~\ref{sec:flash:pgmerase}).
As a result of this TAT effect, the electrons present in the
floating gate (FG) leak away much faster through the intrinsic
electric field. Hence, the threshold voltage of the flash
cell decreases over time. As the flash cell wears out with
increasing P/E cycles, the amount of trapped charge also
increases~\cite{R37, R57}, and so does the TAT effect. At high P/E
cycles, the amount of trapped charge is large enough to form
percolation paths that significantly hamper the insulating
properties of the gate dielectric~\cite{R30, R37}, resulting in retention
failure. \emph{Charge detrapping}, where charge previously
trapped in the tunnel oxide is freed spontaneously, can also
occur over time~\cite{R30, R37, R57, R59}. The charge polarity can
be either negative (i.e., electrons) or positive (i.e., holes).
Hence, charge detrapping can either decrease or increase the
threshold voltage of a flash cell, depending on the polarity of
the detrapped charge.

\begin{figure}[h]
  \centering
  \includegraphics[width=0.7\columnwidth]{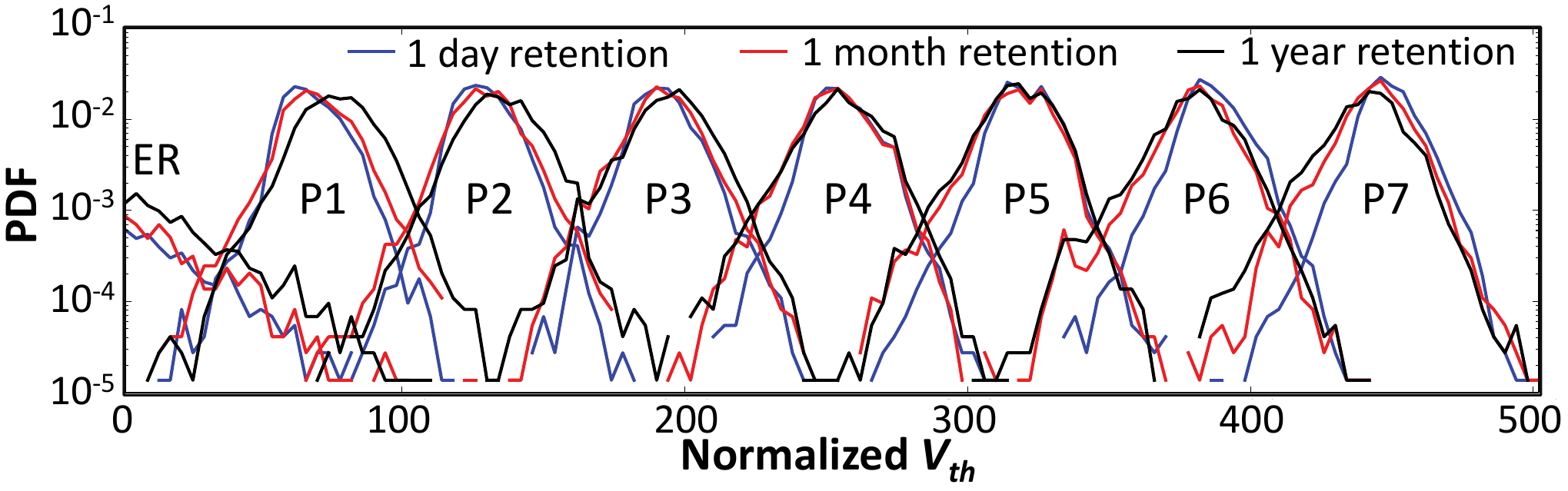}%
  \vspace{-5pt}%
  \caption{Threshold voltage distribution for TLC NAND flash memory
after one day, one month, and one year of retention time. \chI{Reproduced from \cite{cai.arxiv17}.}}%
  \label{fig:F18}%
\end{figure}

Figure~\ref{fig:F18} illustrates how the voltage distribution shifts
for data we program into TLC NAND flash, as the data sits
untouched over a period of one day, one month, and one year.
The mean and standard deviation are provided in Table~\ref{tbl:T5} in
the Appendix (which includes data for other retention ages
as well). These results are obtained from real flash memory
chips we tested. We distill three major findings from these
results, which are similar to our previously reported findings
for retention behavior on MLC NAND flash memory~\cite{R37}.

First, as the \emph{retention age} (i.e., the length of time after
programming) of the data increases, the threshold voltage distributions
of the higher-voltage states shift to lower voltages,
while the threshold voltage distributions of the lower-voltage
states shift to higher voltages. As the intrinsic electric field
strength is higher for the cells in higher-voltage states, TAT
is the dominant failure mechanism for these cells, which can
\emph{only} decrease the threshold voltage, as the resulting SILC can
flow only in the direction of the intrinsic electric field generated
by the electrons in the FG. Cells at the lowest-voltage
states, where the intrinsic electric field strength is low, do not
experience high TAT, and instead contain many \emph{holes} (i.e.,
positive charge) that leak away as the retention age grows,
leading to increase in threshold voltage.

Second, the threshold voltage distribution of each state
becomes wider with retention age. Charge detrapping can
cause cells to shift in either direction (i.e., toward lower or
higher voltages), contributing to the widening of the distribution.
The rate at which TAT occurs can also vary from cell
to cell, as a result of process variation, which further widens
the distribution.

Third, the threshold voltage distributions of higher-voltage
states shift by a larger amount than the distributions
of lower-voltage states. This is again a result of TAT. Cells
at higher-voltage states have greater intrinsic electric field
intensity, which leads to larger SILC. A cell where the SILC
is larger experiences a greater drop in its threshold voltage
than a cell where the SILC is smaller.

More findings on the nature of data retention and the
impact of data retention behavior on NAND flash memory
errors and lifetime can be found in our prior work~\cite{R32, R34, R37, R39, R177}.

\subsection{Read Disturb Errors}
\label{sec:errors:readdisturb}

Read disturb is a phenomenon in NAND flash memory
where reading data from a flash cell can cause the threshold
voltages of other (unread) cells in the same block to shift to
a higher value~\cite{R20, R32, R38, R54, R61, R62, R64}. While a
single threshold voltage shift is small, such shifts can accumulate
over time, eventually becoming large enough to alter the
state of some cells and hence generate \emph{read disturb errors}.

The failure mechanism of a read disturb error is similar
to the mechanism of a normal program operation. A program
operation applies a high programming voltage (e.g.,
+\SI{15}{\volt}) to the cell to change the cell's threshold voltage to
the desired range. Similarly, a read operation applies a \emph{high
pass-through voltage} (e.g., +\SI{6}{\volt}) to \emph{all other cells} that share
the same bitline with the cell that is being read. Although
the pass-through voltage is not as high as the programming
voltage, it still generates a \emph{weak programming effect} on the
cells it is applied to~\cite{R38}, which can unintentionally change
these cells' threshold voltages.

\begin{figure}[h]
  \centering
  \includegraphics[width=0.7\columnwidth]{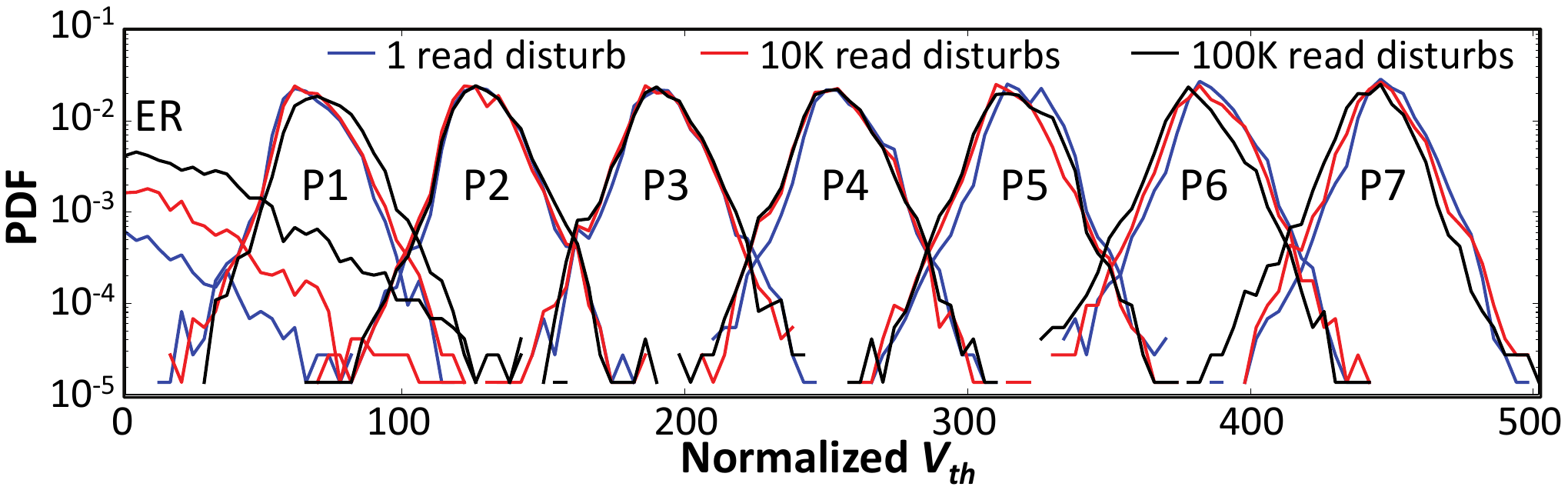}%
  \vspace{-5pt}%
  \caption{Threshold voltage distribution for TLC NAND flash memory
after 1, 10K, and 100K read disturb operations. \chI{Reproduced from \cite{cai.arxiv17}.}}%
  \label{fig:F19}%
\end{figure}

Figure~\ref{fig:F19} shows how read disturb errors impact threshold
voltage distributions in real TLC NAND flash memory
chips. We use blocks that have endured 2,000 P/E cycles,
and we experimentally study the impact of read disturb on
a single wordline in each block. We then read from a second
wordline in the same block 1, 10K, and 100K times to
induce different levels of read disturb. The mean and standard
deviation of each distribution are provided in Table~\ref{tbl:T6} in
the Appendix. We derive three major findings from these
results, which are similar to our previous findings for read
disturb behavior in MLC NAND flash memory~\cite{R38}.

First, as the read disturb count increases, the threshold
voltages increase (i.e., the voltage distribution shifts to the
right). In particular, we find that the distribution shifts are
greater for lower-voltage states, indicating that read disturb
impacts cells in the ER and P1 states the most. This is because
we apply the same pass-through voltage ($V_{pass}$) to \emph{all} unread
cells during a read operation, \emph{regardless} of the threshold voltages
of the cells. A lower threshold voltage on a cell induces
a larger voltage difference ($V_{pass} - V_{th}$) through the tunnel
oxide layer of the cell, and in turn generates a stronger tunneling
current, making the cell more vulnerable to read disturb
(as described in detail in our prior work~\cite{R38}).

Second, cells whose threshold voltages are closer to the
point at which the voltage distributions of the ER and P1
states intersect are more vulnerable to read disturb errors.
This is because process variation causes different cells to have
different degrees of vulnerability to read disturb. We find that
cells that are \emph{prone} to read disturb end up at the right tail
of the threshold voltage distribution of the ER state, as these
cells' threshold voltages increase more rapidly, and that cells
that are relatively \emph{resistant} to read disturb end up at the left
tail of the threshold voltage distribution of the P1 state, as
their threshold voltages increase more slowly. We can exploit
this divergent behavior of cells that end up at the left and
right distribution tails to perform error recovery in the event
of an uncorrectable error, as we discuss in Section~\ref{sec:correction:recovery}.

Third, unlike with the other states, the threshold voltages
of the cells at the left tail of the highest-voltage state
(P7) in TLC NAND flash memory actually \emph{decreases} as the
read disturb count increases. This occurs for two reasons:
(1)~applying $V_{pass}$ causes electrons to move from the floating
gate to the control gate for a cell at high voltage (i.e., a cell
containing a large number of electrons), thus \emph{reducing} its
threshold voltage~\cite{R38, R89}; and (2)~some retention time elapses
while we sweep the voltages during our read disturb experiments,
inducing trap-assisted tunneling (see Section~\ref{sec:errors:retention})
and leading to retention errors that decrease the voltage.

More findings on the nature of read disturb and the
impact of read disturb on NAND flash memory errors and
lifetime can be found in our prior work~\cite{R38}.

\subsection{Large-Scale Studies on SSD Errors}
\label{sec:errors:largescale}

The error characterization studies we have discussed so
far examine the susceptibility of real NAND flash memory
devices to specific error sources, by conducting controlled
experiments on individual flash devices in controlled environments.
To examine the \chII{\emph{aggregate}} effect of these error
sources on flash devices that operate in the field, several
recent studies have analyzed the reliability of SSDs deployed
at a large scale (\chII{e.g., hundreds} of thousands of SSDs)
in production data centers~\cite{R48, R49, R50}. Unlike the controlled
low-level error characterization studies discussed in
Sections~\ref{sec:errors:pe} through \ref{sec:errors:readdisturb}, these large-scale studies analyze
the observed errors and error rates in an \emph{uncontrolled}
manner, i.e., based on real data center workloads operating
at field conditions \chII{(as opposed to carefully controlling access patterns
and operating conditions)}. As such, these large-scale studies
can study flash memory behavior and reliability using only
a black-box approach, where they are able to access only the
registers used by the SSD to record select statistics. 
\chII{Because of this, their conclusions are usually correlational in nature,
as opposed to identifying the underlying causes behind the observations.}
On the
other hand, these studies incorporate the effects of a real
system, including the system software stack and real workloads~\cite{R48}
\chII{and real operational conditions in data centers}, 
on the flash memory devices, which is not present
in the controlled small-scale studies.

These \chII{recent} large-scale studies have made a number of observations
across large sets of SSDs \chII{employed in the data centers of large
internet companies: Facebook~\cite{R48}, Google~\cite{R49}, and Microsoft~\cite{R50}}.
We highlight \chII{six} key observations from these studies \chII{about the
\emph{SSD failure rate}, which is the fraction of SSDs that have experienced at least one
uncorrectable error.}

\begin{figure}[h]
  \centering
  \includegraphics[width=0.41\columnwidth]{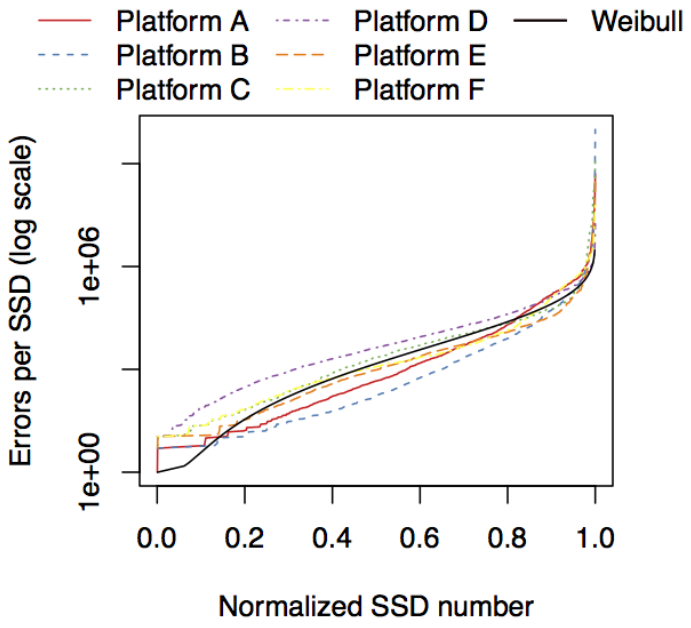}%
  \vspace{-5pt}%
  \caption{\chII{Distribution of uncorrectable errors across SSDs used in 
  Facebook's data centers.  Reproduced from \cite{R48}.}}%
  \label{fig:large-scale-distribution}%
\end{figure}

\chII{First, the number of uncorrectable errors observed varies significantly 
for each SSD.  Figure~\ref{fig:large-scale-distribution} shows the
distribution of uncorrectable errors per SSD across a large set of SSDs
used by Facebook.  The
distributions are grouped into six different \emph{platforms}
that are deployed in Facebook's data center.\chIII{\footnote{\chIII{Each platform has a different combination of SSDs, host controller
interfaces, and workloads. The six platforms are described in detail in~\cite{R48}.}}}
For every platform, we observe that the top 10\% of SSDs, when sorted by
their uncorrectable error count, account for over 80\% of the total uncorrectable
errors observed across all SSDs for that platform.  We find that the
distribution of uncorrectable errors across all SSDs belonging to a platform
follows a Weibull distribution, which we show using a solid black line in
Figure~\ref{fig:large-scale-distribution}.}

\begin{figure}[h]
  \centering
  \includegraphics[width=0.35\columnwidth]{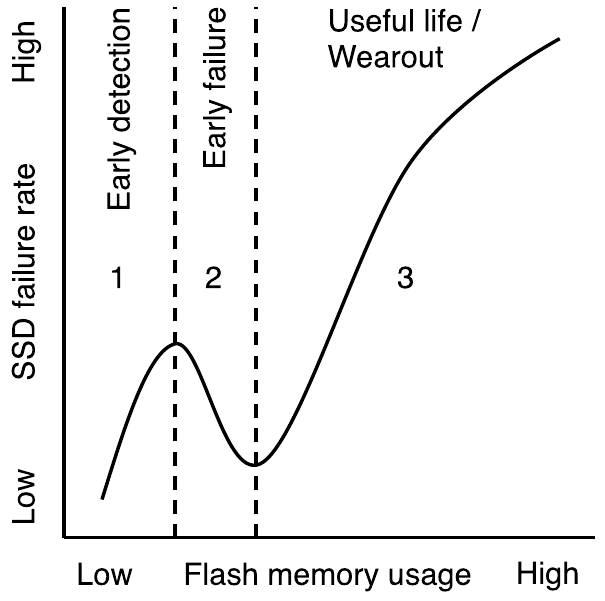}%
  \vspace{-5pt}%
  \caption{\chI{\chII{\chIII{Pictorial and abstract \chIV{depiction} of the pattern} of SSD failure rates observed in real SSDs operating
  in a modern data center.}  An SSD fails at different rates
  during distinct periods throughout the SSD lifetime. Reproduced from \cite{R48}.}}%
  \label{fig:lifecycle}%
\end{figure}

\begin{figure}[h]
  \centering
  \includegraphics[width=0.32\columnwidth]{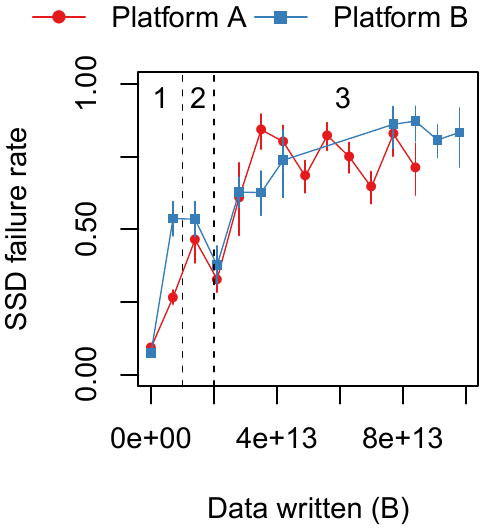}%
  \hfill
  \includegraphics[width=0.32\columnwidth]{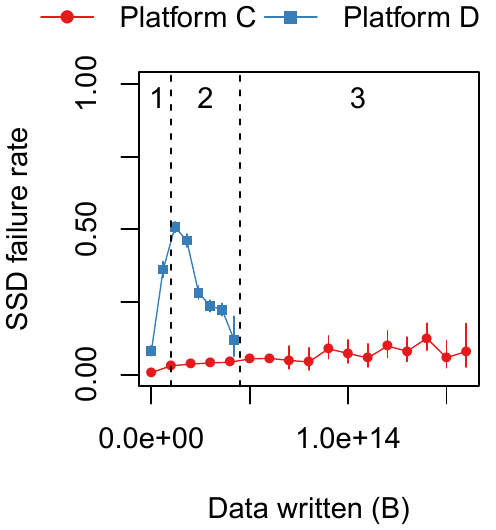}%
  \hfill
  \includegraphics[width=0.32\columnwidth]{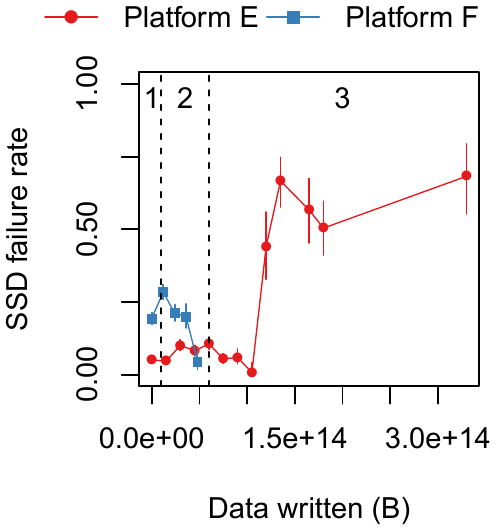}%
  \vspace{-5pt}%
  \caption{\chI{SSD failure rate vs.\ the amount of data written to the SSD.
  The three \chII{periods of failure rates, shown \chIII{pictorially and abstractly}
  in Figure~\ref{fig:lifecycle},} are annotated on each graph:
  (1)~early detection, (2)~early failure, and (3)~useful life/wearout. 
  Reproduced from \cite{R48}.}}%
  \label{fig:lifecycle-write}%
\end{figure}

\chI{\chII{Second, the SSD failure rate does} \emph{not}
increase monotonically with the P/E cycle count.  Instead, we observe
several \emph{distinct} periods of reliability, as illustrated \chII{pictorially and abstractly} in 
Figure~\ref{fig:lifecycle}, \chII{which is based on data obtained from analyzing
errors in SSDs used in Facebook's data centers~\cite{R48}}.  The failure rate increases when the SSDs are
relatively new (shown as the \emph{early detection} period in
Figure~\ref{fig:lifecycle}), as the SSD controller identifies unreliable
NAND flash cells during the initial read and write operations to the devices
\chII{and removes them from the address space (see \chIV{Section~\ref{sec:ssdarch:ctrl:badblocks}})}.  
As the SSDs are used \chII{more}, they enter the
\emph{early failure} period, where failures are less likely to occur.
When the SSDs approach the end of their lifetime (\emph{useful life/wearout}
in the figure), the failure rate increases again, as more cells become 
unreliable due to wearout.  Figure~\ref{fig:lifecycle-write} shows how the 
measured failure rate changes as more writes are performed to the SSDs
\chII{(i.e., how real data collected from Facebook's SSDs \chIII{corresponds to} the
pictorial depiction in Figure~\ref{fig:lifecycle}) for the
same six platforms shown in Figure~\ref{fig:large-scale-distribution}}. 
\chII{We observe that the failure rates in each platform exhibit the
distinct periods that are illustrated in Figure~\ref{fig:lifecycle}.
For example, \chIII{let us consider} the SSDs in Platforms A and B\chIII{, which} have more data written to
their cells than SSDs in other platforms.  We observe from 
Figure~\ref{fig:lifecycle-write} that for SSDs in Platform~A,
there is an 81.7\% \chIII{increase from} the failure rate during the
early detection period \chIII{to} the failure rate during the wearout period~\cite{R48}.}}

\chII{Third}, the raw bit error rate grows with the age of the
device even if the P/E cycle count is held constant, indicating
that mechanisms such as silicon aging \chII{likely contribute}
to the error rate~\cite{R50}. 

\chII{Fourth}, the observed failure rate of SSDs
has been noted to be significantly higher than the failure rates
specified by the manufacturers~\cite{R49}. 

\begin{figure}[h]
  \centering
  \includegraphics[width=0.32\columnwidth]{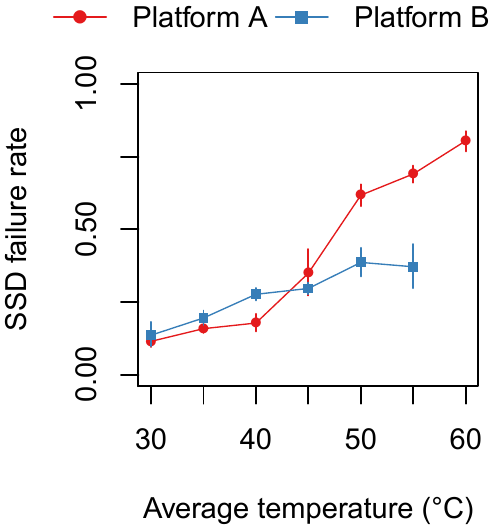}%
  \hfill
  \includegraphics[width=0.32\columnwidth]{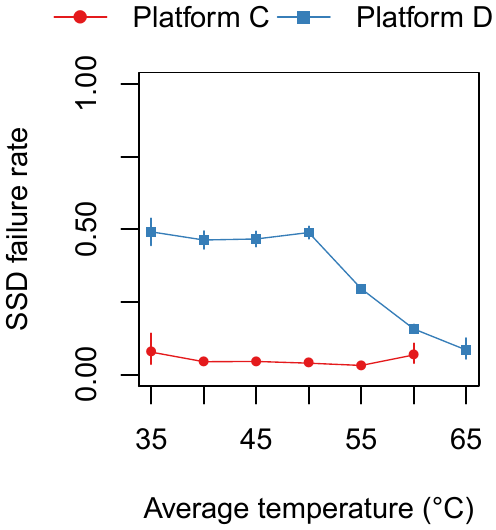}%
  \hfill
  \includegraphics[width=0.32\columnwidth]{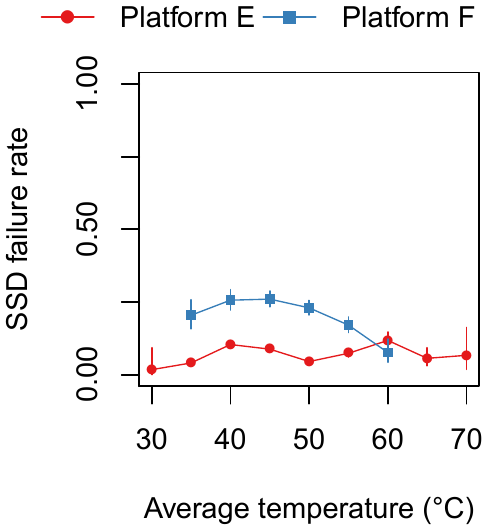}%
  \vspace{-5pt}%
  \caption{\chII{SSD failure rate vs.\ operating temperature.
  Reproduced from~\cite{R48}.}}%
  \label{fig:datacenter-temp}%
\end{figure}

\chII{Fifth}, higher operating
temperatures can lead to higher failure rates, but modern SSDs
employ throttling techniques that reduce the access rates to
the underlying flash chips, which can greatly reduce the negative
reliability impact of higher temperatures~\cite{R48}. 
\chII{For example, Figure~\ref{fig:datacenter-temp} shows the
SSD failure rate as the SSD operating temperature varies, for SSDs from the
same six platforms shown in Figure~\ref{fig:large-scale-distribution}~\cite{R48}.
We observe that at an operating temperature range of \SIrange{30}{40}{\celsius},
SSDs \chIII{either (1)~have} similar failure rates \chIII{across the different temperatures, 
or (2)~experience} slight increases in the failure rate as
the temperature increases.  As the temperature increases beyond \SI{40}{\celsius},
the SSDs fall into three categories:
(1)~temperature-sensitive with increasing failure rate (Platforms~A and B),
(2)~less temperature-sensitive (Platforms~C and E), and
(3)~temperature-sensitive with decreasing failure rate (Platforms~D and F).
There are two factors that affect the temperature sensitivity of each platform:
(1)~some, but not all, of the platforms employ techniques to throttle SSD
activity at high operating temperatures to reduce the failure rate \chIII{(e.g., Platform~D)}; and
(2)~the platform configuration (e.g., the number of SSDs in each machine,
system airflow) can shorten or prolong the effects of higher operating 
temperatures.}

\chII{Sixth}, while
SSD failure rates are higher than specified \chII{by the manufacturers}, the overall occurrence
of \emph{uncorrectable} errors is lower than expected\chIII{~\cite{R48}} because
(1)~effective bad block management policies (see \chIV{Section~\ref{sec:ssdarch:ctrl:badblocks}})
are implemented in SSD controllers; and (2)~certain types of
error sources, such as read disturb~\cite{R48, R50} and incomplete
erase operations~\cite{R50}, have yet to become a major source of
uncorrectable errors at the system level.

\section{Error Mitigation}
\label{sec:mitigation}

Several different types of errors can occur in NAND flash
memory, as we described in Section~\ref{sec:errors}. As NAND flash memory
continues to scale to smaller technology nodes, the magnitude
of these errors has been increasing~\cite{R53, R60, R81}.
This, in turn, uses up the limited error correction capability
of ECC more rapidly than in past flash memory generations
and shortens the lifetime of modern SSDs. To overcome the
decrease in lifetime, a number of error mitigation techniques \chII{have been designed.
These techniques} exploit intrinsic properties of the different types of
errors to reduce the rate at which they lead to raw bit \chII{errors.}
In this section, we discuss how the flash
controller mitigates each of the error types via \chII{various} proposed error
mitigation mechanisms. Table~\ref{tbl:T3} shows the techniques we
overview and which errors (from Section~\ref{sec:errors}) they mitigate.

\begin{table}[h]
\centering
\small
\setlength{\tabcolsep}{0.26em}
\setlength\arrayrulewidth{0.75pt}
\caption{List of different types of errors mitigated by \chII{various} NAND flash
error mitigation mechanisms.}
\label{tbl:T3}
\vspace{-5pt}%
\begin{tabular}{|c||c|c|c|c|c|}
\hline
& \multicolumn{5}{c|}{\cellcolor{Gray!20}\textbf{\emph{Error Type}}} \\
& \cellcolor{Gray!20} & \cellcolor{Gray!20} & \cellcolor{Gray!20} & \cellcolor{Gray!20} & \cellcolor{Gray!20} \\
& \cellcolor{Gray!20} & \cellcolor{Gray!20} & \cellcolor{Gray!20} & \cellcolor{Gray!20} & \cellcolor{Gray!20} \\
& \cellcolor{Gray!20} & \cellcolor{Gray!20} & \cellcolor{Gray!20} & \cellcolor{Gray!20} & \cellcolor{Gray!20} \\
& \cellcolor{Gray!20} & \cellcolor{Gray!20} & \cellcolor{Gray!20} & \cellcolor{Gray!20} & \cellcolor{Gray!20} \\
& \cellcolor{Gray!20} & \cellcolor{Gray!20} & \cellcolor{Gray!20} & \cellcolor{Gray!20} & \cellcolor{Gray!20} \\
& \cellcolor{Gray!20} & \cellcolor{Gray!20} & \cellcolor{Gray!20} & \cellcolor{Gray!20} & \cellcolor{Gray!20} \\
& \cellcolor{Gray!20} & \cellcolor{Gray!20} & \cellcolor{Gray!20} & \cellcolor{Gray!20} & \cellcolor{Gray!20} \\
\textbf{Mitigation} & \cellcolor{Gray!20} & \cellcolor{Gray!20} & \cellcolor{Gray!20} & \cellcolor{Gray!20} & \cellcolor{Gray!20} \\
\textbf{Mechanism}
& \multirow{-9}{*}{\cellcolor{Gray!20}\rotatebox{90}{\parbox[b]{11em}{\makecell[l]{\textbf{\emph{P/E Cycling}}\\\cite{R32, R33, R42} (\S\ref{sec:errors:pe})}}}}
& \multirow{-9}{*}{\cellcolor{Gray!20}\rotatebox{90}{\parbox[b]{11em}{\makecell[l]{\textbf{\emph{Program}}\\\cite{R40, R42, R53} (\S\ref{sec:errors:pgm})}}}}
& \multirow{-9}{*}{\cellcolor{Gray!20}\rotatebox{90}{\parbox[b]{11em}{\makecell[l]{\textbf{\emph{Cell-to-Cell Interference}}\\\cite{R32, R35, R36, R55} (\S\ref{sec:errors:celltocell})}}}}
& \multirow{-9}{*}{\cellcolor{Gray!20}\rotatebox{90}{\parbox[b]{11em}{\makecell[l]{\textbf{\emph{Data Retention}}\\\cite{R20, R32, R34, R37, R39} (\S\ref{sec:errors:retention})}}}}
& \multirow{-9}{*}{\cellcolor{Gray!20}\rotatebox{90}{\parbox[b]{11em}{\makecell[l]{\textbf{\emph{Read Disturb}}\\\cite{R20, R32, R38, R62} (\S\ref{sec:errors:readdisturb})}}}} \\ \hhline{|=#=|=|=|=|=|}
\textbf{Shadow Program Sequencing} & & & \multirow{2}{*}{X} & & \\
\cite{R35, R40} (Section~\ref{sec:mitigation:shadow}) & & & & & \\ \hline
\textbf{Neighbor-Cell Assisted Error} & & & \multirow{2}{*}{X} & & \\
\textbf{Correction} \cite{R36} (Section~\ref{sec:mitigation:nac}) & & & & & \\ \hline
\textbf{Refresh} & & & & \multirow{2}{*}{X} & \multirow{2}{*}{X} \\
\cite{R34, R39, R67, R68} (Section~\ref{sec:mitigation:refresh}) & & & & & \\ \hline
\textbf{Read-Retry} & \multirow{2}{*}{X} & & & \multirow{2}{*}{X} & \multirow{2}{*}{X} \\
\cite{R33, R72, R107} (Section~\ref{sec:mitigation:retry}) & & & & & \\ \hline
\textbf{Voltage Optimization} & \multirow{2}{*}{X} & & & \multirow{2}{*}{X} & \multirow{2}{*}{X} \\
\cite{R37, R38, R74} (Section~\ref{sec:mitigation:voltage}) & & & & & \\ \hline
\textbf{Hot Data Management} & \multirow{2}{*}{X} & \multirow{2}{*}{X} & \multirow{2}{*}{X} & \multirow{2}{*}{X} & \multirow{2}{*}{X} \\
\cite{R41, R63, R70} (Section~\ref{sec:mitigation:hotcold}) & & & & & \\ \hline
\textbf{Adaptive Error Mitigation} & \multirow{2}{*}{X} & \multirow{2}{*}{X} & \multirow{2}{*}{X} & \multirow{2}{*}{X} & \multirow{2}{*}{X} \\
\cite{R43, R65, R77, R78, R82} (Section~\ref{sec:mitigation:adaptive}) & & & & & \\ \hline
\end{tabular}
\end{table}

\subsection{Shadow Program Sequencing}
\label{sec:mitigation:shadow}

As discussed in Section~\ref{sec:errors:celltocell}, cell-to-cell program interference
is a function of the distance between the cells of the
wordline that is being programmed and the cells of the victim
wordline. The impact of program interference is greatest on
a victim wordline when either of the victim's immediately-adjacent
wordlines is programmed (e.g., if we program WL1
in Figure~\ref{fig:F8}, WL0 and WL2 experience the greatest amount of
interference). Early MLC flash memories used one-shot programming,
where both the LSB and MSB pages of a wordline
are programmed at the same time. As flash memory scaled to
smaller process technologies, one-shot programming resulted
in much larger amounts of cell-to-cell program interference.
As a result, manufacturers introduced two-step programming
for MLC NAND flash (see Section~\ref{sec:flash:pgmerase}), where the SSD controller
writes values of the two pages within a wordline in two
independent steps.

\begin{figure}[h]
  \centering
  \includegraphics[width=0.7\columnwidth]{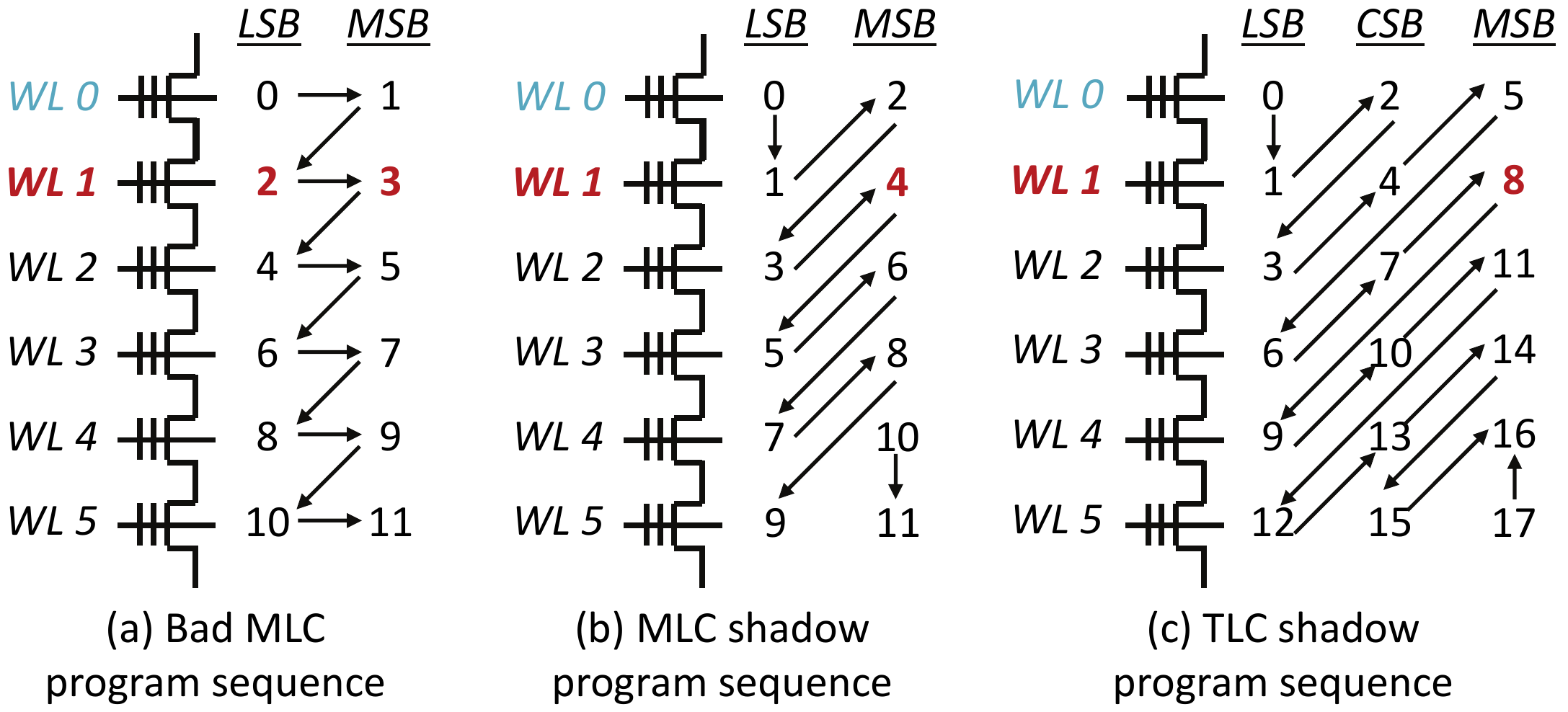}%
  \vspace{-5pt}%
  \caption{Order in which the pages of each wordline (WL) are
programmed using (a)~a bad programming sequence, and using
shadow sequencing for (b)~MLC and (c)~TLC NAND flash. The bold
page programming operations for WL1 induce cell-to-cell program
interference when WL0 is fully programmed. \chI{Reproduced from \cite{cai.arxiv17}.}}%
  \label{fig:F20}%
\end{figure}

The SSD controller minimizes the interference that
occurs during two-step programming by using \emph{shadow program
sequencing}~\cite{R27, R35, R40} to determine the order that
data is written to different pages in a block. If we program
the LSB and MSB pages of the same wordline back to back,
as shown in Figure~\ref{fig:F20}a, both programming steps induce
interference on a \emph{fully-programmed wordline} (i.e., a wordline
where both the LSB and MSB pages are already written).
For example, if the controller programs both pages of WL1
back to back, shown as bold page programming operations
in Figure~\ref{fig:F20}a, the program operations induce a high amount
of interference on WL0, which is fully programmed. The key
idea of shadow program sequencing is to ensure that a fully-programmed
wordline experiences interference minimally,
i.e., \emph{only} during MSB page programming (and \emph{not} during
LSB page programming). In shadow program sequencing, we
assign a unique page number to each page within a block, as
shown in Figure~\ref{fig:F20}b. The LSB page of wordline~$i$ is numbered
page $2i - 1$, and the MSB page is numbered page $2i + 2$. The
only exceptions to the numbering are the LSB page of wordline~0 
(page~0) and the MSB page of the last wordline~$n$ (page
$2n + 1$). Two-step programming writes to pages in \emph{increasing}
order of page number inside a block~\cite{R27, R35, R40}, such that
a \emph{fully-programmed} wordline experiences interference only
from the MSB page programming of the wordline directly
above it, shown as the bold page programming operation in
Figure~\ref{fig:F20}b. With this programming order/sequence, the LSB
page of the wordline above, and both pages of the wordline
below, do \emph{not} cause interference to fully-programmed data~\cite{R27, R35, R40}, 
as these two pages are programmed \emph{before}
programming the MSB page of the given wordline. Foggy-fine
programming in TLC NAND flash (see Section~\ref{sec:flash:pgmerase})
uses a similar ordering to reduce cell-to-cell program interference,
as shown in Figure~\ref{fig:F20}c.

Shadow program sequencing is an effective solution
to minimize cell-to-cell program interference on fully-programmed
wordlines during two-step programming, and
is employed in commercial SSDs today.

\subsection{Neighbor-Cell Assisted Error Correction}
\label{sec:mitigation:nac}

The threshold voltage shift that occurs due to program
interference is highly correlated with the values stored in
the cells of the \emph{immediately-adjacent wordlines}, as we discussed
in Section~\ref{sec:errors:celltocell}. Due to this correlation, knowing
the value programmed in the immediately-adjacent cell
(i.e., a \emph{neighbor cell}) makes it easier to correctly determine
the value stored in the flash cell that is being read~\cite{R36}. We
describe a recently proposed error correction method that
takes advantage of this observation, called \emph{neighbor-cell-assisted
error correction} (NAC). The key idea of NAC is to
use the data values stored in the cells of the immediately-adjacent
wordline to determine a better set of read reference
voltages for the wordline that is being read. Doing so leads
to a more accurate identification of the logical data value
that is being read, as the data in the immediately-adjacent
wordline was \emph{partially responsible} for shifting the threshold
voltage of the cells in the wordline that is being read when
the immediately-adjacent wordline was programmed.

\begin{figure}[h]
  \centering
  \includegraphics[width=0.65\columnwidth]{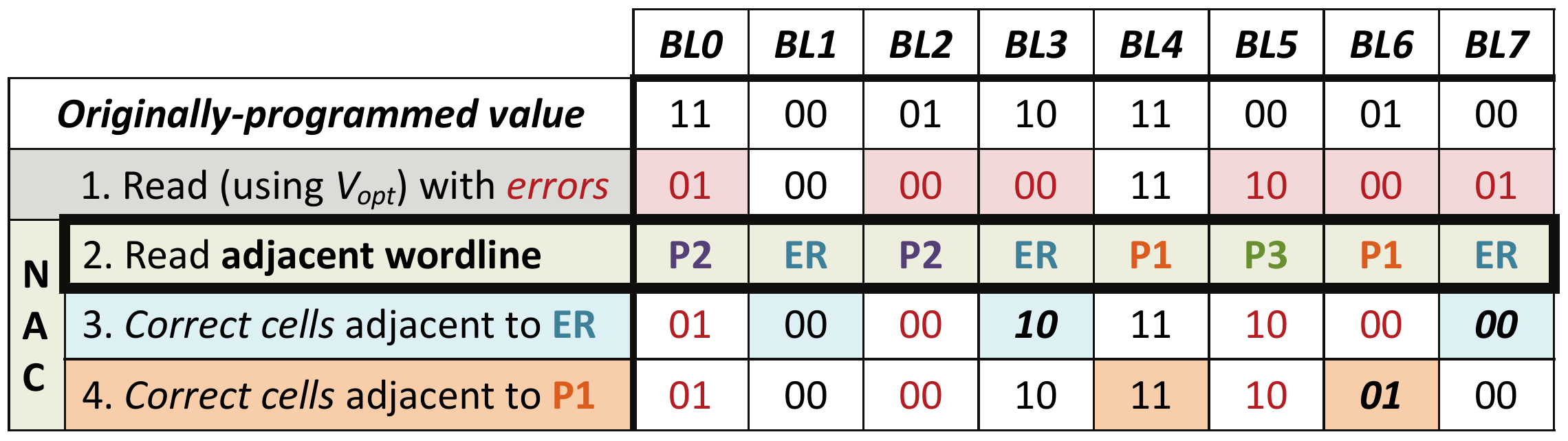}%
  \vspace{-5pt}%
  \caption{Overview of neighbor-cell-assisted error correction (NAC). \chI{Reproduced from \cite{cai.arxiv17}.}}%
  \label{fig:F21}%
\end{figure}

Figure~\ref{fig:F21} shows an operational example of NAC that
is applied to eight bitlines (BL) of an MLC flash wordline.
The SSD controller first reads a flash page from a
wordline using the standard read reference voltages (step~1 in
Figure~\ref{fig:F21}). The bit values read from the wordline are then buffered
in the controller. If there are no errors uncorrectable by ECC,
the read was successful, and nothing else is done. However,
if there are errors that are \emph{uncorrectable} by ECC, we assume
that the threshold voltage distribution of the page shifted due to
cell-to-cell program interference, triggering further correction.
In this case, NAC reads the LSB and MSB pages of the wordline
\emph{immediately above} the requested page (i.e., the \emph{adjacent} wordline
that was programmed \emph{after} the requested page) to classify
the cells of the requested page (step~2). NAC then identifies the
cells adjacent to (i.e., connected to the same bitline as) the ER
cells (i.e., cells in the immediately above wordline that are in
the ER state), such as the cells on BL1, BL3, and BL7 in Figure~\ref{fig:F21}.
NAC rereads these cells using read reference voltages that \emph{compensate
for} the threshold voltage shift caused by programming
the adjacent cell to the ER state (step~3). If ECC can correct the
remaining errors, the controller returns the corrected page to
the host. If ECC fails again, the process is repeated using a different
set of read reference voltages for cells that are adjacent
to the P1 cells (step~4). If ECC continues to fail, the process is
repeated for cells that are adjacent to P2 and P3 cells (steps~5
and 6, respectively, which are not shown in the figure) until
either ECC is able to correct the page or all possible adjacent
values are exhausted.

NAC extends the lifetime of an SSD by reducing the
number of errors that need to be corrected using the limited
correction capability of ECC. With the use of experimental
data collected from real MLC NAND flash memory
chips, we show that NAC extends the NAND flash memory
lifetime by 33\%~\cite{R36}. Our previous work~\cite{R36} provides a
detailed description of NAC, including a theoretical treatment
of why it works and a practical implementation that
minimizes the number of reads performed, even in the case
when the neighboring wordline itself has errors.

\subsection{Refresh Mechanisms}
\label{sec:mitigation:refresh}

As we see in Figure~\ref{fig:F12}, during the time period after a flash
page is programmed, retention (Section~\ref{sec:errors:retention}) and read
disturb (Section~\ref{sec:errors:readdisturb}) can cause an increasing number of
raw bit errors to accumulate over time. This is particularly
problematic for a page that is not updated frequently. Due
to the limited error correction capability, the accumulation
of these errors can potentially lead to data loss for a
page with a \emph{high retention age} (i.e., a page that has not been
programmed for a long time). To avoid data loss, \emph{refresh
mechanisms} have been proposed, where the stored data is
periodically read, corrected, and reprogrammed, in order
to eliminate the retention and read disturb errors that
have accumulated prior to this periodic 
read/correction/reprogramming (i.e., refresh). The concept of refresh in
flash memory is thus conceptually similar to the refresh
mechanisms found in DRAM~\cite{R66, R69, R104, R123}. By
performing refresh and limiting the number of retention
and read disturb errors that can accumulate, the lifetime of
the SSD increases significantly. In this section, we describe
three types of refresh mechanisms used in modern SSDs:
remapping-based refresh, in-place refresh, and read reclaim.

\paratitle{Remapping-Based Refresh}
Flash cells must first be
erased before they can be reprogrammed, due to the fact
the programming a cell via ISPP can only increase the
charge level of the cell but not reduce it (Section~\ref{sec:flash:pgmerase}).
The key idea of \emph{remapping-based refresh} is to periodically
read data from each valid flash block, correct any data
errors, and \emph{remap the data to a different physical location},
in order to prevent the data from accumulating too many
retention errors~\cite{R34, R39, R67, R68, R177}. During each refresh
interval, a block with valid data that needs to be refreshed
is selected. The valid data in the selected block is read out
page by page and moved to the SSD controller. The ECC
engine in the SSD controller corrects the errors in the read
data, including retention errors that have accumulated
since the last refresh. A new block is then selected from
the free list (see \chIV{Section~\ref{sec:ssdarch:ctrl:ftl}}), the error-free data is programmed
to a page within the new block, and the logical
address is remapped to point to the newly-programmed
physical page. By reducing the accumulation of retention
and read disturb errors, remapping-based refresh increases
SSD lifetime by an average of 9x for a variety of disk workloads~\cite{R34, R39}.

Prior work proposes extensions to the basic remapping-based
refresh approach. One work, \emph{refresh SSDs}, proposes a
refresh scheduling algorithm based on an earliest deadline
first policy to guarantee that all data is refreshed in time~\cite{R68}. 
The \emph{quasi-nonvolatile SSD} proposes to use remapping-based
refresh to choose between improving flash endurance
and reducing the flash programming latency (by using
larger ISPP step-pulses)~\cite{R67}. In the quasi-nonvolatile SSD,
refresh requests are deprioritized, scheduled at idle times,
and can be interrupted after refreshing any page within a
block, to minimize the delays that refresh can cause for
the response time of pending workload requests to the
SSD. A refresh operation can also be triggered proactively
based on the data read latency observed for a page, which
is indicative of how many errors the page has experienced~\cite{R87}. 
Triggering refresh \emph{proactively} based on the observed
read latency (as opposed to doing so \emph{periodically}) improves
SSD latency and throughput~\cite{R87}. Whenever the read
latency for a page within a block exceeds a fixed threshold,
the valid data in the block is refreshed, i.e., remapped to a
new block~\cite{R87}.

\paratitle{In-Place Refresh}
A major drawback of remapping-based
refresh is that it performs \emph{additional writes} to the NAND
flash memory, accelerating wearout. To reduce the wearout
overhead of refresh, we propose \emph{in-place refresh}~\cite{R34, R39, R177}. As
data sits unmodified in the SSD, data retention errors dominate~\cite{R32, R39, R56}, 
leading to charge loss and causing the
threshold voltage distribution to shift to the left, as we showed
in Section~\ref{sec:errors:retention}. The key idea of in-place refresh is to incrementally
replenish the lost charge of each page \emph{at its current
location}, i.e., in place, without the need for remapping.

\begin{figure}[h]
  \centering
  \includegraphics[width=0.7\columnwidth]{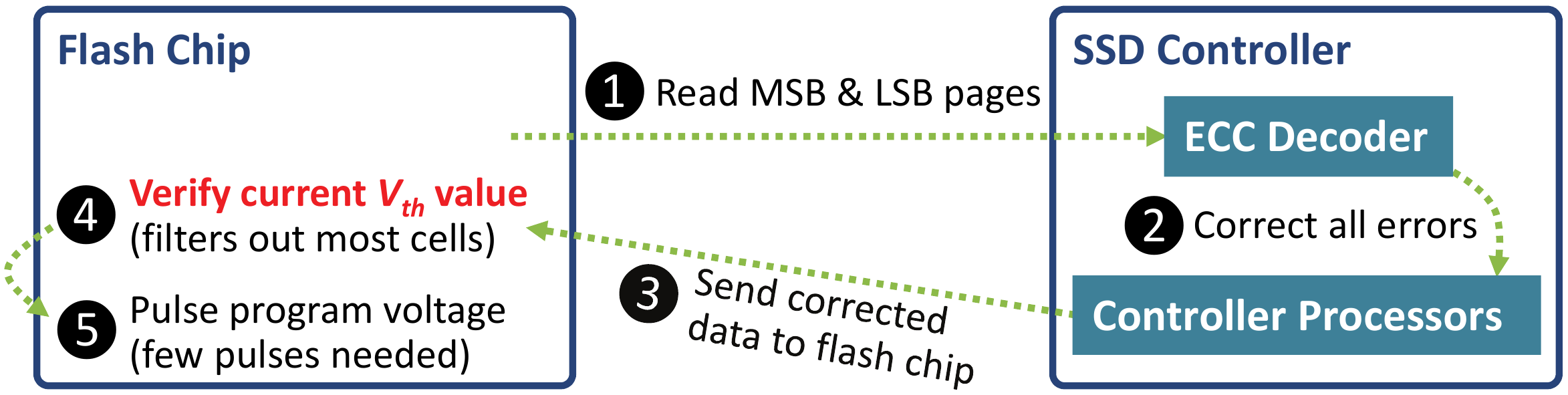}%
  \vspace{-5pt}%
  \caption{Overview of in-place refresh mechanism for MLC NAND
flash memory. \chI{Reproduced from \cite{cai.arxiv17}.}}%
  \label{fig:F22}%
\end{figure}

Figure~\ref{fig:F22} shows a high-level overview of in-place refresh for
a wordline. The SSD controller first reads all of the pages
in the wordline (\incircle{1} in Figure~\ref{fig:F22}). The controller invokes the
ECC decoder to correct the errors within each page (\incircle{2}), and
sends the corrected data back to the flash chips (\incircle{3}). In-place
refresh then invokes a modified version of the ISPP mechanism
(see Section~\ref{sec:flash:pgmerase}), which we call \emph{Verify-ISPP} (V-ISPP),
to compensate for retention errors by restoring the charge
that was lost. In V-ISPP, we first verify the voltage currently
programmed in a flash cell (\incircle{4}). If the current voltage of the
cell is \emph{lower} than the target threshold voltage of the state that
the cell should be in, V-ISPP pulses the programming voltage
in steps, gradually injecting charge into the cell until the
cell returns to the target threshold voltage (\incircle{5}). If the current
voltage of the cell is \emph{higher} than the target threshold voltage,
V-ISPP inhibits the programming pulses to the cell.

When the controller invokes in-place refresh, it is unable
to use shadow program sequencing (Section~\ref{sec:mitigation:shadow}), as all of the
pages within the wordline have already been programmed.
However, unlike traditional ISPP, V-ISPP does not introduce
a high amount of cell-to-cell program interference (Section~\ref{sec:errors:celltocell}) 
for two reasons. First, V-ISPP programs \emph{only} those cells
that have retention errors, which typically account for less
than 1\% of the total number of cells in a wordline selected
for refresh~\cite{R34}. Second, for the small number of cells that
are selected to be refreshed, their threshold voltage is usually
only slightly lower than the target threshold voltage,
which means that only a few programming pulses need to
be applied. As cell-to-cell interference is linearly correlated
with the threshold voltage change to immediately-adjacent
cells~\cite{R35, R36}, the small voltage change on these in-place
refreshed cells leads to only a small interference effect.

One issue with in-place refresh is that it is unable to
correct retention errors for cells in lower-voltage states.
Retention errors cause the threshold voltage of a cell in a
lower-voltage state to \emph{increase} (e.g., see Section~\ref{sec:errors:retention}, ER and
P1 states in Figure~\ref{fig:F18}), but V-ISPP \emph{cannot} decrease the threshold
voltage of a cell. To achieve a balance between the wearout
overhead due to remapping-based refresh and errors that
increase the threshold voltage due to in-place refresh, we propose
\emph{hybrid in-place refresh}~\cite{R34, R39, R177}. The key idea is to use
in-place refresh when the number of program errors (caused
due to reprogramming) is within the correction capability of
ECC, but to use remapping-based refresh if the number of
program errors is too large to tolerate. To accomplish this, the
controller tracks the number of \emph{right-shift errors} (i.e., errors
that move a cell to a higher-voltage state)~\cite{R34, R39}. If the
number of right-shift errors remains under a certain threshold,
the controller performs in-place refresh; otherwise, it
performs remapping-based refresh. Such a hybrid in-place
refresh mechanism increases SSD lifetime by an average of
31x for a variety of disk workloads~\cite{R34, R39}.

\paratitle{Read Reclaim to Reduce Read Disturb Errors}
We can
also mitigate read disturb errors using an idea similar to
remapping-based refresh, known as \emph{read reclaim}. The key
idea of read reclaim is to remap the data in a block to a new
flash block, if the block has experienced a high number of
reads~\cite{R63, R70, R173}. To bound the number of read disturb
errors, some flash vendors specify a maximum number of
tolerable reads for a flash block, at which point read reclaim
rewrites the data to a new block (just as is done for remapping-
based refresh).

\paratitle{Adaptive Refresh and Read Reclaim Mechanisms}
For
the refresh and read reclaim mechanisms discussed above,
the SSD controller can (1)~invoke the mechanisms at fixed
regular intervals; or (2)~\emph{adapt} the rate at which it invokes the
mechanisms, based on various conditions that impact the
rate at which data retention and read disturb errors occur.
By adapting the mechanisms based on the current conditions
of the SSD, the controller can reduce the overhead
of performing refresh or read reclaim. The controller can
adaptively adjust the rate that the mechanisms are invoked
based on (1)~the wearout (i.e., the current P/E cycle count) of
the NAND flash memory~\cite{R34, R39}; or (2)~the temperature
of the SSD~\cite{R32, R37}.

As we discuss in Section~\ref{sec:errors:retention}, for data with a given
retention age, the number of retention errors grows as the
P/E cycle count increases. Exploiting this P/E cycle dependent
behavior of retention time, the SSD controller can perform
refresh less frequently (e.g., once every year) when
the P/E cycle count is low, and more frequently (e.g., once
every week) when the P/E cycle count is high, as proposed
and described in our prior works~\cite{R34, R39}. Similarly, for
data with a given read disturb count, as the P/E cycle count
increases, the number of read disturb errors increases as
well~\cite{R38}. As a result, the SSD controller can perform read
reclaim less frequently (i.e., it increases the maximum number
of tolerable reads per block before read reclaim is triggered)
when the P/E cycle count is low, and more frequently
when the P/E cycle count is high.

Prior works demonstrate that for a given retention time,
the number of data retention errors increases as the NAND
flash memory's operating temperature increases~\cite{R32, R37}.
To compensate for the increased number of retention errors
at high temperature, a state-of-the-art SSD controller adapts
the rate at which it triggers refresh. The SSD contains sensors
that monitor the current environmental temperature
every few milliseconds~\cite{R48, R192}. The controller then
uses the Arrhenius equation~\cite{R68, R186, R187} to estimate
the rate at which retention errors accumulate at the current
temperature of the SSD. Based on the error rate estimate,
the controller decides if it needs to increase the rate
at which it triggers refresh to ensure that the data is not lost.

By employing adaptive refresh and/or read reclaim mechanisms,
the SSD controller can successfully reduce the mechanism
overheads while effectively mitigating the larger number
of data retention errors that occur under various conditions.

\subsection{Read-Retry}
\label{sec:mitigation:retry}

In earlier generations of NAND flash memory, the read
reference voltage values were fixed at design time~\cite{R20, R33}.
However, several types of errors cause the threshold voltage
distribution to shift, as shown in Figure~\ref{fig:F13}. To compensate for
threshold voltage distribution shifts, a mechanism called \emph{read-retry}
has been implemented in modern flash memories (typically
those below \SI{30}{\nano\meter} for planar flash~\cite{R33, R71, R72, R107}).

The read-retry mechanism allows the read reference
voltages to dynamically adjust to changes in distributions.
During read-retry, the SSD controller first reads the data out
of NAND flash memory with the default read reference voltage.
It then sends the data for error correction. If ECC successfully
corrects the errors in the data, the read operation
succeeds. Otherwise, the SSD controller reads the memory
again with a \emph{different} read reference voltage. The controller
repeats these steps until it either successfully reads the data
using a certain set of read reference voltages or is unable to
correctly read the data using all of the read reference voltages
that are available to the mechanism.

While read-retry is widely implemented today, it can
significantly increase the overall read operation latency due
to the multiple read attempts it causes~\cite{R37}. Mechanisms
have been proposed to reduce the number of read-retry
attempts while taking advantage of the effective capability
of read-retry for reducing read errors, and read-retry has
also been used to enable mitigation mechanisms for various
other types of errors, as we describe in Section~\ref{sec:mitigation:voltage}. As a
result, read-retry is an essential mechanism in modern SSDs
to mitigate read errors (i.e., errors that manifest themselves
during a read operation).

\subsection{Voltage Optimization}
\label{sec:mitigation:voltage}

Many raw bit errors in NAND flash memory are affected
by the various voltages used within the memory to enable
reading of values. We give two examples. First, a suboptimal
\emph{read reference voltage} can lead to a large number of read
errors (Section~\ref{sec:errors}), especially after the threshold voltage distribution
shifts. Second, as we saw in Section~\ref{sec:errors:readdisturb}, the \emph{pass-through
voltage} can have a significant effect on the number
of read disturb errors that occur. As a result, optimizing these
voltages such that they minimize the total number of errors
that are induced can greatly mitigate error counts. In this section,
we discuss mechanisms that can discover and employ
the optimal\footnote{Or, more precisely, near-optimal, if the read-retry steps are too
coarse grained to find the optimal voltage.} read reference and pass-through voltages.

\begin{figure}[h]
  \centering
  \includegraphics[width=0.7\columnwidth]{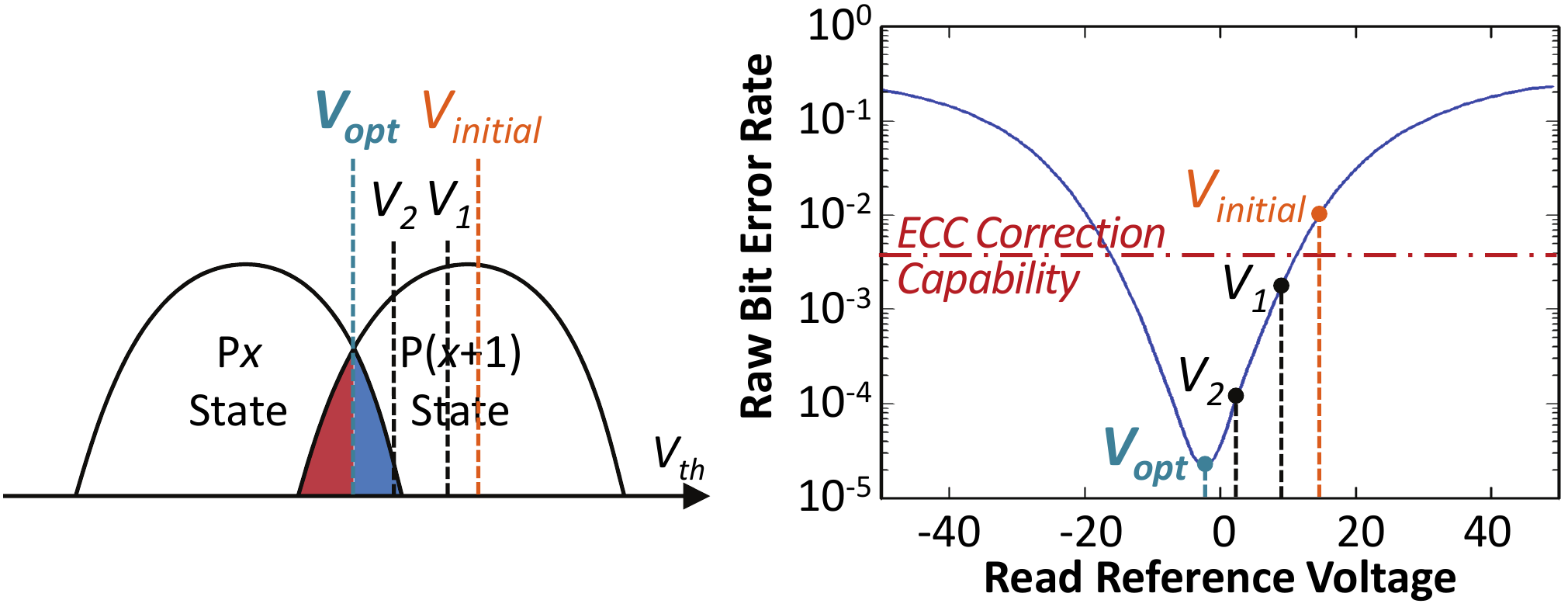}%
  \vspace{-5pt}%
  \caption{Finding the optimal read reference voltage after the
threshold voltage distributions overlap (left), and raw bit error rate
as a function of the selected read reference voltage (right). \chI{Reproduced from \cite{cai.arxiv17}.}}%
  \label{fig:F23}%
\end{figure}

\sloppypar
\paratitle{Optimizing Read Reference Voltages Using Disparity-Based Approximation and Sampling}
As we discussed in
Section~\ref{sec:mitigation:retry}, when the threshold voltage distribution shifts,
it is important to move the read reference voltage to the
point where the number of read errors is minimized. After
the shift occurs and the threshold voltage distribution of
each state widens, the distributions of different states may
overlap with each other, causing many of the cells within
the overlapping regions to be misread. The number of errors
due to misread cells can be minimized by setting the read
reference voltage to be exactly at the point where the distributions
of two neighboring states intersect, which we call
the \emph{optimal read reference voltage} ($V_{opt}$)~\cite{R35, R36, R37, R42, R54}, 
illustrated in Figure~\ref{fig:F23}. Once the optimal read reference
voltage is applied, the raw bit error rate is minimized,
improving the reliability of the device.

One approach to finding $V_{opt}$ is to adaptively learn and
apply the optimal read reference voltage for each flash block
through sampling~\cite{R37, R90, R165, R166}. The key idea is
to periodically (1)~use \emph{disparity} information (i.e., the ratio
of 1s to 0s in the data) to attempt to find a read reference
voltage for which the error rate is lower than the ECC correction
capability; and to (2)~use \emph{sampling} to efficiently tune
the read reference voltage to its optimal value to reduce the
read operation latency. Prior characterization of real NAND
flash memory~\cite{R37, R54} found that the value of $V_{opt}$ does
\emph{not} shift greatly over a short period of time (e.g., a day), and
that all pages within a block experience \emph{similar} amounts of
threshold voltage shifts, as they have the same amount of
wearout and are programmed around the same time~\cite{R37, R54}. 
Therefore, we can invoke our $V_{opt}$ learning mechanism
periodically (e.g., daily) to efficiently tune the \emph{initial
read reference voltage} (i.e., the first read reference voltage
used when the controller invokes the read-retry mechanism,
described in Section~\ref{sec:mitigation:retry}) for each flash block, ensuring that
the initial voltage used by read-retry stays close to $V_{opt}$ even
as the threshold voltage distribution shifts.

The SSD controller searches for $V_{opt}$ by counting the
number of errors that need to be corrected by ECC during
a read. However, there may be times where the initial
read reference voltage ($V_{initial}$) is set to a value at which the
number of errors during a read exceeds the ECC correction
capability, such as the raw bit error rate for $V_{initial}$ in Figure~\ref{fig:F23}
(right). When the ECC correction capability is exceeded, the
SSD controller is unable to count how many errors exist in
the raw data. The SSD controller uses \emph{disparity-based read
reference voltage approximation}~\cite{R90, R165, R166} for each
flash block to try to bring $V_{initial}$ to a region where the number
of errors does not exceed the ECC correction capability.
Disparity-based read reference voltage approximation takes
advantage of data scrambling. Recall from \chIV{Section~\ref{sec:ssdarch:ctrl:scrambling}} that
to minimize data value dependencies for the error rate, the
SSD controller scrambles the data written to the SSD to
probabilistically ensure that an equal number of 0s and 1s
exist in the flash memory cells. The key idea of disparity-based
read reference voltage approximation is to find the
read reference voltages that result in approximately 50\%
of the cells reading out bit value 0, and the other 50\% of
the cells reading out bit value 1. To achieve this, the SSD
controller employs a binary search algorithm, which tracks
the ratio of 0s to 1s for each read reference voltage it tries.
The binary search tests various read reference voltage values,
using the ratios of previously tested voltages to narrow
down the range where the read reference voltage can have
an equal ratio of 0s to 1s. The binary search algorithm continues
narrowing down the range until it finds a read reference
voltage that satisfies the ratio.

\begin{figure}[h]
  \centering
  \includegraphics[width=0.65\columnwidth]{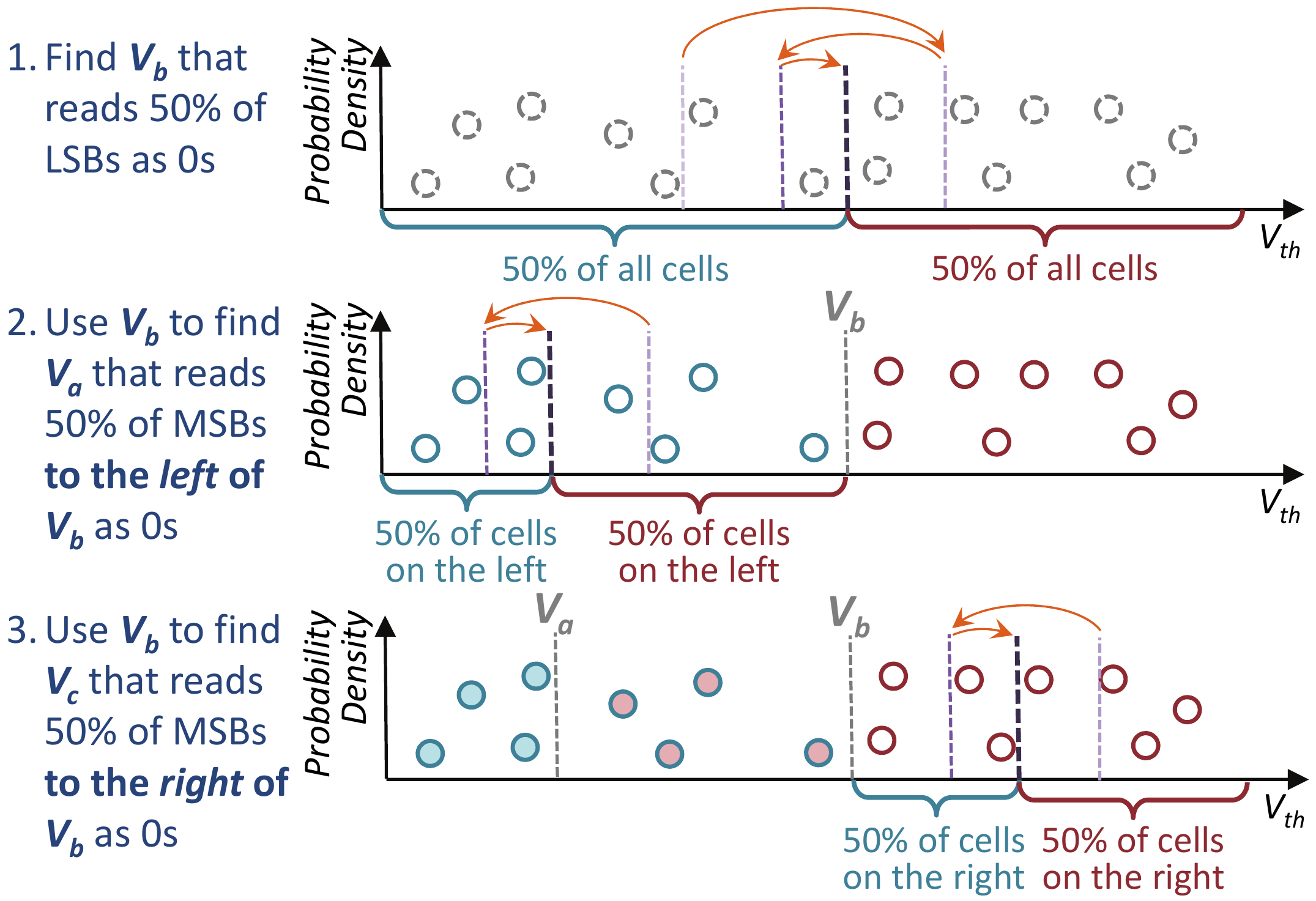}%
  \vspace{-5pt}%
  \caption{Disparity-based read reference voltage approximation to
find $V_{initial}$ for MLC NAND flash memory. Each circle represents a
cell, where a dashed border indicates that the LSB is undetermined,
a solid border indicates that the LSB is known, a hollow circle
indicates that the MSB is unknown, and a filled circle indicates that
the MSB is known. \chI{Reproduced from \cite{cai.arxiv17}.}}%
  \label{fig:F24}%
\end{figure}

The usage of the binary search algorithm depends on the
type of NAND flash memory used within the SSD. For SLC
NAND flash, the controller searches for only a single read
reference voltage. For MLC NAND flash, there are three read
reference voltages: the LSB is determined using $V_b$, and the
MSB is determined using both $V_a$ and $V_c$ (see Section~\ref{sec:flash:read}).
Figure~\ref{fig:F24} illustrates the search procedure for MLC NAND flash.
First, the controller uses binary search to find $V_b$, choosing a
voltage that reads the LSB of 50\% of the cells as data value 0
(step~1 in Figure~\ref{fig:F24}). For the MSB, the controller uses the discovered
$V_b$ value to help search for $V_a$ and $V_c$. Due to scrambling,
cells should be equally distributed across each of the
four voltage states. The controller uses binary search to set
$V_a$ such that 25\% of the cells are in the ER state, by ensuring
that half of the cells \emph{to the left of $V_b$} are read with an MSB of
0 (step~2). Likewise, the controller uses binary search to set
$V_c$ such that 25\% of the cells are in the P3 state, by ensuring
that half of the cells \emph{to the right of $V_b$} are read with an
MSB of 0 (step~3). This procedure is extended in a similar
way to approximate the voltages for TLC NAND flash.

If disparity-based approximation finds a value for $V_{initial}$
where the number of errors during a read can be counted by
the SSD controller, the controller invokes \emph{sampling-based
adaptive $V_{opt}$ discovery}~\cite{R37} to minimize the error count, and
thus reduce the read latency. Sampling-based adaptive $V_{opt}$
discovery learns and records $V_{opt}$ for the \emph{last-programmed
page} in each block. We sample only the last-programmed
page because it is the page with the lowest data retention
age in the flash block. As retention errors cause the higher-voltage
states to shift to the left (i.e., to lower voltages), the
last-programmed page usually provides an \emph{upper bound} of
$V_{opt}$ for the entire block.

During sampling-based adaptive $V_{opt}$ discovery, the SSD
controller first reads the last-programmed page using $V_{initial}$,
and attempts to correct the errors in the raw data read from
the page. Next, it records the number of raw bit errors as
the current lowest error count $N_{ERR}$, and sets the applied
read reference voltage ($V_{ref}$) as $V_{initial}$. Since $V_{opt}$ typically
decreases over retention age, the controller first attempts
to lower the read reference voltage for the last-programmed
page, decreasing the voltage to $V_{ref} - \Delta V$ and reading the
page. If the number of corrected errors in the new read is
less than or equal to the old $N_{ERR}$, the controller updates
$N_{ERR}$ and $V_{ref}$ with the new values. The controller continues
to lower the read reference voltage until the number
of corrected errors in the data is greater than the old $N_{ERR}$
or the lowest possible read reference voltage is reached.
Since the optimal threshold voltage might increase in rare
cases, the controller also tests increasing the read reference
voltage. It increases the voltage to $V_{ref} + \Delta V$ and reads
the last-programmed page to see if $N_{ERR}$ decreases. Again, it
repeats increasing $V_{ref}$ until the number of corrected errors
in the data is greater than the old $N_{ERR}$ or the highest possible
read reference voltage is reached. The controller sets the
initial read reference voltage of the block as the value of $V_{ref}$
at the end of this process so that the next time an uncorrectable
error occurs, read-retry starts at a $V_{initial}$ that is hopefully
closer to the optimal read reference voltage ($V_{opt}$).

During the course of the day, as more retention errors
(the dominant source of errors on already-programmed
blocks) accumulate, the threshold voltage distribution shifts
to the left (i.e., voltages decrease), and our initial read reference
voltage (i.e., $V_{initial}$) is now an upper bound for the read-retry
voltages. Therefore, whenever read-retry is invoked,
the controller now needs to only decrease the read reference
voltages (as opposed to traditional read-retry, which
tries \emph{both} lower and higher voltages~\cite{R37}). Sampling-based
adaptive $V_{opt}$ discovery improves the \emph{endurance} (i.e., the
number of P/E cycles before the ECC correction capability is
exceeded) of the NAND flash memory by 64\% and reduces
error correction latency by 10\%~\cite{R37}, and is employed in
some modern SSDs today.

\paratitle{Other Approaches to Optimizing Read Reference Voltages}
One drawback of the sampling-based adaptive technique is
that it requires time and storage overhead to find and record
the per-block initial voltages. To avoid this, the SSD controller
can employ an accurate \emph{online threshold voltage distribution
model}~\cite{R33, R42, R177}, which can efficiently track and
predict the shift in the distribution over time. The model
represents the threshold voltage distribution of each state as
a probability density function (PDF), and the controller can
use the model to calculate the intersection of the different
PDFs. The controller uses the PDF in place of the threshold
voltage sampling, determining $V_{opt}$ by calculating the intersection
of the distribution of each state in the model. The
endurance improvement from our state-of-the-art model-based
$V_{opt}$ estimation technique~\cite{R42} is within 2\% of the
improvement from an ideal $V_{opt}$ identification mechanism~\cite{R42}. 
An online threshold voltage distribution model can be
used for a number of other purposes, such as estimating the
future growth in the raw bit error rate and improving error
correction~\cite{R42}.

Other prior work examines adapting read reference voltages
based on P/E cycle count, retention age, or read disturb.
In one such work, the controller periodically learns
read reference voltages by testing three read reference voltages
on six pages per block, which the work demonstrates
to be sufficiently accurate~\cite{R54}. Similarly, error correction
using LDPC soft decoding (see Section~\ref{sec:correction:ldpc}) requires reading
the same page using multiple sets of read reference
voltages to provide fine-grained information on the probability
of each cell representing a bit value 0 or a bit value 1.
Another prior work optimizes the read reference voltages to
increase the ECC correction capability without increasing
the coding rate~\cite{R73}.

\paratitle{Optimizing Pass-Through Voltage to Reduce Read Disturb Errors}
As we discussed in Section~\ref{sec:errors:readdisturb}, the vulnerability of a
cell to read disturb is directly correlated with the voltage difference
($V_{pass} - V_{th}$) through the cell oxide~\cite{R38}. Traditionally,
a single $V_{pass}$ value is used \emph{globally} for the entire flash memory,
and the value of $V_{pass}$ must be higher than \emph{all} potential
threshold voltages within the chip to ensure that unread
cells along a bitline are turned on during a read operation
(see Section~\ref{sec:flash:read}). To reduce the impact of read disturb,
we can tune $V_{pass}$ to reduce the size of the voltage difference
($V_{pass} - V_{th}$). However, it is difficult to reduce $V_{pass}$ \emph{globally},
as any cell with a value of $V_{th} > V_{pass}$ introduces an
error during a read operation (which we call a \emph{pass-through
error}).

\begin{figure}[h]
  \centering
  \includegraphics[width=0.65\columnwidth]{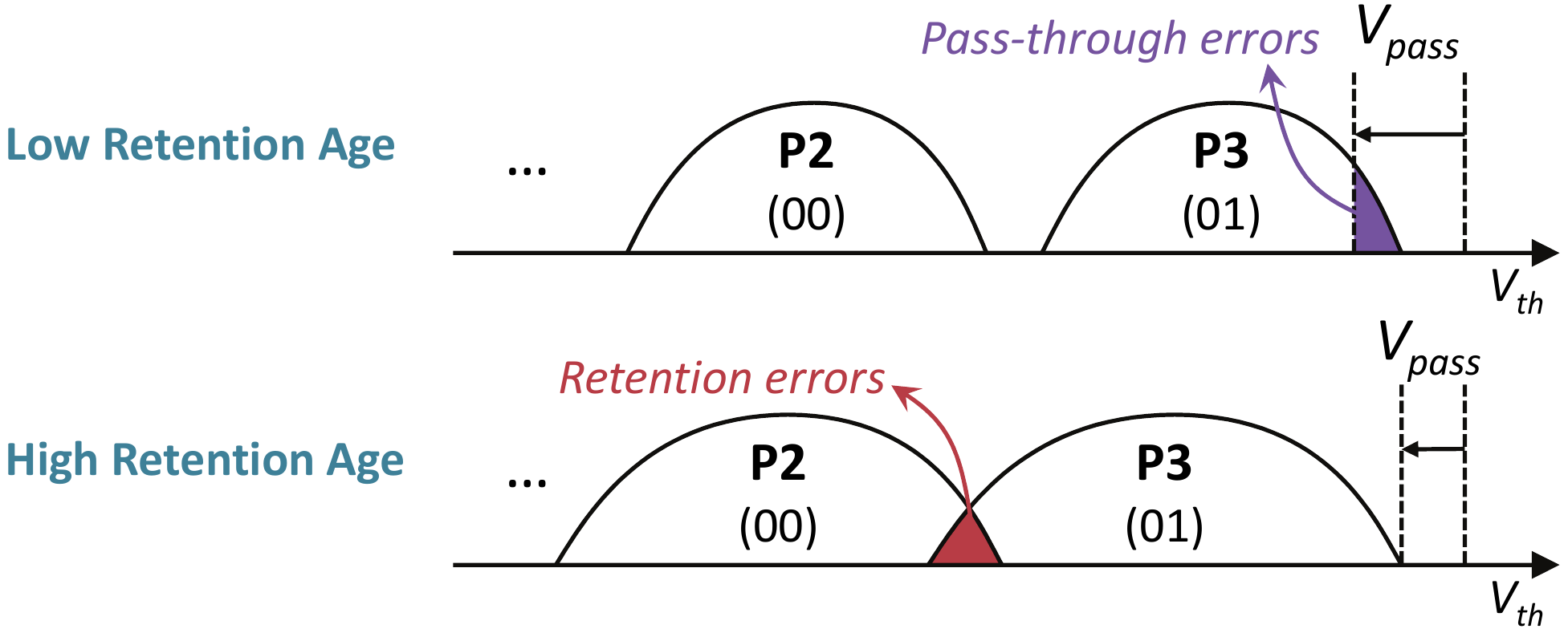}%
  \vspace{-5pt}%
  \caption{Dynamic pass-through voltage tuning at different
retention ages. \chI{Reproduced from \cite{cai.arxiv17}.}}%
  \label{fig:F25}%
\end{figure}

We propose a mechanism that can dynamically lower
$V_{pass}$ while ensuring that it can correct any new pass-through
errors introduced. The key idea of the mechanism is to lower
$V_{pass}$ only for those blocks where ECC has enough leftover
error correction capability (see \chIV{Section~\ref{sec:ssdarch:ctrl:ecc}}) to correct the
newly introduced pass-through errors. When the retention
age of the data within a block is low, we find that the raw
bit error rate of the block is much lower than the rate for
the block when the retention age is high, as the number of
data retention and read disturb errors remains low at low
retention age~\cite{R38, R70}. As a result, a block with a low retention
age has significant \emph{unused} ECC correction capability,
which we can use to correct the pass-through errors we
introduce when we lower $V_{pass}$, as shown in Figure~\ref{fig:F25}. Thus,
when a block has a low retention age, the controller lowers
$V_{pass}$ aggressively, making it much less likely for read disturbs
to induce an uncorrectable error. When a block has
a high retention age, the controller also lowers $V_{pass}$, but
does not reduce the voltage aggressively, since the limited
ECC correction capability now needs to correct retention
errors, and might not have enough unused correction capability
to correct many new pass-through errors. By reducing
$V_{pass}$ aggressively when a block has a low retention age, we
can extend the time before the ECC correction capability is
exhausted, improving the flash lifetime.

Our read disturb mitigation mechanism~\cite{R38} learns the
minimum pass-through voltage for each block, such that
all data within the block can be read correctly with ECC.
Our learning mechanism works online and is triggered
periodically (e.g., daily). The mechanism is implemented in
the controller, and has two components. It first finds the
size of the ECC margin $M$ (i.e., the unused correction capability)
that can be exploited to tolerate additional read errors
for each block. Once it knows the available margin $M$, our
mechanism calibrates $V_{pass}$ on a per-block basis to find the
lowest value of $V_{pass}$ that introduces no more than $M$ additional
raw errors (i.e., there are no more than M cells where
$V_{th} > V_{pass}$). Our findings on MLC NAND flash memory
show that the mechanism can improve flash endurance by
an average of 21\% for a variety of disk workloads~\cite{R38}.

\paratitle{Programming and Erase Voltages}
Prior work also examines
tuning the programming and erase voltages to extend
flash endurance~\cite{R74}. By decreasing the two voltages when
the P/E cycle count is low, the accumulated wearout for
each program or erase operation is reduced, which, in turn,
increases the overall flash endurance. Decreasing the programming
voltage, however, comes at the cost of increasing
the time required to perform ISPP, which, in turn, increases
the overall SSD write latency~\cite{R74}.

\subsection{Hot Data Management}
\label{sec:mitigation:hotcold}

\begin{figure}[h]
  \centering
  \includegraphics[width=\columnwidth]{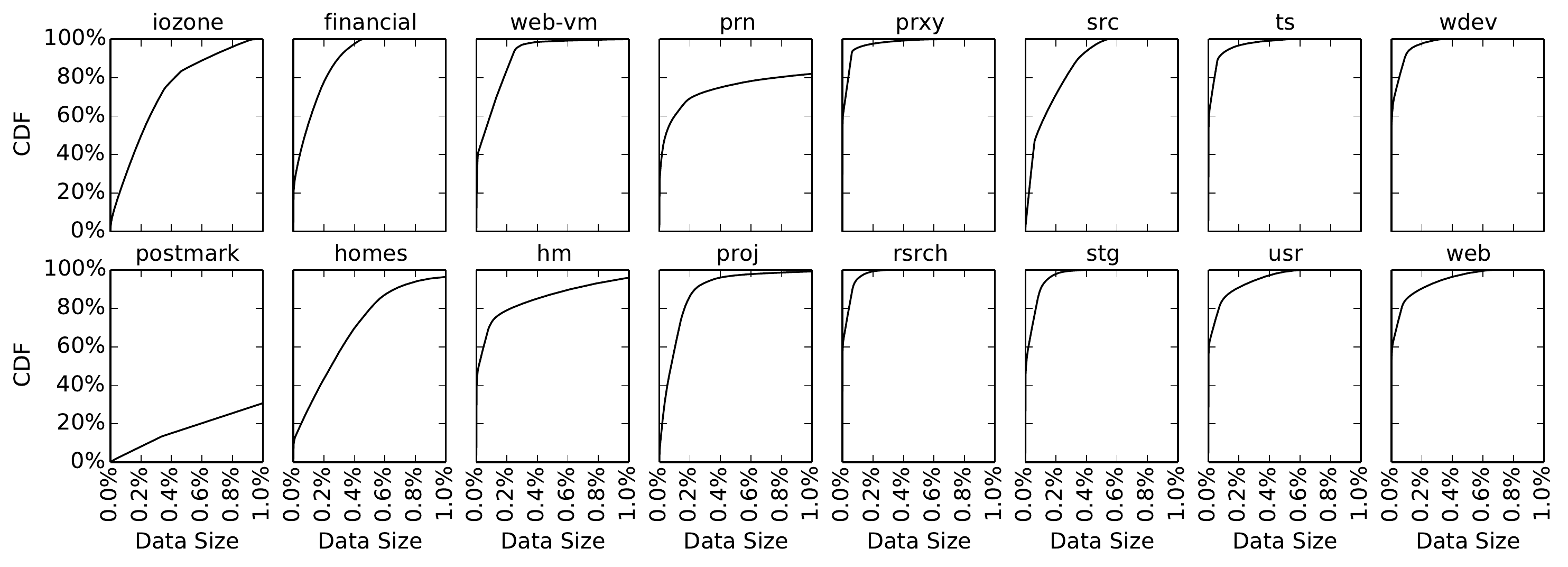}%
  \vspace{-5pt}%
  \caption{\chI{Cumulative distribution function of \chIII{the fraction of writes performed by a workload to NAND flash memory pages,} for
    16~evaluated workloads.  \chIII{For every workload except \emph{postmark}, over 95\% of all writes performed by the workload are destined for less than 1.0\% of the workload's pages.} Total data footprint \chIII{of each workload is} \SI{217.6}{\giga\byte}, i.e., 1.0\% on the x-axis represents \chIII{\SI{2.176}{\giga\byte}} of data.
    Reproduced from \cite{R41}.}}%
  \label{fig:writefreq}%
\end{figure}

The data stored in \chIII{different locations of an} SSD can be accessed by the host at
different rates. For example, we find that across a wide range
of disk workloads, almost 100\% of the write operations target
less than 1\% of the pages within an SSD~\cite{R41}, \chII{as shown in
Figure~\ref{fig:writefreq}.  These pages exhibit high temporal write
locality, and are called \emph{write-hot} pages.}
Likewise, pages with a high
amount of temporal read locality \chII{(i.e., pages that are accessed
by a large fraction of the read operations)} are called \emph{read-hot} pages.
A number of issues can arise when an SSD does not distinguish
between write-hot pages and \emph{write-cold} pages (i.e.,
pages with low temporal write locality), or between read-hot
pages and \emph{read-cold} pages (i.e., pages with low temporal
read locality). For example, if write-hot pages and write-cold
pages are \chII{stored} within the same block, refresh
mechanisms \chIII{(which operate at the block level; see Section~\ref{sec:mitigation:refresh}) \emph{cannot}} avoid refreshes to pages that were overwritten
recently. \chIII{This increases} not only \chIII{the} energy consumption
but also \chIII{the} write amplification due to remapping-based refresh~\cite{R41}.
Likewise, if read-hot and read-cold pages are \chII{stored}
within the same block, read-cold pages are unnecessarily
exposed to a high number of read disturb errors~\cite{R63, R70}.
\emph{Hot data management} refers to a set of mechanisms that can
identify \chII{and exploit} write-hot or read-hot pages in the SSD. 
The key idea \chII{common to such mechanisms}
is to apply special SSD management policies by placing hot
pages and cold pages into \emph{separate} flash blocks.

\chI{\chII{A state-of-the-art hot data management mechanism is
\emph{write-hotness aware refresh management} (WARM)~\cite{R41}, which}
efficiently identifies write-hot pages and uses this information to \chII{carefully}
place pages within blocks.  WARM \chIII{aims to ensure} that every block in the NAND flash
memory contains either \emph{only} write-hot pages or \emph{only} write-cold
pages.  A small pool of blocks in the SSD are designated to exclusively store
\chII{the small amount of write-hot data (as shown in Figure~\ref{fig:writefreq})}.
This block-level segregation between write-hot pages and
write-cold pages allows WARM to apply separate \chII{specialized} management policies
based on the write-hotness of the pages in each block.}

\chI{Two examples of policies for write-hot blocks in WARM
are the \chII{write-hotness-aware refresh policy (see 
Section~\ref{sec:mitigation:refresh} for baseline refresh policies)}
and the \chIII{write-hotness-aware} garbage collection algorithm (\chII{see} \chIV{Section~\ref{sec:ssdarch:ctrl:gc}}).
\chII{In write-hotness-aware refresh}, since write-hot data is overwritten more frequently than the
refresh interval, the SSD controller skips refresh operations to the
write-hot blocks.  \chII{As the retention time for write-hot data never
exceeds the refresh interval, performing refresh to this data
does \chIII{\emph{not}} reduce the error rate.  By skipping refresh for write-hot data,
WARM reduces the total number of writes performed on the SSD, which in turn
increases the SSD lifetime, without introducing uncorrectable errors.}
\chII{In write-hotness-aware garbage collection}, the SSD controller performs \chII{\chIV{\emph{oldest-block-first} garbage
collection}.  WARM sizes the pool of write-hot blocks such that when a write-hot
block becomes the oldest block in the \chIII{pool of write-hot blocks}, all of the data that was in the
block \chIII{is likely to} already have been overwritten.  As a result, all of the pages within the
oldest write-hot block \chIII{is likely to} be invalid, and the block can be erased without the
need to migrate any remaining valid pages to a new block.  By always selecting the
oldest block \chIII{in the pool of write-hot blocks} for garbage collection, the write-hotness-aware garbage collection
algorithm \chIII{(1)}~does not spend time searching for a block to select (as traditional garbage
collection algorithms do)}, and \chIII{(2)}~rarely needs to migrate pages from the
selected block. \chIII{Both of these lead to a reduction in} the performance overhead of
garbage collection.}

\chI{WARM continues to \chIII{use the} traditional controller policies (i.e., the policies
described in Section~\ref{sec:ssdarch:ctrl}) and refresh \chII{mechanisms} for the
write-cold blocks.
\chIII{WARM reduces fragmentation within write-cold blocks (i.e., each write-cold
block is likely to have few, if any, invalid pages), because each page
within the block does \chIV{\emph{not}} get updated frequently by the application.}
Due to the \chIII{write-hotness-aware} policies and
reduced fragmentation, WARM reduces \chII{write amplification significantly,}
which translates to an
average lifetime improvement of 21\% over an SSD that employs a state-of-the-art
refresh mechanism\chII{~\cite{R34} (see \emph{Adaptive Refresh and Read Reclaim Mechanisms} 
in Section~\ref{sec:mitigation:refresh})}, across a \chIII{wide} \chIV{variety} of disk workloads~\cite{R41}.}

Another work~\cite{R75} proposes to reuse the correctly functioning
flash pages within bad blocks (see \chIV{Section~\ref{sec:ssdarch:ctrl:badblocks}}) to store
write-cold data. This technique increases the total number
of usable blocks available for overprovisioning, and extends
flash lifetime by delaying the point at which each flash chip
reaches the upper limit of bad blocks it can tolerate.

RedFTL identifies and replicates read-hot pages across
multiple flash blocks, allowing the controller to evenly
distribute read requests to these pages across the replicas~\cite{R63}. 
Other works reduce the number of read reclaims
(see Section~\ref{sec:mitigation:refresh}) that need to be performed by mapping
read-hot data to particular flash blocks and lowering the
maximum possible threshold voltage for such blocks~\cite{R45, R70}.
By lowering the maximum possible threshold voltage
for these blocks, the SSD controller can use a lower $V_{pass}$
value (see Section~\ref{sec:mitigation:voltage}) on the blocks without introducing
any additional errors during a read operation. To lower the
maximum threshold voltage in these blocks, the width of
the voltage window for each voltage state is decreased, and
each voltage window shifts to the left~\cite{R45, R70}. Another
work applies stronger ECC encodings to \emph{only} read-hot
blocks based on the total read count of the block, in order
to increase SSD endurance without significantly reducing
the amount of overprovisioning~\cite{R88} (see Section~\ref{sec:ssdarch:reliability} for
a discussion on the tradeoff between ECC strength and
overprovisioning).

\subsection{Adaptive Error Mitigation Mechanisms}
\label{sec:mitigation:adaptive}

Due to the many different factors that contribute to raw
bit errors, error rates in NAND flash memory can be highly
variable. Adaptive error mitigation mechanisms are capable of
adapting error tolerance capability to the error rate. They provide
stronger error tolerance capability when the error rate is
higher, improving flash lifetime significantly. When the error
rate is low, adaptive error mitigation techniques reduce error
tolerance capability to lower the cost of the error mitigation
techniques. In this section, we examine two types of adaptive
techniques: (1)~multi-rate ECC and (2)~dynamic cell levels.

\begin{figure}[h]
  \centering
  \includegraphics[width=0.65\columnwidth]{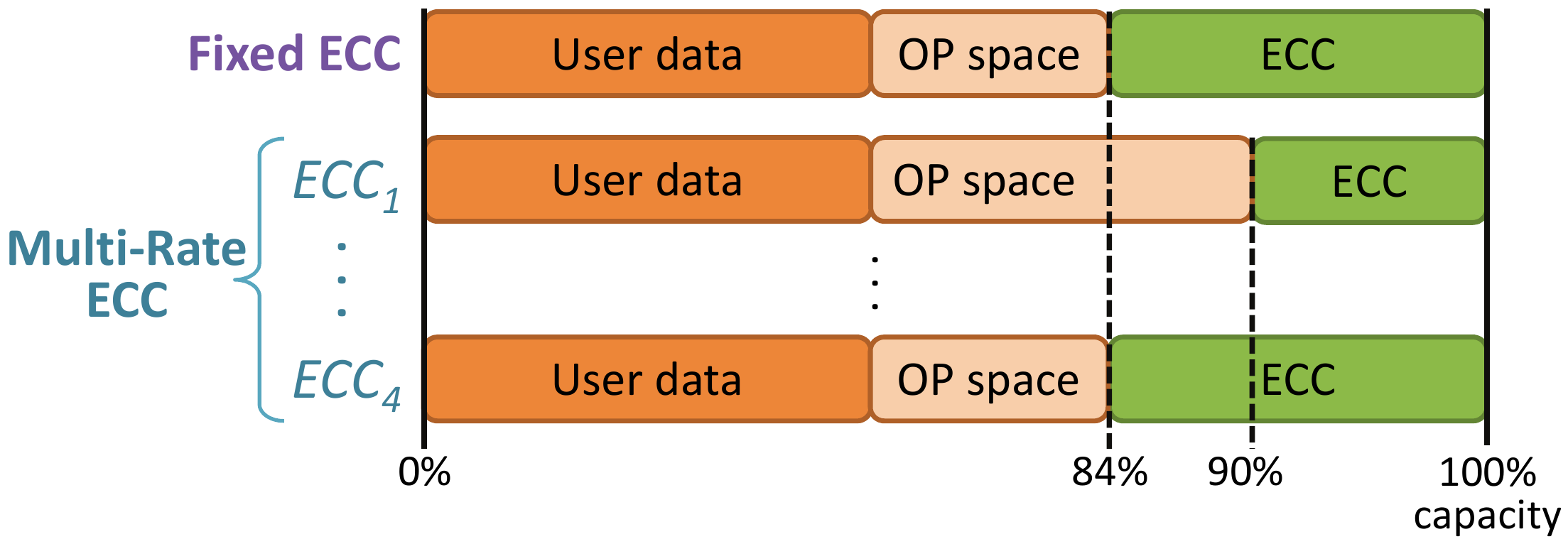}%
  \vspace{-5pt}%
  \caption{Comparison of space used for user data, overprovisioning,
and ECC between a fixed ECC and a multi-rate ECC mechanism. \chI{Reproduced from \cite{cai.arxiv17}.}}%
  \label{fig:F26}%
\end{figure}

\paratitle{Multi-Rate ECC}
Some works propose to employ
multiple ECC algorithms in the SSD controller~\cite{R43, R65, R76, R77, R82}. 
Recall from Section~\ref{sec:ssdarch:reliability} that there is a
tradeoff between ECC strength (i.e., the coding rate; see
\chIV{Section~\ref{sec:ssdarch:ctrl:ecc}}) and overprovisioning, as a codeword (which
contains a data chunk \emph{and} its corresponding ECC information)
uses more bits when stronger ECC is employed. The
key idea of multi-rate ECC is to employ a weaker codeword
(i.e., one that uses fewer bits for ECC) when the SSD is relatively
new and has a smaller number of raw bit errors, and
to use the saved SSD space to provide additional overprovisioning,
as shown in Figure~\ref{fig:F26}.

\begin{figure}[h]
  \centering
  \includegraphics[width=0.4\columnwidth]{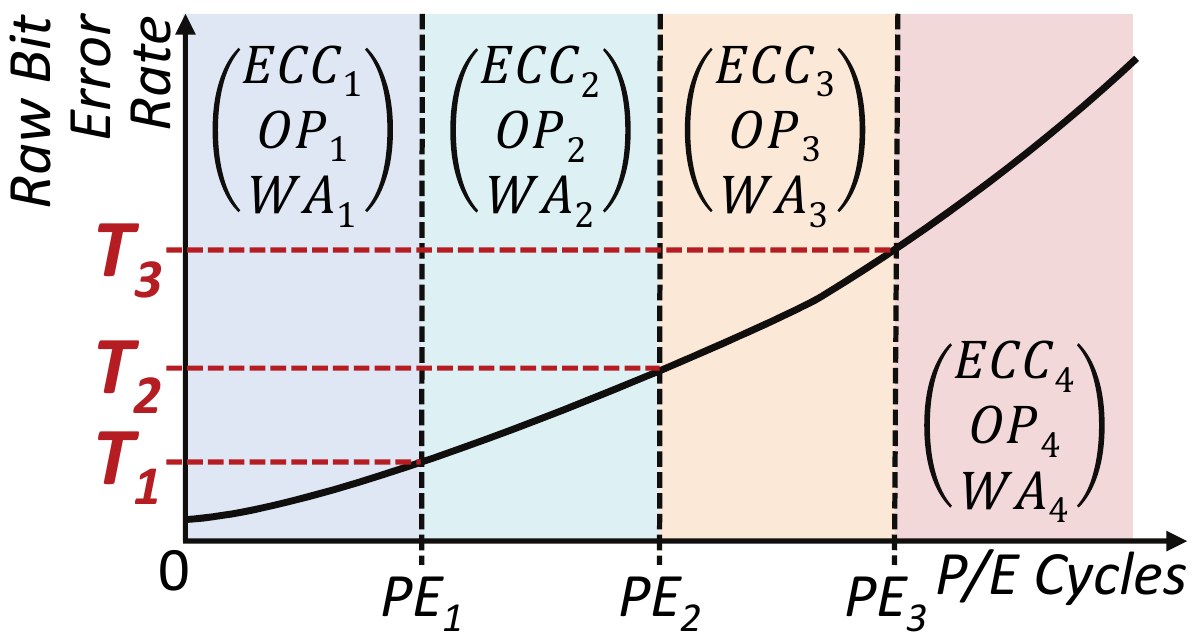}%
  \vspace{-5pt}%
  \caption{Illustration of how multi-rate ECC switches to different
ECC codewords (i.e., ECC$_i$) as the RBER grows. OP$_i$ is the
overprovisioning factor used for engine ECC$_i$, and WA$_i$ is the
resulting write amplification value. \chI{Reproduced from \cite{cai.arxiv17}.}}%
  \label{fig:F27}%
\end{figure}

Let us assume that the controller contains a configurable
ECC engine that can support $n$ different types of ECC
codewords, which we call $\text{ECC}_i$. Figure~\ref{fig:F26} shows an example
of multi-rate ECC that uses four ECC engines, where $\text{ECC}_1$
provides the weakest protection but has the smallest codeword,
while $\text{ECC}_4$ provides the strongest protection with
the largest codeword. We need to ensure that the NAND
flash memory has enough space to fit the largest codewords,
e.g., those for $\text{ECC}_4$ in Figure~\ref{fig:F26}. Initially, when the raw bit
error rate (RBER) is low, the controller employs $\text{ECC}_1$,
as shown in Figure~\ref{fig:F27}. The smaller codeword size for $\text{ECC}_1$
provides additional space for overprovisioning, as shown
in Figure~\ref{fig:F26}, and thus reduces the effects of write amplification.
Multi-rate ECC works on an interval-by-interval
basis. Every interval (in this case, a predefined number
of P/E cycles), the controller measures the RBER. When
the RBER exceeds the threshold set for transitioning from
a weaker ECC to a stronger ECC, the controller switches
to the stronger ECC. For example, when the SSD exceeds
the first RBER threshold for switching ($T_1$ in Figure~\ref{fig:F27}),
the controller starts switching from $\text{ECC}_1$ to $\text{ECC}_2$. When
switching between ECC engines, the controller uses the
$\text{ECC}_1$ engine to decode data the next time the data is read
out, and stores a new codeword using the $\text{ECC}_2$ engine.
This process is repeated during the lifetime of flash memory
for each stronger engine $\text{ECC}_i$, where each engine has
a corresponding threshold that triggers switching~\cite{R43, R65, R82}, 
as shown in Figure~\ref{fig:F27}.

Multi-rate ECC allows the same maximum P/E cycle
count for each block as if $\text{ECC}_n$ was used throughout the
lifetime of the SSD, but reduces write amplification and
improves performance during the periods where the lower
strength engines are employed, by providing additional
overprovisioning (see Section~\ref{sec:ssdarch:reliability}) during those times.
As the lower-strength engines use smaller codewords
(e.g., $\text{ECC}_1$ versus $\text{ECC}_4$ in Figure~\ref{fig:F26}), the resulting free space
can instead be employed to further increase the amount of
overprovisioning within the NAND flash memory, which in
turn increases the total lifetime of the SSD. We compute
the lifetime improvement by modifying Equation~\ref{eq:E4} (Section~\ref{sec:ssdarch:reliability})
to account for each engine, as follows:
\begin{equation}
\text{Lifetime} = \sum_{i = 1}^{n} \frac{\text{PEC}_i \times (1 + \text{OP}_i)}{365 \times \text{DWPD} \times \text{WA}_i \times R_{compress}}
\label{eq:E9}
\end{equation}
In Equation~\ref{eq:E9}, $\text{WA}_i$ and $\text{OP}_i$ are the write amplification and overprovisioning
factor for $\text{ECC}_i$, and $\text{PEC}_i$ is the number of P/E
cycles that $\text{ECC}_i$ is used for. Manufacturers can set parameters
to maximize SSD lifetime in Equation~\ref{eq:E9}, by optimizing the values
of $\text{WA}_i$ and $\text{OP}_i$.

\begin{figure}[h]
  \centering
  \includegraphics[width=0.65\columnwidth]{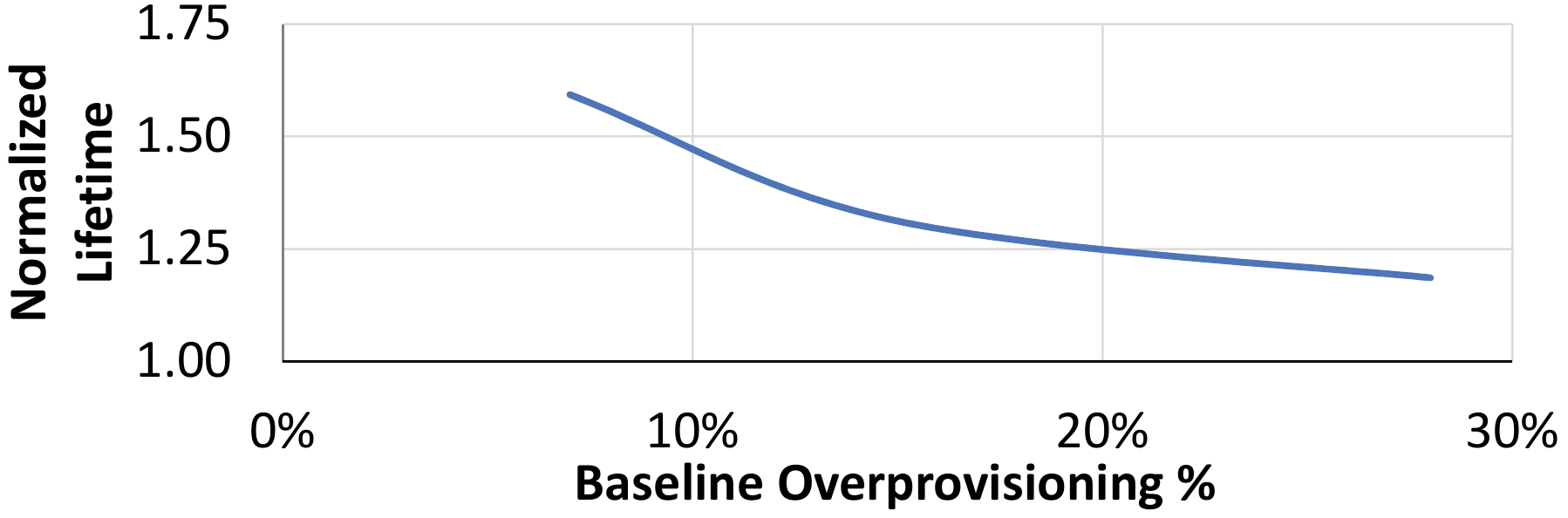}%
  \vspace{-5pt}%
  \caption{Lifetime improvements of using multi-rate ECC over using
a fixed ECC coding rate. \chI{Reproduced from \cite{cai.arxiv17}.}}%
  \label{fig:F28}%
\end{figure}

Figure~\ref{fig:F28} shows the lifetime improvements for a four-engine
multi-rate ECC, with the coding rates for the four
ECC engines ($\text{ECC}_1$--$\text{ECC}_4$) set to 0.90, 0.88, 0.86, and 0.84
(recall that a \emph{lower} coding rate provides stronger protection;
see Section~\ref{sec:ssdarch:reliability}), over a fixed ECC engine that employs a
coding rate of 0.84. We see that the lifetime improvements
of using multi-rate ECC are: (1)~significant, with a 31.2\%
increase if the baseline NAND flash memory has 15\% overprovisioning;
and (2)~greater when the SSD initially has a
smaller amount of overprovisioning.

\paratitle{Dynamic Cell Levels}
A major reason that errors occur
in NAND flash memory is because the threshold voltage distribution
of each state overlaps more with those of neighboring
states as the distributions widen over time. Distribution
overlaps are a greater problem when more states are encoded
within the same voltage range. Hence, TLC flash has a much
lower endurance than MLC, and MLC has a much lower
endurance than SLC (assuming the same process technology
node). If we can increase the margins between the
states' threshold voltage distributions, the amount of overlap
can be reduced significantly, which in turn reduces the
number of errors.

\begin{figure}[h]
  \centering
  \includegraphics[width=0.75\columnwidth]{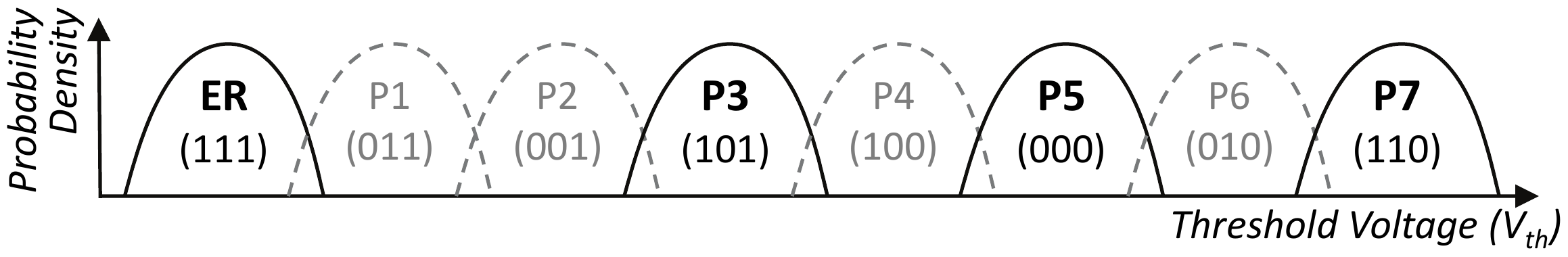}%
  \vspace{-5pt}%
  \caption{States used when a TLC cell (with 8 states) is downgraded
to an MLC cell (with 4 states). \chI{Reproduced from \cite{cai.arxiv17}.}}%
  \label{fig:F29}%
\end{figure}

Prior work proposes to increase margins by \emph{dynamically}
reducing the number of bits stored within a cell, e.g.,
by going from three bits that encode eight states (TLC)
to two bits that encode four states (equivalent to MLC),
or to one bit that encodes two states (equivalent to SLC)~\cite{R45, R78}. 
Recall that TLC uses the ER state and states
P1--P7, which are spaced out approximately equally.
When we \emph{downgrade} a flash block (i.e., reduce the number
of states its cells can represent) from eight states to
four, the cells in the block now employ only the ER state
and states P3, P5, and P7. As we can see from Figure~\ref{fig:F29}, this
provides large margins between states P3, P5, and P7, and
provides an even larger margin between ER and P3. The
SSD controller maintains a list of all of the blocks that
have been downgraded. For each read operation, the SSD
controller checks if the target block is in the downgraded
block list, and uses this information to interpret the data
that it reads out from the wordline of the block.

A cell can be downgraded to reduce various types of
errors (e.g., wearout, read disturb). To reduce wearout, a
cell is downgraded when it has high wearout. To reduce read
disturb, a cell can be downgraded if it stores \emph{read-hot} data
(i.e., the most frequently read data in the SSD). By using
fewer states for a block that holds read-hot data, we can
reduce the impact of read disturb because it becomes harder
for the read disturb mechanism to affect the distributions
enough for them to overlap. As an optimization, the SSD
controller can employ various hot-cold data partitioning
mechanisms (e.g., \cite{R41, R45, R63, R88}) to keep read-hot
data in specially designated blocks~\cite{R45, R63, R70, R88},
allowing the controller to reduce the size of the downgraded
block list and isolate the impact of read disturb from \emph{read-cold}
(i.e., infrequently read) data.

Another approach to dynamically increasing the distribution
margins is to perform program and erase operations
more slowly when the SSD write request throughput is low~\cite{R45, R74}. 
Slower program/erase operations allow the final
voltage of a cell to be programmed more precisely, and
reduce the amount of oxide degradation that occurs during
programming. As a result, the distribution of each state is
initially much narrower, and subsequent widening of the
distributions results in much lower overlap for a given P/E
cycle count. This technique improves the SSD lifetime by
an average of 61.2\% for a variety of disk workloads~\cite{R74}.
Unfortunately, the slower program/erase operations come
at the cost of higher SSD latency, and are thus not applied
during periods of high write traffic. One way to mitigate
the impact of the higher write latency is to perform slower
program/erase operations only during garbage collection,
which ensures that the higher latency occurs only when the
SSD is idle~\cite{R45}. As a result, read and write requests from
the host do not experience any additional delays.


\section{Error Correction and Data Recovery Techniques}
\label{sec:correction}

Now that we have described a variety of error mitigation
mechanisms that can target various types of error sources,
we turn our attention to the error correction flow that is
employed in modern SSDs as well as \emph{data recovery techniques}
that can be employed when the error correction flow
fails to produce correct data.
\chI{In this section, we briefly overview the major error correction steps
an SSD performs when reading data. 
We first discuss two ECC encodings that are typically used by modern SSDs:
Bose--Chaudhuri--Hocquenghem (BCH) codes~\cite{R9, R10, R92, R93} and 
low-density parity-check (LDPC) codes\chV{~\cite{R9, R94, R95, gallager.tit62}}
(Section~\ref{sec:correction:ecc}).
Next, we go through example error correction flows for an SSD that uses
either BCH codes or LDPC codes (Section~\ref{sec:correction:flow}).
Then, we compare the error correction strength (i.e., the
number of errors that ECC can correct) when we employ
BCH codes or LDPC codes in an SSD (Section~\ref{sec:correction:strength}). Finally,}
we discuss techniques that can rescue data from an SSD
when the BCH/LDPC decoding fails to correct all errors
(Section~\ref{sec:correction:recovery}).

\subsection{\chI{Error-Correcting Codes Used in SSDs}}
\label{sec:correction:ecc}

Modern SSDs typically employ one of two types of
ECC. Bose--Chaudhuri--Hocquenghem (BCH) codes
allow for the correction of multiple bit errors~\cite{R9, R10, R92, R93}, 
and are used to correct the errors observed during
a \emph{single} read from the NAND flash memory~\cite{R10}. Low-density
parity-check (LDPC) codes employ information
accumulated over \emph{multiple} read operations to determine
the likelihood of each cell containing a bit value 1 or a bit
value 0\chV{~\cite{R9, R94, R95, gallager.tit62}}, providing stronger protection at
the cost of greater decoding latency and storage overhead~\cite{R11, R73}.
\chII{Next, we describe the basics of BCH and LDPC codes.}

\subsubsection{\chI{Bose--Chaudhuri--Hocquenghem (BCH) Codes}}
\label{sec:correction:ecc:bch}

\chI{BCH codes\chV{~\cite{R9, R10, R92, R93}} \chIII{have been} widely used in modern SSDs \chII{during} the past decade due to \chII{their}
ability to detect and correct multi-bit errors while keeping \chII{the latency
and hardware cost of encoding and decoding low}~\cite{R10, micheloni.isscc06, chen.tit81,
marelli.chapter16}. 
\chIII{For SSDs, BCH codes are designed to be \emph{systematic}, which means that the 
\chIV{original data message is embedded \chV{\emph{verbatim}} within the codeword}.}
\chII{Within an $n$-bit codeword} \chII{(see 
\chIV{Section~\ref{sec:ssdarch:ctrl:ecc}}), error-correcting codes use} the first $k$ bits
\chII{of the codeword, \chV{called \emph{data bits},} to hold the \chIV{data} message bits,}
and the remaining $(n-k)$ bits, \chIV{called \emph{check bits},} \chII{to hold}
error correction information \chII{\chIII{that} protects} the data bits. 
BCH codes are designed \chIII{to \emph{guarantee} that they}
correct up to a certain number of raw bit errors (e.g., \chV{$t$} error bits)
within each codeword, \chII{which depends on the values chosen for $n$ and $k$}.
A stronger error correction strength (i.e., a larger \chV{$t$})
requires more redundant \chIV{check} bits (i.e., $(n-k)$) or a longer codeword length (i.e.,
$n$).}

\chII{A BCH code\chIV{~\cite{R9, R10, R92, R93}} is a linear block code that consists of check bits generated by
an algorithm.  The codeword generation algorithm ensures that the check bits are selected such that
the check bits can be used during a parity check to detect and correct up to
\chV{$t$} bit errors in the codeword.  A BCH code is defined by (1)~a generator matrix 
$G$, which informs the generation algorithm of how to generate each check bit
using the data bits; and (2)~a parity check matrix $H$, which can be applied to
the codeword to detect if any errors exist.  In order for a BCH code to guarantee
that it can correct \chV{$t$} errors within each codeword, the minimum separation \chIII{$d$}
(i.e., the Hamming distance) between valid codewords must be at least
\chV{$d = 2t+1$}~\cite{R9}.}

\paratitle{\chII{BCH Encoding}}
\chI{\chII{The codeword generation algorithm encodes a $k$-bit \chIV{data} message $m$ into an
$n$-bit BCH codeword $c$, by computing the dot product of \chIII{$m$} and the 
generator matrix $G$ (i.e., $c = m \cdot G$).  $G$ is defined
within a finite Galois field \chIII{$GF(2^d) = \{0, \alpha^0, \alpha^1, \ldots,
\alpha^{2^d-1}\}$, where $\alpha$ is a \emph{primitive element} of the field
and $d$ is a positive integer}~\cite{dolecek.fms14}.
An SSD manufacturer constructs $G$} from a set of
polynomials \chII{$g_1(x), g_2(x), \ldots g_{2t}(x)$}, where $g_i(\alpha^i) = 0$. 
\chIII{Each polynomial generates a \emph{parity bit}, which is 
used during decoding to determine if any errors were introduced.}
The
$i$-th row of $G$ encodes the $i$-th polynomial $g_i(x)$. When decoding, the
codeword $c$ can be viewed as a polynomial $c(x)$. Since $c(x)$ is generated by
$g_i(x)$ which has a root $\alpha^i$, $\alpha^i$ should also be a root of
$c(x)$. The parity check matrix $H$ is constructed such that $cH^t$ calculates
$c(\alpha_i)$. Thus, the element \chIII{in the $i$-th row and $j$-th column of $H$} is
$H_{ij} = \alpha^{(j-1) (i+1)}$. \chIII{This allows the decoder to use $H$ to
quickly determine if any of the parity bits do not match, which indicates that
there are errors within the codeword.}  \chIII{BCH codes in SSDs are
typically designed to be \emph{systematic}, which \chIV{guarantees that a
verbatim copy of the \chIV{data} message is embedded within}
the codeword. To form a systematic BCH code,} the generator
matrix and the parity check matrix are transformed such that \chIII{they
contain the identity matrix.}

\paratitle{\chI{BCH Decoding}}
\chIII{When the SSD controller is servicing a read request, it must extract
the data bits (i.e., the $k$-bit \chIV{data} message $m$) from the BCH codeword that is 
stored in the NAND flash memory chips.
Once the controller retrieves the codeword, which we call $r$, from NAND flash 
memory, it sends $r$ to a BCH decoder.  The decoder performs five steps, as 
illustrated in Figure~\ref{fig:bch-decoding-flow}, which correct the retrieved 
codeword $r$ to obtain the originally-written codeword $c$, and then extract 
the \chIV{data} message $m$ from $c$.
In Step~1, the decoder uses \emph{syndrome calculation} to detect if any errors exist
within the retrieved codeword $r$.  If no errors are detected, the decoder uses
the retrieved codeword as the original codeword, $c$, and skips
to Step~5.  Otherwise, the decoder continues on to correct the errors and
recover $c$.
In Step~2, the decoder uses the syndromes from Step~1 to construct an
\emph{error location polynomial}, which encodes the locations of each
detected bit error within $r$.
In Step~3, the decoder extracts the specific location of each detected bit error from the
error location polynomial.
In Step~4, the decoder corrects each detected bit error in the retrieved codeword $r$
to recover the original codeword $c$.
In Step~5, the decoder extracts the \chIV{data} message from the original codeword $c$.
We describe the algorithms \chIV{most commonly used by BCH decoders in 
SSDs~\cite{R10, choi.tvlsi09, liu.sips06}} for each step in detail below.}

\begin{figure}[h]
  \centering
  \includegraphics[width=0.85\linewidth]{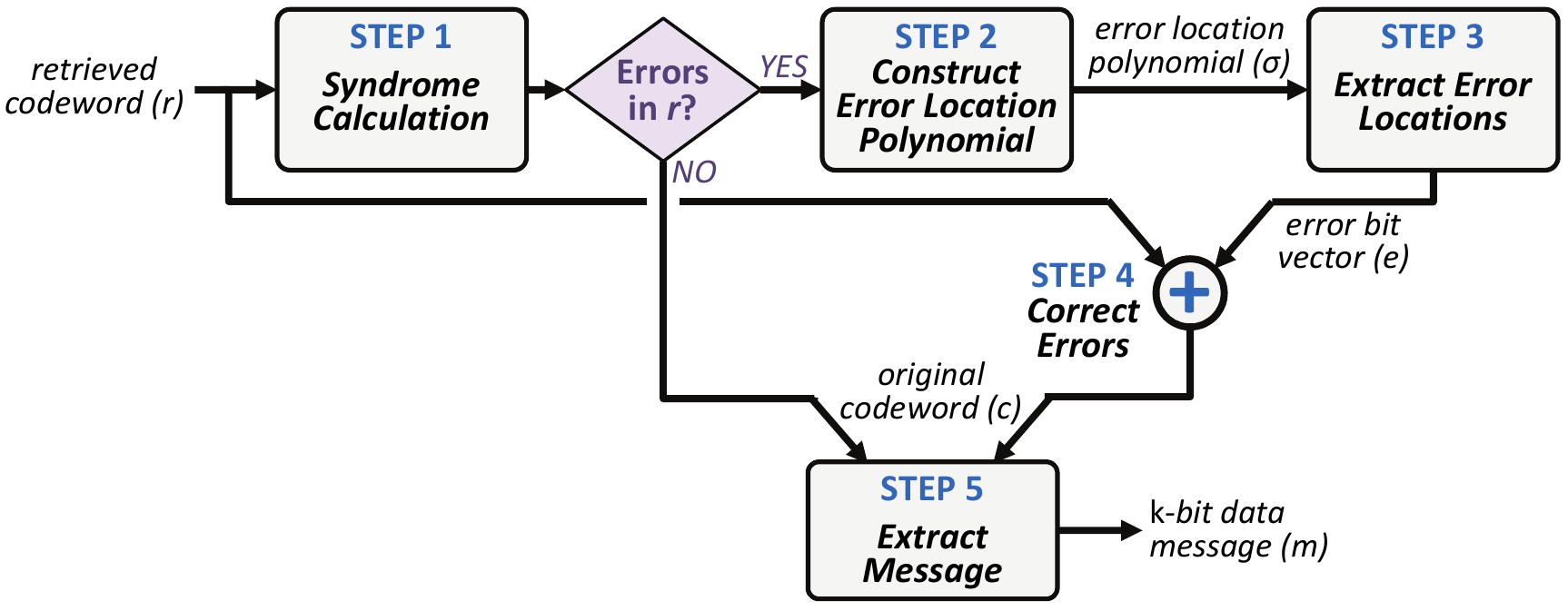}%
  \vspace{-5pt}%
  \caption{\chI{BCH decoding steps.}}%
  \label{fig:bch-decoding-flow}%
\end{figure}

\emph{Step~1---Syndrome Calculation:}  
\chIII{To determine whether the retrieved codeword $r$ contains any errors,
the decoder computes the \emph{syndrome vector}, $S$, which indicates
how many of the parity check polynomials no longer match with the parity  
bits originally computed during encoding.} 
The $i$-th syndrome, $S_i$, is set to one if \chIII{parity bit $i$ does
\chIV{\emph{not}} match its corresponding polynomial, and to zero otherwise.}
To calculate $S$, \chIII{the decoder calculates the dot product of $r$ and
the parity check matrix $H$ (i.e., $S = r \cdot H$).  If every syndrome in
S is set to 0, the decoder does not detect any errors within the codeword,
and skips to Step~5.  Otherwise, the decoder proceeds to Step~2.}

\vspace{3pt}%
\chIII{\emph{Step~2---Constructing the Error Location Polynomial:}}
\chIII{A state-of-the-art BCH decoder uses the} Berlekamp--Massey
algorithm\chIV{~\cite{berlekamp.isit67, massey.tit69, chen.tit81, ryan.cup09}} to construct an error location
polynomial, $\sigma(x)$, whose roots encode the error locations of the 
\chIII{codeword}:
\begin{equation}
  \sigma(x) =
    1 + \sigma_1 \cdot x + \sigma_2 \cdot x^2 + \ldots + \sigma_b \cdot x^b
  \label{eqn:error-location-polynomial}
\end{equation}
In Equation~\ref{eqn:error-location-polynomial}, \chIV{$b$} is the number of raw bit
errors in the codeword.

The polynomial is constructed \chIII{using} an iterative
process.  Since $b$ is not known \chIII{initially}, the algorithm 
\chIII{initially assumes that} $b = 0$ (i.e., $\sigma(x) = 1$). Then, it
updates $\sigma(x)$ by adding \chIII{a \emph{correction term} to the
equation in each iteration,} until $\sigma(x)$ successfully
encodes all \chIII{of the errors that were detected during syndrome calculation}. 
In each iteration, a \chIII{new} correction
term is calculated \chIII{using both the syndromes from Step~1 and the $\sigma(x)$
equations from prior iterations of the algorithm, as long as these prior values 
of $\sigma(x)$ satisfy certain conditions.}
\chIII{This algorithm successfully finds $\sigma(x)$ after $n = (t+b) / 2$ iterations,
where \chV{$t$} is the maximum number of bit errors correctable by the BCH 
code~\cite{dolecek.fms14}.}

Note that (1)~the highest order of the polynomial, $b$, is \chIV{directly correlated with} the
number of errors in the codeword; (2)~the number of iterations, $n$, is also
proportional to the number of errors; (3)~each iteration is compute-intensive, as
it involves several multiply and add operations; and (4)~this algorithm cannot
be parallelized across iterations, as the computation in each iteration is
dependent on the previous ones.

\vspace{3pt}%
\emph{Step~3---Extracting Bit Error Locations from the Error Polynomial:}
\chIII{A state-of-the-art decoder applies the Chien 
search~\cite{chien.tit64, R9} on} the error
location polynomial to find \chIII{the location of \chIV{\emph{all}} raw bit errors that have
been detected \chIV{during Step~1} in the retrieved codeword $r$}. Each
\chIII{bit error} location is encoded with a known function $f$~\cite{ryan.cup09}.
\chIII{The error polynomial from Step~2 is constructed such that if the $i$-th 
bit of the codeword has an error, the error location polynomial
$\sigma(f(i)) = 0$; otherwise, if the $i$-th bit does \chIV{\emph{not}} have an error,
$\sigma(f(i)) \neq 0$.}
\chIV{The} Chien search simply uses trial-and-error
(i.e., tests if $\sigma(f(i))$ is zero), \chIV{testing each bit in the codeword
starting at bit~0.
As the decoder needs to correct only the first $k$ bits of the codeword that
contain the data message $m$, the Chien search
needs to evaluate only $k$ different values of $\sigma(f(i))$.
The algorithm builds a bit vector $e$, which is the same length as the 
retrieved codeword $r$, where the $i$-th bit of $e$ is set to one if bit~$i$
of $r$ contains a bit error, and is set to zero if bit~$i$ of $r$ does \emph{not}
contain an error, or if $i \geq k$ (since \chV{there is no} need to correct the
parity bits).}

Note that (1)~\chIII{the calculation of $\sigma(f(i))$ is compute-intensive}, but
can be parallelized \chIII{because the calculation of each bit $i$ is independent
of the other bits}, and
(2)~the complexity of Step~3 is \chIII{linearly} correlated with the number of 
\chIII{detected errors in the codeword}.

\vspace{3pt}%
\chIII{\emph{Step~4---Correcting the Bit Errors:}
The decoder corrects each detected bit error location
by flipping the bit at that location in the
retrieved codeword $r$.  
This simply involves XORing $r$ with the error vector $e$
created in Step~3.
After the errors are corrected, the decoder now has 
the estimated value of the originally-written codeword $c$
(i.e., $c = r \oplus e$).
\chIV{The decoded version of $c$ is only an \emph{estimate}
of the original codeword, since if $r$ contains more bit errors
than the maximum number of errors (\chV{$t$}) that the BCH can correct,
there may be some uncorrectable errors that were \chIV{\emph{not}} detected 
during syndrome calculation (Step~1).}  In such cases, the decoder cannot
\emph{guarantee} that it has determined the actual original codeword.
In a modern SSD, the bit error rate of a codeword \emph{after} BCH correction 
is \chIV{expected to be} less than $10^{-15}$~\cite{R12}.}

\vspace{3pt}%
\chIII{\emph{Step~5---Extracting the Message from the Codeword:}
As we discuss above, during BCH codeword encoding, the
generator matrix $G$ contains the identity matrix, to ensure that 
\chIV{\chV{the $k$-bit message} $m$ is embedded verbatim into the codeword $c$.}
Therefore, the decoder recovers \chV{$m$} by simply
truncating the last $(n-k)$ bits from the $n$-bit codeword $c$.}

\paratitle{\chIII{BCH Decoder Latency Analysis}}
\chIII{We can model the latency of the state-of-the-art BCH decoder 
($T_{BCH}^{dec}$) that we described above as:}
\begin{equation}
  T_{BCH}^{dec} = T_{Syndrome} + N \cdot T_{Berlekamp}
    + \frac{k}{p} \cdot T_{Chien}
  \label{eqn:bch-decoding-latency}
\end{equation}
In Equation~\ref{eqn:bch-decoding-latency}, $T_{Syndrome}$ is the latency for
calculating the syndrome, which is determined by the size of the parity check
matrix $H$; $T_{Berlekamp}$ is the latency \chIII{of} one iteration of the
Berlekamp--Massey algorithm; $N$ is the total number of iterations that the
Berlekamp--Massey algorithm \chIII{performs}; $T_{Chien}$ is the latency for deciding
whether or not \chIII{a single bit location contains an error, using the
Chien search; $k$ is the length of the \chIV{data} message $m$; and 
$p$} is the number of bits that are processed in parallel
in Step~3.
In this equation, $T_{Syndrome}$, $T_{Berlekamp}$, $k$, and $p$ are constants
\chIII{for a BCH decoder implementation}, while $N$ and $T_{Chien}$ are proportional to the
raw bit error count of the \chIII{codeword}.
\chIV{Note that Steps~4 and 5 can typically be implemented such that they take less than one
\chV{clock} cycle in modern hardware, and thus their latencies are \chV{\emph{not}} included in
Equation~\ref{eqn:bch-decoding-latency}.}

\subsubsection{\chI{Low-Density Parity-Check (LDPC) Codes}}
\label{sec:correction:ecc:ldpc}

\chI{LDPC codes\chV{~\cite{R9, R94, R95, gallager.tit62}} are now used widely in modern SSDs, as LDPC codes provide 
a stronger error correction capability than BCH codes, albeit at a greater
storage cost~\cite{R11, R73}.
LDPC codes are one type of \emph{capacity-approaching codes}, \chIV{which
are error-correcting codes that come close to the \emph{Shannon limit}, \chV{i.e.,} the 
maximum number of data message bits ($k_{max}$) that can be delivered without errors 
for a certain codeword size ($n$) under a given error 
rate~\cite{shannon.bell48a, shannon.bell48b}.}
Unlike BCH codes, LDPC codes cannot \emph{guarantee} that they will correct a
minimum number of raw bit errors.
Instead, \chIV{a good LDPC code guarantees} that the \emph{failure rate} (i.e., the fraction 
of all reads where \chV{the LDPC code} cannot successfully correct the data) is less than a
target rate for a given number of bit errors.}
\chIII{Like BCH codes, LDPC codes for SSDs are designed to be \emph{systematic}, 
\chV{i.e., to} contain \chIV{the data message verbatim}
within the codeword.}

\chIII{An LDPC code\chV{~\cite{R9, R94, R95, gallager.tit62}} is a linear code that, like a BCH code, consists of check bits generated by
an algorithm.  For an LDPC code, these check bits are used to form a bipartite graph,
where one side of the graph contains nodes that represent each bit in the codeword, and the other 
side of the graph contains nodes that represent the parity check equations used 
to generate each parity bit.  When a codeword containing errors is retrieved from memory, 
an LDPC decoder applies \emph{belief propagation}~\cite{pearl.aaai82} to iteratively identify the
bits within the codeword that are \emph{most likely} to contain a bit error.}

\chI{An LDPC code is defined using a binary \chIII{parity} check matrix $H$, where $H$ is very
sparse (i.e., there are few ones in the matrix).  Figure~\ref{fig:ldpc}a shows an
example $H$ matrix for a \chIII{seven}-bit codeword $c$ (see \chIV{Section~\ref{sec:ssdarch:ctrl:ecc}}). 
\chIII{For an $n$-bit codeword that encodes a $k$-bit \chIV{data} message, $H$ is sized to
be an $(n-k) \times n$ matrix.}
Within the matrix, each \emph{row} 
represents a \emph{parity check equation}, while each \emph{column} represents one of the
\chIII{seven} bits in the codeword.  As our example matrix has three rows, this means that
our error correction uses three parity check equations (denoted as $f$).  A bit value 1 in row $i$,
column $j$ indicates that parity check equation $f_i$ contains bit $c_j$.
Each parity check equation XORs all of the codeword bits in the equation to see
whether the output is zero.
For example, parity check equation \chIII{$f_1$} from the $H$ matrix in
Figure~\ref{fig:ldpc}a is:}
\begin{equation}
f_1 = c_1 \oplus c_2 \oplus c_4 \oplus c_5 = 0
\end{equation}
\chIII{This means that $c$ is a valid codeword only if $H \cdot c^T = 0$,
where $c^T$ is the transpose matrix of \chIV{the codeword} $c$.}

\begin{figure}[h]
  \centering
  \includegraphics[width=.75\columnwidth]{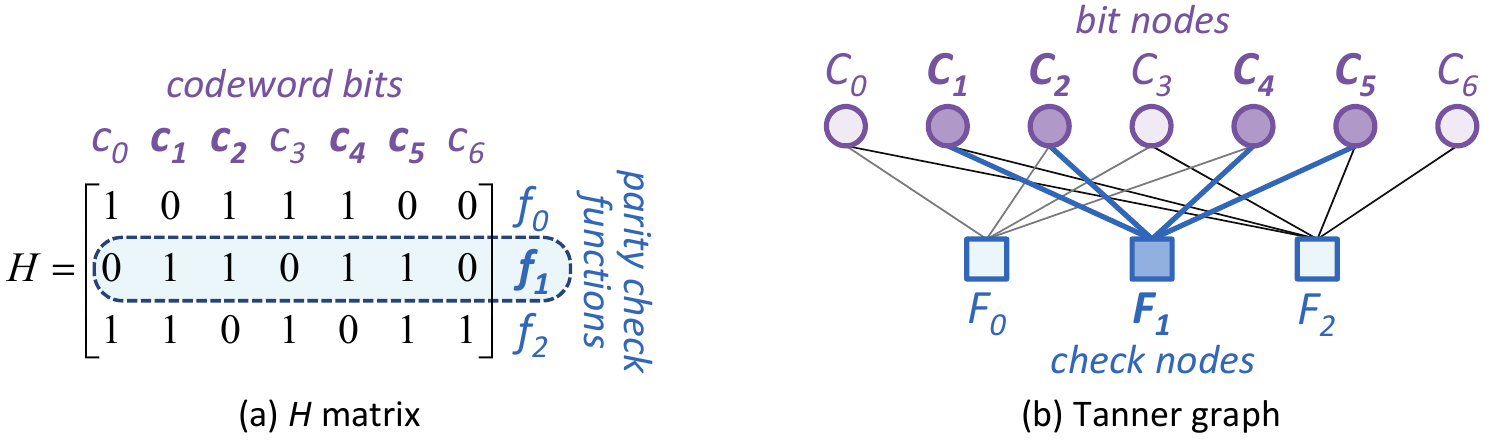}%
  \vspace{-5pt}%
  \caption{\chI{Example LDPC code for a \chIII{seven}-bit codeword with
  \chIII{a four-bit \chIV{data} message \chIV{(stored in bits $c_0$, $c_1$, $c_2$, and $c_3$)} 
  and} three parity check equations
  \chIII{(i.e., $n=7$, $k=4$)},
  represented as (a)~an \emph{H} matrix and (b)~a Tanner graph.}}%
  \label{fig:ldpc}%
\end{figure}

\chI{\chIII{In order to perform belief propagation, $H$} can be represented 
using a \emph{Tanner graph}~\cite{tanner.tit81}.  \chIII{A Tanner graph is a} bipartite graph
that contains \emph{check nodes}, which represent the parity check equations,
and \emph{bit nodes}, which represent the bits in the codeword.
An edge connects a check node $F_i$ to a bit node $C_j$ only if parity check 
equation $f_i$ contains bit $c_j$.  Figure~\ref{fig:ldpc}b shows the Tanner 
graph that corresponds to the $H$ matrix in Figure~\ref{fig:ldpc}a.
\chIII{For example, since parity check equation $f_1$ uses codeword bits
$c_1$, $c_2$, $c_4$, and $c_5$, the $F_1$ check node in Figure~\ref{fig:ldpc}b
is connected to bit nodes $C_1$, $C_2$, $C_4$, and $C_5$.}}

\paratitle{\chIII{LDPC Encoding}}
\chIII{As was the case with BCH, the LDPC codeword generation algorithm encodes
a $k$-bit \chIV{data} message $m$ into an $n$-bit LDPC codeword $c$ by computing the dot
product of $m$ and a generator matrix $G$ (i.e., $c = m \cdot G$).
For an LDPC code, the generator matrix is designed to (1)~preserve 
\chIV{$m$ verbatim within the codeword,}
and (2)~generate the parity bits for each parity check equation in $H$.
Thus, $G$ is defined using the parity check matrix $H$.
With linear algebra \chIV{based} transformations, $H$ can be expressed in the form
$H=[A, I_{(n-k)}]$, where $H$ is composed of $A$, an $(n-k) \times k$ binary matrix,
and $I_{(n-k)}$, an $(n-k) \times (n-k)$ identity matrix~\cite{johnson.ldpc}.
The generator matrix $G$ can then be created using the composition
$G = [I_k, A^T]$, where $A^T$ is the transpose matrix of $A$.}

\paratitle{\chIII{LDPC Decoding}}
\chIII{When the SSD controller is servicing a read request, it must extract 
the $k$-bit \chIV{data} message from the LDPC codeword $r$ that is stored in NAND flash memory.
\chV{In an SSD, an LDPC decoder performs multiple \emph{levels} of 
decoding~\cite{R11, dolecek.fms14, varnica.fms13},}
which correct the retrieved codeword $r$ to obtain the originally-written 
codeword $c$ and extract the \chIV{data} message $m$ from $c$.
\chV{Initially, the decoder performs a single level of \emph{hard decoding}, 
where it uses the information from a single read operation on the codeword to 
attempt to correct the codeword bit errors.  If the decoder cannot correct all 
errors using hard decoding, it then initiates the first level of \emph{soft decoding}, where a 
\emph{second} read operation is performed on the \emph{same} codeword using a 
\emph{different} set of read reference voltages.  The \chVI{second} read 
provides \chVI{\emph{additional}} information on the \emph{probability} 
that each bit in the codeword is a zero or a one.  An LDPC \chVI{decoder} typically uses 
multiple levels of soft decoding, where each new level \chVI{performs} an additional 
read operation to \chVI{calculate a more accurate probability for each bit 
value}.  We discuss multi-level soft decoding in detail in 
Section~\ref{sec:correction:ldpcflow}.}

\chVI{For each level, the} decoder performs \chV{five} steps, as illustrated in
Figure~\ref{fig:ldpc-decoding-flow}.
\chIII{\chV{\chVI{At each level}, the
decoder uses \chVI{two pieces of information to determine which bits are 
\emph{most likely} to contain errors: (1)~the \emph{probability} that each bit 
in $r$ is a zero or a one, and
(2)~the parity check equations.}}
In \chV{Step~1} \chVI{(Figure~\ref{fig:ldpc-decoding-flow})}, the decoder 
computes an initial \emph{log likelihood ratio} (LLR) for each bit of
the stored codeword.  We refer to the initial codeword LLR values as $L$,
where $L_j$ is the LLR value for bit~$j$ of the codeword.
$L_j$ expresses the likelihood (i.e., \emph{confidence})
that bit~$j$ \emph{should be} a zero 
or a one,
based on the current 
threshold voltage of the NAND flash cell where bit~$j$ is stored.
The decoder uses $L$ as the initial \emph{LLR message} generated using the bit nodes.
\chIV{An LLR message consists of the LLR values for each bit, which are updated by and 
communicated between the check nodes and bit nodes during each step of belief
propagation.\footnote{\chIV{Note that an LLR message is \chV{\emph{not}} the same as the $k$-bit
data message.  The \emph{data message} refers to the actual data stored
within the SSD, which, \chV{when read,} is modeled in information theory as a message 
\chVI{that is transmitted} across a noisy communication channel.
\chV{In contrast, an} \emph{LLR message} refers to the updated LLR values for each bit of
the codeword that are exchanged between the check nodes and the bit nodes 
during belief propagation. \chV{Thus, there is no relationship between a data 
message and an LLR message.}}}}
In \chV{Steps~2 through 4}, the belief propagation algorithm~\cite{pearl.aaai82} 
iteratively updates the LLR message, using the Tanner graph to identify those bits that
are most likely to be incorrect (i.e., the codeword bits whose \chIV{(1)~}\chV{bit nodes
are connected to the largest number of check nodes that currently contain a parity error,}
and \chIV{(2)~}LLR values
indicate low confidence).
Several decoding algorithms exist to perform belief propagation
for LDPC codes.  The most commonly-used belief propagation
algorithm is the \emph{min-sum algorithm}~\cite{fossorier.tcomm99, chen.tcomm02},
a simplified version of the original sum-product algorithm 
for LDPC\chV{~\cite{R94, gallager.tit62}} with near-equivalent error correction capability~\cite{anastasopoulos.globecom01}.
During each iteration of the min-sum algorithm, the decoder
\chIV{identifies a set of codeword bits that likely contain errors and \chV{thus} need to be
flipped.  \chV{The decoder accomplishes this} by}
(1)~\chIV{having each check node use its parity check information to determine
how much the LLR value of each bit should be updated by,} 
using the most recent LLR messages from the bit nodes;
(2)~\chIV{having each bit node gather the LLR updates from each bit to
generate a new LLR value for the bit,} 
using the most recent LLR messages from the check nodes; and
(3)~\chIV{using} the parity check equations to see if the values predicted by the new
LLR message for each node are correct.
\chV{The min-sum algorithm terminates \chVI{under} one of two conditions:
(1)~the predicted bit values after the most recent iteration are all correct,
which means that the decoder now has an estimate of the original codeword $c$,
and can advance to Step~5; or
(2)~the algorithm exceeds a predetermined number of iterations, at which point
the decoder moves onto the next decoding level, \chVI{or returns a decoding failure
if the maximum number of decoding levels have been performed}.
In Step~5, once the errors are corrected, and the decoder has the original codeword $c$, 
the decoder extracts the $k$-bit data message $m$ from the codeword.}
\chIII{We describe the steps used by a state-of-the-art decoder in detail below, which uses
an optimized version of the min-sum algorithm that 
can be implemented efficiently in hardware~\cite{gunnam.icc07, gunnam.fms14}.}

\begin{figure}[h]
  \centering
  \includegraphics[width=0.92\linewidth]{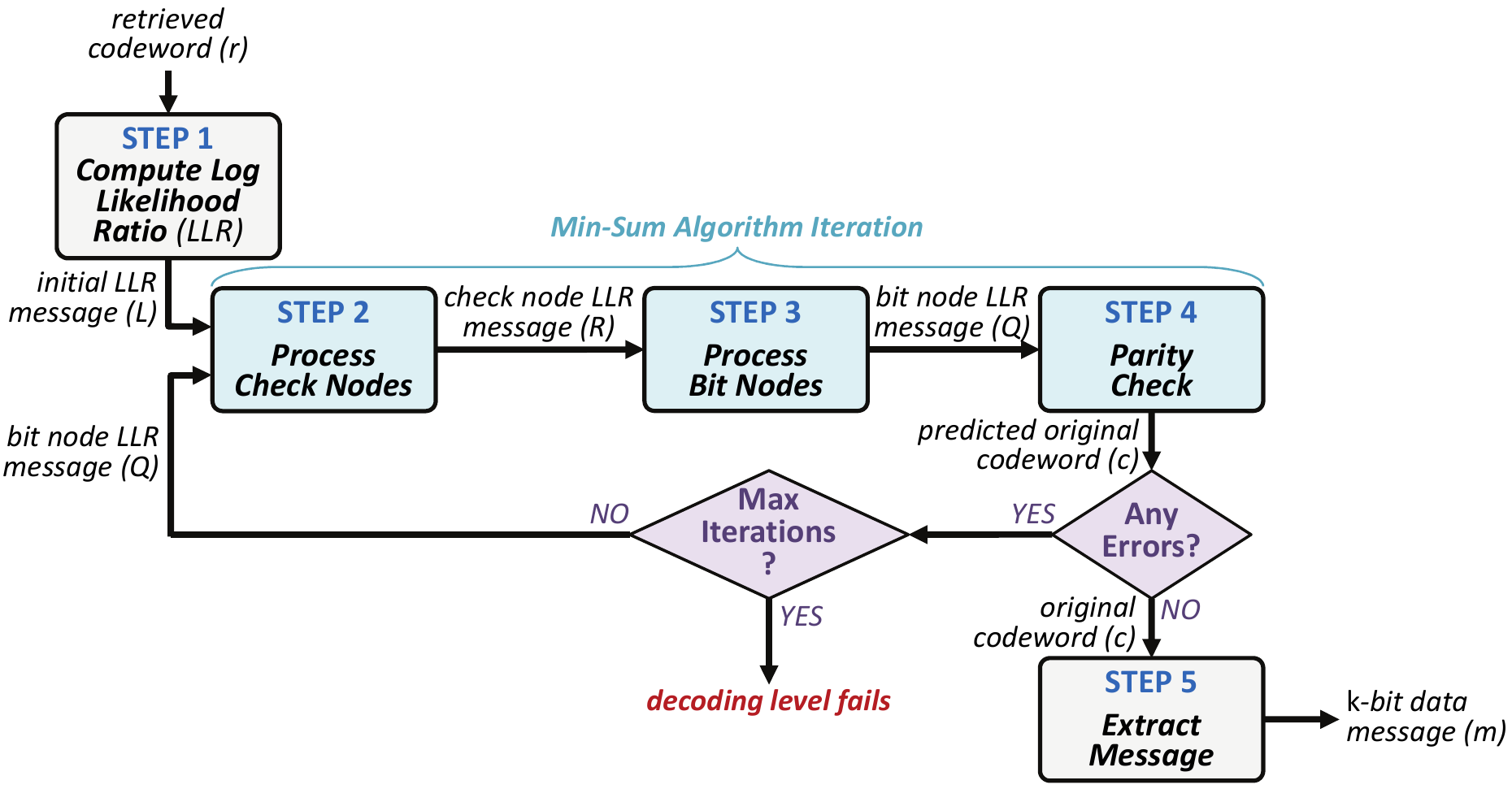}%
  \vspace{-5pt}%
  \caption{\chI{LDPC decoding steps \chV{for a single level of hard or soft decoding}.}}%
  \label{fig:ldpc-decoding-flow}%
\end{figure}

\vspace{3pt}%
\chIII{\emph{\chV{Step~1}---Computing the Log Likelihood Ratio (LLR):}}
\chV{The LDPC decoder uses the \emph{probability} (i.e.,
\emph{likelihood}) that a bit is a zero or a one to identify errors, instead of
using the bit values directly.}
\chI{The \chV{\emph{log likelihood ratio} (LLR)} is the probability \chVI{that a certain
bit is} zero, i.e., $P(x = 0| V_{th})$, over the probability \chVI{that
the bit is} one, i.e., $P(x = 1| V_{th})$, given a certain threshold
voltage range ($V_{th}$) bounded by two threshold voltage values
(i.e., the maximum and the minimum voltage of the threshold
voltage range)~\cite{R11, R73}:}
\begin{equation}
\text{LLR} = \log \frac{P(x = 0| V_{th})}{P(x = 1| V_{th})}
\label{eq:E10}
\end{equation}
\chIII{The sign of the LLR value indicates whether the bit is likely to be a 
zero (when the LLR value is positive) or a one (when the LLR value is negative).
A \emph{larger} magnitude (i.e., absolute value) of the LLR value indicates 
a \emph{greater} confidence
that a bit should be zero or one, while an LLR value closer to zero indicates
low confidence.  The bits whose LLR values have the smallest magnitudes are the
ones \chIV{that are} most likely to contain errors.}

%
\chI{There are several alternatives
for how to compute the LLR values. A common approach for
LLR computation is to treat a flash cell as a communication
channel, where the channel takes an input program signal
(i.e., the target threshold voltage for the cell) and outputs
an observed signal (i.e., the current threshold voltage of the
cell)~\cite{R33}. The observed signal differs from the input signal
due to the various types of NAND flash memory errors. The
communication channel model allows us to break down
the threshold voltage of a cell into two components: (1)~the
expected signal; and (2)~the additive signal noise due to
errors. By enabling the modeling of these two components
separately, the communication channel model allows us to
estimate the current threshold voltage distribution of each
state~\cite{R33}. The threshold voltage distributions can be used to
predict how likely a cell within a certain voltage region is to
belong to a particular voltage state.}

\chI{One popular variant of the communication channel
model assumes that the threshold voltage distribution of
each state can be modeled as a Gaussian distribution~\cite{R33}.
If we use the mean observed threshold voltage of each state
(denoted as $\mu$) to represent the signal, we find that the P/E
cycling noise (i.e., the shift in the distribution of threshold
voltages due to the accumulation of charge from repeated
programming operations; see Section~\ref{sec:errors:pe}) can be modeled
as \emph{additive white Gaussian noise} (AWGN)~\cite{R33}, which
is represented by the standard deviation of the distribution
(denoted as $\sigma$). The closed-form AWGN-based model can be
used to determine the LLR value for a cell with threshold
voltage $y$, as follows:}
\begin{equation}
\text{LLR}(y) = \frac{\mu_1^2 - \mu_0^2}{2 \sigma^2} + \frac{y(\mu_0 - \mu_1)}{\sigma^2}
\label{eq:E11}
\end{equation}
\chI{where $\mu_{0}$ and $\mu_{1}$ are the mean threshold voltages for the distributions
of the threshold voltage states for bit value 0 and
bit value 1, respectively, and $\sigma$ is the standard deviation of
both distributions (assuming that the standard deviation
of each threshold voltage state distribution is equal). Since
\chI{the SSD controller} uses threshold voltage ranges to categorize
a flash cell, we can substitute $\mu_{R_j}$, the mean threshold
voltage of the threshold voltage range $R_j$, in place of $y$ in Equation~\ref{eq:E11}.}

\chI{The AWGN-based LLR model in Equation~\ref{eq:E11} provides only an
estimate of the LLR, because (1)~the actual threshold voltage
distributions observed in NAND flash memory are \emph{not} perfectly
Gaussian in nature~\cite{R33, R42}; (2)~the controller uses
the mean voltage of the threshold voltage range to \emph{approximate}
the actual threshold voltage of a cell; and (3)~the standard
deviations of each threshold voltage state distribution
are \emph{not} perfectly equal (see Tables \ref{tbl:T4}--\ref{tbl:T6} in the Appendix).
A number of methods have been proposed to improve
upon the AWGN-based LLR estimate by: (1)~using nonlinear
transformations to convert the AWGN-based LLR into a
more accurate LLR value~\cite{R188}; (2)~scaling and rounding the
AWGN-based LLR to compensate for the estimation error~\cite{R189}; 
(3)~initially using the AWGN-based LLR to read the
data, and, if the read fails, using the ECC information from
the failed read attempt to optimize the LLR and to perform
the read again with the optimized LLR~\cite{R190}; and (4)~using
online and offline training to empirically determine the
LLR values under a wide range of conditions (e.g., P/E cycle
count, retention time, read disturb count)~\cite{R191}. The SSD
controller can either compute the LLR values at runtime, or
statically store precomputed LLR values in a table.}

\chIII{Once the decoder calculates the LLR values for each bit of the codeword,
which we call the initial LLR message $L$, the decoder starts the first 
iteration of the min-sum algorithm (\chV{Steps 2--4} below).}

\vspace{3pt}%
\chIII{\emph{\chV{Step~2}---Check Node Processing:}
In every iteration of the min-sum algorithm, each} \chI{check node $i$ \chIII{(see Figure~\ref{fig:ldpc})}
generates a revised check node LLR message $R_{ij}$ to send to each bit node $j$ \chIII{(see Figure~\ref{fig:ldpc})} that is
connected to check node $i$.  \chIII{The decoder computes} $R_{ij}$ as:}
\begin{equation}
R_{ij} = \delta_{ij} \kappa_{ij}
\end{equation}
\chI{where $\delta_{ij}$ is the sign of the \chIV{LLR} message, and $\kappa_{ij}$ is the
magnitude of the \chIV{LLR} message.}
\chIII{The decoder determines the values of both $\delta_{ij}$ and $\kappa_{ij}$
using the bit node LLR message $Q'_{ji}$.
\chIV{At a high level, each check node collects LLR values sent from each bit
node ($Q'_{ji}$), and then determines how much each bit's LLR value should be
adjusted using the parity information available at the check node.  These LLR 
value updates are then bundled together into the LLR message $R_{ij}$.}
During the first iteration of the min-sum algorithm,
the decoder sets $Q'_{ji} = L_j$,
the initial LLR value from \chV{Step~1}.  In subsequent iterations, the decoder uses the 
value of $Q'_{ji}$
that was generated in \chV{Step~3} of the \emph{previous} iteration.
The decoder calculates $\delta_{ij}$, the sign of the check node LLR
message, as:}
\begin{equation}
\delta_{ij} = \displaystyle\prod_{J} \text{sgn}(Q'_{Ji})
\end{equation}
\chI{where $J$ represents all bit nodes connected to check node $i$ \emph{except}
for bit node $j$.  
\chIV{The sign of a bit node indicates whether the value of a bit is predicted to be a zero
(if the sign is positive) or a one (if the sign is negative).}
\chIII{The decoder calculates $\kappa_{ij}$, the 
magnitude of the check node LLR message, as:}
\begin{equation}
\kappa_{ij} = \displaystyle\min_J |Q'_{Ji}|
\end{equation}
\chI{In essence, the smaller the magnitude of $Q'_{ji}$ is, the more
uncertain we are about whether the bit should be a \chIV{zero or a one}.  \chIV{At} each check node,
\chIV{the decoder updates the LLR value of each bit node $j$, adjusting the LLR by} the smallest value of
$Q'$ for \chIII{any of the other bits connected to the check node} 
(i.e., the LLR value of the most uncertain bit \chIII{aside from bit~$j$}).}

\vspace{3pt}%
\chIII{\emph{\chV{Step~3}---Bit Node Processing:}}
\chI{Once each check node generates the
LLR messages for each bit node, we combine the LLR messages received by each 
bit node to update the LLR value of the bit.  \chIII{The decoder first generates} the LLR messages
to be used by the check nodes in the next iteration \chIII{of the min-sum algorithm}.  \chIII{The decoder calculates} the bit node
LLR message $Q_{ji}$ to send from bit node $j$ to check node $i$ as follows:}
\begin{equation}
Q_{ji} = L_j + \displaystyle\sum_{I} R_{Ij}
\end{equation}
\chI{where $I$ represents all check nodes connected to bit node $j$ \emph{except}
for check node $i$, \chIV{and $L_j$ is the original LLR value for bit~$j$ generated 
in \chV{Step~1}}.  In essence, for each check node, the bit node LLR message combines the LLR 
messages from the \emph{other check nodes} to ensure that all \chIII{of the LLR value updates}
are propagated globally \chIII{across all of the check nodes}.}

\vspace{3pt}%
\chIII{\emph{\chV{Step~4}---Parity Check:}}
\chI{After the bit node processing is
complete, the decoder uses the revised \chV{LLR} information to predict the value of
each bit.  For bit node $j$, the predicted bit value $P_j$ is calculated as:}
\begin{equation}
P_j = L_j + \displaystyle\sum_{i} R_{ij}
\end{equation}
\chII{where $i$ represents \emph{all} check nodes connected to bit node $j$,
\chIV{\emph{including} check node $i$, and $L_j$ is the original LLR value for bit~$j$ generated 
in \chV{Step~1}}.}
\chI{If $P_j$ is positive, bit~$j$ \chIII{of the original codeword $c$} is predicted to be a \chIV{zero}; otherwise, bit~$j$ is 
predicted to be a \chIV{one}.
Once the predicted values have been computed for all bits
\chIII{of $c$}, the $H$ matrix is
used to check the parity, by \chIII{computing $H \cdot c^T$.  If $H \cdot c^T = 0$},}
then the predicted bit values are correct, \chIII{the min-sum algorithm 
terminates, and the decoder goes to \chV{Step~5}}.  Otherwise,
at least one bit is still incorrect, and the decoder goes back to \chV{Step~2} to 
perform the next iteration \chIII{of the min-sum algorithm}.  
\chIV{In the next iteration, the min-sum algorithm uses the updated LLR values
from the current iteration to identify the next set of bits that are most likely 
incorrect and need to be flipped.}

\chV{The current decoding level fails to correct the data when the decoder 
cannot determine the correct codeword \chVI{bit values} after a predetermined number of 
min-sum algorithm iterations.  If the decoder has more soft decoding levels 
left to perform, it
advances to the next soft decoding level.  For the new level, the SSD controller 
performs an additional read operation using a different set of read reference 
voltages than \chVI{the ones} it used for the prior decoding levels.  The decoder then goes 
back to Step~1 to generate the new LLR information, using the output of 
\emph{all} of the read operations performed for each decoding level so far.
We discuss how the number of decoding levels and the read reference voltages
are determined, as well as what happens if \emph{all} soft decoding levels fail, 
in Section~\ref{sec:correction:ldpcflow}.}

\vspace{3pt}%
\chIII{\emph{\chV{Step~5}---Extracting the Message from the Codeword:}
As we discuss above, during LDPC codeword encoding, the
generator matrix $G$ contains the identity matrix, to ensure that the
codeword $c$ includes \chIV{a verbatim} version of $m$.
Therefore, the decoder recovers the $k$-bit \chIV{data} message $m$ by simply
truncating the last $(n-k)$ bits from the $n$-bit codeword $c$.}

\subsection{\chI{Error Correction Flow}}
\label{sec:correction:flow}

\chI{For both BCH and LDPC codes, the SSD controller performs several stages
of error correction to retrieve the data, known as the \emph{error correction
flow}.  The error correction flow is invoked when the SSD performs a read operation.
The SSD starts the read operation by using the initial read reference
voltages ($V_{initial}$; see Section~\ref{sec:mitigation:voltage}) to read the raw
data stored within a page of NAND flash memory into the
controller.  Once the raw data is read, the controller starts
error correction.}

\begin{figure}[t]
\begin{algorithm}[H]
\vspace{2pt}%
\caption{Example BCH/LDPC Error Correction Procedure}
\label{alg:A1}
\includegraphics[width=0.49\columnwidth]{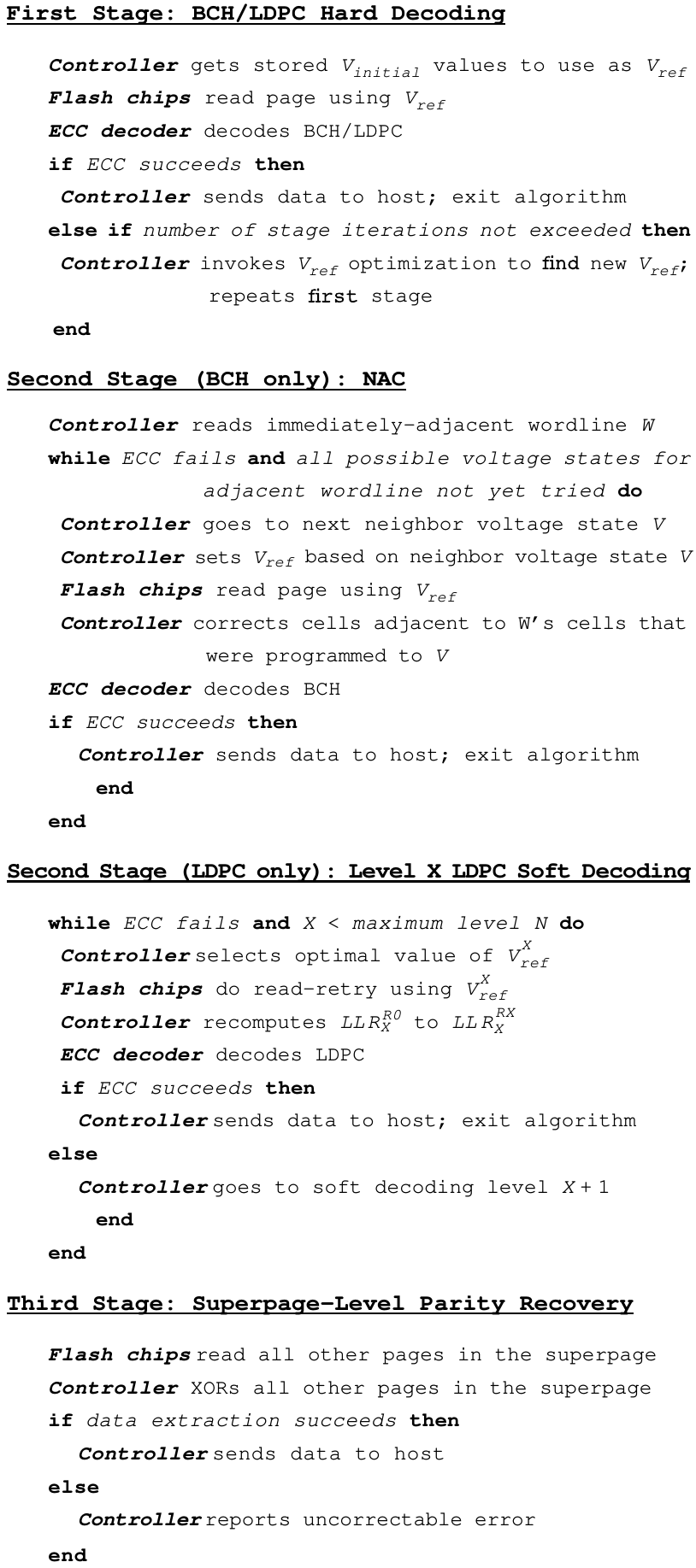}
\end{algorithm}
\vspace{-25pt}%
\end{figure}

\chI{Algorithm~\ref{alg:A1} lists the three stages of an example error correction 
flow, which can be used to decode either BCH codes or LDPC codes.
In the first stage, the ECC engine performs \emph{hard decoding} on the raw
data.  In hard decoding, the ECC engine uses only the
\emph{hard} bit value information (i.e., either a 1 or a 0) read for a
cell using a \emph{single} set of read reference voltages.
If the first stage succeeds (i.e., the controller detects that the error rate
of the data after correction is lower than a predetermined threshold),
the flow finishes.  If the first stage fails, then the flow moves on to the
second stage of error correction.  The second stage differs significantly for
BCH and for LDPC, which we discuss below.  If the second stage succeeds,
the flow terminates; otherwise, the flow moves to the third stage of error
correction.  In the third stage, the controller tries to
correct the errors using the more expensive superpage-level
parity recovery (\chIV{see Section~\ref{sec:ssdarch:ctrl:parity}}). The steps for superpage-level
parity recovery are shown in the third stage of Algorithm~\ref{alg:A1}.
If the data can be extracted successfully from the other pages
in the superpage, the data from the target page can be recovered.
Whenever data is successfully decoded or recovered,
the data is sent to the host (and it is also reprogrammed into
a new physical page to ensure that the \emph{corrected} data values
are stored for the logical page). Otherwise, the SSD controller
reports an uncorrectable error to the host.}

\chI{Figure~\ref{fig:F30} compares the error correction flow with BCH
codes to the flow with LDPC codes.
Next, we discuss \chIII{the flows used with} both BCH codes
(Section~\ref{sec:correction:bchflow}) and LDPC codes
(Section~\ref{sec:correction:ldpcflow}).}

\begin{figure}[h]
  \centering
  \includegraphics[width=0.8\columnwidth]{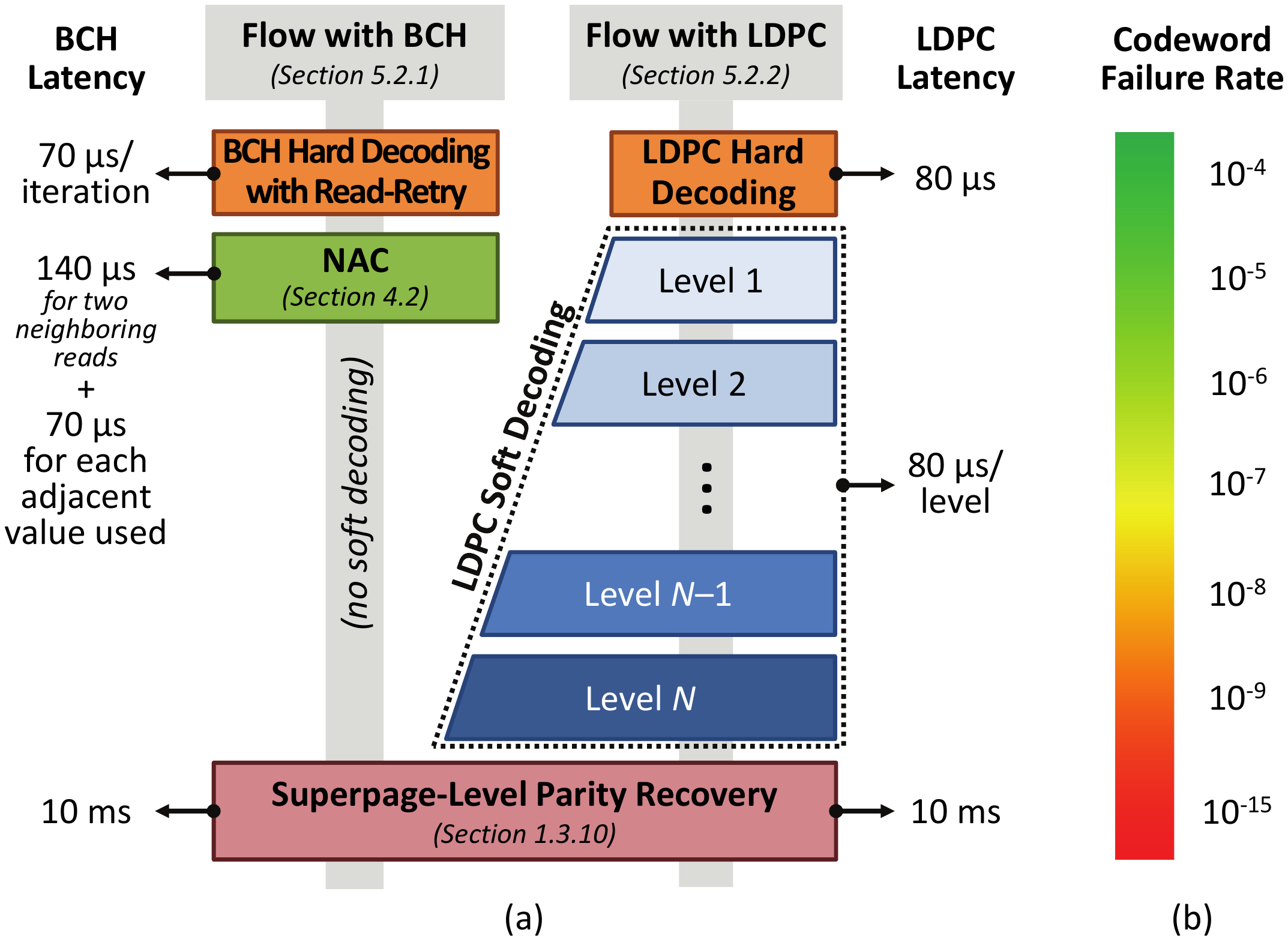}%
  \vspace{-5pt}%
  \caption{(a)~Example error correction flow using BCH codes and
LDPC codes, \chVI{with average latency of each BCH/LDPC stage.} 
(b)~The corresponding codeword failure rate for each LDPC stage. 
\chI{\chVI{Adapted} from \cite{cai.arxiv17}.}}%
  \label{fig:F30}%
\end{figure}

\subsubsection{\chI{Flow Stages for BCH Codes}}
\label{sec:correction:bchflow}
\label{sec:correction:bch}

\chI{An example flow of
the stages for BCH decoding is
shown on the left-hand side of Figure~\ref{fig:F30}a. In the first stage,
the ECC engine performs BCH hard decoding on the raw data,
which reports the total number of bit errors in the data.
If the data cannot be corrected by the implemented BCH
codes, many controllers invoke read-retry (Section~\ref{sec:mitigation:retry}) or
read reference voltage optimization (Section~\ref{sec:mitigation:voltage}) to find a
new set of read reference voltages ($V_{ref}$) that lower the raw
bit error rate of the data from the error rate when using
$V_{initial}$. The controller uses the new $V_{ref}$ values to read the
data again, and then repeats the BCH decoding.}
\chV{We discuss the algorithm used to perform decoding for
BCH codes in Section~\ref{sec:correction:ecc:bch}.}

If the controller exhausts the maximum number of read
attempts (specified as a parameter in the controller), it
employs correction techniques such as neighbor-cell-assisted
correction (NAC; see Section~\ref{sec:mitigation:nac}) to further reduce the error
rate, as shown in the second BCH stage of Algorithm~\ref{alg:A1}. If NAC
cannot successfully read the data, the controller then tries to
correct the errors using the more expensive superpage-level
parity recovery \chIII{(see \chIV{Section~\ref{sec:ssdarch:ctrl:parity}})}.

\subsubsection{\chI{Flow Stages for LDPC Codes}}
\label{sec:correction:ldpcflow}
\label{sec:correction:ldpc}

\chI{An example flow of
the stages for LDPC decoding is
shown on the right-hand side of Figure~\ref{fig:F30}a.}
LDPC decoding consists of three major steps. First,
the SSD controller performs LDPC hard decoding, where
the controller reads the data using the optimal read reference
voltages. The process for LDPC hard decoding is similar
to that of BCH hard decoding (as shown in the first stage
of Algorithm~\ref{alg:A1}), but does not typically invoke read-retry if
the first read attempt fails. Second, if LDPC hard decoding
cannot correct all of the errors, the controller uses LDPC
\emph{soft decoding} to decode the data (which we describe in detail
below). Third, if LDPC soft decoding also cannot correct all
of the errors, the controller invokes superpage-level parity.
\chV{We discuss the algorithm used to perform hard and soft decoding for
LDPC codes in Section~\ref{sec:correction:ecc:ldpc}.}

\paratitle{Soft Decoding}
Unlike BCH codes, which require
the invocation of expensive superpage-level parity recovery
immediately if the hard decoding attempts (\chIII{i.e.,} BCH hard
decoding with read-retry or NAC) fail to return correct data,
LDPC decoding fails more gracefully: it can perform multiple
levels of \emph{soft decoding} (\chIII{shown in the second stage of} Algorithm~\ref{alg:A1})
after hard decoding fails before invoking superpage-level
parity recovery~\cite{R11, R73}. The key idea of soft decoding is \chIII{to}
use \emph{soft} information for each cell (i.e., the \emph{probability} that
the cell contains a 1 or a 0) obtained from \emph{multiple} reads of
the cell via the use of different sets of read reference voltages\chV{~\cite{R9, R11, R94, R95, gallager.tit62, dolecek.fms14}}. 
\chI{Soft information is typically represented by
the \emph{log likelihood ratio} (LLR; see Section~\ref{sec:correction:ecc:ldpc}).}

Every additional level of soft decoding (i.e., the use of
a new set of read reference voltages, which we call $V_{ref}^X$ for
level $X$) increases the strength of the error correction, as the
level \emph{adds} new information about the cell (as opposed to
hard decoding, where a new decoding step simply \emph{replaces}
prior information about the cell). The new read reference
voltages, unlike the ones used for hard decoding, are
optimized such that the amount of useful information (or
\emph{mutual information}) provided to the LDPC decoder is maximized~\cite{R73}. 
Thus, the use of soft decoding reduces the frequency
at which superpage-level parity needs to be invoked.

Figure~\ref{fig:F31} illustrates the read reference voltages used during
\chV{LDPC hard decoding and during}
the first \chV{two} levels of LDPC soft decoding. At each level, a
new read reference voltage is applied, which divides an existing
threshold voltage range into two ranges. Based on the bit
values read using the various read reference voltages, the SSD
controller bins each cell into a certain $V_{th}$ range, and sends
the bin categorization of all the cells to the LDPC decoder.
For each cell, the decoder applies an LLR value, precomputed
by the SSD manufacturer, which corresponds to the cell's bin
and decodes the data. For example, as shown in the bottom
of Figure~\ref{fig:F31}, the three read reference voltages in \chV{Level~2} soft
decoding form four threshold voltage ranges (i.e., R0--R3).
Each of these ranges corresponds to a different LLR value
(i.e., \chV{$\text{LLR}_2^{R0}$ to  $\text{LLR}_2^{R3}$},
where $\text{LLR}_i^{Rj}$ is the LLR value for range
$R_j$ in \chV{soft decoding} level $i$). Compared with \chV{hard} decoding (shown
at the top of Figure~\ref{fig:F31}), which \chVI{has only} two LLR values, 
\chV{Level~2} soft decoding provides more accurate information to the
decoder, and thus has stronger error correction capability.

\begin{figure}[h]
  \centering
  \includegraphics[width=0.72\columnwidth]{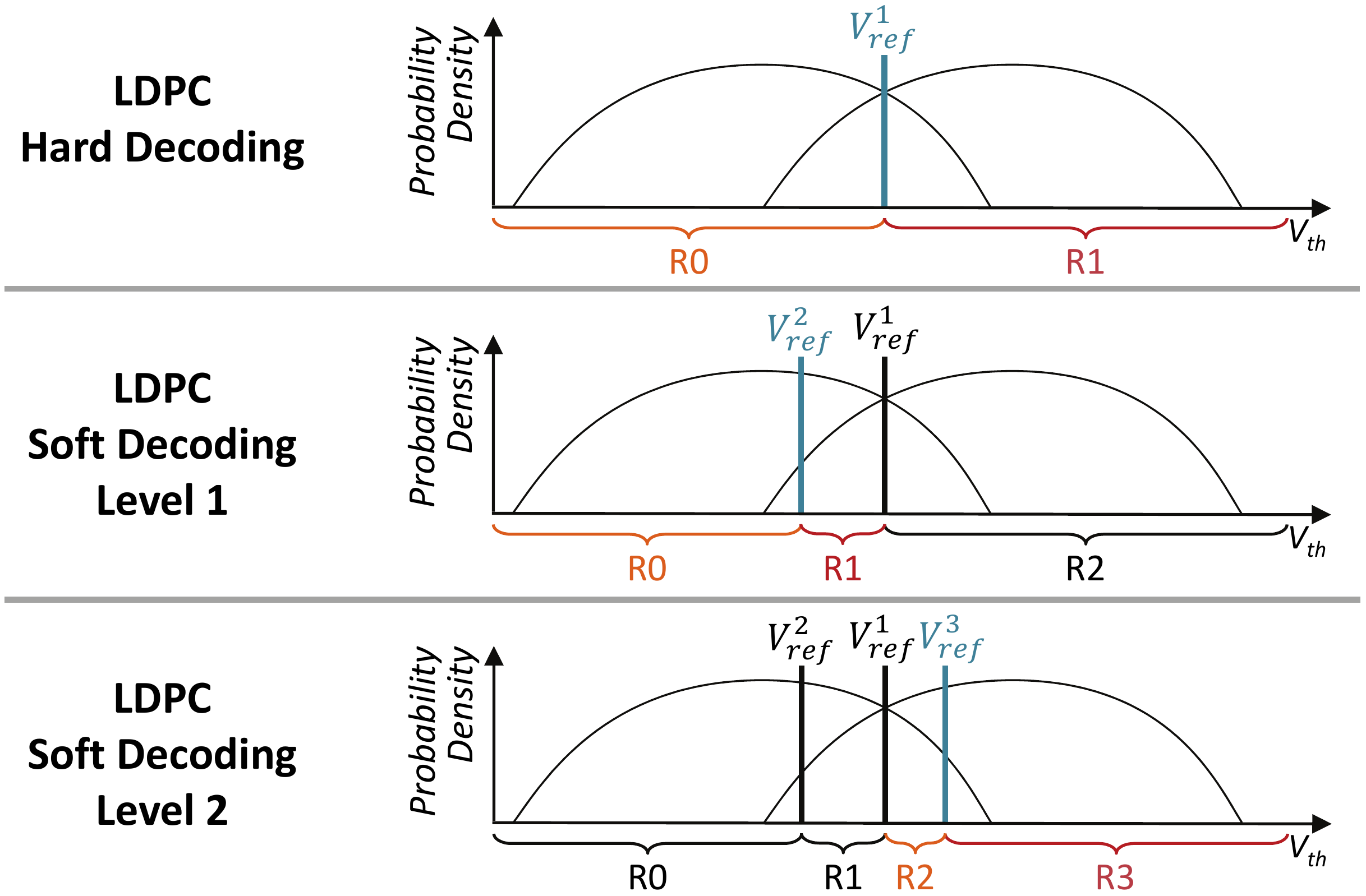}%
  \vspace{-5pt}%
  \caption{\chV{LDPC hard decoding and the first two} levels of LDPC soft decoding, showing the $V_{ref}$
value added at each level, and the resulting threshold voltage
ranges (R0--R3) used for flash cell categorization. \chI{\chV{Adapted} from \cite{cai.arxiv17}.}}%
  \label{fig:F31}%
\end{figure}

\paratitle{Determining the Number of Soft Decoding Levels}
If the
final level of soft decoding, i.e., level $N$ in Figure~\ref{fig:F30}a, fails,
the controller attempts to read the data using superpage-level
parity (\chIV{see Section~\ref{sec:ssdarch:ctrl:parity}}). The number of levels used for
soft decoding depends on the improved reliability that each
additional level provides, taking into account the latency of
performing additional decoding. Figure~\ref{fig:F30}b shows a rough
estimation of the average latency and the codeword failure
rate for each stage. There is a tradeoff between the number
of levels employed for soft decoding and the expected
read latency. For a smaller number of levels, the additional
reliability can be worth the latency penalty. For example,
while a five-level soft decoding step requires up to \SI{480}{\micro\second}, it
effectively reduces the codeword failure rate by five orders
of magnitude. This not only improves overall reliability,
but also reduces the frequency of triggering expensive
superpage-level parity recovery, which can take around
\SI{10}{\milli\second}~\cite{R65}. However, manufacturers limit the number of
levels, as the benefit of employing an additional soft decoding
level (which requires more read operations) becomes
smaller due to diminishing returns in the number of additional
errors corrected.

\subsection{BCH and LDPC Error Correction Strength}
\label{sec:correction:strength}

BCH and LDPC codes provide different strengths of
error correction. While LDPC codes can offer a stronger
error correction capability, soft LDPC decoding can lead
to a greater latency for error correction. Figure~\ref{fig:F32} compares
the error correction strength of BCH codes, hard LDPC
codes, and soft LDPC codes~\cite{R113}. The x-axis shows the raw
bit error rate (RBER) of the data being corrected, and the
y-axis shows the \emph{uncorrectable bit error rate} (UBER), or the
error rate after correction, once the error correction code
has been applied. The UBER is defined as the ECC codeword
(see \chIV{Section~\ref{sec:ssdarch:ctrl:ecc}}) failure rate divided by the codeword
length~\cite{R110}. To ensure a fair comparison, we choose a similar
codeword length for both BCH and LDPC codes, and use
a similar coding rate (0.935 for BCH, and 0.936 for LDPC)~\cite{R113}. 
We make two observations from Figure~\ref{fig:F32}.

\begin{figure}[h]
  \centering
  \includegraphics[width=0.65\columnwidth]{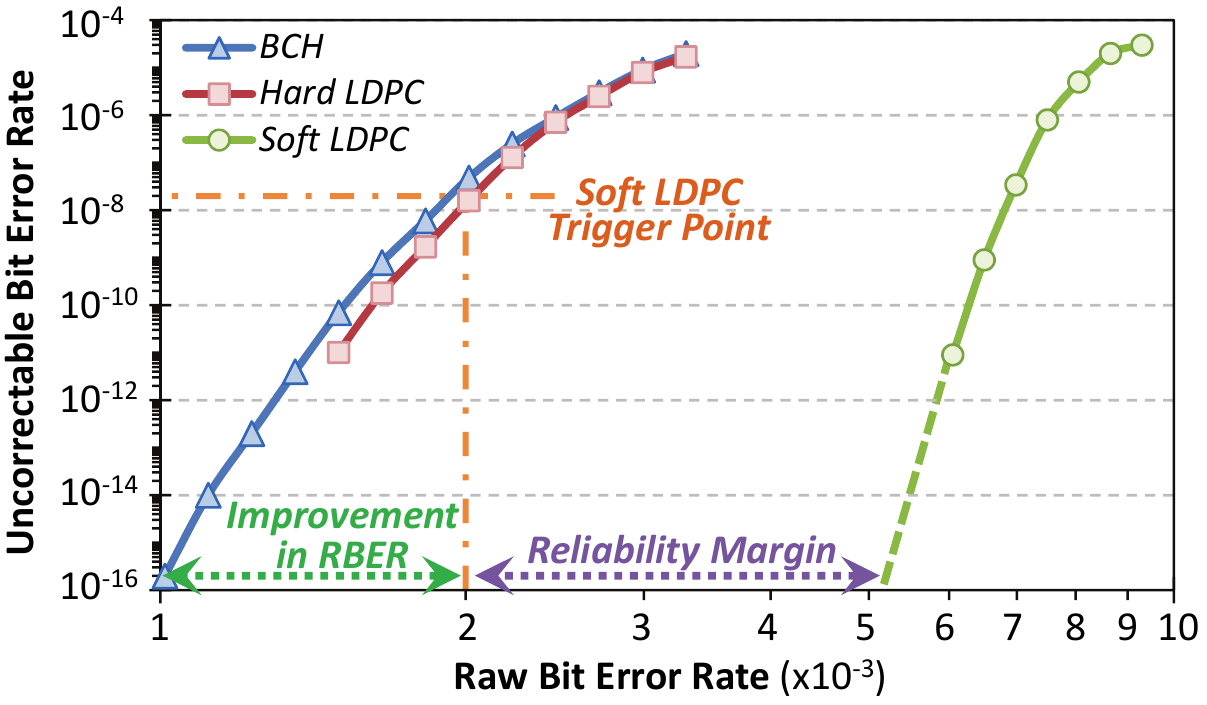}%
  \vspace{-7pt}%
  \caption{Raw bit error rate versus uncorrectable bit error rate for
BCH codes, hard LDPC codes, and soft LDPC codes. \chI{Reproduced from \cite{cai.arxiv17}.}}%
  \label{fig:F32}%
\end{figure}

First, we observe that the error correction strength of
the hard LDPC code is similar to that of the BCH codes.
Thus, on its own, hard LDPC does not provide a significant
advantage over BCH codes, as it provides an equivalent
degree of error correction with similar latency (i.e.,
one read operation). Second, we observe that soft LDPC
decoding provides a significant advantage in error correction
capability. Contemporary SSD manufacturers target a
UBER of $10^{-16}$~\cite{R110}. The example BCH code with a coding
rate of 0.935 can successfully correct data with an RBER
of $1.0 \times 10^{-3}$ while remaining within the target UBER. The
example LDPC code with a coding rate of 0.936 is more
successful with soft decoding, and can correct data with an
RBER as high as $5.0 \times 10^{-3}$ while remaining within the target
UBER, based on the error rate extrapolation shown in
Figure~\ref{fig:F32}. While soft LDPC can tolerate up to five times the
raw bit errors as BCH, this comes at a cost of latency (not
shown on the graph), as soft LDPC can require several additional
read operations after hard LDPC decoding fails, while
BCH requires only the original read.

To understand the benefit of LDPC codes over BCH
codes, we need to consider the combined effect of hard
LDPC decoding and soft LDPC decoding. As discussed in
Section~\ref{sec:correction:ldpcflow}, soft LDPC decoding is invoked \emph{only when hard
LDPC decoding fails}. To balance error correction strength
with read performance, SSD manufacturers can require that
the hard LDPC failure rate cannot exceed a certain threshold,
and that the overall read latency (which includes the
error correction time) cannot exceed a certain target~\cite{R65, R113}. 
For example, to limit the impact of error correction
on read performance, a manufacturer can require 99.99\% of
the error correction operations to be completed after a single
read. To meet our example requirement, the hard LDPC
failure rate should not be greater than $10^{-4}$ (i.e., 99.99\%),
which corresponds to an RBER of $2.0 \times 10^{-3}$ and a UBER
of $10^{-8}$ (shown as \emph{Soft LDPC Trigger Point} in Figure~\ref{fig:F32}). For
only the data that contains one or more failed codewords,
soft LDPC is invoked (i.e., soft LDPC is invoked only 0.01\%
of the time). For our example LDPC code with a coding
rate of 0.936, soft LDPC decoding is able to correct these
codewords: for an RBER of $2.0 \times 10^{-3}$, using soft LDPC
results in a UBER well below $10^{-16}$, as shown in Figure~\ref{fig:F32}.

To gauge the combined effectiveness of hard and soft
LDPC codes, we calculate the overhead of using the combined
LDPC decoding over using BCH decoding. If 0.01\%
of the codeword corrections fail, we can assume that in
the worst case, each failed codeword resides in a different
flash page. As the failure of a single codeword in a flash
page causes soft LDPC to be invoked for the entire flash
page, our assumption maximizes the number of flash pages
that require soft LDPC decoding. For an SSD with four
codewords per flash page, our assumption results in up
to 0.04\% of the data reads requiring soft LDPC decoding.
Assuming that the example soft LDPC decoding requires
seven additional reads, this corresponds to 0.28\% more
reads when using combined hard and soft LDPC over BCH
codes. Thus, with a 0.28\% overhead in the number of reads
performed, the combined hard and soft LDPC decoding
provides twice the error correction strength of BCH codes
(shown as \emph{Improvement in RBER} in Figure~\ref{fig:F32}).

In our example, the lifetime of an SSD is limited by
both the UBER and whether more than 0.01\% of the codeword
corrections invoke soft LDPC, to ensure that the
overhead of error correction does not significantly increase
the read latency~\cite{R113}. In this case, when the lifetime
of the SSD ends, we can still read out the data correctly
from the SSD, albeit at an increased read latency. This is
because even though we capped the SSD lifetime to an
RBER of $2.0 \times 10^{-3}$ in our example shown in Figure~\ref{fig:F32}, soft
LDPC is able to correct data with an RBER as high as
$5.0 \times 10^{-3}$ while still maintaining an acceptable UBER
($10^{-16}$) based on the error rate extrapolation shown.
Thus, LDPC codes have a margin, which we call the \emph{reliability
margin} and show in Figure~\ref{fig:F32}. This reliability margin
enables us to trade off lifetime with read latency.

We conclude that with a combination of hard and soft
LDPC decoding, an SSD can offer a significant improvement
in error correction strength over using BCH codes.

\subsection{SSD Data Recovery}
\label{sec:correction:recovery}

When the number of errors in data exceeds the ECC
correction capability and the error correction techniques in
Sections~\ref{sec:correction:bchflow} and \ref{sec:correction:ldpcflow}
are unable to correct the read data,
then data loss can occur. At this point, the SSD is considered
to have reached the end of its lifetime. In order to avoid such
data loss and \emph{recover} (or, \emph{rescue}) the data from the SSD, we
can harness our understanding of data retention and read
disturb behavior. The SSD controller can employ two conceptually
similar mechanisms, \emph{Retention Failure Recovery}
(RFR)~\cite{R37} and \emph{Read Disturb Recovery} (RDR)~\cite{R38}, to undo
errors that were introduced into the data as a result of data
retention and read disturb, respectively. The key idea of
both of these mechanisms is to exploit the wide variation of
different flash cells in their susceptibility to data retention
loss and read disturbance effects, respectively, in order to
correct some of the errors \emph{without} the assistance of ECC so
that the remaining error count falls within the ECC error
correction capability.

When a flash page read fails (i.e., uncorrectable errors
exist), RFR and RDR record the current threshold voltages
of each cell in the page using the read-retry mechanism (see
Section~\ref{sec:mitigation:retry}), and identify the cells that are \emph{susceptible} to
generating errors due to retention and read disturb (i.e.,
cells that lie at the tails of the threshold voltage distributions
of each state, where the distributions overlap with
each other), respectively. We observe that some flash cells
are more likely to be affected by retention leakage and read
disturb than others, as a result of process variation~\cite{R37, R38}. 
We call these cells retention/read disturb \emph{prone}, while
cells that are less likely to be affected are called retention/read 
disturb \emph{resistant}. RFR and RDR classify the susceptible
cells as retention/read disturb prone or resistant by inducing
\emph{even more} retention and read disturb on the failed flash
page, and then recording the new threshold voltages of the
susceptible cells. We classify the susceptible cells by observing
the magnitude of the threshold voltage shift due to the
additional retention/read disturb induction.

\begin{figure}[h]
  \centering
  \includegraphics[width=0.65\columnwidth]{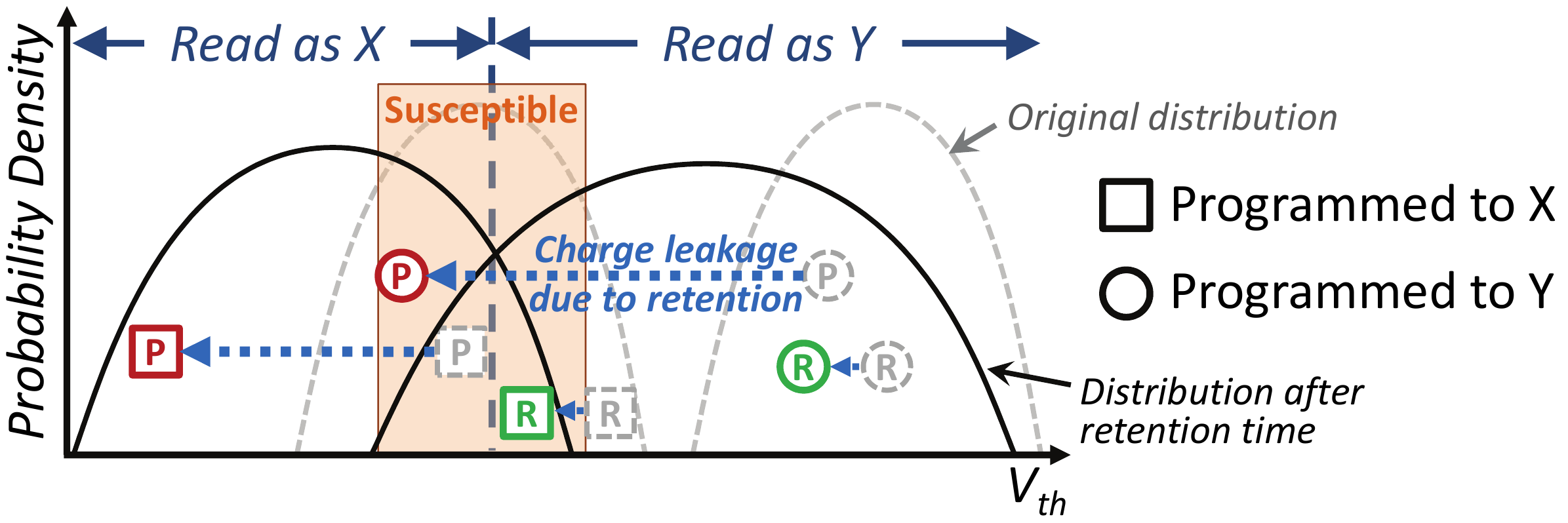}%
  \vspace{-5pt}%
  \caption{Some retention-prone (P) and retention-resistant (R) cells
are incorrectly read after charge leakage due to retention time.
RFR identifies and corrects the incorrectly read cells based on their
leakage behavior. \chI{Reproduced from \cite{cai.arxiv17}.}}%
  \label{fig:F33}%
\end{figure}

Figure~\ref{fig:F33} shows how the threshold voltage of a retention-prone
cell (i.e., a \emph{fast-leaking} cell, labeled P in the figure)
decreases over time (i.e., the cell shifts to the left) due to
retention leakage, while the threshold voltage of a retention-
resistant cell (i.e., a \emph{slow-leaking} cell, labeled R in the
figure) does not change significantly over time. Retention
Failure Recovery (RFR) uses this classification of retention-
prone versus retention-resistant cells to correct the
data from the failed page \emph{without} the assistance of ECC.
Without loss of generality, let us assume that we are studying
susceptible cells near the intersection of two threshold
voltage distributions X and Y, where Y contains higher voltages
than X. Figure~\ref{fig:F33} highlights the region of cells considered
susceptible by RFR using a box, labeled \emph{Susceptible}.
A susceptible cell within the box that is retention prone
likely belongs to distribution Y, as a retention-prone cell
shifts rapidly to a lower voltage (see the circled cell labeled
P within the \emph{susceptible} region in the figure). A retention-resistant
cell in the same \emph{susceptible} region likely belongs
to distribution X (see the boxed cell labeled R within the
\emph{susceptible} region in the figure).

Similarly, Read Disturb Recovery (RDR) uses the classification
of read disturb prone versus read disturb resistant
cells to correct data. For RDR, disturb-prone cells shift more
rapidly to higher voltages, and are thus likely to belong to
distribution X, while disturb-resistant cells shift little and
are thus likely to belong to distribution Y. Both RFR and
RDR correct the bit errors for the susceptible cells based on
such \emph{expected} behavior, reducing the number of errors that
ECC needs to correct.

RFR and RDR are highly effective at reducing the error
rate of failed pages, reducing the raw bit error rate by 50\%
and 36\%, respectively, as shown in our prior works~\cite{R37, R38}, 
where more detailed information and analyses can
be found.

\section{Emerging Reliability Issues for 3D NAND Flash Memory}
\label{sec:3d}

\chI{While the demand for NAND flash memory capacity continues to grow,
manufacturers have found it increasingly difficult to rely on manufacturing
process technology scaling to achieve increased capacity~\cite{park.jssc15}.
Due to a combination of limitations in manufacturing process technology
and the increasing reliability issues as manufacturers move to smaller
process technology nodes, planar (i.e., 2D) NAND flash scaling has become
difficult for manufacturers to sustain.  This has led manufacturers to seek
alternative approaches to increase NAND flash memory capacity.}

\chI{Recently, manufacturers have begun to produce SSDs that
contain \emph{three-dimensional} (3D) NAND flash memory\chIII{~\cite{R79, park.jssc15, kang.isscc16,
im.isscc15, micheloni.procieee17, micheloni.sn16}}.  In 3D NAND flash memory, 
\emph{multiple layers} of flash cells are stacked vertically to increase the density
and to improve the scalability of the memory~\cite{R79}. 
In order to achieve this stacking, manufacturers have changed a number of 
underlying properties of the flash memory design.  In this section, we examine
these changes, and discuss how they affect the reliability of the \chII{flash} memory devices.
In Section~\ref{sec:3d:org}, we discuss the flash \chII{memory} cell design commonly used
in contemporary 3D NAND flash memory, and how these cells are organized 
across the multiple layers.
In Section~\ref{sec:3d:errors}, we discuss how the reliability of
3D NAND flash memory compares to the reliability of the planar NAND flash
memory that we have studied so far in this work.
\chIII{Table~\ref{tbl:3dchanges} summarizes the differences observed in
3D NAND flash memory reliability.}
In Section~\ref{sec:3d:mitigation}, we briefly discuss \chII{error} mitigation mechanisms
that cater to emerging reliability issues in 3D NAND flash memory.}

\begin{table}[h]
\centering
\small
\setlength{\tabcolsep}{0.26em}
\setlength\arrayrulewidth{0.75pt}
\caption{\chIII{Changes in behavior of different types of errors in 3D NAND 
flash memory, compared to planar (i.e., two-dimensional) NAND flash memory.
See Section~\ref{sec:3d:errors} for a detailed discussion.}}
\label{tbl:3dchanges}
\vspace{-5pt}
\begin{tabular}{|c||c|}
\hline
\rowcolor{Gray!20} \textbf{Error Type} & \textbf{Change in 3D vs.\ Planar} \\
\hhline{|=#=|}
\textbf{P/E Cycling} & 3D is \emph{less susceptible}, \\
(Section~\ref{sec:errors:pe}) & due to \chIV{current use of charge trap transistors for flash cells} \\
\hline
\textbf{Program} & 3D is \emph{less susceptible for now}, \\
(Section~\ref{sec:errors:pgm}) & due to use of one-shot programming (see Section~\ref{sec:flash:pgmerase}) \\
\hline
\textbf{Cell-to-Cell Interference} & 3D is \emph{less susceptible for now}, \\
(Section~\ref{sec:errors:celltocell}) & due to larger manufacturing process technology \\
\hline
\textbf{Data Retention} & 3D is \emph{more susceptible}, \\
(Section~\ref{sec:errors:retention}) & due to early retention loss \\
\hline
\textbf{Read Disturb} & 3D is \emph{less susceptible for now}, \\
(Section~\ref{sec:errors:readdisturb}) & due to larger manufacturing process technology \\
\hline
\end{tabular}
\end{table}

\subsection{\chI{3D NAND Flash Design and Operation}}
\label{sec:3d:org}

\chI{As we discuss in Section~\ref{sec:flash:data}, NAND flash memory stores
data as the threshold voltage of each flash cell.  In planar NAND flash memory,
we achieve this using a floating-gate transistor as a flash cell, as shown in
Figure~\ref{fig:F6}.  The floating-gate transistor stores charge in the floating
gate of the cell, which consists of a conductive material.  The floating gate is
surrounded on both sides by an oxide layer.  When high voltage is applied to
the control gate of the transistor, charge can migrate through the oxide layers
into the floating gate due to Fowler-Nordheim (FN) tunneling~\cite{R21} (see 
Section~\ref{sec:flash:pgmerase}).}

\chI{Most manufacturers use a \emph{charge trap transistor}\chVII{~\cite{R105, R172}} as the flash cell
in 3D NAND flash memories, instead of using a floating-gate transistor.
Figure~\ref{fig:3dcell} shows \chII{the} cross section of a charge trap transistor.
Unlike a floating-gate transistor, which stores data in the form of charge 
within a \emph{conductive} material, a charge trap transistor stores data as
charge within an \emph{insulating} material, known as the \emph{charge trap}.
In a 3D circuit, the charge trap wraps around a cylindrical transistor substrate, 
which contains the source (labeled \emph{S} in Figure~\ref{fig:3dcell}) and 
drain (labeled \emph{D} in the figure), and a control gate wraps around the charge trap.
This arrangement allows the channel between the source and drain to form
\emph{vertically} within the transistor.
As is the case with a floating-gate transistor, a \chII{tunnel oxide layer} exists between
the charge trap and the substrate, and a gate oxide \chII{layer} exists between the charge trap and the control 
gate.}

\begin{figure}[h]
\centering
\includegraphics[width=.35\linewidth]{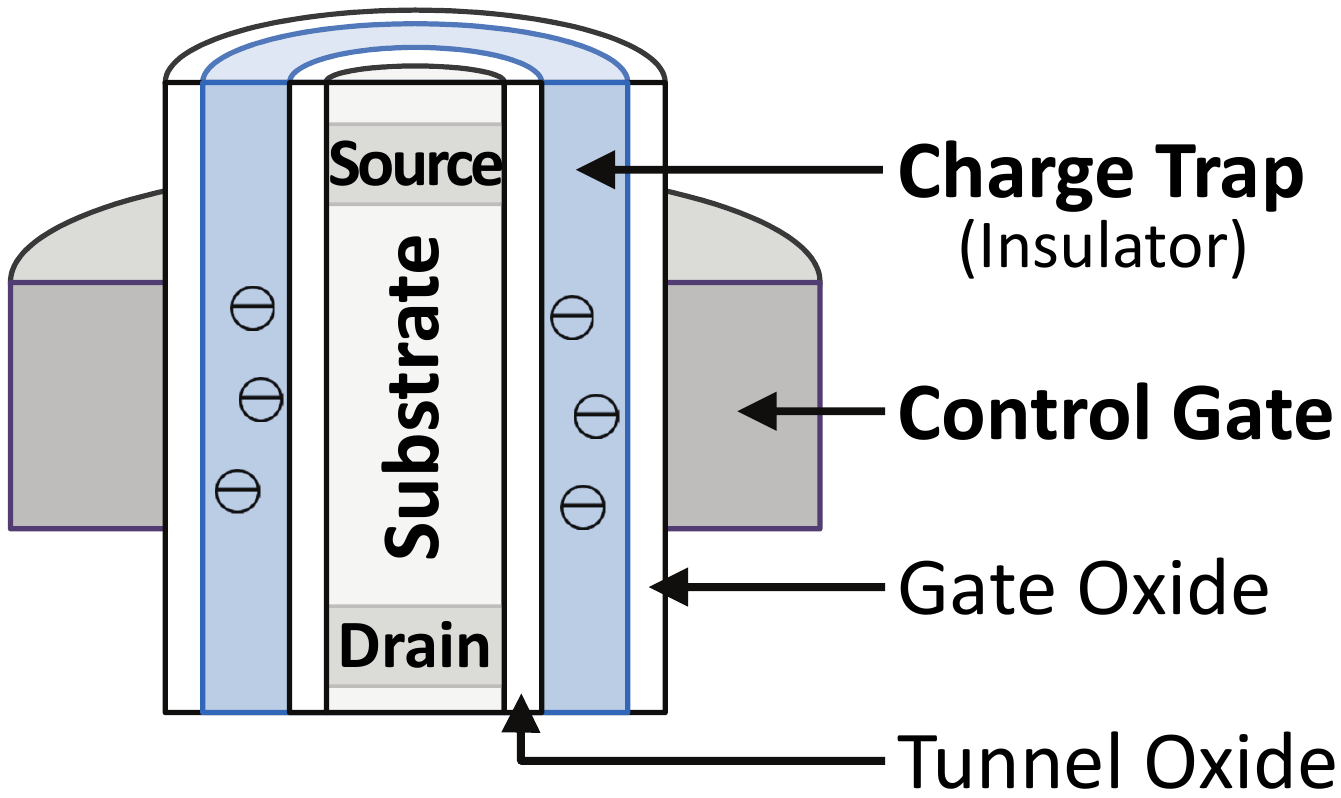}
\caption{\chI{Cross section of a charge trap transistor, used as a flash cell in 3D \chII{charge trap} NAND flash memory.}}
\label{fig:3dcell}
\end{figure}

\chI{Despite the change in cell structure, the mechanism for transferring charge
into and out of the charge trap is similar to the mechanism for transferring
charge into and out of the floating gate.  In 3D NAND flash memory, the charge
trap transistor typically employs FN tunneling to change the threshold voltage 
of the charge trap~\cite{park.jssc15, katsumata.vlsit09}.\footnote{\chI{Note that \chV{\emph{not}} all 
charge trap transistors rely on FN tunneling.  Charge trap transistors used for
NOR flash memory change their threshold voltage using \emph{channel hot
electron injection}, also known as \emph{hot carrier injection}~\cite{luryi.ted84}.}}
When high voltage is applied to the control gate, electrons are injected into
the charge trap from the substrate.  As this behavior is similar to how 
electrons are injected into a floating gate, read, program, and erase operations
remain the same for both planar and 3D NAND flash memory.}

\chI{Figure~\ref{fig:3dorganization} shows how multiple charge trap transistors are
physically organized within 3D NAND flash memory \chIV{to form flash blocks, 
wordlines, and bitlines (see Section~\ref{sec:flash:block})}.  \chII{As mentioned above}, the
channel within a charge trap transistor forms vertically, as opposed to the
horizontal channel that forms within a floating-gate transistor.
The vertical orientation of the channel allows us to stack multiple transistors
\emph{on top of each other} (i.e., along the z-axis) within the chip,
using 3D-stacked circuit integration.
The vertically-connected channels form one bitline of a flash block in 3D NAND
flash memory.  Unlike in planar NAND flash memory, where only the substrates
of flash cells on the same bitline are connected together, flash cells along the
same bitline in 3D NAND flash memory share a common substrate and a common
insulator \chII{(i.e., charge trap)}. The FN tunneling induced by the control gate of the transistor
forms a tunnel only in a local region of the insulator, and, thus, electrons are
injected only into that local region.}
Due to the strong insulating properties of the material used for
the insulator, different regions of a single insulator can have different voltages.
\chII{This means that each region of the insulator can store a different
data value, and thus, the data of \emph{multiple} 3D NAND flash memory cells
can be stored reliably in a \emph{single} insulator.}  This is because the FN tunneling
induced by the control gate of the transistor forms a tunnel only in a \chIV{\emph{local}} region
of the insulator, and, thus, electrons are injected only into that local region.

\begin{figure}[h]
\centering
\includegraphics[width=.6\linewidth]{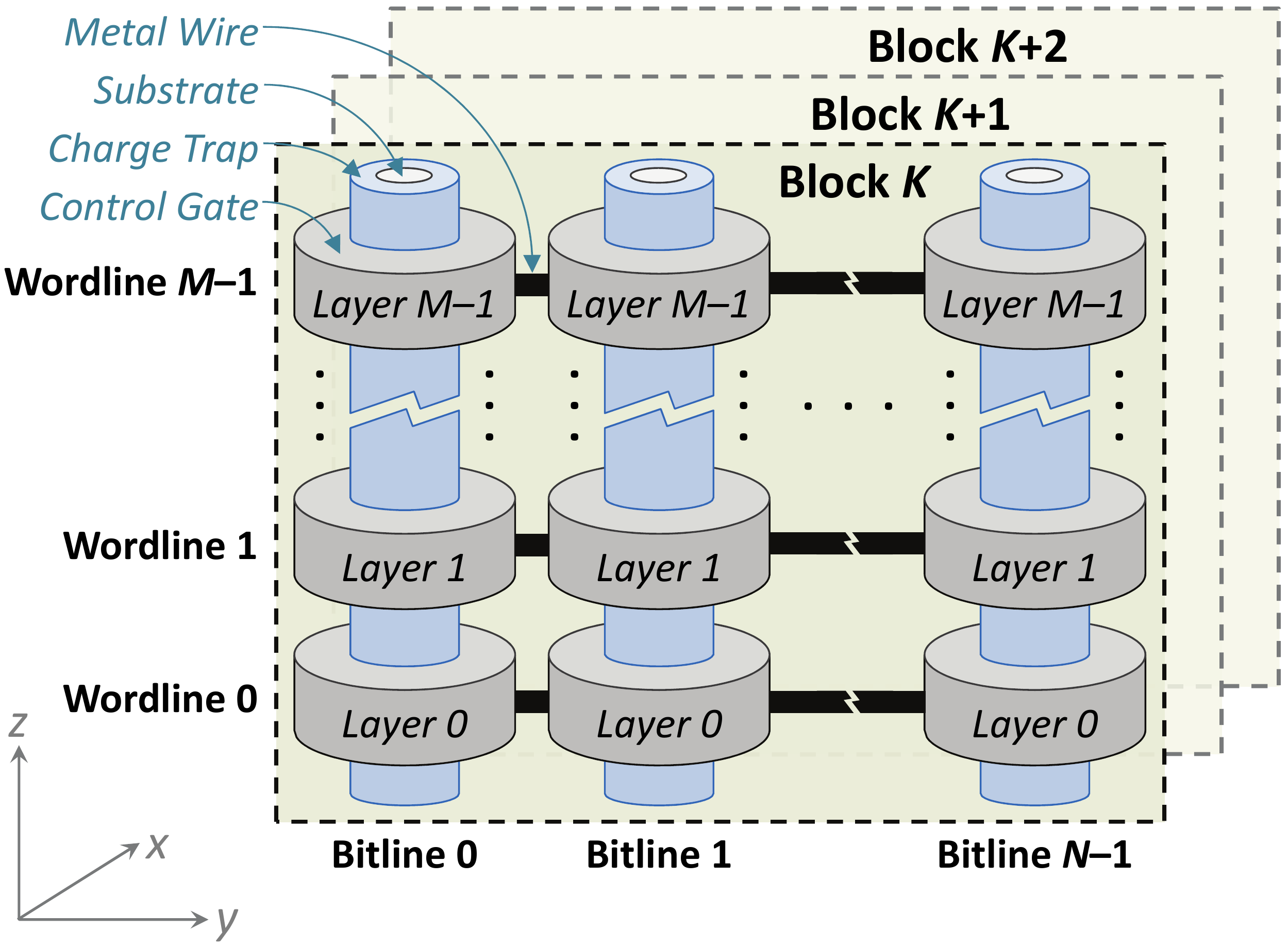}
\caption{\chI{Organization of flash cells in an \emph{M}-layer 3D charge trap NAND flash memory chip,
where each block consists of \emph{M}~wordlines and \emph{N}~bitlines.}}
\label{fig:3dorganization}
\end{figure}

\chI{Each cell along a bitline belongs to a different \emph{layer} of the flash 
memory chip.  Thus, a bitline crosses \chIV{\emph{all}} of the layers within the chip.  
Contemporary 3D NAND flash \chIII{memory contains} 24--\chIII{96 layers~\cite{{R79, park.jssc15,
kang.isscc16, kim.isscc17, elliott.fms17, toshiba.3dnand}}}.
Along the y-axis, the control gates of cells within a \emph{single layer} are 
connected together to form one wordline of a flash block.  As we show in
Figure~\ref{fig:3dorganization}, a block in 3D NAND flash memory consists of 
all of the flash cells within the same y-z plane (i.e., all cells that have the
same coordinate along the x-axis).
\chIV{Note that, while not depicted in Figure~\ref{fig:3dorganization},
each bitline within a 3D NAND flash block includes a sense 
amplifier and two selection transistors used to select the bitline (i.e., the SSL and GSL
transistors; see Section~\ref{sec:flash:block}).  The sense amplifier and 
selection transistors are connected in series with the charge trap transistors
that belong to the same bitline, in a similar manner to
the connections shown for a planar NAND flash block in Figure~\ref{fig:F8}.}}
\chV{More detail on the circuit-level design of 3D NAND flash memory can be
found in \cite{katsumata.vlsit09, komori.iedm08, tanaka.vlsit07, jang.vlsit09}.}

\chI{Due to the use of multiple layers of flash cells within a single NAND flash
memory chip, \chII{which greatly increases capacity per unit area,} manufacturers can achieve high cell density \chIII{\emph{without}} the need to
use small manufacturing process technologies. \chIII{For example, state-of-the-art planar NAND flash
memory uses the \SIrange{15}{19}{\nano\meter} feature size~\cite{R42, R53}.
In contrast, contemporary} 3D NAND flash memory uses larger feature sizes (e.g.,
\SIrange{30}{50}{\nano\meter})~\cite{R79, samsung.whitepaper14}. 
The larger feature sizes reduce manufacturing costs, as their corresponding
manufacturing process technologies are much more mature and have a higher yield
than the \chIII{process} technologies used for small feature sizes.  As we discuss in Section~\ref{sec:3d:errors},
the larger feature size also has an effect on the reliability of 3D NAND flash
memory.}

\subsection{\chI{Errors in 3D NAND Flash \chIII{Memory}}}
\label{sec:3d:errors}

\chI{While the high-level behavior of 3D NAND flash memory is similar to the
behavior of 2D planar NAND flash memory, there are a number of differences 
between the reliability of 3D NAND flash and planar NAND flash,
\chIII{which we summarize in Table~\ref{tbl:3dchanges}}.
There are two reasons for the \chII{differences} in reliability:
(1)~the use of charge trap transistors instead of
floating-gate transistors, and
(2)~moving to a larger manufacturing process technology.
We categorize the changes based on the reason for the change below.}

\paratitle{\chIII{Effects of Charge Trap Transistors}}
\chI{Compared to the reliability \chII{issues} discussed in Section~\ref{sec:errors} for 
planar NAND flash memory, the use of charge trap transistors introduces
two key differences:
(1)~\emph{early retention loss}~\cite{choi.vlsit16, R79, mizoguchi.imw17}, and
(2)~a \emph{reduction in P/E cycling \chII{errors}}~\cite{park.jssc15, R79}.}

\chI{\chII{First,} early retention loss refers to the rapid leaking of electrons from a
flash cell soon after the cell is programmed~\cite{choi.vlsit16, R79}.
Early retention loss occurs in 3D NAND flash memory because charge can now
migrate out of the charge trap in \emph{three} dimensions.  In planar NAND flash
memory, charge leakage due to retention occurs across the tunnel oxide,
\chIII{which \chIV{occupies} two dimensions} (see
Section~\ref{sec:errors:retention}).  In 3D NAND flash memory, charge can 
leak across \emph{both} the tunnel oxide \emph{and} the insulator \chII{that is} used 
for the charge trap, \chIV{i.e., across \emph{three} dimensions}.  The additional charge leakage takes place for only a few
seconds after cell programming.  After a few seconds have passed, the impact
of leakage through the charge trap decreases, and the long-term cell retention
behavior is similar to that of flash cells in planar NAND flash memory~\cite{choi.vlsit16, R79, mizoguchi.imw17}.}

\chI{\chII{Second, P/E cycling errors} (see Section~\ref{sec:errors:pe}) 
reduce \chIII{with 3D NAND flash memory} because the tunneling oxide in charge trap transistors is 
\emph{less} susceptible to breakdown than the oxide in floating-gate transistors
during high-voltage operation~\cite{R79, mizoguchi.imw17}.  As a result, the oxide is less likely
to contain trapped electrons once a cell is erased, which in turn makes it less
likely that the cell is subsequently programmed to an incorrect threshold 
voltage.
One benefit of the reduction in P/E cycling errors is that the
endurance (i.e., the maximum P/E cycle count) for a \chII{3D flash memory} cell has increased by
\chIII{more than} an order of magnitude\chIII{~\cite{R80, parnell.fms17}}.}

\paratitle{\chIII{Effects of Larger Manufacturing Process Technologies}}
\chI{Due to the use of larger manufacturing process technologies for 3D NAND 
flash memory, many of the errors that we observe in 2D planar NAND flash (see
Section~\ref{sec:errors}) are not as prevalent in 3D NAND flash memory.
For example, while read disturb is a prominent source of errors at small feature
sizes (e.g., \SIrange{20}{24}{\nano\meter}), its effects are small at larger feature
sizes~\cite{R38}. Likewise, there are much fewer errors due to cell-to-cell 
program interference (see Section~\ref{sec:errors:celltocell}) in 3D NAND flash
memory, as the physical distance between neighboring cells is much larger due to
the increased feature size.
As a result, both cell-to-cell program interference and read disturb are 
\chIV{\emph{currently} not} major issues in 3D NAND flash memory 
reliability\chIII{~\cite{park.jssc15, R79, parnell.fms17}}.}

\chIII{One advantage of the lower cell-to-cell program interference is that 
3D NAND flash memory uses the older \emph{one-shot programming} 
algorithm~\cite{R80, yoon.fms17, parnell.fms17} (see Section~\ref{sec:flash:pgmerase}).
In planar NAND flash memory, one-shot programming was replaced by two-step
programming (for MLC) and foggy-fine programming (for TLC) in order to reduce 
the impact of cell-to-cell program interference on fully-programmed cells 
\chIV{(as we describe in Section~\ref{sec:flash:pgmerase})}.
The lower interference in 3D NAND flash memory makes two-step and foggy-fine
programming unnecessary.  As a result, none of the cells in 3D NAND flash memory
are partially-programmed, significantly reducing the number of program errors (see 
Section~\ref{sec:errors:pgm}) that occur~\cite{parnell.fms17}.}

\chI{Unlike \chII{the effects on reliability} due to the use of a charge trap transistor, \chII{which are likely longer-term},
\chII{the effects on reliability} due to the use of larger manufacturing process technologies
are expected to \chII{be shorter-term}.  As manufacturers seek to further
increase the density of 3D NAND flash memory, they will reach an upper limit \chII{for}
the number of layers that can be integrated within a 3D-stacked \chII{flash memory} 
chip\chIII{, which is currently projected to be in the range of 300--512 layers~\cite{lapedus.semieng16, lee.eetimes17}}.  At that
point, manufacturers will once again need to scale down the chip to \chIII{\emph{smaller}}
manufacturing process technologies~\cite{R79}, \chII{which, in turn,} will reintroduce high amounts
of read disturb and cell-to-cell program interference (just as it happened for
planar NAND flash memory~\cite{R26, kim.irps10, R38, R35, R36}).}

\subsection{\chI{Changes in Error Mitigation for 3D NAND Flash Memory}}
\label{sec:3d:mitigation}

\chI{Due to the reduction in a number of sources of errors, fewer error mitigation
mechanisms are currently needed for 3D NAND flash memory.  For example,
because the number of errors introduced by cell-to-cell program interference is 
\chII{currently} low, manufacturers have \chIII{\emph{reverted}} to using one-shot programming (see
Section~\ref{sec:flash:pgmerase}) for 3D NAND flash\chIII{~\cite{R80, yoon.fms17, parnell.fms17}}.
As a result of the \chII{currently small} effect of read disturb errors, mitigation and recovery
mechanisms for read disturb (e.g., pass-through voltage optimization in
Section~\ref{sec:mitigation:voltage}, Read Disturb Recovery in
Section~\ref{sec:correction:recovery}) \chIII{may} not \chIII{be} \chII{needed,}
for the time being.  We expect that once 3D NAND flash memory begins to scale
down to smaller manufacturing process technologies, approaching the current
feature sizes used for planar NAND flash memory, there will be a \chII{significant} need for 3D
NAND flash \chIII{memory} to use \chIII{many, if not all, of} the error mitigation mechanisms we discuss in
Section~\ref{sec:mitigation}.}

\chI{To our knowledge, no mechanisms have been designed yet to reduce
the impact of early retention loss,
\chII{which is a new error mechanism in 3D NAND flash memory}.  This is in part due to the reduced overall impact
of retention errors in 3D NAND flash memory compared to planar NAND flash
memory~\cite{choi.vlsit16}, \chII{since} a larger cell contains a greater
number of electrons than a smaller cell at the same threshold voltage.
As a result, existing refresh mechanisms (see Section~\ref{sec:mitigation:refresh})
can be used to tolerate errors introduced by early retention loss with little
modification.} \chII{However, as 3D NAND flash memory scales into future smaller
technology nodes, the early retention loss problem may require new mitigation techniques.}

\chI{\chVII{At the time of writing, only a few rigorous studies examine 
error characteristics of and error mitigation techniques for 3D NAND flash
memories.  An example of
such a study is by Luo et al.~\cite{luo.hpca18}, which (1)~examines the
\emph{self-recovery effect} in 3D NAND flash memory, where the damage caused by 
wearout due to P/E cycling (see Section~\ref{sec:errors:pe}) can be repaired by
\chVIII{\emph{detrapping} electrons that are \emph{inadvertently} trapped in flash cells}; 
(2)~examines how the operating temperature of 3D NAND flash memory affects
the raw bit error rate; 
(3)~comprehensively models the impact of wearout, data retention, self-recovery,
and temperature on 3D NAND flash reliability; and 
\chVIII{(4)}~proposes a new technique to
mitigate errors in 3D NAND flash memory using this comprehensive model.}
\chVII{Other such} studies \chII{(1)}~may expose additional sources of errors that have not yet been
observed, and \chII{that} may be unique to 3D NAND flash memory; \chII{and (2)~can enable
a solid understanding of current error mechanisms in 3D NAND flash memory so
that appropriate specialized mitigation mechanisms can be developed}.
We expect that future works will \chII{experimentally} examine such sources of errors, and will
potentially introduce novel mitigation mechanisms for these errors.}
\chII{Thus, the field (both academia and industry) is currently in much need of
rigorous experimental characterization \chIII{and analysis} of 3D NAND flash
memory devices.}


\section{Similar Errors in Other Memory Technologies}
\label{sec:othermem}

As we discussed in Section~\ref{sec:errors}, there are five major sources
of errors in flash-memory-based SSDs. Many of these error
sources can also be found in other types of memory and
storage technologies. In this section, we take a brief look
at the major reliability issues that exist within DRAM and
in emerging nonvolatile memories. In particular, we focus
on DRAM in our discussion, as modern SSD controllers
have access to dedicated DRAM of considerable capacity
(e.g., \SI{1}{\giga\byte} for every \SI{1}{\tera\byte} of SSD capacity), which exists
within the SSD package (see Section~\ref{sec:ssdarch}). Major sources
of errors in DRAM include data retention, cell-to-cell
interference, and read disturb. There is a wide body of
work on mitigation mechanisms for the \chII{DRAM and
emerging memory technology} errors we describe
in this section, but we explicitly discuss only a select
number of them here, \chII{since a full treatment of
such mechanisms is out of the scope of this current chapter}.

\subsection{Cell-to-Cell Interference Errors in DRAM}
\label{sec:othermem:celltocell}

One similarity
between the capacitive DRAM cell and the floating-gate
cell in NAND flash memory is that they are both vulnerable
to cell-to-cell interference. In DRAM, one important way
in which cell-to-cell interference exhibits itself is the data-dependent
retention behavior, where the retention time of
a DRAM cell is dependent on the values written to \emph{nearby}
DRAM cells\chIII{~\cite{R104, R126, R127, R147, R149, R206}}. This phenomenon
is called \emph{data pattern dependence} (DPD)~\cite{R104}. Data pattern
dependence in DRAM is similar to the data-dependent
nature of program interference that exists in NAND flash
memory (see Section~\ref{sec:errors:celltocell}). Within DRAM, data dependence
occurs as a result of parasitic capacitance coupling (between
DRAM cells). Due to this coupling, the amount of charge
stored in one cell's capacitor can inadvertently affect the
amount of charge stored in an adjacent cell's capacitor\chIII{~\cite{R104, R126, R127, R147, R149, R206}}. 
As DRAM cells become smaller with
technology scaling, cell-to-cell interference worsens because
parasitic capacitance coupling between cells increases~\cite{R104, R126}. 
More findings on cell-to-cell interference and the
data-dependent nature of cell retention times in DRAM,
along with experimental data obtained from modern DRAM
chips, can be found in our prior works~\cite{R104, R125, R126, R127, R147, R149, R174, R206}.

\subsection{Data Retention Errors in DRAM}
\label{sec:othermem:retention}

DRAM uses the charge
within a capacitor to represent one bit of data. Much like the
floating gate within NAND flash memory, charge leaks from
the DRAM capacitor over time, leading to data retention
issues. Charge leakage in DRAM, if left unmitigated, can lead
to much more rapid data loss than the leakage observed in a
NAND flash cell. While leakage from a NAND flash cell typically
leads to data loss after several days to years of retention
time (see Section~\ref{sec:errors:retention}), leakage from a DRAM cell leads to
data loss after a retention time on the order of \emph{milliseconds} to
\emph{seconds}~\cite{R104}.

\chII{The retention time of a DRAM cell depends upon several factors,
including (1)~manufacturing process variation and 
(2)~temperature~\cite{R104}.
Manufacturing process variation affects the amount of current that leaks
from each DRAM cell's capacitor and access transistor~\cite{R104}.
As a result, the retention time of the cells within a single DRAM chip vary
significantly, resulting in \emph{strong cells} that have high retention
times and \emph{weak cells} that have low retention times within each chip.
The operating temperature affects the rate at which charge leaks from the
capacitor.  As the operating temperature increases, the retention time of a
DRAM cell decreases exponentially~\cite{R104, hamamoto.ted98}.
Figure~\ref{fig:dram-retention-temperature} shows the change in retention
time as we vary the operating temperature, as measured from real DRAM 
chips~\cite{R104}.  In Figure~\ref{fig:dram-retention-temperature}, we
normalize the retention time of each cell to its retention time at an
\chIII{operating} temperature of \SI{50}{\celsius}.  As the number of cells is
\chIII{large}, we group the normalized retention times into bins, and plot the
density of each bin.  \chIV{We} draw two \chIII{exponential-fit} curves:
(1)~the \emph{peak} curve, which is drawn through the most populous
bin at each temperature measured; and
(2)~the \emph{tail} curve, which is drawn through the lowest non-zero
bin for each temperature measured.}
\chIII{Figure~\ref{fig:dram-retention-temperature} provides us with three major
conclusions about the relationship between DRAM cell retention time and
temperature.
First, both of the exponential-fit curves fit well, which confirms the exponential decrease
in retention time as the operating temperature increases in modern DRAM devices.
Second, the retention times of different DRAM cells are affected very differently
by changes in temperature.
Third, the variation in retention time across cells increases greatly as temperature
increases.
More analysis of factors that affect DRAM retention times can be found in our
recent works\chV{~\cite{R104, R126, R147, R125, R127, R149, R206}}.}

\begin{figure}[h]
  \centering
  \includegraphics[width=0.65\columnwidth]{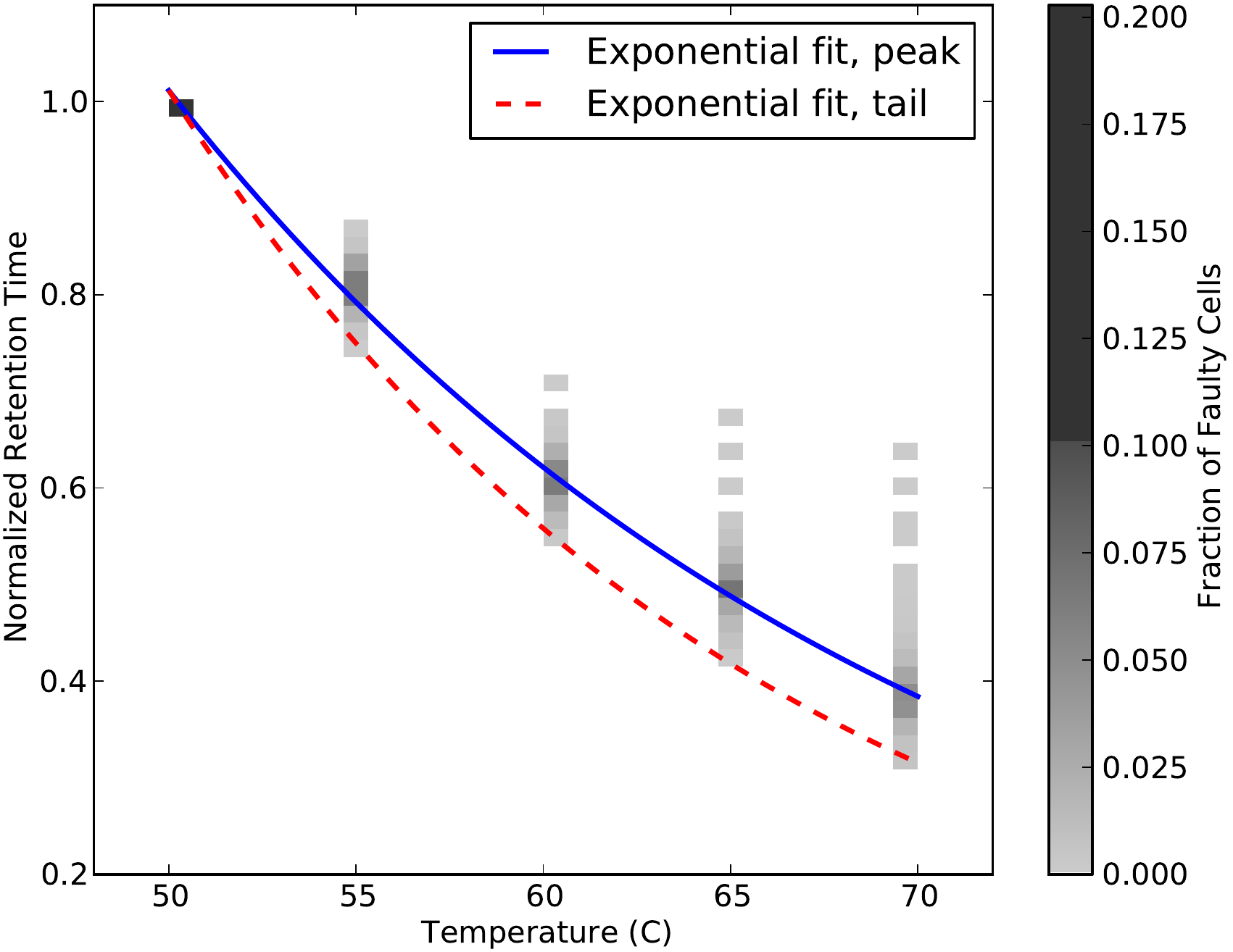}%
  \vspace{-5pt}%
  \caption{\chII{DRAM retention time vs.\ operating temperature, normalized to 
  the retention time of each DRAM cell at \SI{50}{\celsius}.  Reproduced from \cite{R104}.}}%
  \label{fig:dram-retention-temperature}%
\end{figure}

Due to the rapid charge leakage from DRAM
cells, a DRAM controller periodically refreshes all DRAM cells
in place~\cite{R66, R69, R104, R123, R125, R126, R147} (similar to
the techniques discussed in Section~\ref{sec:mitigation:refresh}, but at a much smaller
time scale). DRAM standards require a DRAM cell to be
refreshed once every \SI{64}{\milli\second}~\cite{R123}. As the density of DRAM continues
to increase over successive product generations (e.g., by
128x between 1999 and 2017~\cite{R120, R174}),
\chIII{enabled by the scaling of DRAM to smaller manufacturing process
technology nodes\chV{~\cite{mandelman.ibmjrd02}},} the performance
and energy overheads required to refresh an entire DRAM
module have grown significantly\chIII{~\cite{R66, R69}}.
\chII{It is expected that the refresh problem will get worse and \chIII{limit}
DRAM density scaling, \chIII{as described in a recent work by
Samsung and Intel~\cite{R153} and by our group~\cite{R66}}.
\chIII{Refresh operations in DRAM cause both
(1)~performance loss and (2)~energy waste, both of which together lead to a
difficult technology scaling challenge.
Refresh operations degrade performance due to three major reasons.
First, refresh operations increase the memory latency, as a request to a DRAM bank that is
refreshing must wait for the refresh latency before it can be serviced.
Second, they reduce the amount of bank-level parallelism available to
requests, as a DRAM bank cannot service requests during refresh.
Third, they decrease the row buffer hit rate, as a refresh operation causes all
open rows in a bank to be closed.}
When a DRAM chip scales to a greater capacity, there are more DRAM rows that
need to be refreshed.  As Figure~\ref{fig:dram-refresh-scaling}a shows,
the amount of time spent on \chIII{each refresh operation} scales linearly with the capacity of the
DRAM chip.  The additional time spent on refresh causes the \chIII{DRAM data throughput
loss due to refresh} to become more severe in denser DRAM chips, as shown in 
Figure~\ref{fig:dram-refresh-scaling}b.  For a chip with a density of \SI{64}{\giga\bit},
nearly 50\% of the \chIII{data} throughput is lost due to the high amount of time spent on
refreshing all of the rows in the chip.  The increased refresh time \chIII{also increases the
effect of refresh on
power consumption.}  As we observe from Figure~\ref{fig:dram-refresh-scaling}c,
the fraction of DRAM power spent on refresh is expected to be the dominant
component of \chIII{the} total DRAM power consumption, \chIII{as DRAM chip 
capacity scales to become larger.  For a chip with a density of \SI{64}{\giga\bit},
nearly 50\% of the DRAM chip power is spent on refresh operations.  Thus,} 
refresh poses a clear challenge to DRAM scalability.}

\begin{figure}[t]
  \centering
  \begin{subfigure}[b]{0.32\columnwidth}%
    \includegraphics[width=\textwidth]{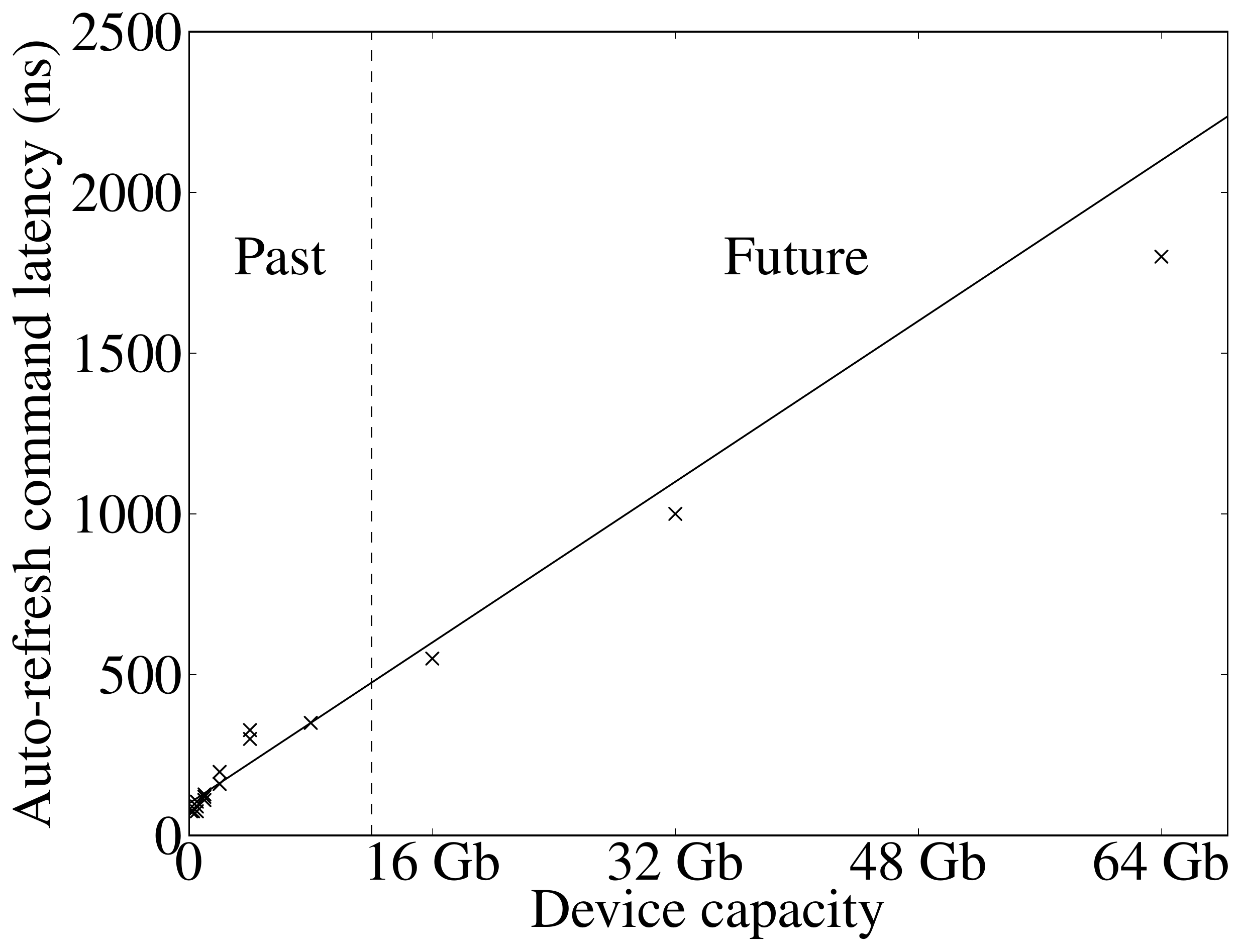}%
    \vspace{-5pt}%
    \caption{Refresh latency}%
  \end{subfigure}%
  \hfill%
  \begin{subfigure}[b]{0.32\columnwidth}%
    \includegraphics[width=\textwidth]{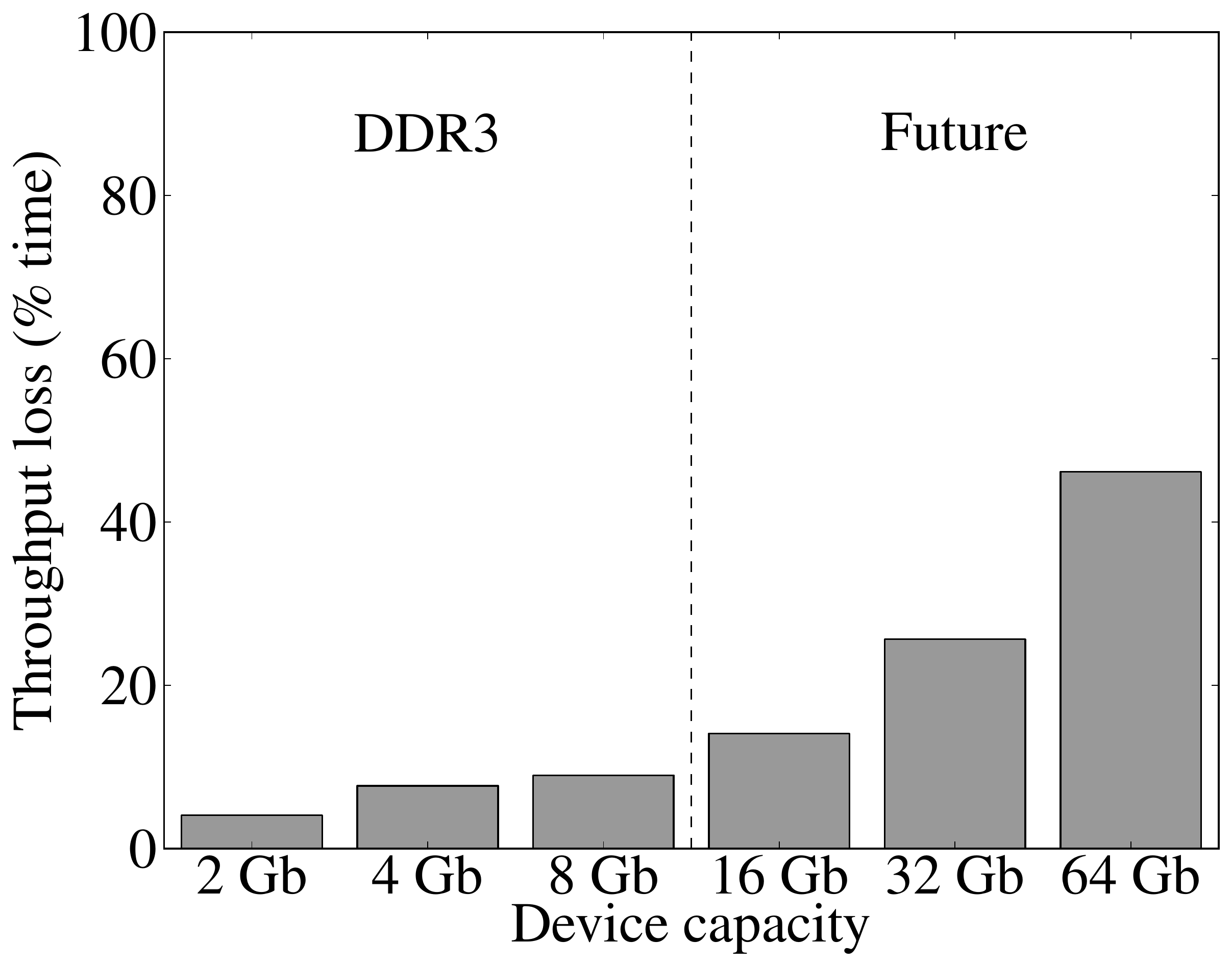}%
    \vspace{-5pt}%
    \caption{Throughput loss}%
  \end{subfigure}%
  \hfill%
  \begin{subfigure}[b]{0.32\columnwidth}%
    \includegraphics[width=\textwidth]{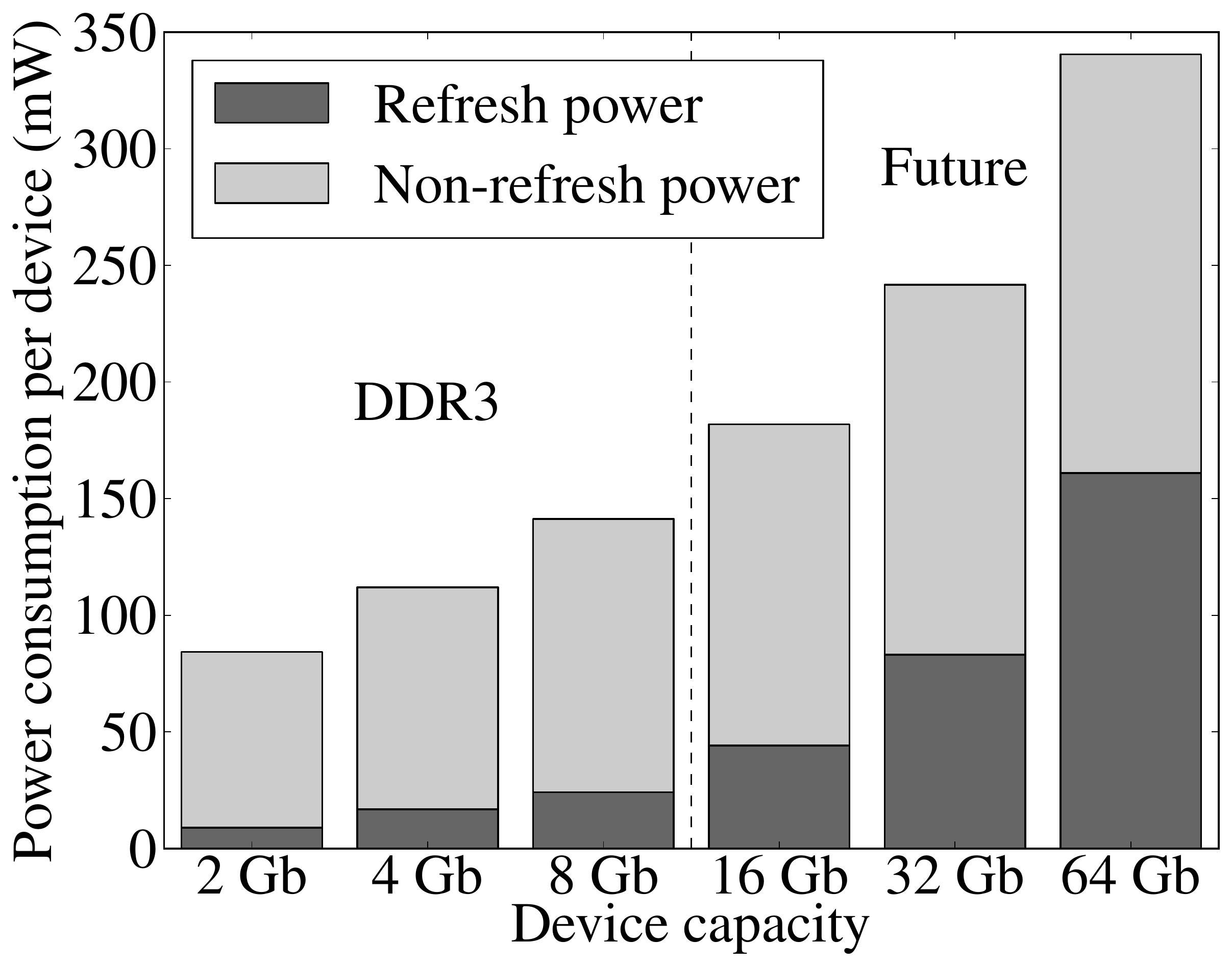}%
    \vspace{-5pt}%
    \caption{Power consumption}%
  \end{subfigure}%
  \vspace{-5pt}%
  \caption{\chII{\chIII{Negative performance and power consumption} effects of refresh in contemporary and future DRAM devices.  
  \chIII{We expect that as the capacity of each DRAM chip increases, (a)~the refresh latency, 
  (b)~the DRAM throughput lost during refresh operations, and (c)~the power consumed by refresh will all increase.}
  Reproduced from \cite{R66}.}}%
  \label{fig:dram-refresh-scaling}%
\end{figure}

To combat the growing performance and energy overheads
of refresh, two classes of techniques have been
developed. The first class of techniques reduce the \emph{frequency}
of refresh operations without sacrificing the reliability
of data stored in DRAM (e.g., \chIII{\cite{R66, R125, R126, R145, R146, R147, R149, baek.tc14, R206}}). 
\chI{Various experimental studies of real DRAM chips (e.g., \chIII{\cite{R66, R104, R119, R125, R126, R127, R147, R148, R157}})
have studied the \chII{data} retention time \chII{of DRAM cells in modern chips}.
Figure~\ref{fig:dram-retention}
shows the retention time measured from seven different \chII{real} DRAM modules
\chII{(by manufacturers A, B, C, D, and E) at an operating temperature of
\SI{45}{\celsius}}, as a
cumulative distribution (CDF) of the fraction of cells that have a retention time
less than the x-axis value~\cite{R104}.  We observe from the figure that even
for the DRAM module whose cells have the worst retention time (i.e., the
CDF is the highest), \chIII{fewer than only} 0.001\% of the total cells have a retention time
\chIII{smaller} than \SI{3}{\second} \chII{at \SI{45}{\celsius}}.  
\chII{As shown in Figure~\ref{fig:dram-retention-temperature}, the retention 
time decreases exponentially as the temperature increases.  We can extrapolate
our observations from Figure~\ref{fig:dram-retention} to the worst-case
operating conditions by using the tail curve from Figure~\ref{fig:dram-retention-temperature}.
DRAM standards specify that the operating temperature of DRAM should not exceed
\SI{85}{\celsius}~\cite{R123}.  Using the tail curve, we find that a retention
time of \SI{3}{\second} at \SI{45}{\celsius} is equivalent to a retention time
of \SI{246}{\milli\second} at the worst-case temperature of \SI{85}{\celsius}.}
Thus, the vast majority of DRAM cells can retain
data without loss for much longer than the \SI{64}{\milli\second} retention
time specified by DRAM standards.  \chIII{The} other experimental studies 
\chIII{of DRAM chips have validated this observation} as well\chIII{~\cite{R66, R119, R125, R126, R127, R147, R148, R157}}.}

\begin{figure}[h]
  \centering
  \includegraphics[width=0.65\columnwidth]{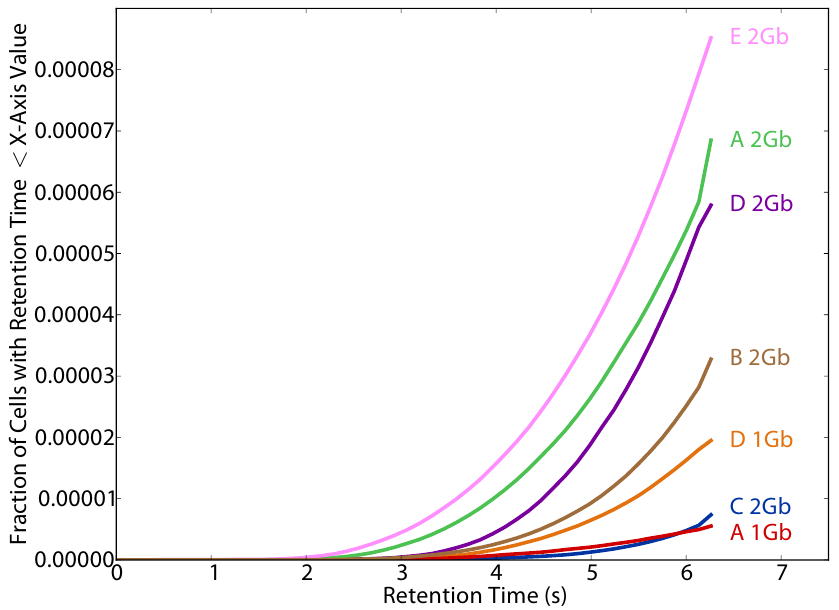}%
  \vspace{-5pt}%
  \caption{\chI{Cumulative distribution of the number of cells in a DRAM module
  with a retention time less than the value on the x-axis, plotted for seven
  different DRAM modules.  Reproduced from \cite{R104}.}}%
  \label{fig:dram-retention}%
\end{figure}

A number of works take advantage of this \chIII{variability in data retention time behavior
across DRAM cells,
by introducing heterogeneous refresh rates, i.e., different refresh rates for
different DRAM rows.  Thus, these works can reduce} the frequency at which the vast majority of DRAM \chIII{rows} within a
module are refreshed \chII{(e.g., \chIII{\cite{R66, R104, R125, R126, R145, R146, R147, R149, baek.tc14}})}.
\chII{For example, the key idea of RAIDR~\cite{R66} is to
refresh the \emph{strong} DRAM rows (i.e., those rows that can retain data
for \chIII{much} longer than the minimum \SI{64}{\milli\second} retention time in the
DDR4 standard~\cite{R123}) less frequently, and refresh the \emph{weak}
DRAM rows (i.e., those rows that can retain data only for the minimum retention
time) more frequently.  The major challenge  in such works is how to accurately
identify the retention time of each DRAM row.  To solve this challenge,
many recent works examine (online) DRAM retention time profiling 
techniques~\chIII{\cite{R147, R125, R127, R126, R104, R206}}.}

The second
class of techniques reduce the interference caused by
refresh requests on demand requests (e.g., \cite{R69, R114, R163}). 
These works either change the scheduling order of
refresh requests~\cite{R69, R114, R163} or slightly modify the
DRAM architecture to enable the servicing of refresh and
demand requests in parallel~\cite{R69}.

\chIII{One \chIV{critical challenge} in developing techniques to reduce 
refresh overheads is that it is getting significantly more difficult to
determine the minimum retention time of a DRAM cell,
as we have shown experimentally on modern DRAM
chips\chIV{~\cite{R104, R126, R127, R125, R147}}.
Thus, determining the correct rate
at which to refresh DRAM cells has become more difficult, as also
indicated by industry~\cite{R153}. This is due to two major phenomena, both
of which get worse (i.e., become more prominent) with manufacturing 
process technology scaling. 
The first phenomenon is \emph{variable retention time} (VRT), where the
retention time of some DRAM cells can change drastically over time,
due to a memoryless random process that
results in very fast charge loss via a phenomenon called \emph{trap-assisted
gate-induced drain leakage}\chIV{~\cite{R104, R125, yaney.iedm87, restle.iedm92}}. 
VRT, as far as we know, is very
difficult to test for, because there seems to be no way of determining
that a cell exhibits VRT until that cell is observed to exhibit VRT,
and the time scale of a cell exhibiting VRT does not seem to
be bounded, based on the current experimental data \chIV{on modern
DRAM devices~\cite{R104, R147}}. 
The second phenomenon is \emph{data pattern dependence} (DPD), which we discuss
in Section~\ref{sec:othermem:celltocell}.
Both of these phenomena greatly
complicate the accurate determination of minimum data retention
time of DRAM cells.  Therefore, data retention in DRAM continues to be 
a vulnerability that can greatly affect \chIV{DRAM technology scaling (and
thus performance and energy consumption) as well as the} reliability and
security of current and future DRAM generations.}

More findings on the
nature of DRAM data retention and associated errors, as
well as relevant experimental data from modern DRAM
chips, can be found in our prior works\chIII{~\cite{R66, R69, R104, R119, R125, R126, R127, R147, R157, R174, R206, R149, R131}}.

\subsection{Read Disturb Errors in DRAM}
\label{sec:othermem:readdisturb}

Commodity DRAM
chips that are sold and used in the field today exhibit read
disturb errors~\cite{R116}, also called \emph{RowHammer}-induced errors~\cite{R131}, 
which are \emph{conceptually} similar to the read disturb
errors found in NAND flash memory (see Section~\ref{sec:errors:readdisturb}).
Repeatedly accessing the same row in DRAM can cause
bit flips in data stored in adjacent DRAM rows. In order to
access data within DRAM, the row of cells corresponding
to the requested address must be \emph{activated} (i.e., opened for
read and write operations). This row must be \emph{precharged}
(i.e., closed) when another row in the same DRAM bank
needs to be activated. Through experimental studies on a
large number of real DRAM chips, we show that when a
DRAM row is activated and precharged repeatedly (i.e.,
\emph{hammered}) enough times within a DRAM refresh interval,
one or more bits in physically-adjacent DRAM rows can be
flipped to the wrong value~\cite{R116}.

\chII{We tested 129~DRAM modules manufactured by three major manufacturers
(A, B, and C) between 2008 and 2014, using an FPGA-based experimental DRAM
testing infrastructure~\cite{R157} (more detail on our experimental
setup, along with a list of all modules and their characteristics, can be found 
in our original RowHammer paper~\cite{R116}).  
Figure~\ref{fig:rowhammer-date} shows the
rate of RowHammer errors that we found, with the 129~modules that we tested
categorized based on their manufacturing date.
We find that 110 of our tested modules exhibit RowHammer errors, with the
earliest such module dating back to 2010.  In particular, we find that
\emph{all} of the modules manufactured in 2012--2013 that we tested are
vulnerable to RowHammer. Like with many NAND flash \chIII{memory error 
mechanisms, especially read disturb}, RowHammer
is a recent phenomenon that \chIII{especially affects DRAM chips} manufactured with more advanced
manufacturing process technology generations.}

\begin{figure}[h]
  \centering
  \includegraphics[width=0.5\columnwidth]{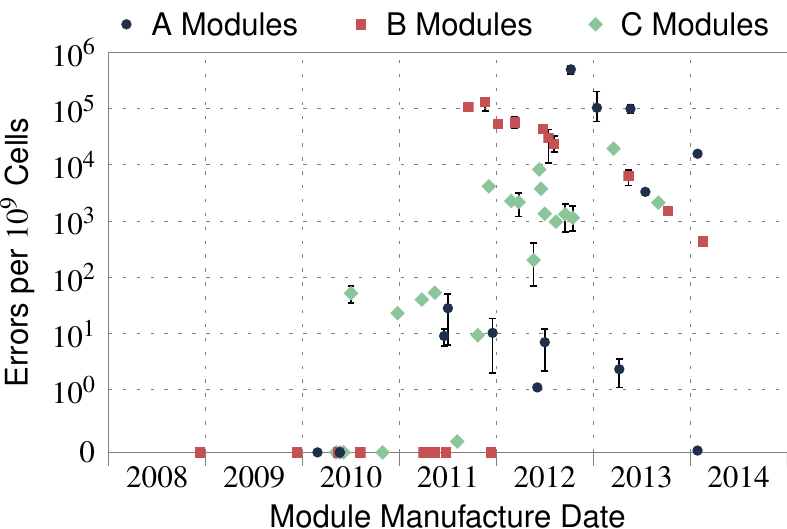}%
  \vspace{-5pt}%
  \caption{\chII{RowHammer error rate vs.\ manufacturing dates of 129~DRAM
  modules we tested.  Reproduced from \cite{R116}.}}%
  \label{fig:rowhammer-date}%
\end{figure}

\chI{Figure~\ref{fig:rowhammer-interference} shows the distribution of
the number of rows (plotted in log scale on the y-axis) within a DRAM module that flip the
number of bits along the x-axis, as measured for example DRAM 
modules from three different DRAM manufacturers~\cite{R116}.
We make two observations from the figure.  First, the number of bits
flipped when we hammer a row (known as the \emph{aggressor row}) can vary
significantly within a module.  \chII{Second, each module has a different 
distribution of the number of rows.}
Despite these differences, we find that this DRAM failure mode}
affects more than 80\% of the DRAM chips we tested~\cite{R116}.
As indicated above, this read disturb error mechanism in
DRAM is popularly called RowHammer~\cite{R131}.

\begin{figure}[h]
  \centering
  \includegraphics[width=0.5\columnwidth]{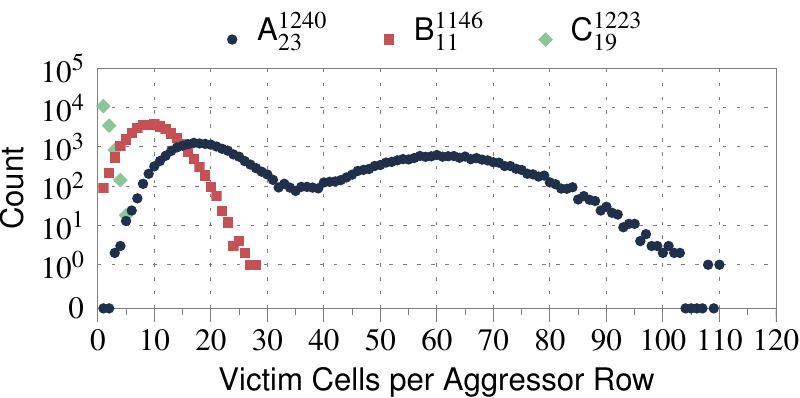}%
  \vspace{-5pt}%
  \caption{\chI{Number of victim cells (i.e., number of bit errors) when an
  aggressor row is repeatedly activated, for three representative DRAM modules
  from three major manufacturers.  \chII{We label the modules in the format $X^{yyww}_n$, 
  where $X$ is the manufacturer (A, B, or C), $yyww$ is the manufacture year ($yy$) and
  week of the year ($ww$), and $n$ is the number of the selected module.} 
  Reproduced from \cite{R116}.}}%
  \label{fig:rowhammer-interference}%
\end{figure}

Various recent works show that RowHammer can be
maliciously exploited by user-level software programs to
(1)~induce errors in existing DRAM modules~\cite{R116, R131}
and (2)~launch attacks to compromise the security of various
systems\chIV{~\cite{R115, R131, R132, R135, R136, R137, R138, R158, R179, gruss.arxiv17}}.
For example, by exploiting the RowHammer read disturb
mechanism, a user-level program can gain kernel-level
privileges on real laptop systems~\cite{R115, R132}, take over a
server vulnerable to RowHammer~\cite{R135}, take over a victim
virtual machine running on the same system~\cite{R136}, and
take over a mobile device~\cite{R138}. Thus, the RowHammer
read disturb mechanism is a prime (and perhaps the
first) example of how a circuit-level failure mechanism in
DRAM can cause a practical and widespread system security
vulnerability.  We believe similar (yet \chII{likely} more difficult to
exploit) vulnerabilities exist in MLC NAND flash memory
as well, as described in our recent work~\cite{R40}.

\chII{Note that various solutions to RowHammer exist~\cite{R116, R131, R176},
but we do not discuss them in detail here.  Our recent work~\cite{R131} 
provides a comprehensive overview.  A very promising proposal is to modify
either the memory controller or the DRAM chip such that it probabilistically
refreshes the physically-adjacent rows of a recently-activated row, with very
low probability.  This solution is called \emph{Probabilistic Adjacent Row
Activation} (PARA)\chIII{~\cite{R116}}.  Our prior work shows that this low-cost, low-complexity
solution, which does not require any storage overhead, greatly closes the
RowHammer vulnerability~\cite{R116}.}

The RowHammer effect in DRAM worsens as the manufacturing
process scales down to smaller node sizes~\cite{R116, R131}. 
More findings on RowHammer, along with extensive
experimental data from real DRAM devices, can be found in
our prior works~\cite{R116, R131, R176}.

\subsection{Large-Scale DRAM Error Studies}
\label{sec:othermem:largescale}

Like flash memory,
DRAM is employed in a wide range of computing systems,
at scale. Thus, there is a similar need to study the aggregate
behavior of errors observed in a large number of DRAM
chips deployed in the field. Akin to the large-scale flash
memory SSD reliability studies discussed in Section~\ref{sec:errors:largescale}, a
number of experimental studies characterize the reliability
of DRAM at large scale in the field (e.g., \cite{R117, R118, R124, R150, R151}). 
\chII{We highlight three notable results from these studies.}

\chII{First, as we saw for large-scale studies of SSDs (see 
Section~\ref{sec:errors:largescale}), the number of errors
observed varies significantly for each DRAM module~\cite{R117}.
Figure~\ref{fig:dram-large-scale}a shows the distribution of correctable
errors across the \emph{entire fleet} of servers at Facebook over a
fourteen-month period, omitting the servers that did not exhibit any
correctable DRAM errors.  The x-axis shows the normalized device
number, with devices sorted based on the number of errors they
experienced in a month.  As we saw in the case of SSDs,
a small number of servers accounts for the majority of errors.
As we see from Figure~\ref{fig:dram-large-scale}a, the top 1\%
of servers account for 97.8\% of all observed correctable DRAM errors.
The distribution of the number of errors among servers follows a
power law model.  We show the probability density distribution of
correctable errors in Figure~\ref{fig:dram-large-scale}b, \chIII{which indicates}
that the distribution of errors across servers follows a Pareto 
distribution, with a decreasing hazard rate\chIII{~\cite{R117}}.  This means that a
server that has experienced more errors in the past is likely to 
experience more errors in the future.}

\begin{figure}[h]
  \centering
  \hfill%
  \begin{subfigure}[b]{0.34\columnwidth}%
    \includegraphics[width=\textwidth]{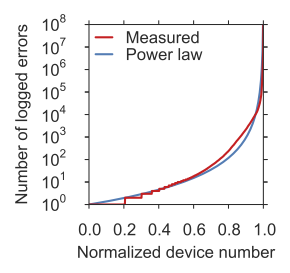}%
    \vspace{-5pt}%
    \caption{}%
  \end{subfigure}%
  \hfill%
  \begin{subfigure}[b]{0.34\columnwidth}%
    \includegraphics[width=\textwidth]{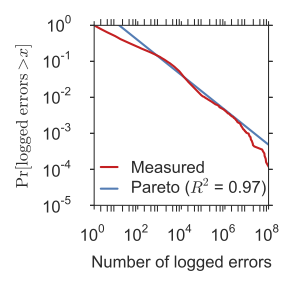}%
    \vspace{-5pt}%
    \caption{}%
  \end{subfigure}%
  \hfill%
  \vspace{-5pt}%
  \caption{\chII{Distribution of memory errors among servers with errors (a),
  which resembles a power law distribution.  Memory errors follow a Pareto
  distribution among servers with errors (b).  
  Reproduced from \cite{R117}.}}%
  \label{fig:dram-large-scale}%
\end{figure}

\chII{Second,} unlike SSDs, DRAM does \chIII{\emph{not} seem to} show any clearly
discernible trend where higher utilization and age lead to
a greater raw bit error rate~\cite{R117}.

\chII{Third,} the increase in the
density of DRAM chips with technology scaling leads to
higher error rates~\cite{R117}.
\chII{The latter is illustrated in Figure~\ref{fig:dram-density},
which shows how different DRAM chip densities are related to
device failure rate.  We can see that there is a clear trend of
increasing failure rate with increasing chip density.
We find that the failure rate increases because despite small
improvements in the reliability of an \emph{individual} cell, the quadratic
increase in the number of cells per chip greatly increases the probability
of observing a single error in the whole chip\chIII{~\cite{R117}}.}

\begin{figure}[h]
  \centering
  \includegraphics[width=0.37\columnwidth]{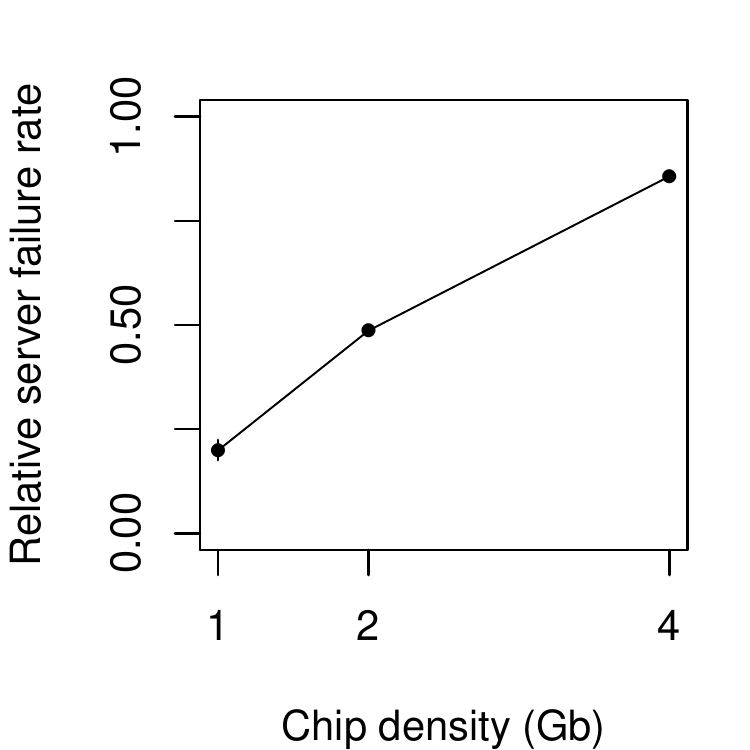}%
  \vspace{-5pt}%
  \caption{\chII{Relative failure rate for servers with different chip densities.
  Higher densities (related to newer technology nodes) show a trend of 
  higher failure rates.  Reproduced from \cite{R117}.} \chIII{See Section~II-E
  of \cite{R117} for the complete definition of the metric plotted on the y-axis,
  i.e., \emph{relative server failure rate}.}}%
  \label{fig:dram-density}%
\end{figure}

\subsection{Latency-Related Errors in DRAM}
\label{sec:othermem:latency}

\chIII{Various} experimental
studies examine the tradeoff between DRAM reliability and
latency\chIV{~\cite{R119, R120, R128, R152, R157, R174, R175, R178, kim.hpca18}}.
These works perform extensive experimental studies on
real DRAM chips to identify the effect of (1)~temperature,
(2)~supply voltage, and (3)~manufacturing process variation
that exists in DRAM on the latency and reliability characteristics
of different DRAM cells and chips. The temperature,
supply voltage, and manufacturing process variation
all dictate the amount of time that each cell needs to safely
complete its operations.
\chI{Several of our works~\cite{R119, R120, R128, R178} examine how one can 
\emph{reliably} exploit different effects of variation to improve DRAM 
performance or energy consumption.}

\chI{Adaptive-Latency DRAM (AL-DRAM)~\cite{R119} \chII{shows} that significant
variation exists in the access latency of \chII{(1)}~different DRAM modules, as a result of 
\chII{manufacturing process variation; and (2)~the same DRAM module over time,
as a result of varying operating temperature, since at low temperatures DRAM
can be accessed faster}.
\chII{The key idea of AL-DRAM is to adapt the DRAM latency to the operating
temperature and the DRAM module that is being accessed.
Experimental results show that AL-DRAM can reduce DRAM read latency by 32.7\%
and write latency by 55.1\%, averaged across 115~DRAM modules operating at
\SI{55}{\celsius}\chIV{~\cite{R119}}.}}

\chI{Voltron~\cite{R178} identifies the \chIII{relationship} between the DRAM supply voltage
and access latency variation.  Voltron uses this \chIII{relationship} to identify the
combination of voltage and access latency that minimizes system-level
energy consumption \chII{without exceeding a user-specified threshold for
the maximum acceptable performance loss.
For example, at an average performance loss of only 1.8\%, Voltron reduces
the DRAM energy consumption by 10.5\%, which translates to a
reduction in the overall system energy consumption of 7.3\%,
averaged over seven memory-intensive quad-core workloads\chIV{~\cite{R178}}.}}

\chI{Flexible-Latency DRAM (FLY-DRAM)~\cite{R120} captures access latency
variation across DRAM cells \emph{within} a single DRAM chip due to manufacturing
process variation.
\chII{For example, Figure~\ref{fig:dram-trcd} shows how the bit error rate (BER)
changes if we reduce one of the timing parameters used to control the DRAM 
access latency below the minimum value specified by the manufacturer~\cite{R120}.
We use an FPGA-based experimental DRAM testing infrastructure~\cite{R157} to
measure the BER \chIII{of} 30~real DRAM modules, over a total of 7500~rounds of
tests, as we lower the $t_{RCD}$ timing parameter (i.e., how long it takes to
open a DRAM row) below its standard value of
\SI{13.125}{\nano\second}.\chIII{\footnote{\chIII{More detail on our 
experimental setup, along with a list of all modules and their 
characteristics, can be found in our original FLY-DRAM paper~\cite{R120}.}}}
In this figure, we use a box plot to summarize the bit error rate measured during
each round.  For each box, the bottom, middle, and
top lines indicate the 25th, 50th, and 75th percentile of the
population.  The ends of the whiskers indicate the minimum
and maximum BER of all modules for a given $t_{RCD}$ value.
Each round of BER measurement is represented as a single point 
overlaid upon the box.  From the figure, we make three observations.
First, the BER decreases exponentially as we reduce $t_{RCD}$.
Second, there are no errors when $t_{RCD}$ is at \SI{12.5}{\nano\second}
or at \SI{10.0}{\nano\second}, indicating that manufacturers provide a
significant latency \emph{guardband} to provide additional protection against
process variation.
Third, the BER variation \chIV{across different models} becomes smaller as $t_{RCD}$ decreases.  \chIII{The}
reliability of a module operating at $t_{RCD}=$~\SI{7.5}{\nano\second}
varies significantly based on the DRAM manufacturer and model.
This variation occurs because the number of DRAM cells that experience
an error within a DRAM chip varies significantly from module to module.
\chIV{Yet, the BER variation across different modules operating at $t_{RCD}=$~\SI{2.5}{\nano\second}
is much smaller, as most modules fail when the latency is reduced so significantly.}}}

\chI{From other experiments that we describe in our FLY-DRAM paper~\cite{R120}, 
we find that there is spatial locality in the slower cells, resulting in \emph{fast regions} (i.e.,
regions where all DRAM cells can operate at significantly-reduced access 
latency without \chIV{experiencing} errors) and \emph{slow regions} (i.e., regions where \emph{some} of the DRAM
cells \emph{cannot} operate at significantly-reduced access latency without \chIV{experiencing} errors)
within each chip.
\chIII{To take advantage of this heterogeneity in the reliable access latency of
DRAM cells within a chip,}
FLY-DRAM (1)~categorizes the \chIII{cells} into fast and slow regions; and
(2)~lowers the overall DRAM latency by accessing fast regions with a lower 
latency.
FLY-DRAM lowers the timing parameters used for the fast region by as much as
42.8\%\chIII{~\cite{R120}}.}
\chIV{FLY-DRAM improves system performance for a wide variety of real workloads,
with the average improvement for an eight-core system ranging between 
13.3\% and 19.5\%, depending on the amount of variation that exists in each 
module~\cite{R120}.}

\begin{figure}[h]
  \centering
  \includegraphics[width=0.55\columnwidth]{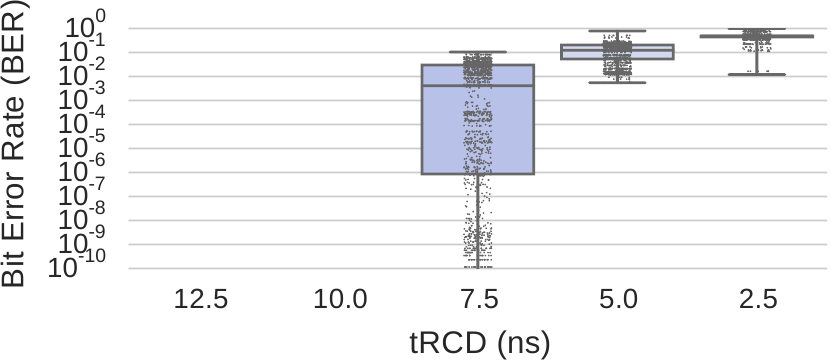}%
  \vspace{-5pt}%
  \caption{\chII{Bit error rates of tested DRAM modules as we reduce the
  DRAM access latency (i.e., the $t_{RCD}$ timing parameter).  Reproduced from \cite{R120}.}}%
  \label{fig:dram-trcd}%
\end{figure}

\chI{Design-Induced Variation-Aware DRAM (DIVA-DRAM)~\cite{R128} identifies the
latency variation within a single DRAM chip that occurs due to the architectural
design of the chip.  For example, a cell that is further away from the
row decoder requires a longer access time than a cell that is close to the \chIII{row
decoder.  Similarly, a cell that is farther away from the wordline driver
requires a larger access time than a cell that is close to the wordline driver.}
DIVA-DRAM uses design-induced variation to reduce the access latency
to different parts of the chip. One can further reduce latency by
sacrificing some amount of reliability and performing error
correction to fix the resulting errors~\cite{R128}.
\chII{Experimental results show that DIVA-DRAM can reduce DRAM read latency by 40.0\%
and write latency by 60.5\%\chIII{~\cite{R128}}.}}
\chIV{In an eight-core system running a wide variety of real workloads, 
DIVA-DRAM improves system performance by an average of 13.8\%~\cite{R128}.}

More information
about the errors caused by reduced latency \chII{and reduced voltage} operation
in DRAM chips and the tradeoff between reliability and
latency \chII{and voltage} can be found in our prior works~\cite{R119, R120, R128, R130, R157, R174, R175, R178}.

\subsection{Error Correction in DRAM}
\label{sec:othermem:ecc}

In order to protect the data
stored within DRAM from various types of errors, some
(but not all) DRAM modules employ ECC~\cite{R130}. The ECC
employed within DRAM is much weaker than the ECC
employed in SSDs (see Section~\ref{sec:correction}) for various reasons. First,
DRAM has a much lower access latency, and error correction
mechanisms should be designed to ensure that DRAM
access latency does not increase significantly. Second, the
error rate of a DRAM chip tends to be lower than that of a
flash memory chip. Third, the granularity of access is much
smaller in a DRAM chip than in a flash memory chip, and
hence sophisticated error correction can come at a high
cost. The most common ECC algorithm used in commodity
DRAM modules is \emph{SECDED} (single error correction, double
error detection)~\cite{R130}. Another ECC algorithm available for
some commodity DRAM modules is \emph{Chipkill}, which can tolerate
the failure of an \chIII{\emph{entire}} DRAM chip within a module~\cite{R139} 
\chIII{at the expense of higher storage overhead and higher latency}. 
For both SECDED and Chipkill, the ECC information
is stored on one or more extra chips within the DRAM module,
and, on a read request, this information is sent alongside
the data to the memory controller, which performs the
error detection and correction.

As DRAM scales to smaller technology nodes, its error rate
continues to increase\chIV{~\cite{R111, R112, R116, R117, R131, R153, mandelman.ibmjrd02}}.  \chIII{Effects}
like read disturb~\cite{R116}, cell-to-cell interference~\cite{R104, R126, R127, R147, R149, R206}, 
and variable retention time~\cite{R104, R125, R126, R147} 
become more severe\chII{~\cite{R111, R112, R116, R131, R153}}.
As a result, there is an increasing need for (1)~employing ECC
algorithms in \emph{all} DRAM chips/modules; (2)~developing more
sophisticated and efficient ECC algorithms for DRAM chips/modules; 
and (3)~developing error-specific mechanisms for error
correction. To this end, recent work follows various directions.
First, in-DRAM ECC, where correction is performed within the
DRAM module itself (as opposed to in the controller), is proposed~\cite{R153}.
One work shows how exposing this in-DRAM ECC
information to the memory controller can provide Chipkill-like
error protection at much lower overhead than the traditional
Chipkill mechanism~\cite{R144}. Second, various works explore and
develop stronger ECC algorithms for DRAM (e.g., \cite{R140, R141, R154}), 
and explore how to make ECC more efficient based
on the current DRAM error rate (e.g., \chII{\cite{R139, R142, R143, R164, R128}}). 
Third, recent work shows how the cost of ECC protection
can be reduced by (1)~exploiting \emph{heterogeneous reliability memory}~\cite{R130}, 
where different portions of DRAM use different strengths
of error protection based on the error tolerance of different applications
and different types of data~\cite{R130, R180}, and (2)~using the
additional DRAM capacity that is otherwise used for ECC to
improve system performance when reliability is not as important
for the given application and/or data~\cite{R207}.

Many of these works that propose error mitigation
mechanisms \chIII{for DRAM do \emph{not}} distinguish between the characteristics
of different types of errors. We believe that, in addition to
providing sophisticated and efficient ECC mechanisms in
DRAM, there is also significant value in and opportunity
for exploring \emph{specialized} error mitigation mechanisms that
are \emph{customized for different error types}, just as it is
done for flash memory (as we discussed in Section~\ref{sec:mitigation}). One
such example of a specialized error mitigation mechanism
is targeted to fix the RowHammer read disturb mechanism,
and is called \emph{Probabilistic Adjacent Row Activation} 
(PARA)~\cite{R116, R131}, \chII{as we discussed earlier.  Recall that
the} key idea of PARA is to refresh the rows that
are physically adjacent to an activated row, with a very low
probability. PARA is shown to be very effective in fixing the
RowHammer problem at no storage cost and at very low
performance overhead~\cite{R116}.
\chII{PARA is a specialized yet very effective solution for fixing a specific 
error mechanism that is important \chIII{and \chIV{prevalent} in modern DRAM devices}.}

\subsection{Errors in Emerging Nonvolatile Memory Technologies}
\label{sec:othermem:emerging}

DRAM operations are several orders of magnitude faster than
SSD operations, but DRAM has two major disadvantages. First,
DRAM offers orders of magnitude less storage density than
NAND-flash-memory-based SSDs. Second, DRAM is volatile
(i.e., the stored data is lost on a power outage). Emerging
nonvolatile memories, such as \emph{phase-change memory} (PCM)~\cite{R121, R129, R134, R155, R159, R160, R208}, 
\emph{spin-transfer torque
magnetic RAM} (STT-RAM or STT-MRAM)~\cite{R122, R133}, \emph{metal-oxide
resistive RAM} (RRAM)~\cite{R156}, and \emph{memristors}~\cite{R181, R182},
are expected to bridge the gap between DRAM and SSDs, providing
DRAM-like access latency and energy, and at the same
time SSD-like large capacity and nonvolatility (and hence SSD-like
data persistence). 
\chIII{These technologies are also expected to be used as part of
\emph{hybrid memory systems} \chIV{(also called \emph{heterogeneous
memory systems})}, where one part of the memory consists of
DRAM modules and another part consists of modules of emerging
technologies\chIV{~\cite{R129, R208, ramos.ics11, yoon.iccd12, qureshi.micro12, 
zhang.pact09, meza.cal12, li.cluster17, meza.weed13, yu.micro17, chou.isca15, 
chou.micro14,jiang.hpca10, phadke.date11, chatterjee.micro12}}.}
PCM-based devices are expected to have
a limited lifetime, as PCM can only endure a certain number
of writes~\cite{R121, R129, R134}, similar to the P/E cycling errors in
NAND-flash-memory-based SSDs (though PCM's write endurance
is higher than that of SSDs). PCM suffers from \chII{(1)}~\chIV{\emph{resistance
drift}~\cite{R134, pirovano.ted04, ielmini.ted07}}, where the resistance used to represent the value
\chII{becomes} higher over time (and eventually \chIII{can introduce} a bit error),
similar to how charge leakage in NAND flash memory and
DRAM lead to retention errors over time; \chII{and
(2)~\chIV{\emph{write disturb}~\cite{jiang.dsn14}}, where the heat generated during the programming of one
PCM cell dissipates into neighboring cells and \chIII{can change} the value that is
stored within the neighboring cells}. 
\chII{STT-RAM suffers} from \chII{(1)}~\chIV{\emph{retention failures}}, where the value
stored for a single bit \chIII{(as the magnetic orientation of the layer that 
stores the bit)} can flip over time; and \chII{(2)}~\chIV{\emph{read disturb}} (\chIII{a
conceptually different phenomenon}
from the read disturb in DRAM and flash memory), where
reading a bit in STT-RAM can inadvertently induce a write to
that same bit~\cite{R122}. Due to the nascent nature of emerging
nonvolatile memory technologies and the lack of availability of
large-capacity devices built with them, extensive and dependable
experimental studies have yet to be conducted on the reliability
of real PCM, STT-RAM, RRAM, and memristor chips.
However, we believe that \chII{error mechanisms conceptually \chIII{or
abstractly} similar} to those we
discussed in this paper for flash memory and DRAM are likely
to be prevalent in emerging technologies as well
\chIII{(as supported by some recent studies~\cite{R122, jiang.dsn14, 
zhang.iccd12, khwa.isscc16, athmanathan.jetcas16, sills.vlsic15, sills.vlsit14})}, 
albeit with different underlying mechanisms and error rates.


\section{Conclusion}
\label{sec:conclusion}

We provide a survey of the fundamentals of and recent
research in NAND-flash-memory-based SSD reliability. As
the underlying NAND flash memory within SSDs scales to
increase storage density, we find that the rate at which raw bit
errors occur in the memory increases significantly, which in
turn reduces the lifetime of the SSD. We describe the prevalent
error mechanisms that affect NAND flash memory, and
examine how they behave in modern NAND flash memory
chips. To compensate for the increased raw bit error rate with
technology scaling, a wide range of error mitigation and data
recovery mechanisms have been proposed. These techniques
effectively undo some of the SSD lifetime reductions that
occur due to flash memory scaling. We describe the state-of-the-art
techniques for error mitigation and data recovery, and
discuss their benefits. Even though our focus is on MLC and
TLC NAND flash memories, for which we provide data from
real flash chips, we believe that these techniques will be applicable
to emerging 3D NAND flash memory technology as
well, especially when the process technology scales to smaller
nodes. Thus, we hope the tutorial presented in this work on
fundamentals and recent research not only enables practitioners
to get acquainted with flash memory errors and how
they are mitigated, but also helps inform future directions in
NAND flash memory and SSD development as well as system
design using flash memory. We believe future is bright for
system-level approaches that codesign system and memory~\cite{R111, R112, R131}
to enhance overall scaling of platforms, and
we hope that the \chII{many} examples of this approach presented in this
tutorial inspire researchers and developers to enhance future
computing platforms via such system-memory codesign.

\section*{Acknowledgments}
The authors would like to thank Rino Micheloni for his helpful
feedback on earlier drafts of the paper. 
They would also like to \chII{thank Seagate} for their continued 
dedicated support.
Special thanks also goes to our research group SAFARI's industrial sponsors 
over the past six years, especially Facebook, Google, Huawei, Intel,
Samsung, Seagate, VMware. This work was also partially supported by 
ETH Z{\"u}rich, the Intel Science and Technology Center for Cloud Computing,
the Data Storage Systems Center at Carnegie Mellon University, and 
NSF grants 1212962 and 1320531.

\chI{An earlier, shorter version of this book chapter appears on 
arxiv.org~\cite{cai.arxiv17} and in the} \emph{Proceedings 
of the IEEE}~\cite{cai.procieee17}.

\newpage
\bibliographystyle{IEEEtranS}
\bibliography{refs}

\newpage
\appendix

\section*{Appendix: TLC Threshold Voltage Distribution Data}
\label{sec:tlc}

\vspace{-5pt}

\begin{table}[H]
\centering
\small
\setlength{\tabcolsep}{0.26em}
\setlength\arrayrulewidth{0.75pt}
\renewcommand{\arraystretch}{0.85}
\caption{Normalized mean (top) and standard deviation (bottom)
values for threshold voltage distribution of each voltage state at
various P/E cycle counts (Section~\ref{sec:errors:pe}).}
\label{tbl:T4}
\vspace{-5pt}
\begin{tabular}{|R{1.4cm}||R{0.8cm}|R{0.8cm}|R{0.8cm}|R{0.8cm}|R{0.8cm}|R{0.8cm}|R{0.8cm}|R{0.8cm}|}
\hline
\rowcolor{Black!70}\multicolumn{1}{|c||}{\textcolor{white}{\textbf{P/E}}} & & & & & & & & \\ 
\rowcolor{Black!70}\multicolumn{1}{|c||}{\textcolor{white}{\textbf{Cycles}}} 
& \multicolumn{1}{c|}{\textcolor{white}{\multirow{-2}{*}{\textbf{ER}}}} 
& \multicolumn{1}{c|}{\textcolor{white}{\multirow{-2}{*}{\textbf{P1}}}} 
& \multicolumn{1}{c|}{\textcolor{white}{\multirow{-2}{*}{\textbf{P2}}}} 
& \multicolumn{1}{c|}{\textcolor{white}{\multirow{-2}{*}{\textbf{P3}}}}
& \multicolumn{1}{c|}{\textcolor{white}{\multirow{-2}{*}{\textbf{P4}}}} 
& \multicolumn{1}{c|}{\textcolor{white}{\multirow{-2}{*}{\textbf{P5}}}} 
& \multicolumn{1}{c|}{\textcolor{white}{\multirow{-2}{*}{\textbf{P6}}}} 
& \multicolumn{1}{c|}{\textcolor{white}{\multirow{-2}{*}{\textbf{P7}}}} \\
\hline 
\textbf{0} & -110.0 & 65.9 & 127.4 & 191.6 & 254.9 & 318.4 & 384.8 & 448.3 \\ \hline
\rowcolor{Gray!30}\textbf{200} & -110.4 & 66.6 & 128.3 & 192.8 & 255.5 & 319.3 & 385.0 & 448.6 \\ \hline
\textbf{400} & -105.0 & 66.0 & 127.3 & 191.7 & 254.5 & 318.2 & 383.9 & 447.7 \\ \hline
\rowcolor{Gray!30}\textbf{1,000} & -99.9 & 66.5 & 127.1 & 191.7 & 254.8 & 318.1 & 384.4 & 447.8 \\ \hline
\textbf{2,000} & -92.7 & 66.6 & 128.1 & 191.9 & 254.9 & 318.3 & 384.3 & 448.1 \\ \hline
\rowcolor{Gray!30}\textbf{3,000} & -84.1 & 68.3 & 128.2 & 193.1 & 255.7 & 319.2 & 385.4 & 449.1 \\ \hline
\end{tabular}

\vspace{3pt}

\begin{tabular}{|R{1.4cm}||R{0.8cm}|R{0.8cm}|R{0.8cm}|R{0.8cm}|R{0.8cm}|R{0.8cm}|R{0.8cm}|R{0.8cm}|}
\hline
\rowcolor{Black!70}\multicolumn{1}{|c||}{\textcolor{white}{\textbf{P/E}}} & & & & & & & & \\ 
\rowcolor{Black!70}\multicolumn{1}{|c||}{\textcolor{white}{\textbf{Cycles}}} 
& \multicolumn{1}{c|}{\textcolor{white}{\multirow{-2}{*}{\textbf{ER}}}} 
& \multicolumn{1}{c|}{\textcolor{white}{\multirow{-2}{*}{\textbf{P1}}}} 
& \multicolumn{1}{c|}{\textcolor{white}{\multirow{-2}{*}{\textbf{P2}}}} 
& \multicolumn{1}{c|}{\textcolor{white}{\multirow{-2}{*}{\textbf{P3}}}}
& \multicolumn{1}{c|}{\textcolor{white}{\multirow{-2}{*}{\textbf{P4}}}} 
& \multicolumn{1}{c|}{\textcolor{white}{\multirow{-2}{*}{\textbf{P5}}}} 
& \multicolumn{1}{c|}{\textcolor{white}{\multirow{-2}{*}{\textbf{P6}}}} 
& \multicolumn{1}{c|}{\textcolor{white}{\multirow{-2}{*}{\textbf{P7}}}} \\
\hline 
\textbf{0} & 45.9 & 9.0 & 9.4 & 8.9 & 8.8 & 8.9 & 9.3 & 8.5 \\ \hline
\rowcolor{Gray!30}\textbf{200} & 46.2 & 9.2 & 9.8 & 9.0 & 8.8 & 9.0 & 9.1 & 8.5 \\ \hline
\textbf{400} & 46.4 & 9.2 & 9.5 & 9.1 & 8.8 & 8.8 & 9.0 & 8.6 \\ \hline
\rowcolor{Gray!30}\textbf{1,000} & 47.3 & 9.5 & 9.4 & 9.1 & 9.3 & 8.9 & 9.4 & 8.8 \\ \hline
\textbf{2,000} & 48.2 & 9.7 & 9.7 & 9.4 & 9.3 & 9.1 & 9.5 & 9.1 \\ \hline
\rowcolor{Gray!30}\textbf{3,000} & 49.4 & 10.2 & 10.2 & 9.6 & 9.7 & 9.5 & 9.8 & 9.4 \\ \hline
\end{tabular}
\end{table}

\vspace{-7pt}

\begin{table}[H]
\centering
\small
\setlength{\tabcolsep}{0.26em}
\setlength\arrayrulewidth{0.75pt}
\renewcommand{\arraystretch}{0.85}
\caption{Normalized mean (top) and standard deviation (bottom)
values for threshold voltage distribution of each voltage state at
various data retention times (Section~\ref{sec:errors:retention}).}
\label{tbl:T5}
\vspace{-5pt}
\begin{tabular}{|C{1.4cm}||R{0.8cm}|R{0.8cm}|R{0.8cm}|R{0.8cm}|R{0.8cm}|R{0.8cm}|R{0.8cm}|R{0.8cm}|}
\hline
\rowcolor{Black!70} & & & & & & & & \\
\rowcolor{Black!70}\multicolumn{1}{|c||}{\textcolor{white}{\multirow{-2}{*}{\textbf{Time}}}} 
& \multicolumn{1}{c|}{\textcolor{white}{\multirow{-2}{*}{\textbf{ER}}}} 
& \multicolumn{1}{c|}{\textcolor{white}{\multirow{-2}{*}{\textbf{P1}}}} 
& \multicolumn{1}{c|}{\textcolor{white}{\multirow{-2}{*}{\textbf{P2}}}} 
& \multicolumn{1}{c|}{\textcolor{white}{\multirow{-2}{*}{\textbf{P3}}}}
& \multicolumn{1}{c|}{\textcolor{white}{\multirow{-2}{*}{\textbf{P4}}}} 
& \multicolumn{1}{c|}{\textcolor{white}{\multirow{-2}{*}{\textbf{P5}}}} 
& \multicolumn{1}{c|}{\textcolor{white}{\multirow{-2}{*}{\textbf{P6}}}} 
& \multicolumn{1}{c|}{\textcolor{white}{\multirow{-2}{*}{\textbf{P7}}}} \\
\hline 
\textbf{1 day} & -92.7 & 66.6 & 128.1 & 191.9 & 254.9 & 318.3 & 384.3 & 448.1 \\ \hline
\rowcolor{Gray!30}\textbf{1 week} & -86.7 & 67.5 & 128.1 & 191.4 & 253.8 & 316.5 & 381.8 & 444.9 \\ \hline
\textbf{1 month} & -84.4 & 68.6 & 128.7 & 191.6 & 253.5 & 315.8 & 380.9 & 443.6 \\ \hline
\rowcolor{Gray!30}\textbf{3 months} & -75.6 & 72.8 & 131.6 & 193.3 & 254.3 & 315.7 & 380.2 & 442.2 \\ \hline
\textbf{1 year} & -69.4 & 76.6 & 134.2 & 195.2 & 255.3 & 316.0 & 379.6 & 440.8 \\ \hline
\end{tabular}

\vspace{3pt}

\begin{tabular}{|C{1.4cm}||R{0.8cm}|R{0.8cm}|R{0.8cm}|R{0.8cm}|R{0.8cm}|R{0.8cm}|R{0.8cm}|R{0.8cm}|}
\hline
\rowcolor{Black!70} & & & & & & & & \\
\rowcolor{Black!70}\multicolumn{1}{|c||}{\textcolor{white}{\multirow{-2}{*}{\textbf{Time}}}} 
& \multicolumn{1}{c|}{\textcolor{white}{\multirow{-2}{*}{\textbf{ER}}}} 
& \multicolumn{1}{c|}{\textcolor{white}{\multirow{-2}{*}{\textbf{P1}}}} 
& \multicolumn{1}{c|}{\textcolor{white}{\multirow{-2}{*}{\textbf{P2}}}} 
& \multicolumn{1}{c|}{\textcolor{white}{\multirow{-2}{*}{\textbf{P3}}}}
& \multicolumn{1}{c|}{\textcolor{white}{\multirow{-2}{*}{\textbf{P4}}}} 
& \multicolumn{1}{c|}{\textcolor{white}{\multirow{-2}{*}{\textbf{P5}}}} 
& \multicolumn{1}{c|}{\textcolor{white}{\multirow{-2}{*}{\textbf{P6}}}} 
& \multicolumn{1}{c|}{\textcolor{white}{\multirow{-2}{*}{\textbf{P7}}}} \\
\hline 
\textbf{1 day} & 48.2 & 9.7 & 9.7 & 9.4 & 9.3 & 9.1 & 9.5 & 9.1 \\ \hline
\rowcolor{Gray!30}\textbf{1 week} & 46.4 & 10.7 & 10.8 & 10.5 & 10.6 & 10.3 & 10.6 & 10.6 \\ \hline
\textbf{1 month} & 46.8 & 11.3 & 11.2 & 11.0 & 10.9 & 10.8 & 11.2 & 11.1 \\ \hline
\rowcolor{Gray!30}\textbf{3 months} & 45.9 & 12.0 & 11.8 & 11.5 & 11.4 & 11.4 & 11.7 & 11.7 \\ \hline
\textbf{1 year} & 45.9 & 12.8 & 12.4 & 12.0 & 12.0 & 11.9 & 12.3 & 12.4 \\ \hline
\end{tabular}
\end{table}

\vspace{-7pt}

\begin{table}[H]
\centering
\small
\setlength{\tabcolsep}{0.26em}
\setlength\arrayrulewidth{0.75pt}
\renewcommand{\arraystretch}{0.85}
\caption{Normalized mean (top) and standard deviation (bottom)
values for threshold voltage distribution of each voltage state at
various read disturb counts (Section~\ref{sec:errors:readdisturb}).}
\label{tbl:T6}
\vspace{-5pt}
\begin{tabular}{|R{1.4cm}||R{0.8cm}|R{0.8cm}|R{0.8cm}|R{0.8cm}|R{0.8cm}|R{0.8cm}|R{0.8cm}|R{0.8cm}|}
\hline
\rowcolor{Black!70}\multicolumn{1}{|c||}{\textcolor{white}{\textbf{Read}}} & & & & & & & & \\ 
\rowcolor{Black!70}\multicolumn{1}{|c||}{\textcolor{white}{\textbf{Disturbs}}} 
& \multicolumn{1}{c|}{\textcolor{white}{\multirow{-2}{*}{\textbf{ER}}}} 
& \multicolumn{1}{c|}{\textcolor{white}{\multirow{-2}{*}{\textbf{P1}}}} 
& \multicolumn{1}{c|}{\textcolor{white}{\multirow{-2}{*}{\textbf{P2}}}} 
& \multicolumn{1}{c|}{\textcolor{white}{\multirow{-2}{*}{\textbf{P3}}}}
& \multicolumn{1}{c|}{\textcolor{white}{\multirow{-2}{*}{\textbf{P4}}}} 
& \multicolumn{1}{c|}{\textcolor{white}{\multirow{-2}{*}{\textbf{P5}}}} 
& \multicolumn{1}{c|}{\textcolor{white}{\multirow{-2}{*}{\textbf{P6}}}} 
& \multicolumn{1}{c|}{\textcolor{white}{\multirow{-2}{*}{\textbf{P7}}}} \\
\hline 
\textbf{1} & -84.2 & 66.2 & 126.3 & 191.5 & 253.7 & 316.8 & 384.3 & 448.0 \\ \hline
\rowcolor{Gray!30}\textbf{1,000} & -76.1 & 66.7 & 126.6 & 191.5 & 253.6 & 316.4 & 383.8 & 447.5 \\ \hline
\textbf{10,000} & -57.0 & 67.9 & 127.0 & 191.5 & 253.3 & 315.7 & 382.9 & 445.7 \\ \hline
\rowcolor{Gray!30}\textbf{50,000} & -33.4 & 69.9 & 128.0 & 191.9 & 253.3 & 315.4 & 382.0 & 444.1 \\ \hline
\textbf{100,000} & -20.4 & 71.6 & 128.8 & 192.1 & 253.3 & 315.0 & 381.1 & 443.0 \\ \hline
\end{tabular}

\vspace{3pt}

\begin{tabular}{|R{1.4cm}||R{0.8cm}|R{0.8cm}|R{0.8cm}|R{0.8cm}|R{0.8cm}|R{0.8cm}|R{0.8cm}|R{0.8cm}|}
\hline
\rowcolor{Black!70}\multicolumn{1}{|c||}{\textcolor{white}{\textbf{Read}}} & & & & & & & & \\ 
\rowcolor{Black!70}\multicolumn{1}{|c||}{\textcolor{white}{\textbf{Disturbs}}} 
& \multicolumn{1}{c|}{\textcolor{white}{\multirow{-2}{*}{\textbf{ER}}}} 
& \multicolumn{1}{c|}{\textcolor{white}{\multirow{-2}{*}{\textbf{P1}}}} 
& \multicolumn{1}{c|}{\textcolor{white}{\multirow{-2}{*}{\textbf{P2}}}} 
& \multicolumn{1}{c|}{\textcolor{white}{\multirow{-2}{*}{\textbf{P3}}}}
& \multicolumn{1}{c|}{\textcolor{white}{\multirow{-2}{*}{\textbf{P4}}}} 
& \multicolumn{1}{c|}{\textcolor{white}{\multirow{-2}{*}{\textbf{P5}}}} 
& \multicolumn{1}{c|}{\textcolor{white}{\multirow{-2}{*}{\textbf{P6}}}} 
& \multicolumn{1}{c|}{\textcolor{white}{\multirow{-2}{*}{\textbf{P7}}}} \\
\hline 
\textbf{1} & 48.2 & 9.7 & 9.7 & 9.4 & 9.3 & 9.1 & 9.5 & 9.1 \\ \hline
\rowcolor{Gray!30}\textbf{1,000} & 47.4 & 10.7 & 10.8 & 10.5 & 10.6 & 10.3 & 10.6 & 10.6 \\ \hline
\textbf{10,000} & 46.3 & 12.0 & 11.7 & 11.4 & 11.4 & 11.4 & 11.7 & 11.7 \\ \hline
\rowcolor{Gray!30}\textbf{50,000} & 46.1 & 12.3 & 12.1 & 11.7 & 11.6 & 11.7 & 12.0 & 12.4 \\ \hline
\textbf{100,000} & 45.9 & 12.8 & 12.4 & 12.0 & 12.0 & 11.9 & 12.3 & 12.4 \\ \hline
\end{tabular}
\end{table}


\section*{About the Authors}
\label{sec:authors}

\vspace{10pt}

\begin{wrapfigure}[8]{L}{65pt}
  \centering
  \vspace{-12pt}%
  \includegraphics[width=65pt]{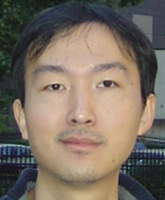}%
\end{wrapfigure}

\textbf{Yu Cai} received the B.S.\ degree from Beijing
University of Posts and Telecommunications
in Telecommunication Engineering, Beijing,
China, the M.S.\ degree in electronic engineering
from Tsinghua University, Beijing, China,
and the Ph.D.\ degree in computer engineering
from Carnegie Mellon University, Pittsburgh,
PA, USA.

He has worked as a solid-state disk system
architect at SK Hynix, Seagate Technology, Avago
Technologies, and LSI Corporation. Prior to that, he worked on wireless
communications at the Hong Kong Applied Science and Technology
Research Institute (ASTRI), Alcatel-Lucent, and Microsoft Research Asia
(MSRA). He has authored over 20 peer-reviewed papers and holds more
than 30 U.S. patents.

Dr.\ Cai received the Best Paper Runner-Up Award from the IEEE International
Symposium on High-Performance Computer Architecture (HPCA)
in 2015. He also received the Best Paper Award from the DFRWS Digital
Forensics Research Conference Europe in 2017.

\vspace{30pt}

\begin{wrapfigure}[8]{L}{65pt}
  \centering
  \vspace{-12pt}%
  \includegraphics[width=65pt, trim = 0 5 0 5, clip]{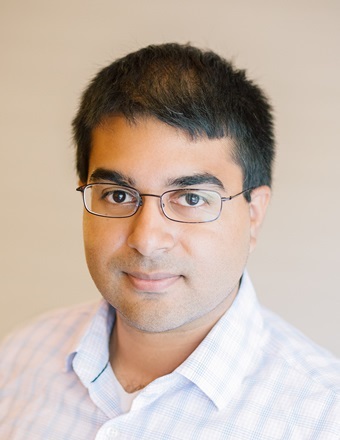}%
\end{wrapfigure}

\noindent\textbf{Saugata Ghose} received dual B.S. degrees in
computer science and in computer engineering
from Binghamton University, State University of
New York, USA, and the M.S.\ and Ph.D.\ degrees
from Cornell University, Ithaca, NY, USA, where he
was the recipient of the NDSEG Fellowship and the
ECE Director's Ph.D.\ Teaching Assistant Award.

He is a Systems Scientist in the Department
of Electrical and Computer Engineering at Carnegie
Mellon University, Pittsburgh, PA, USA. He is
a member of the SAFARI Research Group, led by Dr.\ Onur Mutlu. His current
research interests include application- and system-aware memory and storage
systems, flash reliability, architectural solutions for large-scale systems,
GPUs, and emerging memory technologies.

Dr.\ Ghose received the Best Paper Award from the DFRWS Digital
Forensics Research Conference Europe in 2017. For more information, see
his website at \url{https://ece.cmu.edu/~saugatag/}.

\vspace{30pt}

\noindent\textbf{Erich F.\ Haratsch} is Director of Engineering at Seagate Technology,
where he is responsible for the architecture of flash controllers. He leads
the development of hardware and firmware features that improve the
performance, quality of service, endurance, error correction and media
management capabilities of solid-state drives. Earlier in his career, he
developed signal processing and error correction technologies for hard
disk drive controllers at LSI Corporation and Agere Systems, which
shipped in more than one billion chips. He started his engineering career
at Bell Labs Research, where he invented new chip architectures for Gigabit
Ethernet over copper and optical communications. He is a frequent
speaker at leading industry events, is the author of over 40 peer-reviewed
journal and conference papers, and holds more than 100 U.S.\ patents.

He earned his M.S.\ and Ph.D.\ degrees in electrical engineering from
the Technical University of Munich (Germany).

\vspace{30pt}

\begin{wrapfigure}[8]{L}{65pt}
  \centering
  \vspace{-12pt}%
  \includegraphics[width=65pt, trim=10 0 10 0, clip]{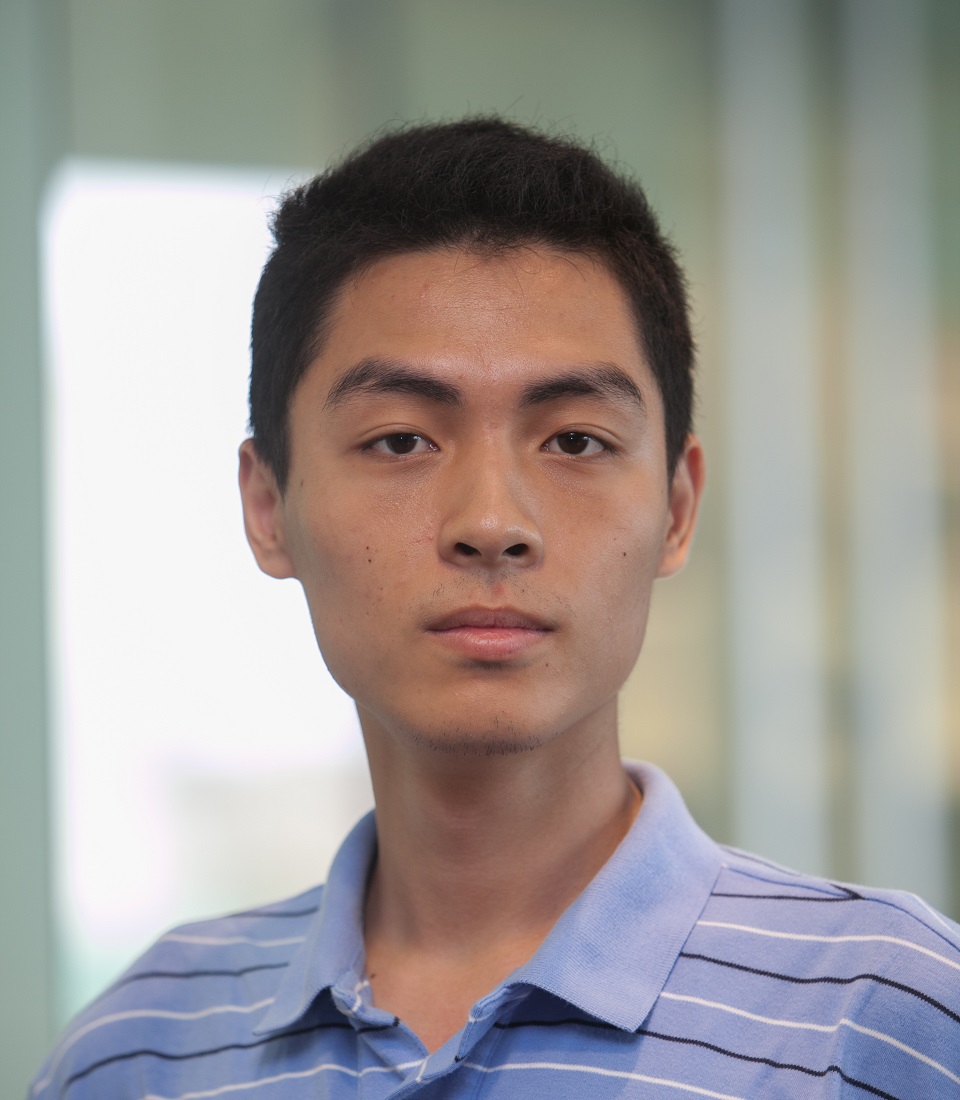}%
\end{wrapfigure}

\noindent\textbf{Yixin Luo} received the B.S.E.\ degree in computer
engineering from the University of Michigan, Ann
Arbor, MI, USA and the B.S.E.\ degree in electrical
engineering from Shanghai Jiao Tong University,
Shanghai, China, in 2012. He is currently working
toward the Ph.D.\ degree in computer science at
Carnegie Mellon University, Pittsburgh, PA, USA.

At Carnegie Mellon, he is involved in
research on DRAM and flash reliability, and on
datacenter reliability and cost optimization.

Mr.\ Luo received the Best Paper Award and the Best Paper Runner-Up
Award from the IEEE International Symposium on High-Performance Computer
Architecture in 2012 and 2015, respectively, and the Best Paper Award
from the DFRWS Digital Forensics Research Conference Europe in 2017.

\vspace{30pt}

\begin{wrapfigure}[8]{L}{65pt}
  \centering
  \vspace{-12pt}%
  \includegraphics[width=65pt]{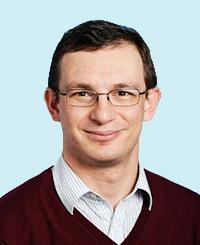}%
\end{wrapfigure}

\noindent\textbf{Onur Mutlu} received B.S.\ degrees in computer
engineering and psychology from the University
of Michigan, Ann Arbor, MI, USA and the
M.S.\ and Ph.D.\ degrees in electrical and computer
engineering from the University of Texas
at Austin, USA.

He is a Professor of Computer Science at ETH
Z{\"u}rich, Switzerland. He is also a faculty member
at Carnegie Mellon University, Pittsburgh, PA,
USA, where he previously held the William D.\ and
Nancy W.\ Strecker Early Career Professorship. His current broader research
interests are in computer architecture, systems, and bioinformatics. He is
especially interested in interactions across domains and between applications,
system software, compilers, and microarchitecture, with a major
current focus on memory and storage systems. His industrial experience
spans starting the Computer Architecture Group at Microsoft Research
(2006--2009), and various product and research positions at Intel Corporation,
Advanced Micro Devices, VMware, and Google. His computer architecture
course lectures and materials are freely available on YouTube, and
his research group makes software artifacts freely available online.

Dr. Mutlu received the inaugural IEEE Computer Society Young
Computer Architect Award, the inaugural Intel Early Career Faculty Award,
faculty partnership awards from various companies, and a healthy number
of best paper and ``Top Pick'' paper recognitions at various computer
systems and architecture venues. For more information, see his webpage
at \url{http://people.inf.ethz.ch/omutlu/}.

\end{document}